\documentstyle[epsfig,pra,aps,eqsecnum]{revtex}

\begin{document}
\draft

\def\be{\begin{eqnarray}}
\def\ee{\end{eqnarray}}
\def\l{\langle}
\def\r{\rangle}

\title
{
Quantum State Reconstruction From Incomplete Data
}

\author{
 V. Bu\v{z}ek$^{1,2}$, G. Drobn\'y$^{1}$, R. Derka$^{2}$, G. Adam$^{3}$, 
and H. Wiedemann$^{4}$
}
\address{
$^{1}$ Institute of Physics, Slovak Academy of Sciences,
D\'ubravsk\'a cesta 9, 842 28 Bratislava, Slovakia\\
$^{2}$	Faculty of Mathematics and Physics, Comenius University,
Mlynsk\'a dolina, 842 15 Bratislava, Slovakia\\
$^{3}$ Institut f\"ur Theoretische Physik, Technische Universit\"at Wien,
    Wiedner Hauptstrasse 8-10, A-1040 Vienna, Austria\\
$^{4}$ Abteilung f\"ur Quantenphysik, Universit\"at Ulm, D-89069 Ulm,
Germany
}

\date{29 April 1998}
\maketitle

\begin{abstract}
Knowing and guessing, these are two essential 
epistemological  pillars in the theory of quantum-mechanical measurement.
As formulated quantum mechanics is a statistical theory. In general,
{\it a priori} unknown states can be completely determined 
only when
measurements on infinite ensembles of identically prepared quantum systems 
are performed. But how one can estimate (guess) quantum state when 
just incomplete data  are available (known)? What is the
most reliable estimation based on a given measured data? What is the optimal
measurement providing only a finite number of identically prepared quantum
objects are available? These are some of the questions we address in the
 article.  

We present  several schemes for a reconstruction of  states
of quantum systems from measured data:\newline
{\bf (1)} We show how the  {\it maximum entropy} (MaxEnt) principle 
 can be efficiently used for 
an estimation of quantum states (i.e. density operators or Wigner functions)
on incomplete observation levels, when just a fraction of system observables
are measured (i.e., the mean values of these observables are known 
from the measurement).
With the extention of observation levels more reliable
estimation of quantum states can be performed. In the limit, when all system
observables (i.e., the quorum of observables) are measured,    the
MaxEnt principle leads to a complete reconstruction of quantum states, i.e.
quantum states are uniquely determined. We analyze the reconstruction
via the MaxEnt principle 
of bosonic systems (e.g. single-mode electromagnetic fields modeled as
harmonic oscillators) as well as spin systems. We present results of
MaxEnt reconstruction of Wigner functions
of various nonclassical states of light in different
observation levels.  
We also present results of numerical 
simulations which illustrate  how the MaxEnt
principle can be efficiently applied for a reconstruction of quantum
states from incomplete tomographic data. 
\newline
{\bf (2)} When only a {\it finite} 
number of identically prepared systems are measured, then 
the measured data contain only information about frequencies
of appearances of  eigenstates of certain observables. We show that in 
 this case  states of quantum systems can be estimated with the help
of quantum Bayesian inference. We analyze the connection between this
reconstruction scheme and the reconstruction via the MaxEnt principle
in the limit of infinite number of measurements.
We discuss how an {\it a priori} knowledge about the state which is going
to be reconstructed can be utilized in the estimation procedure. In
particular, we discuss in detail the difference between the reconstruction
of states which are a priori known to be {\it pure} or {\it impure}. 
\newline
{\bf (3)} We show how to construct the {\it optimal} generalized measurement
of a finite number of identically prepared quantum systems which results
in the estimation of a quantum state with the highest fidelity.
We show how this optimal measurement can in principle be realized. We
analyze two physically interesting examples - a reconstruction of states
of a spin-1/2 and an estimation of phase shifts. 

\end{abstract}

\section{INTRODUCTION: MEASUREMENT OF QUANTUM STATES}
\label{sec1}
The concept of a quantum state represents one of the most fundamental
pillars of the paradigm of quantum theory
\cite{Peres93,Omnes94,Ballentine90}.
Contrary to its
mathematical elegance and convenience in calculations,
the physical interpretation
of a quantum state is not so transparent. The problem is that the quantum
state (described either by a state vector, or density operator or
a phase-space probability density distribution) does not have a well
defined objective status,
i.e. a state vector is not an {\em objective} property
of a particle. According to Peres (see \cite{Peres93}, p. 374):
``There is no physical evidence whatsoever that every physical system
has at every instant a well defined state... In strict interpretation of
quantum theory these mathematical symbols [i.e., state vectors]
represent {\em statistical information} enabling us to compute the
probabilities of occurrence of specific events.'' Once this point of view
is adopted then it becomes clear that any ``measurement'' or a reconstruction
of a density operator (or its mathematical equivalent) can be understood
exclusively as an expression of our knowledge about the quantum
mechanical state based on a certain set of measured data. To be more
specific, any quantum-mechanical
reconstruction scheme  is nothing more than
an  {\em a posteriori}
estimation of the density operator
of a quantum-mechanical (microscopic) system
based on data obtained with the help of a macroscopic measurement
apparatus
\cite{Ballentine90}.
The quality of the reconstruction depends on the ``quality'' of
the measured data and the efficiency of the reconstruction procedure
with the help of which the data analysis is performed.
In particular, we can specify three different situations. Firstly,
when all system observables are precisely measured. In this case
the complete  reconstruction of an initially unknown state can be
performed (we will call this the reconstruction
on the complete observation level).
Secondly, when just part of the system observables is precisely
measured then one cannot perform a complete reconstruction of the
measured state. Nevertheless, the reconstructed density operator
still uniquely determines mean values of the measured observables
(we will denote this scheme as
reconstruction on incomplete observation levels).
 Finally,  when measurement does not provide  us with sufficient
information  to specify the exact mean values (or probability
distributions) but only the frequencies
of appearances of eigenstates of the measured observables, then
one can perform an estimation
(e.g. reconstruction based on quantum Bayesian  inference) which is
the ``best'' with respect to the given measured data and the {\em a
priori} knowledge about  the state of the measured system.

\subsection{Complete observation level}
\label{sec1.A}
Providing all system observables (i.e., the quorum
\cite{Band79,Newton68})
have  been precisely measured, then the density operator
of a quantum-mechanical
system can be completely reconstructed (i.e., the density operator
can be uniquely determined based on the available data).
In principle,
we can consider two different schemes for reconstruction of the
density operator (or, equivalently, the Wigner function)
of the given quantum-mechanical  system. The
difference between these two schemes is based on the way in which
information about the quantum-mechanical system is obtained. The first
type of  measurement is such that on each element of the ensemble
of the measured states
only a {\em single} observable is measured.
In the second type of measurement
 a {\em simultaneous} measurement of conjugate
observables is assumed. We note that
in both cases we will assume ideal, i.e.,  unit-efficiency,
measurements.

\subsubsection{Quantum tomography}
\label{sec1.A.1}
When the single-observable measurement is performed, a {\em distribution}
$w_{|\Psi\rangle}(A)$ for a particular observable $\hat{A}$ of the state
$| \psi\rangle$ is obtained in an unbiased way
\cite{Neumann55},
 i.e., $w_{|\Psi\rangle}(A)=
|\langle\Phi_A| \Psi\rangle|^2$, where $| \Phi_A\rangle$
are eigenstates of the observable $\hat{A}$ such that
$\sum_A|\Phi_A \rangle\langle\Phi_A |=\hat{1}$.  Here a question arises:
What is the {\em smallest} number of distributions $w_{|\Psi\rangle}(A)$
required to determine the state uniquely? If we consider the reconstruction
of the state of a harmonic oscillator, then
this question is directly
related to the so-called Pauli problem
\cite{Pauli80}
 of the reconstruction of the wave-function from
distributions $w_{|\Psi\rangle}(q)$ and $w_{|\Psi\rangle}(p)$ for the
position and
momentum of the state $| \Psi\rangle$. As shown by Gale, Guth and Trammel
\cite{Gale68}
the knowledge of $w_{|\Psi\rangle}(q)$ and $w_{|\Psi\rangle}(p)$
is not in general sufficient for a complete reconstruction of the wave
(or, equivalently, the Wigner)
function. In contrast, one can consider an {\em infinite}
set of distributions $w_{|\Psi\rangle}(x_{\theta},\theta)$ 
of the rotated quadrature
$\hat{x}_{\theta}=\hat{q}\cos\theta +\hat{p}\sin\theta$. Each distribution
$w_{|\Psi\rangle}(x_{\theta},\theta)$	
can be obtained from a measurement of a {\em
single}
observable $\hat{x}_{\theta}$, in which case a detector (filter) is prepared
in an eigenstate $|x_{\theta} \rangle$ of this observable. It has been
shown by Vogel and Risken
\cite{Vogel89}
that from an infinite set (in the case of the harmonic oscillator)
of the measured distributions $w_{|\Psi\rangle}(x_{\theta},\theta)$
for all values of $\theta$ such that $[0<\theta\leq\pi]$, the
Wigner function can be reconstructed uniquely via the inverse Radon
transformation. In other words knowledge of the set of distributions
$w_{|\Psi\rangle}(x_{\theta},\theta)$ 
is equivalent to  knowledge of the Wigner function.
This scheme for reconstruction of the Wigner function (i.e., the
{\em optical homodyne tomography}) has recently been realized experimentally
by Raymer and his coworkers \cite{Smithey93,Munroe95}. In these experiments
the Wigner functions of
a coherent state and a squeezed vacuum state have been
reconstructed from tomographic data. 
Very comprehensive discussion of the quantum homodyne tomography can be
found in the book by Leonhardt \cite{Leonhardt1997} and the review
article by Welsch, Vogel and Opatrn\'y \cite{Welsch1998}. Quantum homodyne
tomography can be efficiently performed not only with the help
of the inverse Radon transformation but also with the help of the so-called
pattern functions \cite{Dariano,Leonhardt96-oc}. Other theoretical
concept of Wigner-function reconstruction has been considered by Royer
\cite{Royer}.  

Quantum-state tomography can be
applied not only to optical fields  but also 
for  reconstruction of other physical systems. In particular,
recently
Janicke and Wilkens  \cite{Wilkens} have suggested that Wigner functions of
atomic waves can be tomographically reconstructed. Kurtsiefer et al.
\cite{Kurtsiefer96} have performed experiments in which
Wigner functions of matter wave packets have been reconstructed.
Yet another example of the tomographic reconstruction
is a reconstruction of Wigner functions of vibrational states
of trapped atomic ions theoretically described by a number of groups
\cite{Vogel95} and experimentally measured by
Leibfried et al. \cite{Leibfried}. Vibrational motional states of molecules
have also been reconstructed by this kind of quantum tomography by
Dunn et al. \cite{Dunn}.

Leonhardt \cite{Leon95b} has recently developed a theory of  quantum
tomography of discrete Wigner functions describing states of quantum
systems with  finite-dimensional Hilbert spaces (for instance,
angular momentum or  spin).
We note that the problem of
reconstruction of states of finite-dimensional systems is closely
related to various aspects of quantum information processing,
such as
 reading of  registers of quantum computers
\cite{Barenco}. This problem also emerges when states
of atoms are reconstructed (see, for instance, \cite{Walser}).

Here we stress once again, that reconstruction
on the complete observation level (such as  quantum tomography)
is a {\em deterministic} inversion procedure which helps us to
``rewrite'' measured data in the more convenient form of a density
operator or a Wigner function of the measured state.

\subsubsection{Filtering with quantum rulers}
\label{sec1.A.2}
For   the case of  simultaneous measurement of two non-commuting
observables (let us say $\hat{q}$ and $\hat{p}$), it is not possible
to construct a joint eigenstate of these two operators, and therefore it is
inevitable that the simultaneous measurement of two non-commuting observables
 introduces additional
noise (of quantum origin) into measured data. This noise is associated
with Heisenberg's  uncertainty relation and it results in
a specific ``smoothing'' (equivalent to a reduction of resolution)
of the original  Wigner function of the system under consideration
(see  \cite{Husimi} and \cite{Arthurs65} and the reviews 
\cite{Stenholm92,Leonhardt95rev}).
To describe the  process of simultaneous measurement of two non-commuting
observables, W\'odkiewicz
\cite{Wod84}
has proposed a formalism
based on an operational probability density distribution which explicitly
takes into account the action of the measurement device modeled as a
``filter''
(quantum ruler). A particular choice of the state of the ruler samples a
specific  type of accessible information concerning the system, i.e.,
information about the system is biased by the filtering 
process\footnote{The quantum filtering, i.e. the measurement with ``unsharp
observables'' belongs to a class of generalized POVM (positive operator value
measure) measurements \cite{Helstrom76,Holevo82}. In Section~\ref{sec10} we
will show that  POVM measurements are in some cases 
the most optimal one when the state
estimation is based on measurements performed on finite ensembles.} 
The quantum-mechanical noise induced by filtering formally results in a
smoothing
of the original Wigner function of the measured state
\cite{Husimi,Arthurs65},
so that the operational probability
density distribution can be expressed as a convolution of the original
Wigner function and the  Wigner function of the filter state.
In particular, if the filter is considered to be in
its  vacuum state then the corresponding operational probability density
distributions is equal to the Husimi ($Q$) function \cite{Husimi}.
The $Q$ function of optical fields has been experimentally
measured using such an approach
by Walker and Carroll \cite{Walker86}. The direct experimental
measurement of the operational probability density distribution
with the filter in an arbitrary state is feasible in an 8-port
experimental  setup of the type
used by Noh, Foug\'eres and Mandel \cite{Noh91}
(see also \cite{Lai89,Leonhardt1997}).

As a consequence of a simultaneous measurement of non-commuting
observables
the measured distributions
are fuzzy (i.e., they are equal to smoothed Wigner functions). Nevertheless,
if  detectors used in the experiment have  unit efficiency (in the
case of an ideal measurement), the noise induced by quantum filtering can be
``separated'' from the measured data and the density operator (Wigner
function) of the measured system
can be
``extracted'' from the operational probability density distribution. In
particular, the Wigner function can be uniquely reconstructed from the
$Q$ function (for more details see \cite{Wuensche97}).
This extraction procedure is
technically quite involved and it suffers significantly if additional
stochastic noise due to imperfect measurement is present in the data.

We note that
propensities, and in particular $Q$-functions, can also be associated
with discrete phase space  and they can in principle be
measured directly \cite{Opat95}. These discrete probability distributions
contain complete information about density operators of measured
systems. Consequently, these density operators can be uniquely
determined from the discrete-phase space propensities.

\subsection{Reduced observation levels and MaxEnt principle}
\label{sec1.B}

As we have already indicated it is  well understood that density operators
(or Wigner functions)
can, in principle, be uniquely reconstructed using either the single
observable measurements (optical homodyne tomography) or the simultaneous
measurement of two non-commuting observables. The completely reconstructed
density operator (or, equivalently, the Wigner function)
 contains information about {\em all} independent moments
of the system operators. For example, in the case of the quantum harmonic
oscillator,
the knowledge of the Wigner function is equivalent to the knowledge of all
moments $\langle(\hat{a}^{\dagger})^m\hat{a}^n\rangle$ of the creation
$(\hat{a}^{\dagger})$ and annihilation $(\hat{a})$ operators.

In many cases it turns out that the state of a harmonic oscillator
 is characterized
by an {\em infinite} number of independent moments
$\langle(\hat{a}^{\dagger})^m\hat{a}^n\rangle$ (for all $m$ and $n$).
Analogously, the state of a quantum system in a finite-dimensional
Hilbert space can be characterized by a very large number
of independent parameters.
A {\em complete} measurement of these moments may  take an
infinite time to perform. This means that even though the Wigner function
can in principle
be reconstructed the collection of a complete set of experimental data points
is (in principle) a  never ending process.
In addition the data processing and numerical reconstruction of the Wigner
function are time consuming. Therefore experimental realization of
the reconstruction of the density operators (or Wigner functions) for many
systems   can be difficult.

In practice, it is possible to	measure  just a finite number
of independent moments of the system operators, so  that only
a subset $\hat{G}_{\nu}$ ($\nu=1,2,...,n)$ of observables from
the quorum (this subset constitutes the so-called observation level
\cite{Jaynes57})
is measured. In this case, when the complete information about the
system is not available, one needs an additional criterion which would
help to reconstruct (or estimate) the density operator uniquely.
Provided mean values of all observables on the given observation
level are measured precisely, then the density operator
(or the Wigner function) of the system under consideration can be
reconstructed with the help of the Jaynes principle of maximum
entropy (the so called {\em MaxEnt} principle)
\cite{Jaynes57} (see also \cite{Fick,Kapur,Katz}).
The MaxEnt principle provides us with a very efficient
prescription  to reconstruct density operators of quantum-mechanical
systems providing the mean values of a given set of observables are known.
It works perfectly well for systems with infinite Hilbert spaces
(such as the quantum-mechanical harmonic oscillator) as well as for
systems with finite-dimensional Hilbert spaces (such as spin systems).
If the observation level is composed of the quorum of the observables
(i.e., the complete observation level), then the
MaxEnt principle represents an alternative to  quantum tomography,
i.e. both schemes are equally suitable for the analysis of the
tomographic data (for details see \cite{Buzek96}). To be specific,
the observation level in this case is composed of all projectors
associated with probability distributions of rotated quadratures.
The power of the MaxEnt principle can be appreciated in analysis
of incomplete tomographic data.
In particular cases {\em MaxEnt} reconstruction  from incomplete
tomographic data can be  several
orders better than a standard tomographic inversion (see Section~\ref{sec6}).
This result suggests that
 the MaxEnt principle is the conceptual
basis underlying incomplete
tomographic reconstruction (irrespective whether this is employed
in continuous or discrete phase spaces).

\subsection{Incomplete measurement and Bayesian inference}
\label{sec1.C}

It has to be stressed that the Jaynes principle of  maximum
entropy can be consistently  applied only when {\em exact}
mean values of the measured observables are available. This
condition implicitly assumes that an infinite number of repeated measurements
on different elements of the ensemble
has to be performed to reveal the exact mean value of the given observable.
In practice only a finite number of measurements can be performed.
What is obtained from these measurements is a specific set
of data indicating the number of times the eigenvalues of given observables
have appeared (which in the limit of an infinite number of measurements
results in the corresponding  quantum probability distributions).
The question is how to obtain the best {\em a posteriori} estimation
of the density operator based on the measured data.
Helstrom \cite{Helstrom76},
Holevo \cite{Holevo82},
 and Jones \cite{Jones91} 
 have shown that the answer to this
question can be given by the Bayesian inference method, providing it is
{\em a priori} known that the quantum-mechanical state which is
to be reconstructed is prepared in a pure (although unknown) state.
When the
purity condition is fulfilled, then the observer can systematically
estimate  an {\em a posteriori}
a  probability distribution in an abstract state space
of the measured system. It is this probability distribution
(conditioned by the assumed Bayesian prior) which characterizes observer's
knowledge of the system after the measurement is performed. Using this
probability distribution one can derive a reconstructed density operator,
which however is subject to certain ambiguity associated with
the choice of
the {\em cost} function (see ref.\cite{Helstrom76}, p. 25). In general,
depending on the choice of the cost function one obtains different
estimators (i.e., different reconstructed density operators). In
this paper we adopt the approach advocated by Jones \cite{Jones91}
when the estimated density operator is equal to
 the mean over all possible {\em pure} states weighted by the
estimated probability distribution (see below in Section~\ref{sec9}).
We note once again that the quantum Bayesian  inference has been developed
for a reconstruction of {\em pure} quantum mechanical states and in this
sense it corresponds to an averaging over a generalized {\em microcanonical}
ensemble. Nevertheless
it can also be applied for a reconstruction of impure states
of quantum systems \cite{Derka96}. The Bayesian inference based on
appropriate {\it a priori} assumptions in the limit of infinite number of
measurements results in the same estimation as the reconstruction via
the MaxEnt principle.

\subsection{The optimal generalized  measurements}
\label{sec1.D}
The quantum Bayesian inference  allows us to estimate reliably the quantum 
state from a given set of measured data obtained in a specific measurement
performed on a finite ensemble of identically prepared quantum objects.
But the measurement itself may be designed very badly, i.e. the chosen
observables do not efficiently reveal the nature of the state. Therefore the
question is: Given the finite ensemble of $N$ identical quantum objects
prepared in an unknown quantum state.
What is the {\it  the optimal measurement} which should provide
us the best possible estimation of this unknown state? Holevo\cite{Holevo82}
(see also \cite{MasPop}) has solved this problem. He has shown that the
so called {\it covariant generalized measurements} are the optimal one.
The problem is that these generalized measurements are associated with
an infinite (continuous) number of observables.  This obviously is
physically unrealizable measurement. On the other hand it has been recently
shown \cite{Derka98} how to find a finite optimal generalized measurements.
This allows to design optimal measurements such that the data obtained
in these measurements allow for best estimation of quantum states.

The purpose of the present paper is to
 show how the various estimation procedures can be applied in
different situations. In particular, we show how 
 the MaxEnt principle 
can be applied for a reconstruction of quantum
states of light fields and spin systems. We show how the quantum Bayesian
inference can be used for a reconstruction of spin systems and how
it is related to the reconstruction via the MaxEnt principle. 
We also present a universal algorithm which allows us to ``construct''
the optimal generalized measurements. 
The paper is organized as follows: In Section~\ref{sec2} we briefly
describe main ideas of the MaxEnt principle. In Section~\ref{sec3} we
set up a scene for a description of reconstruction of quantum states
of light fields. In this section we briefly discuss the phase-space
formalism which can be used for a description of quantum states of light.
In addition we introduce the states of light which are going to be
considered later in the paper. In Section~\ref{sec4} we introduced
of various observation levels suitable for a description of 
light fields. Reconstruction of Wigner functions of light fields on these
observation levels is then discussed in Section~\ref{sec5}. In
Section~\ref{sec6} we present results of numerical reconstruction of quantum
states of light from incomplete tomographic data. We compare two
reconstruction schemes: reconstruction via the MaxEnt principle and the
reconstruction via direct sampling (i.e., the tomography reconstruction via
pattern functions - see below in Section\ref{sec3}). We analyze the
reconstruction of spin systems via the MaxEnt principle in
Section~\ref{sec7}. The Bayesian quantum inference      is
discussed in Section~\ref{sec8}  and its application to spin systems
is presented in Section~\ref{sec9}. Finally, in Section~\ref{sec10} we
discuss how the optimal realizable (i.e. finite) measurements can be designed.

\section{MAXENT PRINCIPLE AND OBSERVATION LEVELS}
\label{sec2}

The state of a quantum system can always be described by a
statistical density operator $\hat \rho$. Depending on the system
preparation, the density operator  represents either a pure
quantum state (complete system preparation) or a statistical
mixture of pure states (incomplete preparation). The degree of
deviation of a statistical mixture from the pure state can be
best described by the {\it uncertainty measure}   $\eta [\hat \rho ]$
(see \cite{Neumann55,Fick,Katz,Wehrl78})
\be
\eta [\hat \rho ]=-\,{\rm Tr}(\hat \rho \ln \hat \rho ).
\label{2.1}
\ee
The uncertainty measure $\eta [\hat \rho ]$
possesses the following properties:\newline
{\bf 1.} In the eigenrepresentation of the density operator $\hat \rho$
\be
\hat \rho \, |r_m\rangle =r_m|r_m\rangle,
\label{2.2}
\ee
we can write
\be
\eta [\hat \rho ] =-\sum\limits_m {r_m\ln r_m}\ge 0 ,
\label{2.3}
\ee
where $r_m$ are eigenvalues and $\vert r_m\rangle$ the
eigenstates of $\hat \rho$.\newline
{\bf 2.} For uncertainty measure $\eta [\hat \rho ]$  the following inequality holds:
\be
0\le \eta [\hat \rho ]\le \ln N ,
\label{2.4}
\ee
where $N$ denotes the dimension of the state space of the
system and  $\eta [\hat \rho]$ takes its  maximum value when
\be
\hat \rho ={\hat1 \over {{\rm Tr} \hat1 }}=\frac{\hat{1}}{N},
\label{2.5}
\ee
In this case all pure states in the mixture appear
with the same probability equal to $1/N$.
If the system is prepared in a pure
state then it holds that $\eta [\hat \rho] = 0.$\newline
{\bf 3.} It can be shown with the help of the  Liouville equation
\be
{\partial  \over {\partial t}}\hat \rho (t)=-{i \over \hbar }
[\hat H,\hat \rho (t)],
\label{2.6}
\ee
that in the case of an isolated system
 the uncertainty measure is a constant of motion, i.e.,
\be
{{d\eta (t)} \over {dt}}=0 .
\label{2.7}
\ee

\subsection{MaxEnt principle}
\label{sec2.A}
When instead of the density operator $\hat \rho$,  expectation
values	$G_\nu$ of a set ${\cal O}$ of operators $\hat G_\nu$
$(\nu = 1,\ldots , n)$
are given, then the uncertainty measure can be determined as well. The set
of linearly independent operators is referred to as the
{\em observation level}  ${\cal O}$ \cite{Fick,Buzek96}.
The operators $\hat{G}_{\nu}$ which belong to a given observation level do
not commutate necessarily. A large number of density operators 
which
fulfill the conditions
\be
{\rm Tr} \,\hat \rho _{\{\hat G\}} = 1,
\label{2.8}\\
{\rm Tr}\,(\hat \rho _{\{\hat G\}}\hat G_\nu )=G_\nu \,,
\,\,\,\,\nu = 1,2,...,n ;
\nonumber
\ee
can be found for a given set of expectation values $G_\nu = \langle \hat
{G_\nu}\rangle$, that is the conditions (\ref{2.8}) specify a set
${\cal C}$ of density operators which has to be considered.
Each of these density operators $\hat \rho _{\{\hat G\}}$ can posses a 
different
value of the uncertainty  measure $\eta [\hat \rho _{\{\hat G\}}]$.
If we wish to use only the expectation
values $G_\nu$ of the chosen observation level for determining the
density operator, we must select a particular density operator $\hat
\rho _{\{\hat G\}} = \hat \sigma _{\{\hat G\}}$  in an unbiased manner.
According to the Jaynes principle of the Maximum Entropy 
\cite{Jaynes57,Fick,Kapur,Katz}
this density operator $\hat \sigma _{\{\hat G\}}$ must
be the one which has the largest uncertainty measure
\be
\eta_{\rm max} \equiv {\rm max}\left\{\eta [\hat \sigma _{\{\hat G\}}]\right\}
\label{2.9}
\ee
and simultaneously fulfills constraints (\ref{2.8}).
As a consequence of Eq.(\ref{2.9}) the following fundamental inequality
holds
\be
\eta [\hat \sigma _{\{\hat G\}}]=-
{\rm Tr}(\hat \sigma _{\{\hat G\}}\ln \hat \sigma _{\{\hat G\}})\ge
  \eta [\hat \rho _{\{\hat G\}}]=-
{\rm Tr}(\hat \rho _{\{\hat G\}}\ln \hat \rho _{\{\hat G\}})
\label{2.10}
\ee
for all possible $\hat \rho _{\{\hat G\}}$ which fulfill  Eqs.(\ref{2.8}).
The variation  determining the maximum of $\eta [\hat \sigma _{\{\hat G\}}]$
under the conditions
(\ref{2.8}) leads to a generalized canonical density operator
\cite{Jaynes57,Fick,Katz,Buzek96}
\be
{\hat \sigma _{\{\hat G\}}={1 \over {Z_{\{\hat G\}}}}
\exp \,(-\sum\limits_\nu  {\lambda _\nu \hat G_\nu })};
\label{2.11}
\ee
\be
{Z_{\{\hat G\}}(\lambda _1,...,\lambda _n)=
{\rm Tr}[\exp (-\sum\limits_\nu  {\lambda _\nu \hat G_\nu })]},
\label{2.12}
\ee
where $\lambda_n$ are the Lagrange multipliers and $Z_{\{\hat G\}}(\lambda_1,
\ldots	\lambda_n)$  is the generalized partition function. By using
the derivatives of the partition function we obtain the expectation
values $G_{\nu}$ as
\be
G_\nu ={\rm Tr}(\hat \sigma _{\{\hat G\}}\hat G_\nu )=
-{\partial  \over {\partial \lambda_\nu }}
\ln Z_{\{\hat G\}}(\lambda _1,...,\lambda _n),
\label{2.13}
\ee
where in the case of noncommuting operators the following
relation has to be used
\be
{\partial  \over {\partial a}}\exp [-\hat X(a)]=
\exp [-\hat X(a)]\int\limits_0^1 {\exp [\mu \hat X(a)]}
{{\partial \hat X(a)} \over {\partial a}}\exp
[-\mu \hat X(a)]\,d\mu.
\label{2.14}         
\ee
By using Eq.(\ref{2.13}) the Lagrange multipliers can, in principle,
be expressed as
functions of the expectation values
\be
\lambda _\nu =\lambda _\nu (G_1,...,G_n) .
\label{2.15}
\ee
We note that Eqs.(\ref{2.13}) for Lagrange multipliers not always
have solutions which lead to physical results (see Section~\ref{sec6.B}),
which means that in these cases states of quantum systems cannot
be reconstructed on a given observation level.

The maximum uncertainty measure regarding an observation level
${\cal O}_{\{ \hat G\}}$
will be referred to  as the entropy $S_{\{\hat G\}}$
\be
S_{\{\hat G\}}\equiv \eta _{\max}=-{\rm Tr}(\hat \sigma _{\{\hat G\}}
\ln \hat \sigma _{\{\hat G\}}).
\label{2.16}
\ee
This means that to different observation levels different entropies are related.
By inserting $\sigma _{\{\hat G\}}$ [cf. Eq.(\ref{2.11})] into
Eq.(\ref{2.16}), we obtain
the following expression for the entropy
\be
S_{\{\hat G\}}=\ln Z_{\{\hat G\}}+\sum\limits_\nu
 {\lambda _\nu }G_\nu.
\label{2.17}
\ee
By making use of Eq.(\ref{2.15}), the parameters $\lambda_\nu$ in the
above equation can be expressed as functions of the expectation values $G_\nu$
and this leads to a new expression for the entropy
\be  
S_{\{\hat G\}}=S(G_1,...,G_n).
\label{2.18}
\ee
We note that using the expression
\be
dS_{\{\hat G\}}=\sum\limits_\nu  {\lambda _\nu d}G_\nu,
\label{2.19}
\ee
which follows from Eqs.(\ref{2.13}) and (\ref{2.17})
the following relation can be obtained
\be
\lambda _\nu ={\partial  \over {\partial G_\nu }}S(G_1,...,G_n).
\label{2.20}
\ee

\subsection{Linear transformations within an observation level}
\label{sec2.B}

An observation level can be defined either by a set of linearly independent
operators $\{\hat G_\nu\}$, or by a set of independent linear combinations
of the same operators
\be
\hat G'_\mu =\sum\limits_\nu  {c_{\mu \nu }\hat G_\nu }.
\label{2.21}
\ee
Therefore, $\hat \sigma$ and $S$ are invariant under a linear
transformation:
\be
\hat \sigma '_{\{\hat G'\}}={{\exp (-\sum\limits_\mu
{\lambda '_\mu \hat G'_\mu })} \over {{\rm Tr}\exp (-\sum\limits_\mu
{\lambda '_\mu \hat G'_\mu })}}=\hat \sigma _{\{\hat G\}}.
\label{2.22}
\ee
As a result, the Lagrange multipliers transform contravariantly to
Eq.(\ref{2.21}), i.e.,
\be
\lambda '_\mu =\sum\limits_\nu  {c'_{\mu \nu }\lambda _\nu \,,}
\label{2.23}
\ee
\be
\sum\limits_\mu  {c'_{\nu \mu }c_{\mu \rho }=\delta _{\nu \rho }\,.}
\label{2.24}
\ee

\subsection{Extension and reduction of the observation level}
\label{sec2.C}
If an observation level ${\cal O}_{\{\hat{G}\}}\equiv
\hat G_1, \ldots , \hat G_n$ is extended by including further
operators $\hat M_1, \ldots , \hat M_l$, then additional
expectation values $M_1 = \langle \hat M_1\rangle , \ldots , M_l =
\langle \hat M_l\rangle$ can only increase amount of available
information about the state of the system. This procedure is called
the {\em extension} of the observation level (from ${\cal O}_{\{\hat{G}\}}$
to ${\cal O}_{\{\hat{G}, \hat{M}\}}$)
and is associated with a decrease of the entropy. More precisely, the
entropy $S_{\{\hat G,\hat M\}}$  of the extended observation level
${\cal O}_{\{\hat G, \hat M\}}$ can be only smaller or equal to the entropy
$S_{\{\hat G\}}$ of the original observation level ${\cal O}_{\{\hat G\}}$,
\be 
S_{\{\hat G,\hat M\}}\le S_{\{\hat G\}}\,.
\label{2.25}
\ee
The generalized canonical density operator of the observation
level ${\cal O}_{\{\hat G, \hat M\}}$
\be
\hat \sigma _{\{\hat G,\hat M\}}={1 \over {Z_{\{\hat G,\hat M\}}}}
\exp \left({-\sum\limits_{\nu=1}^n  {\lambda _\nu \hat G_\nu }-
\sum\limits_{\mu=1}^l
{\kappa _\mu \hat M_\mu }}\right)\,,
\label{2.26}
\ee
with
\be
Z_{\{\hat G,\hat M\}}={\rm Tr}\left[
\exp \left({-\sum\limits_{\nu=1}^n
{\lambda _\nu \hat G_\nu }-
\sum\limits_{\mu=1}^l
{\kappa _\mu \hat M_\mu }}\right)\right],
\label{2.27}
\ee
belongs to the set of density operators $\hat \rho _{\{\hat G\}}$
fulfilling Eq.(\ref{2.8}).
Therefore, Eq.(\ref{2.26}) is a special case of Eq.(\ref{2.11}).
Analogously to Eqs.(\ref{2.13}) and (\ref{2.15}), the Lagrange multipliers can be
expressed by functions of the expectation values
\be
\lambda _\nu =\lambda _\nu (G_1,...,G_n,M_1,...,M_l),
\nonumber
\\
\kappa _\mu =\kappa _\mu (G_1,...,G_n,M_1,...,M_l).
\label{2.28}
\ee
The sign of equality in Eq.(\ref{2.25}) holds only for $\kappa_\mu = 0$.
In this special case the expectation values $M_\mu$ are functions of
the expectation values $G_\nu$. The
measurement of observables $\hat M_\mu$ does not increase information
about the system. Consequently, $\hat \rho _{\{\hat G, \hat M\}}
=\hat \rho _{\{\hat G\}}$ and
$S_{\{\hat G,\hat M\}}=S_{\{\hat G\}}$.

We can also consider
a {\em reduction of the observation level} if we decrease number
of independent	observables which are measured, e.g.,
${\cal O}_{\{ \hat{G}, \hat{M}\}}\rightarrow {\cal O}_{\{ \hat{G}\}}$.
This reduction is accompanied with an increase of the entropy due
to the decrease of the information available about the system.

\subsection{Time dependent entropy of an observation level}
\label{sec2.D}

If the dynamical evolution of the system is governed by the
evolution superoperator
 $\hat{U}(t,t_0)$, such that $\hat \rho (t)=
\hat U(t,t_0)\hat \rho (t_0)$,	then  expectation
values of the operators $\hat G_\nu$ on the given
observation level at time $t$ read
\be
G_\nu (t)={\rm Tr}[\hat G_\nu \hat U(t,t_0)\hat \rho (t_0)].
\label{2.29}
\ee
By using these time--dependent expectation values as constraints
for maximizing the uncertainty measure $\eta [\hat \rho _{\{\hat G\}}(t)]$, we
get the generalized canonical density operator
\be
\hat \sigma _{\{\hat G\}} (t)={{\exp \left( {-\sum\limits_\nu
{\lambda _\nu (t)\hat G_\nu }} \right)} \over {{\rm Tr}\left[
{\exp \left( {-\sum\limits_\nu	{\lambda _\nu (t)\hat G_\nu }}
\right)} \right]}},
\label{2.30}
\ee
and the time--dependent entropy of the corresponding observation
level
\be
S_{\{\hat G\}}(t)=-{\rm Tr}{[\hat \sigma _{\{\hat G\}}(t)\ln \hat \sigma _
{\{\hat G\}}(t)]}=\ln Z_{\{\hat G\}}(t)+\sum\limits_\nu
{\lambda _\nu (t)}\kern 1pt G_\nu (t)\,.
\label{2.31}
\ee
This generalized canonical density operator does not satisfy the
von Neumann equation but it satisfies an integro--differential
equation derived by Robertson \cite{Robertson66} (see also \cite{Seke}).
The time--dependent
entropy is defined for any system being arbitrarily far from
equilibrium. In the case of an isolated system the entropy can
increase or decrease during the time evolution (see, for
example Ref. \cite{Katz}, Sec. 5.6).

\section{STATES OF LIGHT: PHASE-SPACE DESCRIPTION}
\label{sec3}

Utilizing a close analogy between the operator for the electric
component $\hat{E}(r,t)$ of a monochromatic light field and the
quantum-mechanical harmonic oscillator we will
 consider  a dynamical system which is described by a pair of
canonically conjugated Hermitean observables $\hat{q}$ and $\hat{p}$,
\be
[\hat{q},\hat{p}]=i\hbar.
\label{3.1}
\ee
Eigenvalues of these operators
range continuously from $-\infty$ to $+\infty$. The annihilation and
creation operators $\hat{a}$ and $\hat{a}^{\dagger}$ can be expressed as a
complex linear combination of $\hat{q}$ and $\hat{p}$:
\be
\hat{a}=\frac{1}{\sqrt{2\hbar}}\left(\lambda
\hat{q}+i\lambda^{-1}\hat{p}\right);\qquad
\hat{a}^{\dagger}=\frac{1}{\sqrt{2\hbar}}\left(\lambda
\hat{q}-i\lambda^{-1}\hat{p}\right),
\label{3.2}
\ee
 where $\lambda$ is
an arbitrary real parameter. The operators $\hat{a}$ and $\hat{a}^{\dagger}$
obey the Weyl-Heisenberg commutation relation
\be
[\hat{a},\hat{a}^{\dagger}]=1,
\label{3.3}
\ee
 and therefore possess the same
algebraic properties as the operator associated with the complex
amplitude of a harmonic oscillator (in this case
$\lambda=\sqrt{m\omega}$, where $m$ and $\omega$ are the mass and the
frequency of the quantum-mechanical oscillator, respectively) or the photon
annihilation and creation operators of a single mode of the quantum
electromagnetic field. In this case $\lambda=\sqrt{\epsilon_0\omega}$
($\epsilon_0$ is the dielectric constant and $\omega$ is the
frequency of the field mode) and the operator for the electric field
reads (we do not take into account polarization of the field)
\be
\hat{E}(r,t)=\sqrt{2}{\cal E}_0\left( \hat{a} {\rm e}^{-i\omega t}
+\hat{a}^{\dagger} {\rm e}^{i\omega t} \right) u(r),
\label{3.4}
\ee
where $u(r)$ describes the spatial field distribution and is same in both
classical and quantum theories. The constant
${\cal E}_0=(\hbar\omega/2\epsilon_0 V)^{1/2}$	is equal to
the ``electric field per photon'' in the cavity of volume $V$.

A particularly useful set of states is the overcomplete set of
coherent states $|\alpha\rangle$ which are the eigenstates of the
annihilation operator $\hat{a}$:
\be
\hat{a}|\alpha\rangle=\alpha|\alpha\rangle.
\label{3.5}
\ee
 These coherent
states can be generated from the vacuum state $|0\rangle$
[defined as $\hat{a}|0\rangle=0$] by the action of the unitary displacement
operator $\hat{D}(\alpha)$ \cite{Glauber}
\be
\hat{D}(\alpha)\equiv \exp\left[\alpha\hat{a}^{\dagger}-\alpha^*\hat{a}\right];
\qquad |\alpha\rangle=\hat{D}(\alpha)|0\rangle.  
\label{3.6}
\ee
 The parametric space of
eigenvalues, i.e., the {\em phase space} for our dynamical system, is the
{\em infinite} plane of eigenvalues $(q,p)$ of the Hermitean operators
$\hat{q}$ and $\hat{p}$. An equivalent phase space is the complex plane
of eigenvalues
\be
\alpha=\frac{1}{\sqrt{2\hbar}}\left(\lambda
q+i\lambda^{-1} p\right);
\label{3.7}
\ee
of the annihilation operator
$\hat{a}$. We should note here that the coherent state $|\alpha\rangle$
is not an eigenstate of either $\hat{q}$ or $\hat{p}$.	The quantities
$q$ and $p$ in Eq.(\ref{3.7}) can be interpreted as the expectation values of
the operators $\hat{q}$ and $\hat{p}$ in the state $|\alpha\rangle$.  Two
invariant differential elements of the two phase-spaces are related as:
\be
\frac{1}{\pi}d^2\alpha=\frac{1}{\pi}d[{\rm Re}(\alpha)]\,d[{\rm
Im}(\alpha)]= \frac{1}{2\pi\hbar}dq\,dp. 
\label{3.8}
\ee
The phase-space description of the quantum-mechanical oscillator which
is in the state described by the density operator $\hat{\rho}$ (in what
follows we will consider mainly pure states such that
$\hat{\rho}=|\Psi\rangle\langle \Psi|$) is
based on the definition of the Wigner function 
\cite{Wigner} $W_{\hat{\rho}}(\xi)$. Here the subscript $\hat{\rho}$ in
the expression $W_{\hat{\rho}}(\xi)$
 explicitly indicates the
state which is described by the given  Wigner function.

The
Wigner function is related to the characteristic function
$C^{(W)}_{\hat{\rho}}(\eta)$ of the Weyl-ordered moments of the annihilation and
creation operators of the harmonic oscillator as follows \cite{Cahill}
\be
W_{\hat{\rho}}(\xi)
={1\over{\pi}}\int\,C^{(W)}_{\hat{\rho}}(\eta)\,
\exp(\xi\eta^*-\xi^*\eta)\,d^2\eta.
\label{3.9}
\ee
The characteristic function $C^{(W)}_{\hat{\rho}}(\eta)$ of the system
described by the  density operator $\hat{\rho}$ is defined as
\be
C^{(W)}_{\hat{\rho}}(\eta)\equiv {\rm Tr}[\hat{\rho}\hat{D}(\eta)],
\label{3.10}
\ee
where $\hat{D}(\eta)$ is the displacement operator given by Eq.(\ref{3.6}). The
characteristic function $C^{(W)}_{\hat{\rho}}(\eta)$ can be used for
the evaluation of  the Weyl-ordered
products of the annihilation and creation operators:
\be
\left.\langle\{(\hat{a}^{\dagger})^m\hat{a}^n\}\rangle=
{\partial^{(m+n)}\over{\partial\eta^m\partial(-\eta^*)^n}}C^{(W)}_{\hat{\rho}}(\eta)
\right|_{\eta=0},
\label{3.11}
\ee
 On the other hand the mean value of
the Weyl-ordered product $\langle\{(\hat{a}^{\dagger})^m\hat{a}^n\}\rangle$ can
be obtained by using the Wigner function directly
\be 
 {\rm Tr}\left[\{(\hat{a}^{\dagger})^m\hat{a}^n\}\hat{\rho}\right] =
\frac{1}{\pi}\int d^2\xi\,
(\xi^*)^m \xi^n W_{\hat{\rho}}(\xi).
\label{3.12}
\ee
 For instance, the Weyl-ordered
product $\langle\{\hat{a}^{\dagger}\hat{a}^2\}\rangle$ can be evaluated as:
\be
\langle\{\hat{a}^{\dagger}\hat{a}^2\}\rangle=
\frac{1}{3}\langle\hat{a}^{\dagger}\hat{a}^2+
\hat{a}\hat{a}^{\dagger}\hat{a}+\hat{a}^2\hat{a}^{\dagger}\rangle
=\frac{1}{\pi}\int d^2\xi\, |\xi|^2 \xi
W_{\hat{\rho}}(\xi).
\label{3.13}
\ee

In this paper we will several times refer to mean values of
central moments and cumulants of the system operators $\hat{a}$ and
$\hat{a}^{\dagger}$. We will denote central moments as
$\langle...\rangle^{(c)}$ and in what follows we will consider the
Weyl-ordered central moments which are defined as:
\be
\langle\{(\hat{a}^{\dagger})^m\hat{a}^n\}\rangle^{(c)}\equiv
\langle\{(\hat{a}^{\dagger}-\langle\hat{a}^{\dagger}\rangle)^m
(\hat{a}-\langle\hat{a}\rangle)^n\}\rangle.
\label{3.14}
\ee
From this definition it follows that the central moments of the order
$k$ ($k=m+n$) can be expressed by moments of the order less or equal to $k$.
On the other hand we denote cumulants as $\langle\langle...\rangle\rangle$.
The cumulants are usually defined via  characteristic functions.
In particular,	the Weyl-ordered cumulants are defined as
\be
\left.\langle\langle\{(\hat{a}^{\dagger})^m\hat{a}^n\}\rangle\rangle=
{\partial^{(m+n)}\over{\partial\eta^m\partial(-\eta^*)^n}}\ln
C^{(W)}_{\hat{\rho}}(\eta)  \right|_{\eta=0},
\label{3.15}
\ee
where $C^{(W)}_{\hat{\rho}}(\eta)$ is the  characteristic function of the
Weyl-ordered  moments given by Eq.(\ref{3.10}).    The cumulants of the order
$k$ ($k=m+n$) can be expressed in terms moments of the order less or equal
to $k$.

Originally the Wigner function was introduced in a form different from
(\ref{3.9}).
Namely, the Wigner  function was defined as a particular Fourier transform
of the
density operator expressed in the basis of the eigenvectors $|q\rangle$ of the
position
operator $\hat{q}$:
\be
W_{\hat{\rho}}(q,p)\equiv
\int_{-\infty}^{\infty} d\zeta \langle q-\zeta/2|\hat{\rho}|q+
\zeta/2\rangle {\rm e}^{ip\zeta/\hbar},
\label{3.16}
\ee
which for a pure state described by a state vector $|\Psi\rangle$ (i.e.,
$\hat{\rho}=|\Psi\rangle\langle \Psi |$) reads
\be
W_{\hat{\rho}}(q,p)\equiv
\int_{-\infty}^{\infty} d\zeta \psi(q-\zeta/2) \psi^*(q+\zeta/2)
 {\rm e}^{ip\zeta/\hbar},
\label{3.17}
\ee
where $\psi(q)\equiv \langle q|\Psi\rangle$.
It can be shown that both definitions (\ref{3.9}) and (\ref{3.16}) 
of the Wigner
function are identical (see Hillery et al. \cite{Wigner}),
providing the parameters
$\xi$ and $\xi^*$  are related to the coordinates $q$ and $p$
of the phase space as:
\be
\xi=\frac{1}{\sqrt{2\hbar}}\left(\lambda q +i\lambda^{-1}p\right);\qquad
\xi^*=\frac{1}{\sqrt{2\hbar}}\left(\lambda q -i\lambda^{-1}p\right),
\label{3.18}
\ee
i.e.,
\be
 W_{\hat{\rho}}(q,p)=
\frac{1}{2\pi\hbar}\int C^{(W)}_{\hat{\rho}}(q',p')\exp\left[-\frac{i}{\hbar}(qp'-pq')
\right]dq'\,dp',
\label{3.19}
\ee
where the characteristic function $C^{(W)}_{\hat{\rho}}(q,p)$ is given by the relation
\be
C^{(W)}_{\hat{\rho}}(q,p)= {\rm Tr}\left[\hat{\rho}\hat{D}(q,p)\right].
\label{3.20}
\ee
The displacement operator in terms of the position and the momentum operators
reads
\be 
\hat{D}(q,p)= \exp\left[\frac{i}{\hbar}(\hat{q}p-\hat{p}q)\right].
\label{3.21}
\ee
The symmetrically ordered cumulants of the operators $\hat{q}$ and $\hat{p}$
can be evaluated as
\be
\left.\langle\langle\{\hat{p}^m\hat{q}^n\}\rangle\rangle=\hbar^{n+m}
{\partial^{(m+n)}\over{\partial(-iq)^m\partial(ip)^n}}\ln C^{(W)}_{\hat{\rho}}(q,p)
\right|_{q,p=0},
\label{3.22}
\ee
The Wigner function can be interpreted as the quasiprobability (see below)
density distribution through which a probability can be expressed
to find  a quantum-mechanical system  (harmonic oscillator) around the 
``point''
$(q,p)$ of the phase space.

With the help of the Wigner function $W_{\hat{\rho}}(q,p)$ the position and 
momentum
probability distributions $w_{\hat{\rho}}(q)$ and $w_{\hat{\rho}}(p)$ can
be expressed from $W_{\hat{\rho}}(q,p)$
via marginal integration over the conjugated variable (in what follows we
assume $\lambda=1$)
\be
w_{\hat{\rho}}(q)\equiv\frac{1}{\sqrt{2\pi\hbar}}\int dp\,W_{\hat{\rho}}(q,p)
=\sqrt{2\pi\hbar}\langle q|\hat{\rho}|q \rangle,
\label{3.23}
\ee
where $|q \rangle$ is the eigenstate of the position operator $\hat{q}$.
The marginal
probability distribution $W_{\hat{\rho}}(q)$ is normalized to unity, i.e.,
\be
\frac{1}{\sqrt{2\pi\hbar}}\int dq\,w_{\hat{\rho}}(q)=1.
\label{3.24}
\ee

\subsection{Quantum homodyne tomography}
\label{sec3.A}
The relation (\ref{3.23}) 
for the probability distribution $w_{\hat{\rho}}(q)$ of the position
operator $\hat{q}$ can be generalized to the case  of the distribution
of the rotated quadrature operator $\hat{x}_{\theta}$. This operator is
defined as
\be
\hat{x}_{\theta}=\sqrt{\frac{\hbar}{2}}\left[\hat{a}{\rm e}^{-i\theta}
+\hat{a}^{\dagger}{\rm e}^{i\theta}\right],
\label{3.25}
\ee
and the corresponding conjugated operator $\hat{x}_{\theta+\pi/2}$,
such that $[\hat{x}_{\theta},\hat{x}_{\theta+\pi/2}]=i\hbar$,
reads
\be
\hat{x}_{\theta+\pi/2}=\frac{\sqrt{\hbar}}{i\sqrt{2}}
\left[\hat{a}{\rm e}^{-i\theta} -\hat{a}^{\dagger}{\rm e}^{i\theta}\right].
\label{3.26}
\ee
The position and the momentum operators are related to the operator
$\hat{x}_{\theta}$ as, $\hat{q}=\hat{x}_{0}$ and
$\hat{x}_{\pi/2}=\hat{p}$. The rotation (i.e., the linear
homogeneous canonical transformation) given by Eqs.(\ref{3.25})
and (\ref{3.26}) can be performed
by the unitary operator $\hat{U}(\theta)$:
\be
\hat{U}(\theta)=\exp\left[-i\theta\hat{a}^{\dagger}\hat{a}\right],
\label{3.27}
\ee
so that
\be
\hat{x}_{\theta}=\hat{U}^{\dagger}(\theta)\hat{x}_{0}\hat{U}(\theta);
\qquad
\hat{x}_{\theta+\pi/2}=
\hat{U}^{\dagger}(\theta)\hat{x}_{\pi/2}\hat{U}(\theta).
\label{3.28}
\ee
Alternatively, in the vector formalism we can rewrite the transformation 
(\ref{3.28})
as
\be
\left(\begin{array}{c}
\hat{x}_{\theta} \\
\hat{x}_{\theta+\pi/2}
\end{array}\right)={\bf F}
\left(\begin{array}{c}
\hat{q} \\
\hat{p}
\end{array}\right);\qquad {\bf F}=
\left(\begin{array}{cc}
\cos\theta & \sin\theta\\
-\sin\theta & \cos\theta
\end{array}\right).
\label{3.29}
\ee

Eigenvalues $x_\theta$ and $x_{\theta+\pi/2}$ of the operators
$\hat{x}_{\theta}$ and $\hat{x}_{\theta+\pi/2}$ can be expressed
in terms of the eigenvalues $q$ and $p$ of the position and momentum
operators as:
\be
\left(\begin{array}{c}
x_{\theta} \\
x_{\theta+\pi/2}
\end{array}\right)={\bf F}
\left(\begin{array}{c}
q \\
p
\end{array}\right);\qquad
\left(\begin{array}{c}
q \\
p
\end{array}\right)={\bf F}^{-1}
\left(\begin{array}{c}
x_{\theta} \\
x_{\theta+\pi/2}
\end{array}\right);\qquad
{\bf F}^{-1}=
\left(\begin{array}{cc}
\cos\theta & -\sin\theta\\
\sin\theta & \cos\theta
\end{array}\right),
\label{3.30}
\ee
where the matrix ${\bf F}$ is given by Eq.(\ref{3.29})
 and ${\bf F}^{-1}$ is the
corresponding inverse matrix. It has been shown by Ekert and Knight
\cite{Ekert91}
that  Wigner functions are transformed under the action of the
linear canonical transformation (\ref{3.29}) as:
\be
W_{\hat{\rho}}(q,p)\rightarrow W_{\hat{\rho}}
({\bf F}^{-1}(x_{\theta},x_{\theta+\pi/2}))=
W_{\hat{\rho}}(x_{\theta}\cos\theta-x_{\theta+\pi/2}\sin\theta;
x_{\theta}\sin\theta+x_{\theta+\pi/2}\cos\theta),
\label{3.31}
\ee
which means that the probability distribution
$w_{\hat{\rho}}(x_{\theta},\theta)=
\sqrt{2\pi\hbar}\langle x_{\theta}|\hat{\rho}| x_{\theta}\rangle$ can
be evaluated as
\be
w_{\hat{\rho}}(x_{\theta},\theta)
=\frac{1}{\sqrt{2\pi\hbar}}\int_{-\infty}^{\infty}
dx_{\theta+\pi/2}\,
W_{\hat{\rho}}(x_{\theta}\cos\theta-x_{\theta+\pi/2}\sin\theta;
x_{\theta}\sin\theta+x_{\theta+\pi/2}\cos\theta).
\label{3.32}
\ee
As shown by Vogel and Risken \cite{Vogel89} (see also 
\cite{Leonhardt1997,Welsch1998,Dariano,Kuhn})
the knowledge of $w_{\hat{\rho}}(x_{\theta},\theta)$ 
for all values of $\theta$ 
(such that
$[0<\theta\leq\pi]$) is equivalent to the knowledge of the Wigner
function itself. This Wigner function can be obtained from the set
of distributions $w_{\hat{\rho}}(x_{\theta},\theta)$ via the inverse Radon
transformation:
\be
W_{\hat{\rho}}(q,p)=\frac{1}{(2\pi\hbar)^{3/2}}
\int_{-\infty}^{\infty} dx_{\theta}
\int_{-\infty}^{\infty}d\xi\,|\xi|\int_{0}^{\pi} d\theta\,
w_{\hat{\rho}}(x_{\theta},\theta)
\exp\left[\frac{i}{\hbar}\xi(x_{\theta}-q\cos\theta-p\sin\theta)
\right].
\label{3.33}
\ee
It will be shown later in this paper that the optical homodyne tomography is
implicitly based on a measurement of all (in principle, infinite number)
independent moments (cumulants) of the system operators. Nevertheless,
there are states for which the Wigner function can be reconstructed
much easier than via the homodyne tomography. These are Gaussian and
generalized Gaussian states which are completely characterized by the first
two cumulants of the relevant observables while all higher-order cumulants
are equal to zero. On the other hand, if the state under consideration
is characterized by an infinite number of nonzero cumulants
then the homodyne tomography can fail because it does not provide us
with a consistent truncation scheme (see below and \cite{Schack90}).
As we will show later, the MaxEnt principle may help use to
reconstruct reliably the Wigner function from incomplete tomographic
data.

\subsubsection{Quantum tomography via pattern functions}
\label{sec3.A.1}
In a sequence of papers D'Ariano et al. \cite{Dariano},
Leonhardt et al. \cite{Leonhardtt1} and Richter \cite{Richter}
have shown
that Wigner functions can be very efficiently reconstructed from tomographic
data with the help of the so-called pattern functions. This reconstruction
procedure is more effective than the usual Radon transformation 
\cite{Leonhardt96-oc}. To be specific, D'Ariano et al. \cite{Dariano} have
shown that the density matrix $\rho_{mn}$
in the Fock basis\footnote{We note that very analogous procedure for
a reconstruction of density operators in the quadrature basis has been
proposed by K\"{u}hn, Welsch and Vogel \cite{Kuhn}.} 
can be reconstructed
directly from the tomographic data, i.e. from the quadrature-amplitude
``histograms'' (probabilities), $w(x_\theta,\theta)$ via the so-called
{\it direct sampling method} when 
\begin{eqnarray}
\rho_{mn}=\int_{0}^{\pi}\int_{-\infty}^{\infty} w(x_\theta,\theta)
F_{mn}(x_\theta,\theta)\, dx_{\theta} \, d\theta,
\label{3.A.1.1}
\end{eqnarray}
where $F_{mn}(x_\theta,\theta)$ is a set of specific {\it
sampling} functions (see below). Once the density matrix elements are
reconstructed with the help of Eq.(\ref{3.A.1.1}) then the Wigner function
of the corresponding state can be directly obtained using the relation
\begin{eqnarray}
W_{\hat{\rho}}(q,p)=\sum_{m,n}
\rho_{mn} W_{|m\rangle\langle n|}(q,p),
\label{3.A.1.2}
\end{eqnarray}
where $W_{|m\rangle\langle n|}(q,p)$ is the Wigner function of the operator
$|m\rangle\langle n|$.

A serious problem with the direct sampling method as proposed by D'Ariano et
al. \cite{Dariano} is that the sampling functions 
$F_{mn}(x_\theta,\theta)$ are difficult to compute. Later 
D'Ariano, Leonhardt and Paul \cite{Leonhardtt1,Dariano2} have simplified
the expression for the sampling function and have found that it can be
expressed as
\begin{eqnarray}
F_{mn}(x_\theta,\theta)=
f_{mn}(x_\theta) \exp\left[i(m-n)\theta\right],
\label{3.A.1.3}
\end{eqnarray}
where the so-called {\it pattern} function ``picks up'' the pattern in
the quadrature histograms (probability distributions) 
$w_{mn}(x_\theta,\theta)$ which just match the corresponding density-matrix
elements. Recently Leonhardt et al. \cite{Leonhardt96-oc} have shown that
the pattern function $f_{mn}(x_\theta)$ can be expressed as derivatives
\begin{eqnarray}
f_{mn}(x)=\frac{\partial }{\partial x} g_{mn}(x), 
\label{3.A.1.4}
\end{eqnarray}
of functions $g_{mn}(x)$ which are given by the Hilbert transformation
\begin{eqnarray}
g_{mn}(x)=\frac{\cal P}{\pi}
\int_{-\infty}^{\infty}\frac{\psi_m(\zeta)\psi_n(\zeta)}{x-\zeta}
d\zeta,
\label{3.A.1.5}
\end{eqnarray}
where ${\cal P}$ stands for the principal value of the integral and 
 $\psi_n(x)$ are the real energy eigenfunctions of the harmonic
oscillator, i.e. the normalizable solutions of the Schr\"{o}dinger equation
\begin{eqnarray}
\left(-\frac{\hbar^2}{2}\frac{d^2}{dx^2}+ \frac{x^2}{2}\right)
\psi_n(x) = \hbar (n+1/2) \psi_n(x),
\label{3.A.1.6}
\end{eqnarray}
(we assume $m=\omega=1)$. Further details
 of possible applications and discussion
devoted to numerical procedures of the reconstruction of density operators
via the direct sampling method can be found in Ref. \cite{Leonhardt96-oc}.

\subsection{States of light to be considered}
\label{sec3.B}
In this paper we will consider several quantum-mechanical states of a
single-mode  light field. In particular,
we will analyze coherent state, Fock state, squeezed vacuum state,
and superpositions of coherent states.

\subsubsection{ Coherent state}
\label{sec3.B.1}
The coherent state $|\alpha \rangle$ [see Eqs.(\ref{3.5}-\ref{3.6})]
is an eigenstate of the annihilation
operator $\hat{a}$, i.e., $|\alpha \rangle$
is not an eigenstate of an observable \cite{Glauber}. 
The Wigner function [Eq.(\ref{3.9})]
of the coherent state in the complex $\xi$-phase space reads
\be
W_{| \alpha\rangle}(\xi)=2\exp\left(-2|\xi-\alpha|^2\right);\qquad
\alpha=\alpha_x +i \alpha_y,
\label{3.34}
\ee
or alternatively, in the $(q,p)$ phase space  we have:
\be
W_{| \alpha\rangle}(q,p)=\frac{1}{\sigma_q\sigma_p}\exp\left[-\frac{1}{2\hbar}
\frac{(q-\bar{q})^2}{\sigma_q^2}
-\frac{1}{2\hbar}\frac{(p-\bar{p})^2}{\sigma_p^2}\right],
\label{3.35}
\ee
where $\bar{q}=\sqrt{2\hbar}\alpha_x/\lambda$;
$\bar{p}=\sqrt{2\hbar}\alpha_y\lambda$, and
\be
\sigma_q^2=\frac{1}{2\lambda^2}\qquad \mbox{and} \qquad
\sigma_p^2=\frac{\lambda^2}{2}.
\label{3.36}
\ee
The mean photon number in the coherent state is equal to
$\bar{n}=|\alpha|^2$. The variances for the position and momentum operators
are
\be
\langle \alpha|(\Delta\hat{q})^2| \alpha\rangle=\hbar\sigma_q^2;\qquad
\langle \alpha|(\Delta\hat{p})^2| \alpha\rangle=\hbar\sigma_p^2,
\label{3.37}
\ee
from which it is seen that the coherent state belongs to the class of
the minimum uncertainty states for which
\be
\langle (\Delta\hat{q})^2 \rangle
\langle (\Delta\hat{p})^2 \rangle=\hbar^2\sigma_q^2
\sigma_p^2=\frac{\hbar^2}{4}.
\label{3.38}
\ee
Using the expression (\ref{3.35}) 
for the Wigner function in  the $(q,p)$-phase
space we can evaluate the central moments of the Weyl-ordered moments of
the operators $\hat{q}$ and $\hat{p}$ in the coherent state as:
\be
\langle\{\hat{q}^k\hat{p}^l\}\rangle^{(c)}=\left\{
\begin{array}{l}
(2n-1)!!(2m-1)!!(\hbar\sigma_q)^n(\hbar\sigma_p)^m;~~{\rm for}~k=2n, l=2m\\
0;~~~{\rm for~~} k=2n+1 ~{\rm or}~l=2m+1.
\end{array}\right.
\label{3.39}
\ee
We see that all central moments of the order higher than second can be
expressed in terms of the second-order	central moments, so we can conclude
that the coherent state is completely characterized by four mean values
$\langle \hat{q}\rangle$; $\langle \hat{p}\rangle$; $\langle \hat{q}^2\rangle$,
and $\langle \hat{p}^2\rangle$.  With the help of the relation (\ref{3.20}) 
we can
find   the characteristic function $C^{(W)}_{| \alpha\rangle}(q,p)$
of the symmetrically-ordered
moments of the coherent state
\be
C^{(W)}_{| \alpha\rangle}(q,p)=\exp\left[\frac{i}{\hbar}\bar{q}p-\frac{i}{\hbar}\bar{p}q
-\frac{\sigma_q^2}{2\hbar}p^2 -\frac{\sigma_p^2}{2\hbar}q^2\right],
\label{3.40}
\ee
from which the following nonzero cumulants for the coherent state:
\be
\langle\langle \hat{q}\rangle\rangle=\bar{q}; \qquad
\langle\langle \hat{p}\rangle\rangle=\bar{p}; \qquad
\langle\langle \hat{q}^2\rangle\rangle=\hbar\sigma_q^2; \qquad
\langle\langle \hat{p}^2\rangle\rangle=\hbar\sigma_p^2,
\label{3.41}
\ee
can be found. We stress that all other cumulants of the operators
$\hat{q}$ and $\hat{p}$ are equal to zero. This is due to the fact
that the characteristic function of the Weyl-ordered moments is
an exponential of a polynomial of the second order in $q$ and $p$.

\subsubsection{ Fock state}
\label{sec3.B.2}
Eigenstates $|n \rangle$ of the photon number operator $\hat{n}$
\be
\hat{n}=\hat{a}^{\dagger}\hat{a}=
\frac{1}{2\hbar}\left(\hat{q}^2+\hat{p}^2\right) -\frac{1}{2},
\label{3.42}
\ee
are called the Fock states. The Wigner function of the Fock state
$|n \rangle$ is the $\xi$-phase space reads
\be
W_{|n \rangle}(\xi)=2(-1)^n\exp\left(-2|\xi|^2\right){\cal L}_{n}\left(
4|\xi|^2\right),
\label{3.43}
\ee
where ${\cal L}_n(x)$ is the Laguerre polynomial of the order $n$.
In the $(q,p)$ phase space this Wigner function has the form
\be
W_{|n \rangle}(q,p)=2(-1)^n\exp\left(-\frac{q^2+p^2}{\hbar}\right)
{\cal L}_{n}\left(2\frac{q^2+p^2}{\hbar}\right).
\label{3.44}
\ee
The Wigner function (\ref{3.44}) does not have a Gaussian form.
One can find from Eq.(\ref{3.44}) the following expressions
for first few moments of the position and momentum operators:
\be
\langle \hat{q}\rangle & = &  \langle \hat{p}\rangle=0;
\nonumber\\
\langle \hat{q}^2\rangle & =&  \langle \hat{p}^2\rangle=\frac{\hbar}{2}(2n+1);
\nonumber\\
\langle \hat{q}^4\rangle & =&  \langle \hat{p}^4\rangle=
\frac{\hbar^2}{4}(6n^2+6n+3) = \frac{3}{2}\frac{\langle \hat{q}^2\rangle^2
+\langle \hat{p}^2\rangle^2}{2}+\frac{3}{8}\hbar^2;
\label{3.45}
\\
\langle \hat{q}^2\hat{p}^2\rangle & =&  \langle \hat{p}^2\hat{q}^2\rangle=
\frac{\hbar^2}{4}(2n^2+2n-1)=  \frac{1}{2}\frac{\langle \hat{q}^2\rangle^2
+\langle \hat{p}^2\rangle^2}{2}-\frac{3}{8}\hbar^2.
\nonumber
\ee
In addition we find for the characteristic function $C^{(W)}_{| n\rangle}(q,p)$
of the Weyl-ordered moments of the operators $\hat{q}$ and $\hat{p}$
in the Fock state $|n \rangle$ the  expression
\be
C^{(W)}_{| n\rangle}(q,p)=
\exp\left[-\frac{(q^2+p^2)}{4\hbar}\right]
{\cal L}_n\left(\frac{(q^2+p^2)}{2\hbar}\right),
\label{3.46}
\ee
from which it follows that the Fock state is characterized by an infinite
number of nonzero cumulants. On the other hand,
moments of the photon number operator $\hat{n}$ in the
Fock state $|n \rangle$ are
\be
\langle \hat{n}^k\rangle=n^k,
\label{3.47}
\ee
from which it follows higher-order moments of the operator $\hat{n}$
can be expressed in terms of the first-order moment and
that all central moments
$\langle \hat{n}^k\rangle^{(c)}$ are equal to zero.

\subsubsection{Squeezed vacuum state}
\label{sec3.B.3}
The squeezed vacuum state \cite{Loudon87} 
can be expressed in the Fock basis as
\be
| \eta\rangle=\left(1-\eta^2\right)^{1/4}\sum_{n=0}^{\infty}
\frac{[(2n)!]^{1/2}}{2^n n!}\eta^n| 2n\rangle,
\label{3.48}
\ee
where the squeezing parameter $\eta$ (for simplicity we assume $\eta$ to be
real)  ranges from $-1$ to $+1$.
The squeezed vacuum state (\ref{3.48}) can be obtained by the action
of the squeezing operator $\hat{S}(r)$ on the vacuum state $|0 \rangle$
\be
|\eta \rangle=\hat{S}(r)|0 \rangle;\qquad
\hat{S}(r)=\exp\left[-\frac{ir}{2\hbar}(\hat{q}\hat{p}+\hat{p}\hat{q})\right]=
\exp\left[\frac{r}{2}\left(\hat{a}^{\dagger 2}-\hat{a}^2\right)\right],
\label{3.49}
\ee
where the squeezing parameter $r\in(-\infty,+\infty)$ is related to the
parameter $\eta$ as follows, $\eta=\tanh r$.
The mean
photon number in the squeezed vacuum (\ref{3.48}) is given by the relation
\be
\bar{n}=\frac{\eta^2}{1-\eta^2}.
\label{3.50}
\ee
The variances of the position and momentum operators can be expressed in
a form (\ref{3.37}) 
with the parameters $\sigma_q$ and $\sigma_p$ given by the
relations
\be
\sigma_q^2=\frac{1}{2}\left(\frac{1+\eta}{1-\eta}\right);\qquad
\sigma_p^2=\frac{1}{2}\left(\frac{1-\eta}{1+\eta}\right).
\label{3.51}
\ee
If we assume the squeezing parameter to be real and $\eta\in [0,-1)$
then from Eq.(\ref{3.51}) 
it follows that fluctuations in the momentum are reduced
below the vacuum state limit $\hbar/2$ at the expense of increased fluctuations
in the position. Simultaneously it is important to stress that the
product of variances $\langle \left(\Delta \hat{q}\right)^{2}\rangle$
and $\langle \left(\Delta \hat{p}\right)^{2}\rangle$ is equal to
$\hbar^2/4$, which means that the squeezed vacuum state belongs to
the class of the minimum uncertainty states.

The Wigner function of the squeezed vacuum state is of the Gaussian form
\be
W_{| \eta\rangle}(q,p)=\frac{1}{\sigma_q\sigma_p}\exp\left[-\frac{1}{2\hbar}
\frac{q^2}{\sigma_q^2}
-\frac{1}{2\hbar}\frac{p^2}{\sigma_p^2}\right],
\label{3.52}
\ee
with the parameters $\sigma^2_q$ and $\sigma^2_p$ given by Eq.(\ref{3.51}).
From Eq.(\ref{3.52}) it follows that the mean value of the position and the
momentum operators in the squeezed vacuum state are equal to zero,
while the higher-order symmetrically-ordered (central) moments
are given by Eq.(\ref{3.39}) 
with the parameters $\sigma^2_q$ and $\sigma^2_p$
given by Eq.(\ref{3.51}).  We see that higher-order moments can be expressed
in terms of the second-order moments.
We can find the expression for the characteristic function
$C^{(W)}_{| \eta\rangle}(q,p)$ for the squeezed vacuum state which reads
\be
C^{(W)}_{| \eta\rangle}(q,p)=
\exp\left[-\frac{\sigma_q^2}{2\hbar}p^2 -\frac{\sigma_p^2}{2\hbar}q^2\right],
\label{3.53}
\ee
from which it directly follows that the squeezed vacuum state is completely
characterized by to nonzero cumulants
$\langle\langle \hat{q}^2\rangle\rangle=\hbar\sigma_q^2$ and
$\langle\langle \hat{p}^2\rangle\rangle=\hbar\sigma_p^2$
(all other cumulants are equal to zero).

\subsubsection{Even and odd coherent states}
\label{sec3.B.4}
In nonlinear optical processes superpositions of coherent states
can be produced \cite{Buzek95}.
In particular, Brune et al. \cite{Brune}
have shown that an atomic-phase detection quantum non-demolition scheme
can serve for production of superpositions of two coherent states of
a single-mode radiation field. The following superpositions can be produced
via this scheme:
\be
|\alpha_{\rm e}\rangle = N_e^{1/2}\left(| \alpha\rangle+|-\alpha\rangle
\right);\qquad N_e^{-1}=2\left[1+\exp(-2|\alpha|^2)\right],
\label{3.54}
\ee
and
\be
|\alpha_{\rm o}\rangle = N_o^{1/2}\left(| \alpha\rangle-|-\alpha\rangle
\right);\qquad N_o^{-1}=2\left[1-\exp(-2|\alpha|^2)\right],
\label{3.55}
\ee
which are called the even and odd coherent states, respectively. These states
have been introduced by Dodonov et al. \cite{Dodonov} in a formal
group-theoretical analysis of various subsystems of coherent states.
More recently, these states have been analyzed as prototypes of superposition
states of light 
 which exhibit various nonclassical
effects (for the review see \cite{Buzek95}).
In particular, quantum interference between component states
leads to oscillations in the photon number distributions.
Another consequence of this interference is a reduction (squeezing) of
quadrature fluctuations in the even coherent state. On the other hand,
the odd coherent state exhibits reduced fluctuations in the photon number
distribution (sub-Poissonian photon statistics). Nonclassical effects
associated with superposition states can be explained in terms of quantum
interference between the ``points'' (coherent states) in phase space.
The character of quantum interference is very sensitive with respect to
the relative phase between coherent components of superposition states.
To illustrate this effect we write down the expressions for the Wigner
functions of the even and odd coherent states (in what follows we assume
$\alpha$ to be real):
\be
W_{|\alpha_e\rangle}(q,p)
=N_e\left[ W_{| \alpha\rangle}(q,p)+W_{|-\alpha\rangle}(q,p)
+W_{int}(q,p)\right];
\label{3.56}
\ee
\be
W_{|\alpha_o\rangle}(q,p)=
N_o\left[ W_{| \alpha\rangle}(q,p)+W_{|-\alpha\rangle}(q,p)
-W_{int}(q,p)\right],
\label{3.57}
\ee
where the Wigner functions $W_{|\pm\alpha\rangle}(q,p)$ of coherent states
$|\pm\alpha \rangle$ are given by Eq.(\ref{3.35}). 
The interference part of the
Wigner functions (\ref{3.56}) and (\ref{3.57}) is given by the relation
\be
W_{int}(q,p)=\frac{2}{\sigma_q\sigma_p}
\exp\left[-\frac{q^2}{2\hbar\sigma_q^2}-\frac{p^2}{2\hbar\sigma_p^2}\right]
\cos\left(\frac{\bar{q}p}{\hbar\sigma_q\sigma_p}\right),
\label{3.58}
\ee
where $\bar{q}=\sqrt{2\hbar}\alpha$ (we assume real $\alpha$)
and the variances $\sigma_q^2$
and $\sigma_p^2$ are given by Eq.(\ref{3.36}). From Eqs.
(\ref{3.56})-(\ref{3.57})
it follows that the even and odd coherent states  differ by
a sign of the interference part, which results in completely
different quantum-statistical properties of these states.

With the help of the Wigner function (\ref{3.56}) we evaluate  mean values
of moments of the operators $\hat{q}$ and $\hat{p}$. The first
moments are equal to zero, i.e.,
$\langle \hat{q}\rangle  =  \langle \hat{p}\rangle=0$, while for
higher-order moments we find
\be
\begin{array}{rl}
\langle \hat{q}^2\rangle &= \frac{\hbar}{2}\left(1+8N_e\alpha^2\right);\\
\langle \hat{p}^2\rangle &= \frac{\hbar}{2}
\left(1-8N_e\alpha^2{\rm e}^{-2\alpha^2}\right);\\
\langle \hat{q}^4\rangle &= \frac{3\hbar^2}{4}
\left[1+16N_e\alpha^2\left(1+\frac{2}{3}\alpha^2\right) \right];\\
\langle \hat{p}^4\rangle &= \frac{3\hbar^2}{4}
\left[1-16N_e\alpha^2{\rm e}^{-2\alpha^2}
\left(1-\frac{2}{3}\alpha^2\right) \right].
\end{array}
\label{3.59} 
\ee
From Eqs.(\ref{3.59}) 
it follows that the even coherent state exhibits the second
and fourth-order squeezing in the $\hat{p}$-quadrature \cite{Buzek95}.
We do not present explicit expression for higher-order moments, which
in general cannot be expressed in powers of  second-order moments.
In terms of the cumulants it means that the even (and odd) coherent states
are characterized by an infinite number of nonzero cumulants. This can be seen
from the expression for the characteristic function of the even coherent
state which reads
\be
C^{(W)}_{|\alpha_e\rangle}(q,p)=2N_e\exp\left[-\frac{\sigma_p^2}{2\hbar}q^2-
\frac{\sigma_q^2}{2\hbar}p^2\right]\left\{\cos\left(\frac{\bar{q}p}{\hbar}
\right)
+\exp\left(-\frac{\bar{q}^2}{2\hbar\sigma_q^2}\right)
\cosh\left(\frac{\sigma_p}{\hbar\sigma_q}\bar{q}q\right)\right\}.
\label{3.60}
\ee

\section{OBSERVATION LEVELS FOR SINGLE-MODE FIELD}
\label{sec4}
In our paper we will consider two different classes of observation levels.
Namely, we will consider the phase-sensitive and phase-insensitive
observation levels. These two classes do differ by the fact that
phase-sensitive observation levels are related to such operator
which provide some information about off-diagonal matrix elements of the
density operator in the Fock basis (i.e., these observation levels reveal
some information about the phase of states under consideration). On
the contrary, phase-insensitive observation levels are based exclusively on a
measurement of diagonal matrix elements in the Fock basis.
Before we proceed to a detailed description of the phase-sensitive and
phase-insensitive observation levels we introduce two exceptional observation
levels, the complete and thermal observation levels.

\subsection{Two extreme observation levels}
\label{sec4.A}

\subsubsection{ Complete observation level ${\cal O}_{0}\equiv\{
(\hat{a}^{\dagger})^k\hat{a}^l;~\forall k,l\}$}
\label{sec4.A.1}

The set of operators $\vert n \rangle \langle m \vert$ (for all
$n$ and $m$) are referred to as {\em complete}
observation level.
Expectation values  of the operators $\vert n \rangle \langle m \vert$
 are the matrix elements  of the density operator in the Fock basis
\be
\langle m |\hat \rho | n \rangle={\rm Tr}\left[
\hat \rho | n \rangle\langle m |\right];\qquad
\forall n,m ,
\label{4.1}
\ee
and therefore the generalized canonical density operator is identical
with the statistical density operator
\be
\hat \sigma_{0}={1 \over {Z_{0}}}
\exp\left[-\sum\limits_{m,n=0}^\infty
{  {\lambda _{m,n}\left| n
\right\rangle\left\langle m \right|}}\right]=\hat \rho \,;
\label{4.2}
\ee
\be
Z_{0}={\rm Tr}\left\{\exp\left[ -\sum_{m,n=0}^\infty
\lambda _{m,n}\left| n \right\rangle\left\langle m \right|\right]\right\}.
\label{4.3}
\ee
In this case the entropy $S_{0}$ is determined by the density operator
$\hat\rho$ as
\be
S_{0}=-{\rm Tr}\left[\hat \sigma _{0}\ln \hat
\sigma _{0}\right]=-{\rm Tr}
\left[\hat \rho\ln \hat \rho\right]\,.
\label{4.4}
\ee
This entropy is usually called the von Neumann entropy 
\cite{Neumann55,Wehrl78}.

As a consequence of the relation (cf. Sec. 3.3 in \cite{Louisell})
\be
\left| n \right\rangle\left\langle m \right|=\mathop {\lim }
\limits_{\varepsilon \to 1}\sum\limits_{k=0}^\infty  {{{(-\varepsilon )^k}
\over {k ! \sqrt {n !\, m !}}}(\hat a^{\dagger})^{k+n}\hat a^{k+m}}\,,
\label{4.5}
\ee
the complete observation level ${\cal O}_{0}$ can also be given by
a set of operators
$\{(\hat a^{\dagger})^k \hat a^l ;\,\,\, \forall k,l\}$ or
$\{\hat{q}^k \hat p^l ;\,\,\, \forall k,l\}$. The Wigner function
on the complete information level is equal to the Wigner function of
the state itself, i.e.,
$W_{\hat{\rho}}^{(0)}(\xi)= W_{\hat{\rho}}(\xi)$.

\subsubsection{ Thermal observation level ${\cal O}_{\rm th}\equiv\{
\hat{a}^{\dagger}\hat{a}\}$}
\label{sec4.A.2}
The total reduction of the complete observation level ${\cal O}_{0}$
results in a thermal observation level ${\cal O}_{\rm th}$
characterized just by one observable, the photon number operator $\hat{n}$,
i.e., quantum-mechanical states of light on this observation level are
characterized only by their mean photon number $\bar{n}\equiv\langle\hat{n}
\rangle$.
The generalized canonical density
operator of this observation level is the well-known  density operator of
the harmonic oscillator in  the thermal equilibrium
\be
\hat \sigma_{\rm th}={1 \over {Z_{\rm th}}}\exp [-\lambda_{\rm th}
\hat n].
\label{4.6}
\ee
To find an explicit expression for the Lagrange multiplier $\lambda_{\rm th}$
we have to solve the equation
\be
{\rm Tr}\left[\sigma_{\rm th}\hat{n}\right]=\bar{n}, 
\label{4.7}
\ee
from which we find that
\be
\lambda_{\rm th}=\ln\left(\frac{\bar{n}+1}{\bar{n}}\right),
\label{4.8}
\ee
so that the partition function corresponding to the operator
$\hat \sigma_{\rm th}$	reads
\be
Z_{\rm th}=\{1-\exp [-\lambda_{\rm th} ]\}^{-1}=\bar n+1\,.
\label{4.9}
\ee
Now we can rewrite the generalized canonical density operator
$\hat \sigma_{\rm th}$ in the Fock basis in a form
\be
\hat \sigma_{\rm th}=\sum_{n=0}^{\infty}\frac{\bar{n}^n}{(\bar{n}+1)^{n+1}}
|n \rangle\langle n|.
\label{4.10}
\ee
For the entropy of the thermal observation level we find a familiar expression
\be
S_{\rm th}=(\bar n+1)\ln (\bar n+1)-\bar n\ln \bar n \,.
\label{4.11}
\ee
The fact that the entropy $S _{\rm th}$ is larger than zero for any
$\bar{n}>0$ reflects the fact that on the thermal observation level
{\em all} states with the same mean photon number are indistinguishable.
This is the reason why Wigner function of different states on the
thermal information level are identical. The Wigner function of the state
$|\Psi\rangle$ on the
thermal observation level is given by the relation
\be
W^{\rm (th)}_{\hat{\rho}}(\xi)=\frac{2}{1+2\bar{n}}
\exp\left[-\frac{2|\xi|^2}{1+2\bar{n}}\right].
\label{4.12}
\ee
Extending the thermal observation level we can obtain more ``realistic''
Wigner functions which in the limit of the complete observation level are
equal to the Wigner function of the measured state itself, i.e., they are
not biased by the lack of information (measured data) about the state.

\subsection{Phase-sensitive observation levels}
\label{sec4.B}

\subsubsection{Observation level ${\cal O}_{ 1}\equiv\{
\hat{a}^{\dagger}\hat{a},\hat{a}^{\dagger},\hat{a}\}$}
\label{sec4.B.1}
We can extent the thermal observation level if in addition to the observable
$\hat{n}$ we consider also the measurement of mean values of the operators
$\hat{a}$ and $\hat{a}^{\dagger}$ (that is, we consider a measurement
of the observables $\hat{q}$ and $\hat{p}$). If we denote the
(measured) mean values of this operators as $\langle\hat{a}\rangle=\gamma$
and $\langle\hat{a}^{\dagger}\rangle=\gamma^*$, then the generalized
canonical density operator $\hat{\sigma}_1$ can be written as
\be
\hat{\sigma}_1=\frac{1}{Z_1}\exp\left[-\lambda_1(\hat{a}^{\dagger}-\gamma^*)
(\hat{a}-\gamma)\right],
\label{4.13}
\ee
with the partition function $Z_1$ given by the relation
\be
Z_1=\left(1-{\rm e}^{-\lambda_1}\right)^{-1}.
\label{4.14}
\ee
We have chosen the density operator $\hat{\sigma}_1$ in such form that
the conditions
\be
\langle \hat{a}\rangle ={\rm Tr}[\hat{a}\hat{\sigma}_1]=\gamma;\qquad
\langle \hat{a}^{\dagger}\rangle ={\rm Tr}[\hat{a}^{\dagger}
\hat{\sigma}_1]=\gamma^*,
\label{4.15}
\ee
are automatically fulfilled. To see this we rewrite the density operator
$\hat{\sigma}_1$ in the form:
\be
\hat{\sigma}_1=\frac{1}{Z_1}\hat{D}(\gamma)\exp[-\lambda_1\hat{a}^{\dagger}
\hat{a}]\hat{D}^{\dagger}(\gamma),
\label{4.16}
\ee
where we have used the transformation property $\hat{D}(\gamma)\hat{a}
\hat{D}^{\dagger}(\gamma)=\hat{a}-\gamma$, and therefore
\be
{\rm Tr}[\hat{a}\hat{\sigma}_1]=\frac{1}{Z_1}{\rm Tr}\left[
\hat{D}^{\dagger}(\gamma)\hat{a}\hat{D}(\gamma)\exp(-\lambda_1\hat{a}^{\dagger}
\hat{a})\right]
=\gamma+
\frac{1}{Z_1}{\rm Tr}\left[
\hat{a}\exp(-\lambda_1\hat{a}^{\dagger}\hat{a})\right]=\gamma.
\label{4.17}
\ee
To find the Lagrange multiplier $\lambda_1$ we have to solve the equation
${\rm Tr}[\hat{a}^{\dagger}\hat{a}\hat{\sigma}_1]=\bar{n}$ from which we find
\be
{\rm e}^{-\lambda_1}=\frac{\bar{n}-|\gamma|^2}{1+\bar{n}-|\gamma|^2}.
\label{4.18}
\ee
The entropy $S_1$ on the observation level ${\cal O}_1$ can be expressed
in a form very similar to $S_{\rm th}$ [see Eq.(\ref{4.11})]
\be
S_1=[\bar{n}-|\gamma|^2+1]\ln[\bar{n}-|\gamma|^2+1]-
[\bar{n}-|\gamma|^2]\ln[\bar{n}-|\gamma|^2].
\label{4.19}
\ee
The Wigner function $W^{(1)}_{\hat{\rho}}(\xi)$
corresponding to the generalized
canonical density operator $\hat{\sigma}_1$ reads
\be
W^{(1)}_{\hat{\rho}}(\xi)=\frac{2}{1+2(\bar{n}-|\gamma|^2)}\exp\left[
-\frac{2|\xi-\gamma|^2}{1+2(\bar{n}-|\gamma|^2)}\right].
\label{4.20}
\ee
From the expression (\ref{4.19}) for the entropy $S_1$ it follows that
$S_1=0$ for those states for which $\bar{n}=|\gamma|^2$. In fact, there
is only one state with this property. It is a coherent state $| \alpha\rangle$
(\ref{3.6}). In other words, because of the fact that $S_1=0$, 
the coherent state
can be {\em completely} reconstructed on the observation level ${\cal O}_1$.
In this case
\be
W^{(1)}_{|\alpha\rangle}(\xi)=
W^{(0)}_{|\alpha\rangle}(\xi)=2\exp\left[-2|\xi-\alpha|^2\right],
\label{4.21}
\ee
[see Eq.(\ref{3.34})]. For other states $S_1>0$ and therefore to improve our
information about the state we have to perform further measurements,
i.e., we have to extent the observation level ${\cal O}_1$.

\subsubsection{Observation level ${\cal O}_{ 2}\equiv\{
\hat{a}^{\dagger}\hat{a},(\hat{a}^{\dagger})^2,\hat{a}^2,
\hat{a}^{\dagger},\hat{a}\}$}
\label{sec4.B.2}
One of	possible extensions of the observation level ${\cal O}_1$
can be performed with a help of observables $\hat{q}^2$ and $\hat{p}^2$,
i.e., when not only the mean photon number $\bar{n}$ and mean values of
$\hat{q}$ and $\hat{p}$ are known, but also the variances
$\langle \left(\Delta \hat{q}\right)^{2}\rangle$ and
$\langle \left(\Delta \hat{p}\right)^{2}\rangle$ are measured.
On the observation level ${\cal O}_2$ we can express the generalized
canonical operator $\hat{\sigma}_2$ as
\be
\hat{\sigma}_2=\frac{1}{Z_2}\exp\left[-
\frac{\lambda_2}{2}(\hat{a}^{\dagger}-\gamma^*)^2
-\frac{\lambda_2^*}{2}(\hat{a}-\gamma)^2
-\lambda_1(\hat{a}^{\dagger}-\gamma^*)(\hat{a}-\gamma)\right],
\label{4.22}
\ee
where the Lagrange multiplier $\lambda_1$ is real while
$\lambda_2$ can be complex: $\lambda_2 = |\lambda_2| {\rm e}^{-i\theta}$.
We can	rewrite
$\hat{\sigma}_2$ in a form similar to the thermal density operator:
\be
\hat{\sigma}_2=\frac{1}{\tilde{Z}_2}\hat{D}(\gamma)\hat{U}(\theta/2)
\hat{S}(r)\exp\left[-
\left(\lambda_1^2 -|\lambda_2|^2\right)^{1/2}
\hat{a}^{\dagger}\hat{a}\right]\hat{S}^{\dagger}(r)
\hat{U}^{\dagger}(\theta/2)\hat{D}^{\dagger}(\gamma),
\label{4.23}
\ee
where the operators
$\hat{D}(\gamma)$, $\hat{U}(\theta/2)$, and $\hat{S}(r)$ are given by
Eqs.(\ref{3.6}), 
(\ref{3.27}), and (\ref{3.49}), respectively. These operators transform
the annihilation operator $\hat{a}$ as:
\be
\hat{D}^{\dagger}(\gamma)\hat{a}\hat{D}(\gamma) =
\hat{a}+\gamma;
\nonumber
\\
\hat{U}^{\dagger}(\theta/2)\hat{a}\hat{U}(\theta/2) =
\hat{a}{\rm e}^{-i\theta/2};
\label{4.24}
\\
\hat{S}^{\dagger}(r)\hat{a}\hat{S}(r) =
\hat{a}\cosh r +\hat{a}^{\dagger}\sinh r.
\nonumber
\ee
The partition function $\tilde{Z}_2$ in Eq.(\ref{4.23}) can be evaluated in
an explicit form:
\be
\tilde{Z}_2^{-1}=1-\exp[-(\lambda_1^2-|\lambda_2|^2)^{1/2}].
\label{4.25}
\ee
In Eq.(\ref{4.23}) 
we have chosen the parameter $r$ to be given by the relation
$\tanh 2r=-|\lambda_2|/\lambda_1$. The density operator (\ref{4.23})
is defined in such  way that it automatically fulfills the condition
${\rm Tr}[\hat{a}\hat{\sigma}_2]=\gamma$, while the Lagrange multipliers
$\lambda_1$ and $\lambda_2$ have to be found from the relations
${\rm Tr}[\hat{a}^{\dagger}\hat{a}\hat{\sigma}_2]=\bar{n}$
and ${\rm Tr}[\hat{a}^2\hat{\sigma}_2]=\mu$:
\be
{\rm Tr}[\hat{a}^{\dagger}\hat{a}\hat{\sigma}_2]=\bar{n}=
|\gamma|^2-1/2+(\chi+1/2)\cosh 2r;
\nonumber
\\
{\rm Tr}[\hat{a}^2\hat{\sigma}_2]=\mu=\gamma^2+{\rm e}^{-i\theta}
(\chi+1/2)\sinh 2r,
\label{4.26}
\ee
where we have used the notation
\be
\chi=\left\{\exp[(\lambda_1^2-|\lambda_2|^2)^{1/2}]-1\right\}^{-1}.
\label{4.27}
\ee
Instead of finding explicit expressions for the Lagrange multipliers
$\lambda_1$ and $\lambda_2$ we can find solutions for the parameters
$\tanh 2r$ and $\chi$. We express these parameters in terms of the
measured central moments
$\langle \hat{a}^{\dagger}\hat{a}\rangle^{(c)} \equiv N=\bar{n}-|\gamma|^2>0$
and
$\langle \hat{a}^2\rangle^{(c)}\equiv M=|M|{\rm e}^{-i\theta}=\mu-\gamma^2$:
\be
\tanh 2r=\frac{|M|}{N+1/2};
\label{4.28}
\ee
\be
\chi=\left[(N+1/2)^2-|M|^2\right]^{1/2}-1/2.
\label{4.29}
\ee
We remind us that
physical requirements \cite{Gardiner}
 lead to the following restrictions
on the parameters $N$ and $M$:
\be
N\geq 0;\qquad N(N+1)\geq |M|^2.
\label{4.30}
\ee

Once the $\tanh 2r$ and $\chi$ are found we can reconstruct the Wigner function
$W^{(2)}_{\hat{\rho}}(\xi)$
on the observation level ${\cal O}_2$. This Wigner function  reads:
\be
W^{(2)}_{\hat{\rho}}(\xi)=\frac{1}{\left[(N+1/2)^2-|M|^2\right]^{1/2}}
\exp\left[-\frac{(N+1/2)|\xi-\gamma|^2-\frac{M^*}{2}(\xi-\gamma)^2-\frac{M}{2}
(\xi^*-\gamma^*)^2}{\left[(N+1/2)^2-|M|^2\right]}\right].
\label{4.31}
\ee
Analogously
 we can find an expression for the entropy $S_2$:
\be
S_2=(\chi+1)\ln(\chi+1)-\chi\ln\chi.
\label{4.32}
\ee
It has a form of the thermal entropy (\ref{4.11}) with a mean thermal-photon
number equal to $\chi$ [see Eq.(\ref{4.29})].

Using the expression for the Wigner function (\ref{4.31})
we can rewrite the variances of the position and momentum
operators in terms of the parameters $N$ and $M$ as follows
\be 
\langle \left(\Delta \hat{q}\right)^{2}\rangle =
\frac{\hbar}{2}[1+2N+2{\rm Re}M];\qquad
\langle \left(\Delta \hat{p}\right)^{2}\rangle =
\frac{\hbar}{2}[1+2N-2{\rm Re}M].
\label{4.33}
\ee
The product of these variances reads:
\be
\langle \left(\Delta \hat{q}\right)^{2}\rangle
\langle \left(\Delta \hat{p}\right)^{2}\rangle =
\frac{\hbar^2}{4}\left[(1+2N)^2-4({\rm Re}M)^2\right].
\label{4.34}
\ee
From the expression (\ref{4.32}) for the entropy $S_2$ it is seen that
those states for which $N(N+1)=|M|^2$ can be completely reconstructed
of the observation level ${\cal O}_2$, because for these states
$S_2=0$. In fact, it has been shown by Dodonov et al. \cite{Dodonov80}
 that the states
for which $N(N+1)=|M|^2$ are the {\em only} pure states which
have non-negative Wigner functions [see Eq.(4.31)]. For these states
the product of variances (\ref{4.34}) reads
\be
\langle \left(\Delta \hat{q}\right)^{2}\rangle
\langle \left(\Delta \hat{p}\right)^{2}\rangle =
\frac{\hbar^2}{4}\left[1+4({\rm Im}M)^2\right],
\label{4.35}
\ee
which means that if in addition  ${\rm Im}M=0$ (see for instance
squeezed vacuum state with real parameter of squeezing)
then these states also
belong to the class of the minimum uncertainty states.
From our previous discussion it follows that the squeezed vacuum as well as
squeezed coherent states can be completely reconstructed on the observation
level ${\cal O}_2$. More generally, we can say that all  Gaussian  states
for which $N(N+1)=|M|^2$ can be  completely reconstructed on this observation
level.

\subsubsection{Higher-order phase-sensitive observation levels}
\label{sec4.B.3}
There are pure non-Gaussian states (such as the even coherent state) for which
the entropy $S_2$ is larger than zero and therefore in order to reconstruct
 Wigner functions of such states more precisely, we have to extent the
observation level ${\cal O}_2$. Straightforward extension of ${\cal O}_2$
is the observation level ${\cal O}_k\equiv \{ (\hat{a}^{\dagger})^m\hat{a}^n;
\forall m,n; ~~ m+n\leq k\}$, which in the limit $k\rightarrow\infty$ is
extended to the complete observation level.

To perform a reconstruction of the Wigner function on the observation level
${\cal O}_k$ with $k> 2$ an attention has to be paid to the fact that
for a certain choice of possible observables the vacuum-to-vacuum matrix
elements of the generalized canonical density operator
$\langle 0|\hat{\sigma}_k | 0\rangle$ can have divergent Taylor-series
expansion. To be more specific, if we consider an observation level
such that  ${\cal O}_k\equiv \{ (\hat{a}^{\dagger})^k,\hat{a}^k\}$
then for the generalized canonical density operator
\be
\hat{\sigma}_k=\frac{1}{Z_k}\exp\left[-\lambda_k(\hat{a}^{\dagger})^k-
\lambda^*_k\hat{a}^k\right],
\label{4.36}
\ee
the corresponding partition function $Z_k=
{\rm Tr}\exp\left[-\lambda_k(\hat{a}^{\dagger})^k-\lambda^*_k\hat{a}^k\right]$
is divergent \cite{Fischer}. 
This means
that one cannot consistently define an observation level based exclusively on the measurement
of the operators $(\hat{a}^{\dagger})^k$ and $\hat{a}^k$. In general,
to ``regularize''
the problem one has to include the photon number operator $\hat{n}$ into
the observation level. Then the generalized density operator $\hat{\sigma}_k$
\be
\hat{\sigma}_k=\frac{1}{Z_k}\exp\left[
-\lambda_0\hat{a}^{\dagger}\hat{a}
-\lambda_k(\hat{a}^{\dagger})^k-\lambda^*_k\hat{a}^k\right],
\label{4.37}
\ee
can be	properly defined and one may reconstruct the corresponding Wigner
function $W_k(\xi)$.  We note that any observation
has to be chosen in such a way that information about the mean photon number
is available, i.e., knowledge of the mean photon number (the mean energy) of the
system under consideration represents a necessary condition for a reconstruction
of the Wigner function.

\subsection{Phase-insensitive observation levels}
\label{sec4.C}
The choice of the observation level is very important in order to optimize
the strategy for the measurement and the reconstruction of the Wigner
function  of a given quantum-mechanical state of light. For instance, if we
would like to reconstruct the Wigner function of the Fock state $| n\rangle$
at the observation level
  ${\cal O}_k\equiv \{ \hat{a}^{\dagger}\hat{a},(\hat{a}^{\dagger})^m\hat{a}^n;
m+n\leq k ~{\rm and}~~ m\neq n \}$ we find that irrespectively on the number
($k$) of ``measured'' moments $\langle (\hat{a}^{\dagger})^m\hat{a}^n\rangle$
(for $m\neq n$) the reconstructed Wigner function is always equal to the
thermal Wigner function (\ref{4.12}).
So it can
happen that in a very tedious experiment negligible information is obtained.
On the other hand, if a measurement of diagonal elements of the
density operator in the Fock basis is performed  relevant information
can be obtained much easier.

\subsubsection{Observation level ${\cal O}_{\rm A}\equiv
\{\hat{P}_n=|n \rangle\langle n|;~~\forall n\}$}
\label{sec4.C.1}
The most general phase-insensitive observation level corresponds to the case
when {\em all} diagonal elements $P_n=\langle n|\hat{\rho}|n \rangle$
of the density operator $\hat{\rho}$ describing the state under
consideration are measured. The observation level ${\cal O}_{\rm A}$
can be obtained via a reduction of the complete observation
level ${\cal O}_{0}$ and it corresponds to the measurement of
the photon number distribution $P_n$ such that $\sum_n P_n=1$.
Because of the relation [see Eq.(\ref{4.5})]
\be
\left| n \rangle\langle n \right|=
\mathop {\lim }
\limits_{\varepsilon \to 1}\sum\limits_{k=0}^\infty
\frac{(-\varepsilon )^k}{k! n!}(\hat a^{\dagger})^{k+n}\hat a^{k+n}=
\mathop {\lim }
\limits_{\varepsilon \to 1}\sum\limits_{k=0}^\infty
\frac{(-\varepsilon )^k}{k! n!}\frac{\hat{n}!}{(\hat{n}-k-n)!},
\label{4.38}
\ee
we can conclude that the observation level ${\cal O}_{\rm A}$
corresponds to the measurement of all moments of the creation and
annihilation operators of the form
$(\hat{a}^{\dagger})^{k}\hat a^{k}$ or, what is the same, it corresponds
to a measurement of all moments of the photon number operator, i.e.,
\be
{\cal O}_{\rm A}\equiv
\{\hat{P}_n=|n \rangle\langle n|;~~\forall n\}=\{(\hat a^{\dagger})^{k}\hat a^{k};
\forall ~~k\}= \{\hat{n}^k;~~\forall k\}.
\label{4.39}
\ee
The generalized canonical operator $\hat{\sigma}_{\rm A}$ at the observation
level ${\cal O}_{\rm A}$ reads
\be
\hat{\sigma}_{\rm A}=\frac{1}{Z_{\rm A}}\exp\left[-\sum_{n=0}^{\infty}
\lambda_n |n \rangle\langle n| \right];
\label{4.40}
\ee
with the partition function given by the relation
\be
Z_{\rm A}={\rm Tr}\left\{\exp\left[-\sum_{n=0}^{\infty}
\lambda_n |n \rangle\langle n| \right]\right\}
=\sum_{n=0}^{\infty}\exp[-\lambda_n].
\label{4.41}
\ee
The entropy $S_{\rm A}$ at the observation level ${\cal O}_{\rm A}$
can be expressed in the form
\be
S_{\rm A}=\ln Z_{\rm A}+\sum_{n=0}^{\infty}\lambda_n P_n.
\label{4.42}
\ee
The Lagrange multipliers $\lambda_n$ have to be evaluated from an infinite
set of equations:
\be
P_n={\rm Tr}[\hat{\sigma}_{\rm A} \hat{P}_n]=\frac{{\rm e}^{-\lambda_n}}
{Z_{\rm A}};~~~\forall n,
\label{4.43}
\ee
from which we find  $\lambda_n=-\ln[Z_{\rm A} P_n]$.
If we insert $\lambda_n$ into expression (\ref{4.42}) we obtain for the entropy
$S_{\rm A}$ the familiar expression
\be
S_{\rm A}=-\sum_{n=0}^{\infty} P_n\ln P_n,
\label{4.44}
\ee
derived by Shannon \cite{Shannon}. Here it should be briefly noted that
as a consequence of the relation
\be
\sum_{n=0}^{\infty}\hat{P}_n=\hat{1},
\label{4.45}
\ee
the operators $\hat{P}_n$ are not linearly independent, which means
that the Lagrange multipliers $\lambda_n$ and the partition function
$Z_{\rm A}$ are not uniquely defined. Nevertheless, if $Z_{\rm A}$
is chosen to be equal to unity, then the Lagrange multipliers
can be expressed as
\be
\lambda_n=-\ln P_n;
\label{4.46}
\ee
and the generalized canonical density operator reads
\be
\hat{\sigma}_{\rm A}=\sum_{n=0}^{\infty} P_n | n\rangle\langle n|; \qquad
\sum_{n=0}^{\infty}P_n=1.
\label{4.47}
\ee
From here it follows that the Wigner function $W^{\rm (A)}_{\hat{\rho}}(\xi)$
of the state $\hat{\rho}$
at the observation level ${\cal O}_{\rm A}$ can be reconstructed in the
form
\be
W^{\rm (A)}_{\hat{\rho}}(\xi) =\sum_{n=0}^{\infty} P_n W_{| n\rangle}(\xi)  ,
\label{4.48}
\ee
where $W_{| n\rangle}(\xi)$ is the Wigner function of the Fock state
$| n\rangle$ given by Eq.(\ref{3.43}).

The phase-insensitive observation level ${\cal O}_{\rm A}$ can be further
reduced if only a finite number of operators $\hat{P}_n$ [where $n\in {\cal M}$]
is considered. In this case, in general, we have
$\sum_{n\in {\cal M}} P_n<1$
and therefore it is essential that apart of mean values $P_n$ also the
mean photon number $\bar{n}$ is known from the measurement.

\subsubsection{Observation level ${\cal O}_{\rm B}\equiv
\{\hat{n}, \hat{P}_{2n}=|2n \rangle\langle 2n|;~~\forall n\}$}
\label{sec4.C.2}
As an example of the observation level which is reduced with respect
to ${\cal O}_{\rm A}$ we can consider the observation level
${\cal O}_{\rm B}$ which is based on a measurement of the average
photon number $\bar{n}$ and on the photon statistics on the subspace
of the Fock space composed of the even Fock states $| 2n\rangle$.
In this case the generalized canonical density operator $\hat{\sigma}_{\rm B}$
can be written as
\be
\hat{\sigma}_{\rm B}=\frac{1}{Z_{\rm B}}\exp\left[-\lambda \hat{n}
-\sum_{n=0}^{\infty}
\lambda_n \hat{P}_{2n} \right]=
\frac{{\rm e}^{-\lambda\hat{n}}}{Z_{\rm B}}
\left[\left(1-\sum_{n=0}^{\infty}\hat{P}_{2n}\right)+
\sum_{n=0}^{\infty}{\rm e}^{-\lambda_n} \hat{P}_{2n} \right],
\label{4.49}
\ee
where the partition function is given by the relation
\be
Z_{\rm B}={\rm Tr}\left\{ \exp\left[-\lambda \hat{n}
-\sum_{n=0}^{\infty}\lambda_n \hat{P}_{2n} \right]\right\}.
\label{4.50}
\ee
This partition function can be explicitly evaluated with the help
solutions for the Lagrange multipliers
from equations ${\rm Tr}[\hat{P}_{2n}\hat{\sigma}_{\rm B}]=P_{2n}$.
If we introduce the notation
\be
P_{odd}\equiv 1-\sum_{n=0}^{\infty}P_{2n};
\label{4.51}
\ee
\be
\bar{n}_{odd}\equiv \bar{n}-\sum_{n=0}^{\infty}2nP_{2n},
\label{4.52}
\ee
then the partition function $Z_{\rm B}$ can be expressed as
\be
Z_{\rm B}=\frac{\left[\bar{n}_{odd}^2-P_{odd}^2\right]^{1/2}}{2P_{odd}^2}.
\label{4.53}
\ee
Analogously we find for the generalized canonical density operator
the expression
\be
\hat{\sigma}_{\rm B}=\sum_{n=0}^{\infty} P_{2n}| 2n\rangle\langle 2n|
+\sum_{n=0}^{\infty} P_{2n+1}| 2n+1\rangle\langle 2n+1|,
\label{4.54}
\ee
where $P_{2n}$ are measured values of $\hat{P}_{2n}$ and $P_{2n+1}$
are evaluated from the MaxEnt principle:
\be
P_{2n+1}= \frac{2P_{odd}^2}{\bar{n}_{odd}+P_{odd}}
\left(\frac{\bar{n}_{odd}-P_{odd}}{\bar{n}_{odd}+P_{odd}}\right)^n.
\label{4.55}
\ee
From Eq.(\ref{4.55}) we see that on the subspace of odd Fock states
we have obtained from the MaxEnt principle
a ``thermal-like'' photon number distribution. Now, we know all values
of $P_{2n}$ and $P_{2n+1}$ and	using  Eq.(\ref{4.44})
we can easily evaluate the entropy $S_{\rm B}$ and the Wigner function
$W^{\rm (B)}_{\hat{\rho}}(\xi)$
on the observation level ${\cal O}_{\rm B}$ [see Eq.(\ref{4.48})].

\subsubsection{Observation level ${\cal O}_{\rm C}\equiv
\{\hat{n}, \hat{P}_{2n+1}=|2n +1\rangle\langle 2n+1|;~~\forall n\}$}
\label{sec4.C.3}

If  the mean photon number and the probabilities
$P_{2n+1}=\langle 2n+1|\hat{\rho}| 2n+1\rangle$ are known, then we can define
an observation level ${\cal O}_{\rm C}$ which in a sense is a complementary
observation level to ${\cal O}_{\rm B}$. After some algebra one can find for
the generalized canonical density operator $\hat{\sigma}_{\rm C}$
the expression equivalent to Eq.(\ref{4.54}), i.e.,
\be
\hat{\sigma}_{\rm C}=\sum_{n=0}^{\infty} P_{2n}| 2n\rangle\langle 2n|
+\sum_{n=0}^{\infty} P_{2n+1}| 2n+1\rangle\langle 2n+1|,
\label{4.56}
\ee
where the parameters $P_{2n+1}$ are known from measurement and
$P_{2n}$ are evaluated as follows
\be
P_{2n}= \frac{2P_{even}^2}{\bar{n}_{even}+2P_{even}}
\left(\frac{\bar{n}_{even}}{\bar{n}_{even}+2P_{even}}\right)^n.
\label{4.57}
\ee
In Eq.(\ref{4.57}) we have introduced notations
\be
P_{even}\equiv 1-\sum_{n=0}^{\infty}P_{2n+1};
\label{4.58}
\ee
\be
\bar{n}_{even}\equiv \bar{n}-\sum_{n=0}^{\infty}(2n+1)P_{2n+1}.
\label{4.59}
\ee
The explicit expression for the partition function $Z_{\rm C}$ is
\be
Z_{\rm C}=\frac{\bar{n}_{even}+2P_{even}}{2P_{even}^2}.
\label{4.60}
\ee
The reconstruction of the Wigner function $W^{\rm (C)}_{\hat{\rho}}(\xi)$
is now	straightforward [see Eq.(\ref{4.48})].

\subsubsection{Observation level ${\cal O}_{\rm D}\equiv
\{\hat{n}, \hat{P}_{N}=|N\rangle\langle N|\}$}
\label{sec4.C.4}
We can reduce observation levels ${\cal O}_{\rm A,B,C}$ even further
and we can consider only a measurement of the mean photon number $\bar{n}$
and a probability $P_N$ to find the system under consideration in
the Fock state $| N\rangle$. The generalized density operator
$\hat{\sigma}_{\rm D}$ in this case reads
\be
\hat{\sigma}_{\rm D}=\frac{1}{Z_{\rm D}}\exp\left[-\lambda\hat{n}
-\lambda_N\hat{P}_N\right].
\label{4.61}
\ee
Taking into account the fact that the observables under consideration
do commute, i.e., $[\hat{n},\hat{P}_N]=0$, and that the operator
$\hat{P}_N$ is a projector (i.e., $\hat{P}_N^2=\hat{P}_N$) we can rewrite
Eq.(\ref{4.61}) as
\be
\hat{\sigma}_{\rm D}=
\frac{{\rm e}^{-\lambda\hat{n}}}{Z_{\rm D}}\left[(1-\hat{P}_N)
+{\rm e}^{-\lambda_N}\hat{P}_N\right]=
P_N|N \rangle\langle N| +
\sum_{n\neq N}^{\infty}P_n|n \rangle\langle n|,
\label{4.62}
\ee
where $\lambda$ and $\lambda_N$ are Lagrange multipliers associated with
operators $\hat{n}$ and $\hat{P}_N$, respectively, and
$P_n=\exp(-\lambda n)/Z_{\rm D}$ gives the photon number distribution
on the subspace of the Fock space without the vector $| N\rangle$. The
generalized partition function can be expressed as
\be
Z_{\rm D}=\frac{1}{1-x}+x^N(y-1),
\label{4.63}
\ee
where we have introduced notation
\be
x=\exp(-\lambda);\qquad y= \exp(-\lambda_N).
\label{4.64}
\ee
The Lagrange multipliers can be found from equations
\be
P_N=\frac{1}{Z_{\rm D}} x^N y=\frac{(1-x)x^N y}{1+x^N(y-1)(1-x)};
\label{4.65}
\ee
\be
\bar{n}=\frac{1}{Z_{\rm D}}\left[\frac{x}{(1-x)^2}+Nx^N(y-1)\right]=
\frac{x+N x^N(1-x)^2(y-1)}{(1-x)[1+x^N(y-1)(1-x)]}.
\label{4.66}
\ee
Generally, we cannot express the Lagrange multipliers $\lambda$
and $\lambda_N$ as functions of $\bar{n}$ and $P_N$ in an analytical
way for arbitrary $N$ and Eqs.(\ref{4.65}) and (\ref{4.66})
 have to be solved numerically.
Nevertheless, there are two cases when these equations can be solved in a
closed analytical form.

{\bf 1.} If $N=0$ (we will denote this observation level as ${\cal O}_{\rm
D1}$), then we can find for Lagrange multipliers $\lambda$ and
$\lambda_0$ following expressions:
\be
{\rm e}^{-\lambda}=1-\frac{1-P_0}{\bar{n}};\qquad
{\rm e}^{-\lambda_0}=\frac{P_0}{(1-P_0)^2}[\bar{n}-(1-P_0)];
\label{4.67}
\ee
and for the partition function we find
\be
Z_{\rm D1}=\frac{\bar{n}-(1-P_0)}{(1-P_0)^2}.
\label{4.68}
\ee
Then after some straightforward algebra we can evaluate the parameters $P_n$
as
\be
P_n=\left\{
\begin{array}{l}
P_0 ~~~\mbox{ for} ~~ n=0; \\
\frac{(1-P_0)^2}{\bar{n}-(1-P_0)}
\left[\frac{\bar{n}-(1-P_0)}{\bar{n}}\right]^n
~~~\mbox{ for} ~~~ n>0.
\end{array}
\right.
\label{4.69}
\ee
From Eq.(\ref{4.69}) 
which describes the photon number distribution obtained from
the generalized density operator $\hat{\sigma}_{\rm D1}$
it follows that the
reconstructed state on the observation level ${\cal O}_{\rm D1}$ has on
the subspace formed of Fock states except the vacuum a thermal-like
character. Nevertheless,  in this case the reconstructed Wigner function
can be negative (unlike in the case of the thermal observation level).
 This can happen if $P_0$ is close to zero and
$\bar{n}$ is small.
Using explicit expressions for the parameters $P_n$ given by Eq.(\ref{4.69}) 
we can
evaluate the entropy $S_{\rm D1}$ corresponding to the present observation
level:
\be
S_{\rm D1}=- P_0\ln P_0 - (\bar{n} -P) \ln (\bar{n} -P) -
2 P\ln P
+ \bar{n}\ln\bar{n},
\label{4.70}
\ee
where we have used notation $P=1-P_0$. In the limit
$P_0\rightarrow (1+\bar{n})^{-1}$ expression (\ref{4.70})  reads
\be 
\lim_{P_0\rightarrow (1+\bar{n})^{-1}}S_{\rm D1} =
(\bar{n}+1)\ln(\bar{n}+1)
-\bar{n}\ln\bar{n},
\label{4.71}
\ee
which is the entropy on the thermal observation level Eq.(\ref{4.11}).
In this limit the  ${\cal O}_{\rm D1}$ reduces to
the thermal observation level ${\cal O}_{\rm th}$ .
On the other hand, in the limit $P_0\rightarrow 0$ we obtain from
Eq.(\ref{4.70})
\be
\lim_{P_0\rightarrow 0}S_{\rm D1}=
\bar{n}\ln\bar{n}
-(\bar{n}-1)\ln(\bar{n}-1),
\label{4.72}
\ee
from which it directly follows that in this case the mean photon number has
necessary to be larger or equal than unity. Moreover, from Eq.(\ref{4.72})
 we see
that in the limit $\bar{n}\rightarrow 1$ the entropy $S_{\rm D1}=0$
which means that the Fock state $|1 \rangle$ can be completely reconstructed
on the observation level ${\cal O}_{\rm D1}$. This fact can also be seen
from an explicit expression for the photon number distribution (\ref{4.69})
from which it follows that
\be
\lim_{\bar{n}\rightarrow 1}\lim_{P_0\rightarrow 0} P_n = \delta_{n,1}.
\label{4.73}
\ee

 {\bf 2.} If the mean photon number is an integer, then in the case
$N=\bar{n}$ (we will denote this observation level as  ${\cal O}_{\rm D2}$)
we find for the Lagrange multipliers $\lambda$ and $\lambda_{N=\bar{n}}\equiv
\lambda_{\bar{n}}$ the expressions
\be
{\rm e}^{-\lambda}=\frac{\bar{n}}{1+\bar{n}};\qquad
{\rm e}^{-\lambda_{\bar{n}}}=\frac{(1+\bar{n})^{1+\bar{n}}-\bar{n}^{\bar{n}}}
{(1-P_{\bar{n}})\bar{n}^{\bar{n}}} P_{\bar{n}},
\label{4.74}
\ee
and for the partition function we find
\be
Z_{\rm D2}=
\frac{(1+\bar{n})^{1+\bar{n}}-\bar{n}^{\bar{n}}}
{(1-P_{\bar{n}})(1+\bar{n})^{\bar{n}}}.
\label{4.75}
\ee
Taking into account the expression for the reconstructed photon number
distribution
\be
P_n=\langle n|\hat{\sigma}_{\rm D2}| n\rangle
=\frac{{\rm e}^{-n\lambda}}{Z_{\rm D2}}\left[1+\delta_{n,\bar{n}}
\left({\rm e}^{-\lambda_{\bar{n}}}-1\right)\right],
\label{4.76} 
\ee
then with the help of relations (\ref{4.74}) and (\ref{4.75})  we find
\be
P_n=\left\{
\begin{array}{l}
P_{\bar{n}} ~~~; ~~ n=\bar{n} \\
\frac
{(1-P_{\bar{n}})(1+\bar{n})^{\bar{n}}}
{(1+\bar{n})^{1+\bar{n}}-\bar{n}^{\bar{n}}}
\left(\frac{\bar{n}}{1+\bar{n}}\right)^n
~~~;~~~ n\neq \bar{n}.
\end{array}
\right.
\label{4.77}
\ee
We see that the reconstructed photon-number distribution has a thermal-like
character. The corresponding entropy can be evaluated in a closed analytical
form
\be
S_{\rm D2}=- P_{\bar{n}}\ln P_{\bar{n}} -(1-P_{\bar{n}})\ln(1-P_{\bar{n}})
+(1-P_{\bar{n}})\ln\left[\frac{(1+\bar{n})^{1+\bar{n}}}{\bar{n}^{\bar{n}}}
-1\right] .
\label{4.78}
\ee
It is interesting to note that if $P_{\bar{n}}$ is given by its value
in the thermal photon number distribution, i.e.,
\be
P_{\bar{n}}=\frac{\bar{n}^{\bar{n}}}{(1+\bar{n})^{1+\bar{n}}},
\label{4.79}
\ee
then the entropy (\ref{4.78}) reduces to
\be
S_{\rm D2}=
(\bar{n}+1)\ln(\bar{n}+1)
-\bar{n}\ln\bar{n}=
-\ln P_{\bar{n}},
\label{4.80}
\ee
which means that the reconstructed density operator $\hat{\sigma}_{\rm D2}$
on the observation level ${\cal O}_{\rm D2}$ with $P_{\bar{n}}$ given by
Eq.(\ref{4.79}) is equal to the density operator of the thermal field [see
Eq.(\ref{4.10})] and so, in this case the reduction ${\cal O}_{\rm D2}
\rightarrow {\cal O}_{\rm th}$ takes place. Obviously, if $P_{\bar{n}}=1$
then $S_{\rm D2}=0$ and the Fock state $| \bar{n}\rangle$ can be completely
reconstructed on the observation level ${\cal O}_{\rm D2}$.

\subsection{Relations between observation levels}
\label{sec4.D}
Various observation levels considered in this section can be obtained
as a result
of a sequence of mutual reductions. Therefore we can
order observation levels under consideration. This ordering can be done
separately for phase-sensitive and phase-insensitive observation levels.
In particular,
phase-sensitive observation levels are ordered as follows:
\be
{\cal O}_0\supset {\cal O}_2 \supset {\cal O}_1\supset {\cal O}_{\rm th}.
\label{4.81}
\ee
The corresponding entropies are related as
\be
S_0\le S_2\le S_1\le S_{\rm th}.
\label{4.82}
\ee
The ordering of phase-insensitive observation levels ${\cal O}_{\rm A}$,
${\cal O}_{\rm B}$, ${\cal O}_{\rm C}$, ${\cal O}_{\rm D1}$ and
${\cal O}_{\rm D2}$ is more complex. In particular, we find
\be
{\cal O}_0 & \supset {\cal O}_{\rm A} \supset \left\{
\begin{array}{c}
{\cal O}_{\rm B}\\
{\cal O}_{\rm C}
\end{array}\right\}
\supset {\cal O}_{\rm th};
\nonumber
\\
{\cal O}_0 & \supset {\cal O}_{\rm A} \supset \left\{
\begin{array}{c}
{\cal O}_{\rm D1}\\
{\cal O}_{\rm D2}
\end{array}\right\}
\supset {\cal O}_{\rm th},
\label{4.83}
\\
{\cal O}_0 & \supset {\cal O}_{\rm A} \supset
{\cal O}_{\rm B}\supset
{\cal O}_{\rm D1}
\supset {\cal O}_{\rm th}
\nonumber,
\ee
which reflects the fact that observation levels
${\cal O}_{\rm B}$ and	${\cal O}_{\rm C}$ (as well as
${\cal O}_{\rm D1}$ and
${\cal O}_{\rm D2}$) cannot be obtained as a result of mutual reduction
or extension. The corresponding entropies are related as
\be
S_0\le S_{\rm A} & \le \left\{
\begin{array}{c}
S_{\rm B}\\
S_{\rm C}
\end{array}\right\}
\le S_{\rm th};
\nonumber
\\
S_0\le S_{\rm A} & \le \left\{
\begin{array}{c}
S_{\rm D1}\\
S_{\rm D2}
\end{array}\right\}
\le S_{\rm th},
\label{4.84}
\\
S_0\le S_{\rm A} & \le
S_{\rm B}\le
S_{\rm D1}
\le S_{\rm th}.
\nonumber
\ee
For a particular quantum-mechanical state of light observation levels
${\cal O}_{\rm X}$
can be ordered with respect to increasing values of entropies
$S_{\rm X}$. From the above it also follows that if the entropy $S_{\rm X}$
on the observation level ${\cal O}_{\rm X}$ is equal to zero, then the
entropies on the extended observation levels are equal to zero as well.
It this case the complete reconstruction of the Wigner function of
a pure state  can be performed on the observation level which is based on a
measurement of a finite number of observables.

\section{RECONSTRUCTION OF WIGNER FUNCTIONS}
\label{sec5}

\subsection{Coherent states}
\label{sec5.A}
The Wigner function $W_{| \alpha\rangle}(\xi)$ of a coherent state
$| \alpha\rangle$  on the complete observation level
is given by Eq.(\ref{3.34}) [see Fig.\ref{fig1}a]. Coherent states
are uniquely characterized by their amplitude  and phase and therefore
phase-sensitive observation levels have to be considered for a proper
reconstruction of their Wigner functions. In Section~\ref{sec4.A.1} 
we have shown
that the Wigner function of coherent states can be {\em completely}
reconstructed on the observation level ${\cal O}_{1}$ (see Fig.\ref{fig1}a).
Nevertheless it is
interesting to understand how Wigner functions of coherent states can
be reconstructed on phase-insensitive observation levels.

\subsubsection{ Observation level ${\cal O}_{\rm A}$}
\label{sec5.A.1}
The coherent state $|\alpha \rangle$ has a Poissonian photon number
distribution and therefore we obtain for the generalized density
operator of the coherent state on ${\cal O}_{\rm A}$ the expression
\be
\hat{\sigma}_{\rm A}= \sum_{n=0}^{\infty} P_n | n\rangle\langle n |;\qquad
P_n={\rm e}^{-|\alpha|^2}\frac{|\alpha|^{2n}}{n!}.
\label{5.1}
\ee
This density operator describes a phase-diffused coherent state.
Eq.(\ref{5.1})
can be rewritten in the coherent-state basis
\be
\hat{\sigma}_{\rm A}= \frac{1}{2\pi}\int_{-\pi}^{\pi} d\phi\,| \alpha\rangle
\langle \alpha| ;\qquad \alpha=|\alpha|{\rm e}^{i\phi}.
\label{5.2}
\ee
From Eqs.(\ref{5.1}) and ({\ref{5.2}) 
it follows that on the observation level ${\cal O}_{\rm A}$
phase information is completely lost and
the corresponding Wigner function can be written as
\be
W^{\rm (A)}_{|\alpha\rangle}(\xi)=2\exp(-2|\xi|^2-|\alpha|^2)
\sum_{n=0}^{\infty}\frac{(-|\alpha|^2)^n}{n!}{\cal L}_{n}(4|\xi|^2),
\label{5.3}
\ee
or after some algebra we can find
\be
W^{\rm (A)}_{|\alpha\rangle}(\xi)=2\exp(-2|\xi|^2-2|\alpha|^2)
J_0(4i|\alpha|\, |\xi|),
\label{5.4}
\ee
where $J_0(4i|\alpha|\, |\xi|)$ is the Bessel function
\be
J_0(4i|\alpha|\, |\xi|)=\sum_{n=0}^{\infty}\frac{(4|\alpha|^2|\xi|^2)^n}
{(n!)^2},
\label{5.5}
\ee

\begin{figure}[t]
\begin{center}
\epsfig{file=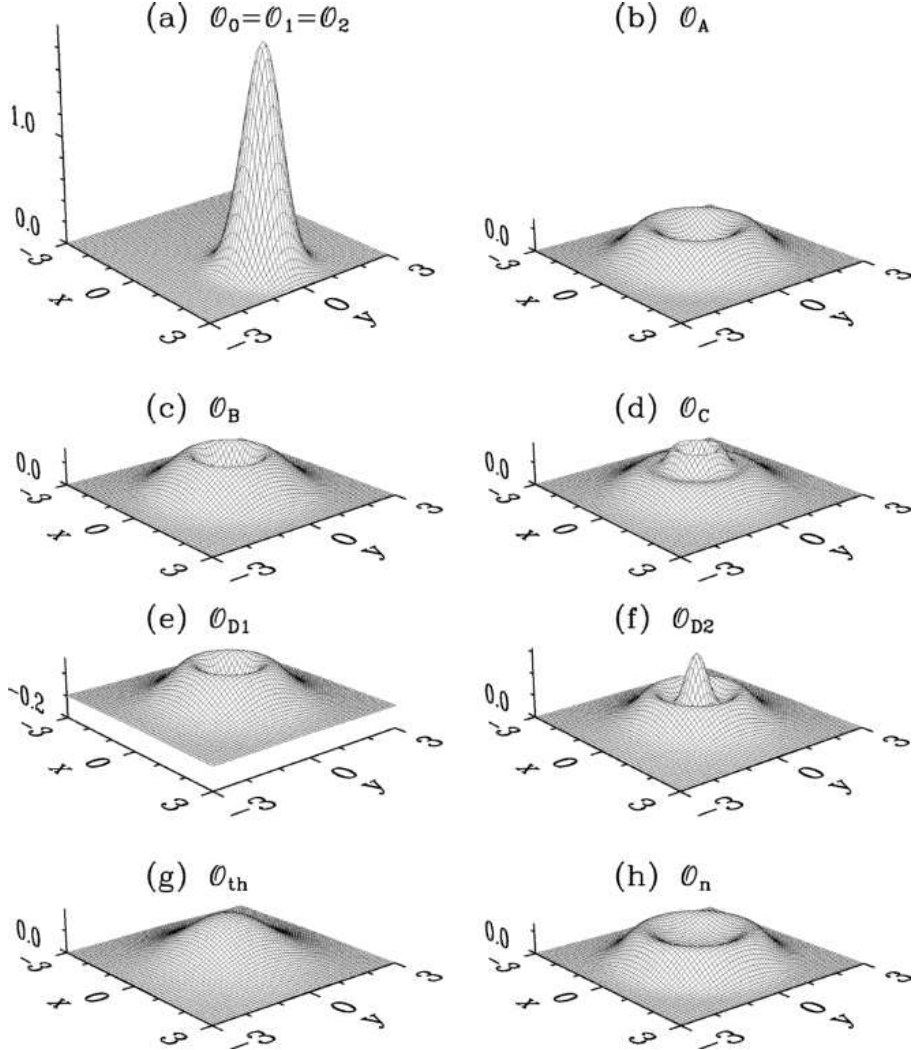, angle=0, width=12cm}
\end{center}
\caption{
The reconstructed Wigner functions of the coherent state
$|\alpha \rangle$ with $\bar{n}=2$. We consider the observation levels
as indicated in the figure. It is clearly seen that the Wigner
function of a coherent state can be easily reconstructed on the most
simple phase-sensitive observation level ${\cal O}_1$. We note that on some
observation levels the reconstructed Wigner function of the coherent
state may take negative values.
}
\label{fig1}
\end{figure}

\noindent
from which we see that the Wigner function (\ref{5.4})
 is positive. We plot
$W^{\rm (A)}_{|\alpha\rangle}(\xi)$
in Fig.~\ref{fig1}b.
We can understand the shape of	$W^{\rm (A)}_{|\alpha\rangle}(\xi)$ if
we imagine phase-averaging of the Wigner function $W_{|\alpha\rangle}(\xi)$
[see Fig.~\ref{fig1}a]. On the other hand we can represent
$W^{\rm (A)}_{|\alpha\rangle}(\xi)$ as a sum of weighted Wigner functions
of Fock states [see Eq.(\ref{5.3})]. 
For the considered coherent state $| \alpha\rangle$
with the mean photon number $\bar{n}=2$ we have $P_1=P_2=2P_0=2\exp(-2)$, so
the Wigner functions of Fock states $|1 \rangle$ and $|2 \rangle$ dominantly
contribute to $W^{\rm (A)}_{|\alpha\rangle}(\xi)$. On the other hand
contribution of the Wigner function of the vacuum state is suppressed and
therefore $W^{\rm (A)}_{|\alpha\rangle}(\xi)$ has a local minimum around
the origin of the phase space while its maximum is at the same distance
from the origin of the phase space as for the Wigner function on the
complete observation level [see Fig.~\ref{fig1}a]. 
We note that the Wigner function
$W^{\rm (A)}_{|\alpha\rangle}(\xi)$ describing the phase-diffused coherent state
has been experimentally reconstructed recently by Munroe et al. 
\cite{Munroe95}.

\subsubsection{ Observation level ${\cal O}_{\rm B}$}
\label{sec5.A.2}

Let us assume that from a measurement the mean photon number $\bar{n}$
and probabilities $P_{2n}$ are know (see Section~\ref{sec4.C.2}). 
If the values of
$P_{2n}$ are given by  Poissonian distribution (\ref{5.1}),
i.e., $P_{2n}=\exp(-\bar{n})\bar{n}^{2n}/(2n)!$,
then using definitions (\ref{4.51}) and (\ref{4.52})
we can find the parameters $P_{odd}$ and
$\bar{n}_{odd}$ to be
\be
P_{odd}={\rm e}^{-\bar{n}}\sinh\bar{n};\qquad
\bar{n}_{odd}=\bar{n}(1-P_{odd}),
\label{5.6}
\ee
The reconstructed probabilities $P_{2n+1}$ are given by Eq.(\ref{4.55})
and in the limit of large $\bar{n}$ (when $P_{odd}\to 1/2$ and
$\bar{n}_{odd}\to\bar{n}/2$) they read
\be
P_{2n+1}\rightarrow\frac{(\bar{n}-1)^{n}}{(\bar{n}+1)^{n+1}}.
\label{5.7}
\ee
With the help of the relation (\ref{4.48}) and explicit expressions for
$P_{2n}$ and $P_{2n+1}$ we can evaluate expression for the Wigner function
of the coherent state on the observation level ${\cal O}_{\rm B}$.
We plot $W^{\rm (B)}_{|\alpha\rangle}(\xi)$ of the coherent state with
the mean photon number equal to two  ($\bar{n}=2$) in Fig.~\ref{fig1}c.
In this case $P_2$ is dominant
 from which it follows that the
Fock state $|2 \rangle$ gives a significant contribution into
$W^{\rm (B)}_{|\alpha\rangle}(\xi)$ [compare with Fig.~\ref{fig1}b].

\subsubsection{ Observation level ${\cal O}_{\rm C}$}
\label{sec5.A.3}

The Wigner function $W^{(C)}_{|\alpha\rangle}(\xi)$
of the coherent state on the observation level ${\cal O}_{\rm C}$
can be reconstructed in exactly same way as on the level ${\cal O}_{\rm B}$.
In Fig.~\ref{fig1}d 
we present a result of this reconstruction. On the observation
level ${\cal O}_{\rm C}$ the contribution of the vacuum state is
more significant than in the case ${\cal O}_{\rm B}$ which is due to the
thermal-like photon number distribution $P_{2n}$ on the even-number
subspace of the Fock space [see Eq.(\ref{4.57})].

\subsubsection{ Observation level ${\cal O}_{\rm D1}$}
\label{sec5.A.4}

We can easily reconstruct the Wigner function of the coherent state
at the observation level ${\cal O}_{\rm D1}$. Using general expressions
from Section~\ref{sec4.C.4} 
we find the following expression for the Wigner function
$W^{\rm (D1)}_{|\alpha\rangle}(\xi)$
[we remind us that for coherent state
the parameter $P_0$ is given by the relation $P_0=\exp(-\bar{n})\,$]:
\be
W^{\rm (D1)}_{|\alpha\rangle}(\xi)=
\left(P_0-\frac{1-P_0}{\tilde{n}}\right)W_{| 0\rangle}(\xi)+
(1-P_0)\frac{\tilde{n}+1}{\tilde{n}} W_{\rm th}(\xi),
\label{5.8} 
\ee
where $W_{| 0\rangle}(\xi)$ is the Wigner function of the vacuum state
given by Eq.(\ref{3.34}) 
and $W_{\rm th}(\xi)$ is the Wigner function of a thermal
state (\ref{4.12}) 
with an effective number of photons equal to $\tilde{n}$, where
\be
\tilde{n}=\frac{\bar{n}}{1-P_0}-1.
\label{5.9}
\ee
In particular, from Eqs.(\ref{5.8}) and (\ref{5.9}) it follows that
\be
\lim_{\bar{n}\rightarrow 0} W^{\rm (D1)}_{|\alpha\rangle}(\xi) =
W_{| 0\rangle}(\xi),
\label{5.10}
\ee
and simultaneously $S_{\rm D1}=0$, which means that the vacuum state
can be completely reconstructed on the present observation level.
Another result which can be derived from Eq.(\ref{5.8}) is that if
$P_0(2\bar{n}+1)<1$, then the
reconstructed Wigner function $W^{\rm (D1)}_{|\alpha\rangle}(\xi)$
of the coherent state
$| \alpha\rangle$ can be negative due to the fact that the contribution
of the Fock state $| 1\rangle$ is more dominant than the contribution
of the vacuum state and then the negativity of the Wigner function
$W_{| 1\rangle}(\xi)$ results into negative values of
$W^{\rm (D1)}_{|\alpha\rangle}(\xi)$.
This means that even though the
Wigner function of the state itself (i.e., the Wigner function at the
complete observation level) is positive, the reconstructed Wigner function
can be negative. This is a clear indication that the observation level
has to be chosen very carefully and that reconstructed Wigner functions
can indicate nonclassical behavior even in those cases when the measured
state itself does not exhibit nonclassical effects. In Fig.~\ref{fig1}e 
we plot the
Wigner function $W^{\rm (D1)}_{|\alpha\rangle}(\xi)$
of the coherent state which illustrates  this effect.

\subsubsection{ Observation level ${\cal O}_{\rm D2}$}
\label{sec5.A.5}

If the mean photon number $\bar{n}$
is an integer, then one may consider the
observation level  ${\cal O}_{\rm D2}$. The Wigner function of the
coherent state at this observation level for which $P_{\bar{n}}=
\exp(-\bar{n})\bar{n}^{\bar{n}}/(\bar{n}!)$  reads
\be
W^{\rm (D2)}_{|\alpha\rangle}(\xi)=\left(1-\frac{1+\bar{n}}{Z_{\rm D2}}\right)
W_{| \bar{n}\rangle}(\xi)+
\frac{\bar{n}+1}{Z_{\rm D2}} W_{\rm th}(\xi),
\label{5.11}
\ee
where $W_{| \bar{n}\rangle}(\xi)$ is the Wigner function of the
Fock state $| \bar{n}\rangle$ and  $W_{\rm th}(\xi)$ is the Wigner function
of the thermal state with the mean photon number equal to $\bar{n}$.
The partition function $Z_{\rm D2}$ is given by the relation (\ref{4.75}).
The Wigner function (\ref{5.11}) 
is plotted in Fig.~\ref{fig1}f. From this figure we see
that the vacuum state $| 0\rangle$
(due to the thermal-like character of the reconstructed photon
number distribution) and the Fock state $|2 \rangle$  (as a consequence
of the measurement)
dominantly contribute to $W^{\rm (D1)}_{|\alpha\rangle}(\xi)$.

\subsection{Squeezed vacuum}
\label{sec5.B}
The Wigner function of the squeezed vacuum state (\ref{3.48}) on  the complete
observation level ${\cal O}_{0}$ is given by Eq.(\ref{3.52}) and is plotted
(in the complex $\xi$ phase space)
in Fig.~\ref{fig2}a.  This is a
Gaussian function, which carries phase information  associated with the phase of
squeezing. On the thermal observation level ${\cal O}_{\rm th}$
which is characterized only by the mean photon number $\bar{n}$ the
reconstructed Wigner function of the squeezed vacuum state is a rotationally
symmetrical
Gaussian function centered at the origin of the phase space [see
Eq.(\ref{4.12})
and Fig.~\ref{fig1}g]. 
On the observation level ${\cal O}_{\rm 1}$ the reconstructed
Wigner function is the same as on the thermal observation level
because the mean amplitudes $\langle \hat{a}\rangle$ and
$\langle \hat{a}^{\dagger}\rangle$ are equal to zero. On the other hand,
 the Wigner function of the squeezed
vacuum can be completely reconstructed on the observation level
${\cal O}_{\rm 2}$. To see this we evaluate the entropy $S_2$
for the squeezed vacuum state [see  Eq.(\ref{4.32})]. The parameters
$M$ and $N$ can be expressed in terms of the squeezing parameter
$\eta$ (we assume $\eta$ to be real) as
\be 
N=\frac{\eta^2}{1-\eta^2};\qquad M=\frac{\eta}{1-\eta^2},
\label{5.12}
\ee
so that $N(N+1)=M^2$. Consequently the parameter $\chi$ given by
Eq.(\ref{4.29}) 
is equal to zero from which it follows that $S_2$ for the squeezed
vacuum is equal to zero.

\subsubsection{ Observation level ${\cal O}_{\rm A}$}
\label{sec5.B.1}
The squeezed vacuum state (\ref{3.48}) is characterized by the oscillatory
photon number distribution $P_n$:
\be
P_{2n} = (1-\eta^2)^{1/2}\frac{(2n)!}{[2^n n!]^2}\eta^{2n}; \qquad
P_{2n+1}  =  0.
\label{5.13}
\ee
Using Eq.(\ref{4.48}) 
we can express the Wigner function $W_{|\eta\rangle}^{(A)}(\xi)$
of the squeezed vacuum on the observation level ${\cal O}_{\rm A}$ as
\be
W_{|\eta\rangle}^{(A)}(\xi)=2(1-\eta^2)^{1/2}{\rm e}^{-2|\xi|^2}
\sum_{n=0}^{\infty}\frac{(2n)!\eta^{2n}}
{2^{2n}(n!)^2}{\cal L}_{2n}(4|\xi|^2).
\label{5.14}
\ee
Taking into account that the Wigner function on the observation level
${\cal O}_{\rm A}$ can be obtained as the phase-averaged Wigner function on the
complete observation level, we can rewrite (\ref{5.14}) as
\be
W_{|\eta\rangle}^{(A)}(\xi)=\frac{1}{2\pi}\int_{-\pi}^{\pi}
W_{|\eta\rangle}(\xi) d\phi;\qquad \xi=|\xi|{\rm e}^{i\phi}.
\label{5.15}
\ee
If we insert the explicit expression for $W_{|\eta\rangle}(\xi)$ 
[see Eq.(\ref{3.52})]
into Eq.(\ref{5.15}) we obtain
\be
W_{|\eta\rangle}^{(A)}(\xi)=2\exp\left[-\left(\frac{|\xi|^2}{2\sigma_q^2}
+\frac{|\xi|^2}{2\sigma_p^2}\right)\right] I_0\left(\frac{|\xi|^2}{2\sigma_q^2}
-\frac{|\xi|^2}{2\sigma_p^2}\right),
\label{5.16}
\ee

\begin{figure}
\begin{center}
\epsfig{file=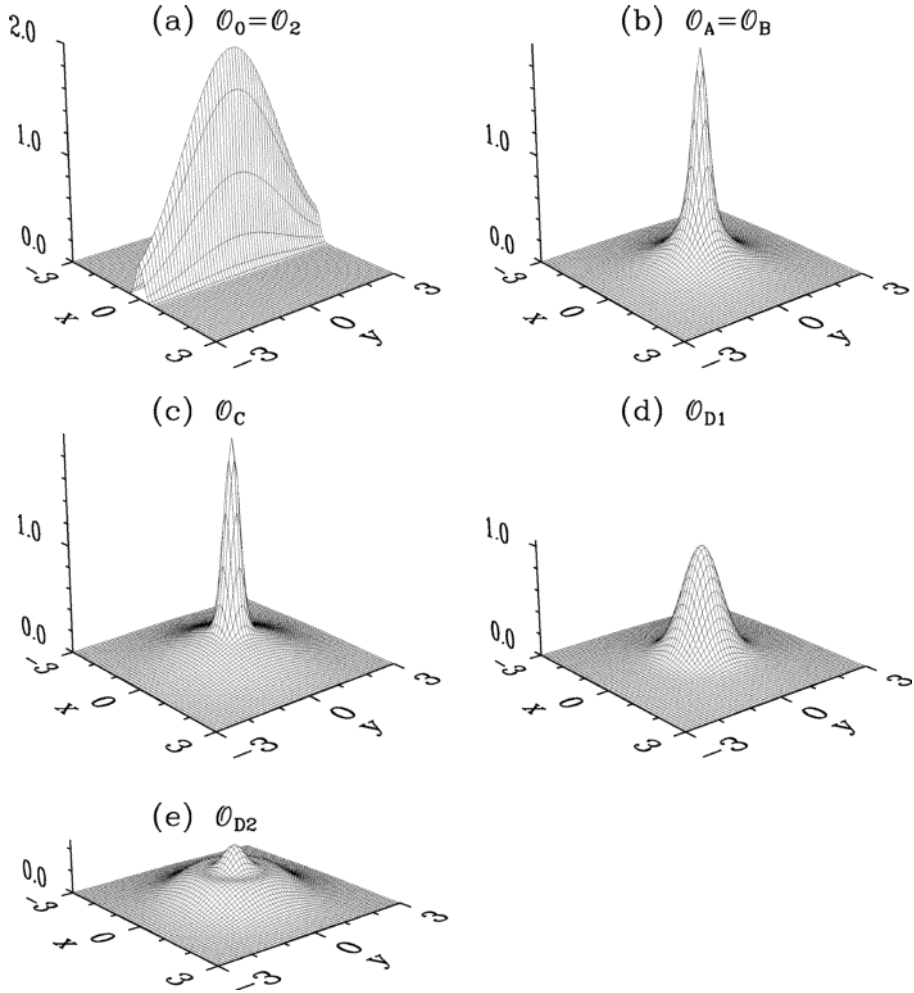, angle=0, width=12cm}
\end{center}
\caption{
The reconstructed Wigner functions of the squeezed vacuum state
$|\eta \rangle$ with $\bar{n}=2$. We consider the observation levels
as indicated in the figure. This Wigner function can be completely
reconstructed on the observation level ${\cal O}_2$.
}
\label{fig2}
\end{figure}
\noindent
where $I_0(x)$ is the modified Bessel function. We plot this Wigner
function in Fig.~\ref{fig2}b. We see that $W_{|\eta\rangle}^{(A)}(\xi)$
is not negative and that it is much narrower in the vicinity of the
origin of the phase space than the Wigner function of the vacuum state
(compare with Fig.~\ref{fig1}a). Nevertheless the total width of
Wigner function $W^{\rm (A)}_{|\eta\rangle}(\xi)$ is much larger
than the width of the Wigner function of the vacuum state.

\subsubsection{ Observation level ${\cal O}_{\rm B}$}
\label{sec5.B.2}
Due to the fact that for the squeezed vacuum state we have $\sum_n P_{2n}=1$,
the Wigner function of the squeezed vacuum state on the observation level
${\cal O}_{\rm B}$ is equal to the Wigner function on the observation
level ${\cal O}_{\rm A}$, i.e., $W_{|\eta\rangle}^{(B)}(\xi)
=W_{|\eta\rangle}^{(A)}(\xi)$.

\subsubsection{ Observation level ${\cal O}_{\rm C}$}
\label{sec5.B.3}
For the squeezed vacuum state all meanvalues $P_{2n+1}$ are equal to zero
and therefore $\sum_n P_{2n+1}=0$. From this fact and from the knowledge
of the mean photon number $\bar{n}$ we can reconstruct the Wigner function
$W_{|\eta\rangle}^{(C)}(\xi)$ in the form [see Section~\ref{sec4.B.3}]
\be
W_{|\eta\rangle}^{(C)}(\xi)=\frac{4{\rm e}^{-2|\xi|^2}}{\bar{n}+2}
\sum_{k=0}^{\infty}\left(\frac{\bar{n}}{\bar{n}+2}\right)^k{\cal L}_{2k}
(4|\xi|^2),
\label{5.17}
\ee
where $\bar{n}$ is the mean photon number in the squeezed vacuum state.
We plot the Wigner function $W_{|\eta\rangle}^{(C)}(\xi)$ in Fig.~\ref{fig2}c.
This Wigner function is very similar to the Wigner function on the
observation level ${\cal O}_{\rm A}$ [see Fig.~\ref{fig2}b] which reflects the
fact that the photon number distribution of the squeezed vacuum state
has a thermal-like character on the even-number subspace of the
Fock space.

\subsubsection{ Observation level ${\cal O}_{\rm D1}$}
\label{sec5.B.4}
With the help of the general formalism presented in Section~\ref{sec4}
 we can express
the Wigner function $W_{|\eta\rangle}^{(D1)}(\xi)$ of the squeezed vacuum state
on the observation level ${\cal O}_{\rm D1}$ in the form [see Eq.(\ref{5.8})]
with
\be
P_0=(1-\eta^2)^{1/2}=(\bar{n}+1)^{-1/2};\qquad \mbox{and}~~~
\tilde{n}=\frac{\bar{n}}{1-(1+\bar{n})^{-1/2}}-1.
\label{5.18}
\ee
We plot the Wigner function $W_{|\eta\rangle}^{(D1)}(\xi)$ in
Fig.~\ref{fig2}d
from which the dominant contribution of the vacuum state is transparent
which is due to the fact that the squeezed vacuum state has a thermal-like
photon number distribution.

\subsubsection{ Observation level ${\cal O}_{\rm D2}$}
\label{sec5.B.5}
If we consider $\bar{n}$ to be an {\em even} integer, then the Wigner
function $W_{|\eta\rangle}^{(D2)}(\xi)$ of the squeezed vacuum state on
${\cal O}_{\rm D2}$ is given by Eq.(\ref{5.11}).
The partition function $Z_{\rm D2}$
is given by Eq.(\ref{4.75}) where
\be
P_{\bar{n}}=\frac{\bar{n}!}{2^{\bar{n}}\left[(
\bar{n}/2)!\right]^2}
\frac{\bar{n}^{\bar{n}/2}}{(1+\bar{n})^{(1+\bar{n})/2}}.
\label{5.19}
\ee
We plot this Wigner
function in Fig.~\ref{fig2}e. It has a thermal-like character
[compare with Fig.~\ref{fig1}g] but contribution of the Fock
state $| \bar n=2\rangle$ is more dominant compared with the proper thermal
distribution.
If $\bar{n}$ is an {\em odd} integer,
then $P_{\bar{n}}=0$  and the corresponding Wigner function can be again
reconstructed with the help of Eqs.(\ref{5.11}) and (\ref{4.75}).

\subsection{Even coherent state}
\label{sec5.C}
We plot the Wigner function of the even coherent state on the complete
observation level in Fig.~\ref{fig3}a. 
Two contributions of coherent component state
$|\alpha \rangle$ and $|-\alpha \rangle$ as well as the interference peak
around the origin of the phase space are transparent in this figure.
As in the case of the squeezed vacuum state, the mean amplitude $\langle
\hat{a}\rangle$ of the even coherent state is equal to zero and therefore the
Wigner function $W_{|\alpha_e\rangle}^{(1)}(\xi)$ of the even coherent state
on the observation level ${\cal O}_1$ is equal to the thermal Wigner function
given by Eq.(\ref{4.12}).

\subsubsection{ Observation level ${\cal O}_{\rm 2}$}
\label{sec5.C.1}

Using general expressions from Section~\ref{sec4.A.2} 
 we can express the Wigner function
$W_{|\alpha_e\rangle}^{(2)}(\xi)$ of the even coherent state on the observation
level ${\cal O}_2$ as
\be
W_{|\alpha_e\rangle}^{(2)}(\xi)=\frac{1}{\left[(N+1/2)^2-M^2\right]^{1/2}}
\exp\left[-\frac{\xi_x^2}{[(N+1/2)+M]} -\frac{\xi_y^2}{[(N+1/2)-M]}\right],
\label{5.20}
\ee
where $\xi=\xi_x +i\xi_y$, and
the parameters $N$ and $M$ read
\be
N=\alpha^2 \tanh\alpha^2;\qquad M=\alpha^2.
\label{5.21}
\ee
We plot the Wigner function $W_{|\alpha_e\rangle}^{(2)}(\xi)$
in Fig.~\ref{fig3}b.  This Wigner function is slightly ``squeezed'' in the
$\xi_y$-direction and stretched in the $\xi_x$-direction. Nevertheless,
the reconstructed Wigner function is different from the Wigner
function of the squeezed vacuum state [compare with Fig.~\ref{fig2}a].

\begin{figure}
\begin{center}
\epsfig{file=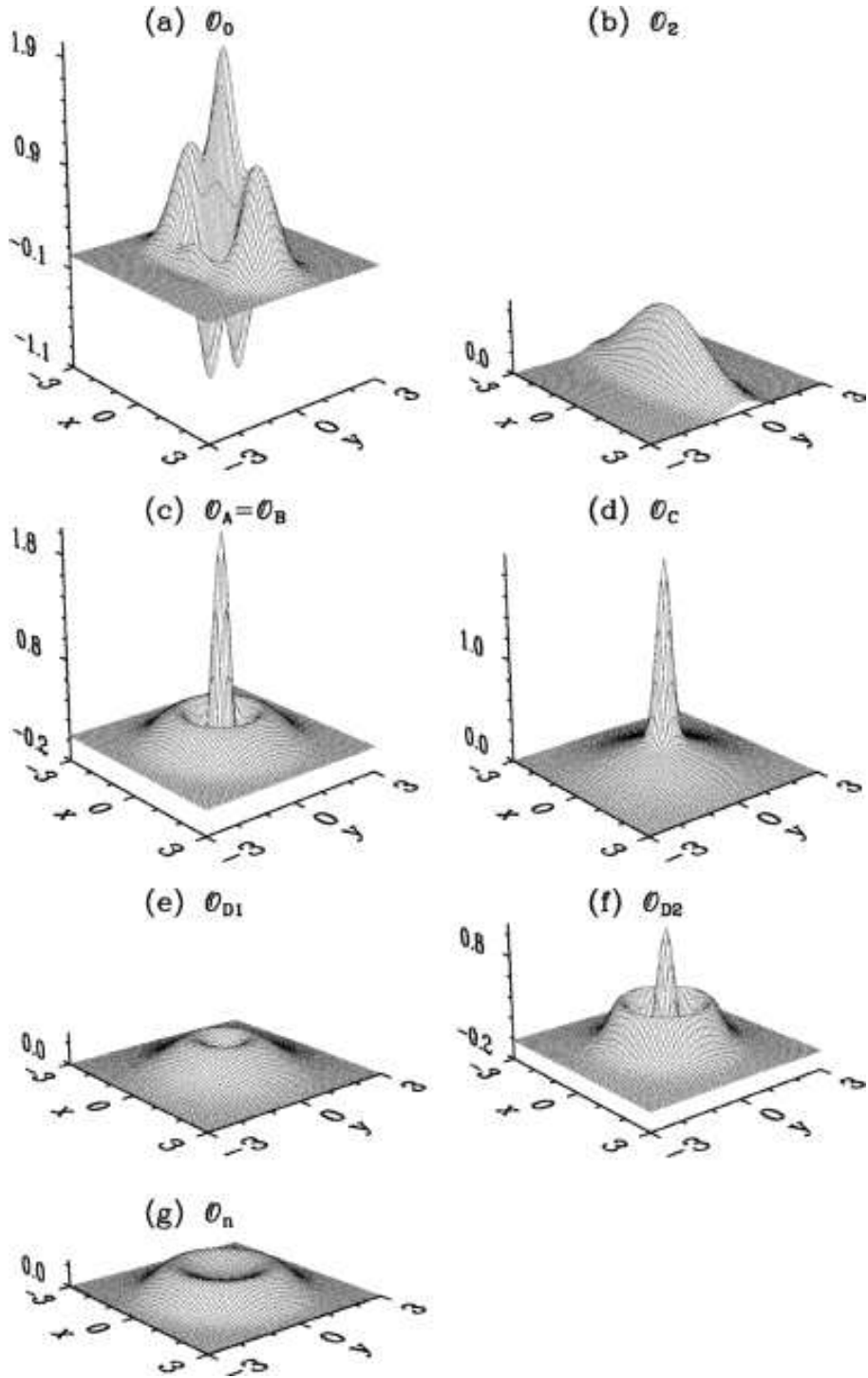, angle=0, width=12cm}
\end{center}
\caption{
The reconstructed Wigner functions of the even coherent state
$|\alpha_e \rangle$ with $\bar{n}=2$. We consider the observation levels
as indicated in the figure. This non-Gaussian Wigner function can be
completely reconstructed only on the complete observation level when
all moments of systems operators are measured.
}
\label{fig3}
\end{figure}

\subsubsection{ Observation level ${\cal O}_{\rm A}$}
\label{sec5.C.2}

The photon number distribution of the even coherent state is given by the
relation (we assume $\alpha$ to be real):
\be
P_{2n} = \frac{1}{\cosh \alpha^2}\frac{\alpha^{4n}}{(2n)!}; \qquad
P_{2n+1}  =  0,
\label{5.22}
\ee
so the corresponding Wigner function can be expressed as Eq.(\ref{4.48}).
We can also express  $W_{|\alpha_e\rangle}^{(A)}(\xi)$ as the phase
averaged Wigner function of the even coherent state
$W_{|\alpha_e\rangle}(\xi)$ given by Eq.(\ref{3.56}).
 After some algebra we find that
$W_{|\alpha_e\rangle}^{(A)}(\xi)$ can be written in a closed form
\be
W_{|\alpha_e\rangle}^{(A)}(\xi)=
\frac{{\rm e}^{-2|\xi|^2}}{\cosh \alpha^2}\left[
{\rm e}^{-\alpha^2} J_0(4i\alpha|\xi|)
+{\rm e}^{\alpha^2} J_0(4\alpha|\xi|)\right].
\label{5.23}
\ee
We plot the Wigner function $W_{|\alpha_e\rangle}^{(A)}(\xi)$ in
Fig.~\ref{fig3}c.
From this figure the dominant contribution of
the Fock state $| 2\rangle$
is transparent (in the present case we have $P_0\simeq 2\exp(-2)$,
$P_2=2P_0$, and $P_4=2P_0/3$,  while all other probabilities $P_n$ are much
smaller) which results in negative Wigner function.

\subsubsection{ Observation level ${\cal O}_{\rm B}$}
\label{sec5.C.3}

Due to the fact that the even coherent state is expressed as a superposition
of only even Fock states, i.e., $\sum_n P_{2n}=1$,
the Wigner functions on the observation levels	${\cal O}_{\rm A}$
and ${\cal O}_{\rm B}$ are equal, i.e.,
$W_{|\alpha_e\rangle}^{(B)}(\xi)=W_{|\alpha_e\rangle}^{(A)}(\xi)$.

\subsubsection{ Observation level ${\cal O}_{\rm C}$}
\label{sec5.C.4}
As a consequence of the fact that
for the even coherent state all meanvalues $P_{2n+1}$ are equal to zero
the information available for the reconstruction of the Wigner function
$W_{|\alpha_e\rangle}^{(C)}(\xi)$ is the same as in the case of the
reconstruction of the Wigner function of the squeezed vacuum state
on the observation level ${\cal O}_{\rm C}$. Therefore,
the Wigner function
$W_{|\alpha_e\rangle}^{(C)}(\xi)$ has exactly the same form as for the
squeezed vacuum state with the same mean photon number $\bar{n}$
[see Fig.~\ref{fig3}d and Fig.~\ref{fig2}c].

\subsubsection{ Observation level ${\cal O}_{\rm D1}$}
\label{sec5.C.5}
The Wigner function $W_{|\alpha_e\rangle}^{(D1)}(\xi)$ of the even coherent state
on the observation level ${\cal O}_{\rm D1}$ is given by Eq.(\ref{5.8})
with
\be
P_0=\frac{1}{\cosh \alpha^2};\qquad
\tilde{n}=\frac{\alpha^2\sinh\alpha^2}{\cosh\alpha^2 -1}-1.
\label{5.24}
\ee
We plot the Wigner function $W_{|\alpha_e\rangle}^{(D1)}(\xi)$ in
Fig.~\ref{fig3}e.
This Wigner function
has a thermal-like character except the fact that the contribution of the
vacuum state is slightly suppressed.

\subsubsection{ Observation level ${\cal O}_{\rm D2}$}
\label{sec5.C.6}

Analogously we can find the Wigner function $W_{|\alpha_e\rangle}^{(D2)}(\xi)$.
If we consider $\bar{n}$ to be an {\em even} integer, then the Wigner
function $W_{|\eta\rangle}^{(D2)}(\xi)$ of the even coherent state on
${\cal O}_{\rm D2}$ is given by Eq.(\ref{5.11}) and Eq.(\ref{4.75})  where
\be 
P_{\bar{n}}=\frac{1}{\cosh \alpha^2}
\frac{\alpha^{2\bar{n}}}{\bar{n}!},
\label{5.25}
\ee
and if $\bar{n}$ is an odd integer then $P_{\bar{n}}=0$. We plot
$W_{|\alpha_e\rangle}^{(D2)}(\xi)$  in Fig.~\ref{fig3}f. From our previous
discussion it is clear that in the present case
the vacuum state and the Fock state
$|2 \rangle$ dominantly contribute to  $W_{|\alpha_e\rangle}^{(D2)}(\xi)$
(similarly as on the observation level ${\cal O}_{\rm A}$ - see
Fig.~\ref{fig3}c).

\subsection{Odd coherent state}
\label{sec5.D}
We present the Wigner function of the odd coherent state with the
mean photon number equal to two in Fig.~\ref{fig4}a.
The mean amplitude $\langle
\hat{a}\rangle$ of the odd coherent state is equal to zero and therefore the
Wigner function $W_{|\alpha_o\rangle}^{(1)}(\xi)$ of this state
on the observation level ${\cal O}_1$ is equal to the thermal Wigner function
given by Eq.(\ref{4.12}) [see Fig.~\ref{fig1}g].

\subsubsection{ Observation level ${\cal O}_{\rm 2}$}
\label{sec5.D.1}
Using general expressions from Section~\ref{sec4.B.2}
 we find that the Wigner function
$W_{|\alpha_o\rangle}^{(2)}(\xi)$ of the odd coherent state on the observation
level ${\cal O}_2$ is the same as for the even coherent state [see
Eq.(\ref{5.20})]
but the parameters $N$ and $M$ in the present case read
\be
N=\alpha^2 \coth\alpha^2;\qquad M=\alpha^2.
\label{5.26}
\ee
We plot the Wigner function $W_{|\alpha_o\rangle}^{(2)}(\xi)$  in
Fig.~\ref{fig4}b.
This is a ``squeezed''-Gaussian function similar to the Wigner
function of the even coherent state on the same observation level
[see Fig.~\ref{fig3}b and discussion in the previous section].

\subsubsection{ Observation level ${\cal O}_{\rm A}$}
\label{sec5.D.2}
The photon number distribution of the odd coherent state is given by the
relation (we assume $\alpha$ to be real):
\be
P_{2n}	=  0;\qquad
P_{2n+1} = \frac{1}{\sinh \alpha^2}\frac{(\alpha^2)^{2n+1}}{(2n+1)!}.
\label{5.27}
\ee
Consequently, the Wigner function $W_{|\alpha_o\rangle}^{(A)}(\xi)$ can be
expressed as (\ref{4.48}).
Alternatively, if we use the fact that
 $W_{|\alpha_o\rangle}^{(A)}(\xi)$ is equal to the phase
averaged Wigner function of the odd coherent state
$W_{|\alpha_o\rangle}(\xi)$ given by Eq.(\ref{3.57}), then we can write
\be
W_{|\alpha_o\rangle}^{(A)}(\xi)=
\frac{{\rm e}^{-2|\xi|^2}}{\sinh \alpha^2}\left[
{\rm e}^{-\alpha^2} J_0(4i\alpha|\xi|)
-{\rm e}^{\alpha^2} J_0(4\alpha|\xi|)\right].
\label{5.28}
\ee
This function is always negative in the origin of the phase space.
We plot the Wigner function $W_{|\alpha_o\rangle}^{(A)}(\xi)$ in
Fig.~\ref{fig4}c.
In the present case $P_0=P_2=0$ and the $P_1$ is the largest probability
therefore the contribution of the Fock state $| 1\rangle$ in
$W_{|\alpha_o\rangle}^{(A)}(\xi)$ is the most dominant which is clearly seen
from Fig.~\ref{fig4}c. 
We also note that, in general, any superposition of odd Fock states
has a negative Wigner function on the observation level ${\cal O}_{\rm A}$.

\subsubsection{ Observation level ${\cal O}_{\rm B}$}
\label{sec5.D.3}

For the odd coherent state all meanvalues $P_{2n}$ are equal to zero.
Taking into account this information and the information about the
mean photon number we reconstruct the Wigner function
$W_{|\alpha_o\rangle}^{(B)}(\xi)$ in the form (for details see Section
~\ref{sec4.C.2})
\be
W_{|\alpha_o\rangle}^{(B)}(\xi) =-\frac{4{\rm e}^{-2|\xi|^2}}{\bar{n}+1}
\sum_{k=0}^{\infty}\left(\frac{\bar{n}-1}{\bar{n}+1}\right)^k
{\cal L}_{2k+1}(4|\xi|^2),
\label{5.29}
\ee
where $\bar{n}=\alpha^2\coth\alpha^2$. We plot this Wigner function in
Fig.~\ref{fig4}d.
In the present case the dominant contribution of the Fock state $|1 \rangle$
is seen ($P_0=P_2=0$ and due to the thermal-like photon number distribution
on the odd-number subspace of the Fock state
$P_3$ is much smaller than $P_1$). We can conclude, that any superposition of
odd Fock states on the observation level ${\cal O}_{\rm B}$ has the
Wigner function given by Eq.(\ref{5.29}), 
i.e., superpositions of odd Fock states
are indistinguishable on ${\cal O}_{\rm B}$.

\subsubsection{ Observation level ${\cal O}_{\rm C}$}
\label{sec5.D.4}

Due to the fact that the odd coherent state is expressed as a superposition
of only odd Fock states, i.e., $\sum_n P_{2n+1}=1$,
the Wigner functions on the observation levels	${\cal O}_{\rm C}$
and ${\cal O}_{\rm A}$ are equal, i.e.,
$W_{|\alpha_o\rangle}^{(C)}(\xi)=W_{|\alpha_o\rangle}^{(A)}(\xi)$.

\begin{figure}
\begin{center}
\epsfig{file=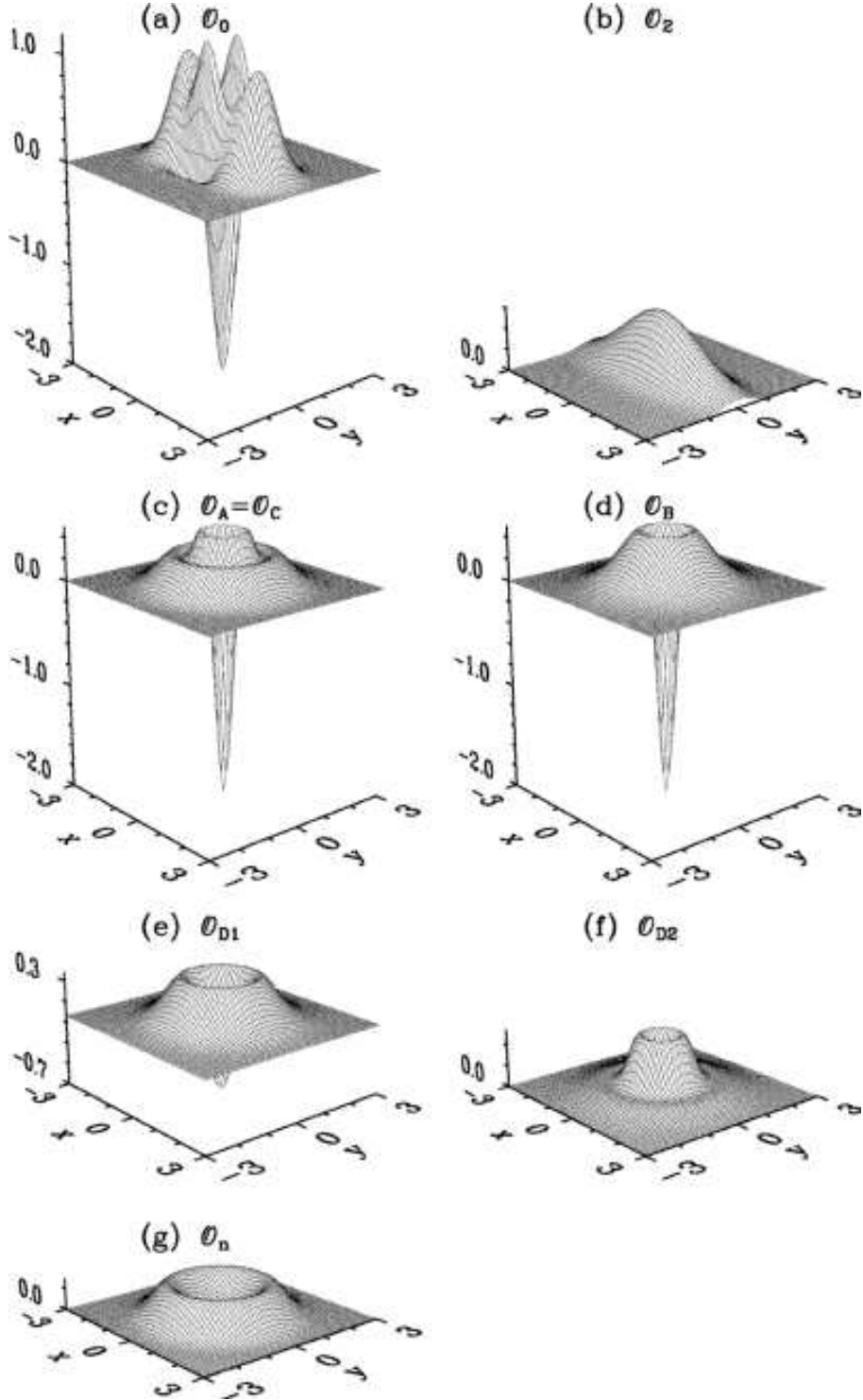, angle=0, width=12cm}
\end{center}
\caption{
The reconstructed Wigner functions of the odd coherent state
$|\alpha_o \rangle$ with $\bar{n}=2$. We consider the observation levels
as indicated in the figure. 
}
\label{fig4}
\end{figure}

\subsubsection{ Observation level ${\cal O}_{\rm D1}$}
\label{sec5.D.5}

The Wigner function $W_{|\alpha_o\rangle}^{(D1)}(\xi)$ of the odd coherent state
on the observation level ${\cal O}_{\rm D1}$ is given by the following relation
[we remind us that for the odd coherent state we have $P_0=0$]
\be
W_{|\alpha_o\rangle}^{(D1)}(\xi)=-\frac{1}{\bar{n}-1}W_{|0\rangle}(\xi)
+\frac{\bar{n}}{\bar{n}-1}W_{\rm th}(\xi),
\label{5.30}
\ee
where $\bar{n}$ is the mean photon number in the odd coherent state;
$W_{|0\rangle}(\xi)$ is the Wigner function of the vacuum state and
$W_{\rm th}(\xi)$ is the thermal Wigner function for the state with
$\bar{n}-1$ photons.
We note that from Eq.(\ref{5.30}) it follows that
\be
\lim_{\bar{n}\rightarrow 1}
W_{|\alpha_o\rangle}^{(D1)}(\xi)=W_{|1\rangle}(\xi),
\label{5.31}
\ee
We plot the Wigner function $W_{|\alpha_o\rangle}^{(D1)}(\xi)$ in
Fig.~\ref{fig4}e.
Compared with Fig.~\ref{fig4}d 
we see that the contribution of the Fock state
$|1 \rangle$ on the observation level ${\cal O}_{\rm D1}$ is smaller
than on ${\cal O}_{\rm B}$. This is due to the fact that on the present
observation level $P_2$ is not equal to zero.

\subsubsection{ Observation level ${\cal O}_{\rm D2}$}
\label{sec5.D.6}
Reconstruction of the Wigner function $W_{|\alpha_o\rangle}^{(D2)}(\xi)$
is straightforward. For the odd coherent state it is valid that
if $\bar{n}$ is an odd integer, then
\be
P_{\bar{n}}=\frac{1}{\sinh \alpha^2}
\frac{\alpha^{2\bar{n}}}{\bar{n}!},
\label{5.32}
\ee
and the Wigner function is given by Eq.(\ref{5.11}). On the other hand if
$\bar{n}$ is an {\em even} integer, then $P_{\bar{n}}=0$ and  we again
use Eq.(\ref{5.11}) for the reconstruction of the Wigner function
$W_{|\eta\rangle}^{(D2)}(\xi)$. We plot this Wigner function in
Fig.~\ref{fig4}f.
Even though on this observation level $P_2=0$ the contribution from the
vacuum state is significant and therefore
$W_{|\alpha_o\rangle}^{(D2)}(\xi)$ is not negative in the present case.

\subsection{Fock state}
\label{sec5.E}
Mean values of the operators $\hat{a}^k$ in the Fock state are equal to zero,
therefore the Wigner functions $W_{|n\rangle}^{(1)}(\xi)$ and
$W_{|n\rangle}^{(2)}(\xi)$ of the Fock state $| n\rangle$ on the observation
levels ${\cal O}_1$ and ${\cal O}_2$, respectively, are equal to the
thermal Wigner function given by Eq.(\ref{4.12}) [see Fig.~\ref{fig5}b].
On the other hand the Shannon entropy
of the Fock state is equal to zero, therefore this state can be completely
reconstructed on the observation level ${\cal O}_{\rm A}$ [see
Fig.~\ref{fig5}a for
the Wigner function of the Fock state $|2 \rangle$].

\subsubsection{ Observation level ${\cal O}_{\rm B}$}
\label{sec5.E.1}
If the Fock state
has an even number of photons then it can also be completely reconstructed on
the observation level ${\cal O}_{\rm B}$. But if the number of photons
of the Fock state is odd then the Wigner function of this Fock state on
 ${\cal O}_{\rm B}$ is given by the relation (\ref{5.29}) with $\bar{n}=n$.

\subsubsection{ Observation level ${\cal O}_{\rm C}$}
\label{sec5.E.2}
If the number of photons on the Fock state is odd than the corresponding Wigner
function can be completely reconstructed on the observation level
${\cal O}_{\rm C}$. If the number of photons is even, then the Wigner function
$W_{|n\rangle}^{(C)}(\xi)$ is given by Eq.(\ref{5.17}) with $\bar{n}=n$.
We plot $W_{|n\rangle}^{(C)}(\xi)$ in Fig.~\ref{fig5}c. This Wigner
function  is the same as for the squeezed vacuum state
$W_{|\eta\rangle}^{(C)}(\xi)$ and the even
coherent state $W_{|\alpha_e\rangle}^{(C)}(\xi)$
with the same mean photon number [see Figs.~\ref{fig2}c and \ref{fig3}d].
More generally, all superpositions of even Fock states with the same
mean photon number are indistinguishable on ${\cal O}_{\rm C}$.

\subsubsection{ Observation level ${\cal O}_{\rm D1}$}
\label{sec5.E.3}
If the Fock state under consideration is the vacuum state then
it can be completely reconstructed on the observation level
${\cal O}_{\rm D1}$. If the number of photons is larger than zero,
then $P_0=0$ and the corresponding Wigner function is given by
Eq.(\ref{5.30})
with $\bar{n}=n$.
We plot $W_{|n=2\rangle}^{(D1)}(\xi)$ in Fig.~\ref{fig5}d.

\subsubsection{ Observation level ${\cal O}_{\rm D2}$}
\label{sec5.E.4}
On this observation level  the Wigner function of the
Fock state $| n\rangle$ can be always completely reconstructed, because
this observation level is defined in such way that $P_n=1$.

\begin{figure}
\begin{center}
\epsfig{file=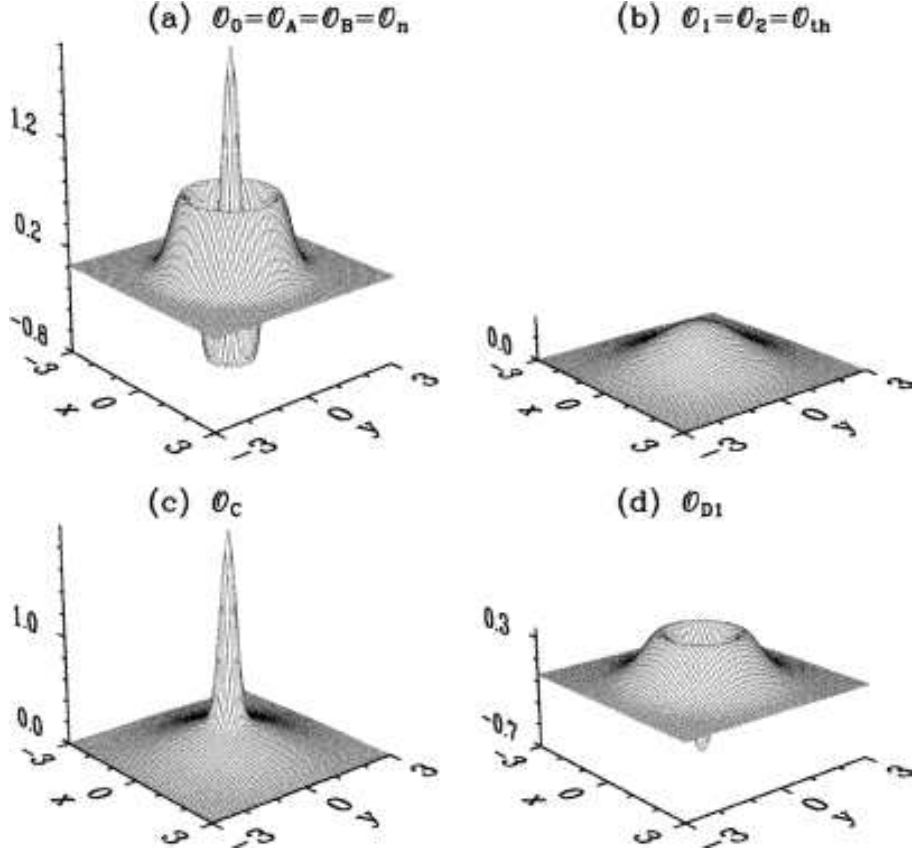, angle=0, width=12cm}
\end{center}
\caption{
The reconstructed Wigner functions of the Fock state
$|n=2 \rangle$. We consider the observation levels
as indicated in the figure. Even though the Fock state is represented
by the non-Gaussian Wigner function it can be completely reconstructed
on the observation level ${\cal O}{\rm n}$.
}
\label{fig5}
\end{figure}

\subsection{Observation level ${\cal O}_{\rm n}\equiv \{\hat{n},\hat{n}^2\}$}
\label{sec5.F}
We will finish this section with a brief discussion about 
the phase-insensitive observation level
${\cal O}_{\rm n}$ which is related to a measurement of the observables
$\hat{n}$ and $\hat{n}^2$.

The generalized canonical density operator $\hat{\sigma}_{\rm}$ on the
observation level ${\cal O}_{\rm n}$ reads:
\begin{eqnarray}
\hat{\sigma}_{\rm n}=\frac{1}{Z_{\rm n}}\exp\left[-\lambda_1\hat{n}
-\lambda_2\hat{n}^2\right]= \frac{1}{Z_{\rm n}}\sum_{n=0}^{\infty}
\exp\left[-\lambda_1 n	-\lambda_2n^2\right] | n\rangle\langle n|.
\label{5.33}
\end{eqnarray}
The Lagrange multipliers are determined by the relations
\begin{eqnarray}
\langle\hat{n}\rangle = -\frac{\partial \ln Z_{\rm n}}{\partial \lambda_1}=
\sum_{m=0}^{\infty} m P_m;
\label{5.34}
\end{eqnarray}
\begin{eqnarray}
\langle\hat{n}^2\rangle = -\frac{\partial \ln Z_{\rm n}}{\partial \lambda_2}=
\sum_{m=0}^{\infty} m^2 P_m,
\label{5.35}
\end{eqnarray}
where
\begin{eqnarray}
P_m= \frac{1}{Z_{\rm n}} \exp\left[-\lambda_1 m  -\lambda_2m^2\right];
\label{5.36}
\end{eqnarray}
and
\begin{eqnarray}
Z_{\rm n}=\sum_{m=0}^{\infty} \exp\left[-\lambda_1 m  -\lambda_2m^2\right].
\label{5.37}
\end{eqnarray}
From Eqs.(\ref{5.34}) and (\ref{5.35})
 it follows that if $\langle \hat{n}\rangle= N$ is an integer,
then in the limit $\sigma_n\rightarrow 0_+$
(where $\sigma_n^2\equiv \langle \hat{n}^2\rangle-\langle \hat{n}\rangle^2$)
$\lambda_1=-2N\lambda_2$ and $\lambda_2$ tends to infinity.
Simultaneously
\begin{eqnarray}
P_m\rightarrow \delta_{m,N},
\label{5.38}
\end{eqnarray}
which means that in this case $\hat{\sigma}_{\rm n}\rightarrow | N\rangle
\langle N |$.  In other words, on the observation level  ${\cal O}_{\rm n}$
the Fock state $| N\rangle$ can be completely reconstructed
(see Fig.\ref{fig5}a) and 
in this case
the corresponding entropy $S_{\rm n}=-k_B\sum_m P_m\ln P_m$ is equal to zero.

The Wigner function of this state
is negative, which in particular reflects the fact that the reconstructed
distribution is narrower than the Poissonian (coherent-state) photon number
distribution, i.e., the state under consideration exhibits sub-Poissonian
photon number distribution. To quantify the degree of the sub-Poissonian
photon statistics one can utilize the Mandel $Q$ parameter 
defined as:
\begin{eqnarray}
Q=\frac{\langle\hat{n}^2\rangle-\langle\hat{n}\rangle^2-\langle\hat{n}\rangle}
{\langle\hat{n}\rangle},
\label{5.39}
\end{eqnarray}
which for  Fock states is equal to -1 while for coherent states is equal
to 0. The state is said to have sub-Poissonian photon statistics providing
$Q<0$. One can easily 
reconstruct  sub-Poissonian states on the
observation level ${\cal O}_{\rm n}$. In addition states with the Poissonian
photon statics $Q=0$ can be partially reconstructed on this observation level
as well. For instance in Fig.\ref{fig1}h
we represent a result of numerical
reconstruction of the Wigner function $W^{(\rm n)}_{|\alpha\rangle}(\xi)$
of the coherent state with a Poissonian photon number distribution
on the observation level ${\cal O}_{\rm n}$.
In this case the reconstructed photon number distribution
 $P_n$ [see Eq.(\ref{5.36})] does not have a Poissonian character,
and therefore the reconstructed Wigner functions of the coherent state
on the observation levels ${\cal O}_{\rm A}$ and ${\cal O}_{\rm n}$ are
different (compare Figs.\ref{fig1}b and \ref{fig1}h, respectively)
 even though the reconstructed states have the same mean photon
number $\langle \hat{n} \rangle$ and the same variance $\sigma_n^2$
in the photon number distribution.

On the observation level ${\cal O}_{\rm n}$ we can reconstruct also the
odd coherent state given by Eq.(\ref{3.55})
which is a sub-Poissonian state with the $Q$ parameter
given by the relation (we assume $\alpha$ to be real):
\begin{eqnarray}
Q=-\frac{4\alpha^2{\rm e}^{-2\alpha^2}}{1-{\rm e}^{-4\alpha^2}}
=-\frac{\bar{n}}{(\cosh\alpha^2)^2} <0 ,
\label{5.40}
\end{eqnarray}
where the mean photon number $\bar{n}$ in the odd coherent state
is given by the relation $\bar{n}=\alpha^2\coth\alpha^2$. We have plotted
the result of the numerical reconstruction of the Wigner function
of the odd coherent state with $\bar{n}=2$
on the given observation level in Fig.\ref{fig4}g.
Due to the fact, that for the given mean photon number
the odd coherent state
does not exhibit a significant degree  of sub-Poissonian photon statistics,
the corresponding Wigner function $W^{(\rm n)}_{|\alpha_o\rangle}(\xi)$
is not negative (compare with Fig.\ref{fig4}c).

The even coherent state (\ref{3.54}) is characterized by the super-Poissonian
photon statistics with the Mandel $Q$ parameter given by the relation
\begin{eqnarray}
Q=\frac{4\alpha^2{\rm e}^{-2\alpha^2}}{1-{\rm e}^{-4\alpha^2}}
=\frac{\bar{n}}{(\sinh\alpha^2)^2} >0 ,
\label{5.41}
\end{eqnarray}
with the mean photon number given by the relation
$\bar{n}=\alpha^2\tanh\alpha^2$. From Eq.(\ref{5.41}) it follows that
for large enough values of $\alpha$ 
(i.e., for large enough values of $\bar{n}$)
the Mandel $Q$ parameter is smaller than $\bar{n}$ (it tends to zero).
In this case the Wigner
function of the even coherent state on the observation level ${\cal O}_{\rm n}$
can be easily reconstructed (see Fig.\ref{fig3}g). 
We can also reconstruct on this
observation level a thermal mixture for which the Mandel $Q$ parameter is
equal to $\bar{n}$ (i.e., $\langle \hat{n}^2\rangle = 2\bar{n}^2+\bar{n}$)
In this case the Lagrange multiplier $\lambda_2$ in expression (\ref{5.33}) is
equal to zero and consequently the results of the reconstruction
on the observation levels ${\cal O}_{\rm n}$ and ${\cal O}_{\rm th}$
(thermal observation level) are equal.

It is important to stress that all those states for which the Mandel
$Q$ parameter is less than $\bar{n}$ (in analogy with sub-Poissonian states
we can call these states as the sub-thermal states) can be reconstructed on
${\cal O}_{\rm n}$. For all these states the Lagrange multiplier
$\lambda_2$ is greater than zero and consequently the generalized partition
function (\ref{5.37}) does exist. Nevertheless there are states for which
$Q>\bar{n}$ (we will call these states as super-thermal states). For these
state the Lagrange multiplier $\lambda_2$ is smaller than zero and
$Z_{\rm n}$ given by Eq.(\ref{5.37}) is diverging. Consequently, these
states cannot be reconstructed on the observation level ${\cal O}_{\rm n}$.
In particular, the Mandel $Q$ parameter for the squeezed vacuum state
(\ref{3.48}) 
reads $Q=2\bar{n}+1$ (for $\bar{n}>0$) and therefore we are not able
to reconstruct the Wigner function of the squeezed vacuum state
on ${\cal O}_{\rm n}$. Analogously, the
even coherent state for small values of $\alpha$ such that $\sinh\alpha^2<1$
has a super-thermal photon number distribution and it cannot be reconstructed
on this observation level.

The mathematical reason behind the fact  that super-thermal states
cannot be reconstructed on ${\cal O}_{\rm n}$ is closely related to
the semi-infiniteness of the Fock state space of the harmonic oscillator,
i.e., the photon number distribution of these states cannot be approximated
by discrete Gaussian distributions $P_m$ (\ref{5.36}) on the interval
$m\in [0,\infty)$. In principle, there exist two ways how to regularize
the problem:  one can either expand the Fock space and to introduce
``negative'' Fock states, i.e., $m\in (-\infty,\infty)$. Alternatively,
one can assume finite-dimensional Fock space such that $m\in [0,s]$.
In both these cases $Z_{\rm n}$ for super-thermal states  is finite
and in principle $\hat{\sigma}_{\rm n}$ can be reconstructed
(but it may depend on the regularization procedure).

\section{OPTICAL HOMODYNE TOMOGRAPHY AND MAXENT PRINCIPLE}
\label{sec6}
From the point of view of the formalism presented in this paper
it follows that from
 the probability density distribution $w_{\hat{\rho}}
(x_{\theta})$ [see Eq.(\ref{3.23})] which corresponds to a measurement
of {\it all} moments $\langle \hat{x}_{\theta}^n\rangle$, the generalized
canonical density operators $\hat{\sigma}_{x_{\theta}}$
[see also Eq.(2.11)]:
\be
\hat{\sigma}_{x_{\theta}}=\frac{1}{Z_{x_{\theta}}}
\exp\left[-\int_{-\infty}^{\infty}dx_{\theta}\,
|x_{\theta}\rangle\langle x_{\theta}| \lambda(x_{\theta})\right]
\label{6.1}
\ee
can be constructed. The Lagrange multipliers $\lambda(x_{\theta})$
are given by an infinite set of equations
\be
w_{\hat{\rho}}(x_{\theta})=\sqrt{2\pi\hbar}
\langle x_{\theta}| \hat{\sigma}_{x_{\theta}}|x_{\theta}\rangle;
\qquad \forall x_{\theta} \in (-\infty,\infty).
\label{6.2}
\ee
If probability distributions $w_{\hat{\rho}}(x_{\theta})$ for all values
of $\theta\in [0,\pi]$ are known then the density operator
on the complete observation level can be obtained in the form
\be
\hat{\rho}=\frac{1}{Z_{0}}
\exp\left[-\int_{0}^{\pi} d\theta\,
\int_{-\infty}^{\infty}dx_{\theta}\,
|x_{\theta}\rangle\langle x_{\theta}| \lambda(x_{\theta})\right],
\label{6.3}
\ee
and the corresponding Wigner function can be reconstructed.
The optical homodyne tomography can be understood as a method
how to find a relation between measured distributions
$w_{\hat{\rho}}(x_{\theta})$ and the Lagrange multipliers
$\lambda(x_{\theta})$ for all values of $x_{\theta}$ and $\theta$.
As we have shown earlier in this section,
the Gaussian and the generalized Gaussian states can be completely
reconstructed on reduced observation levels based on a measurement of just
finite number of moments of system observables, and therefore the
optical homodyne tomography is essentially not needed as a method for
reconstruction of Wigner functions in these cases. On the other hand,
the non-Gaussian states can in principle reconstructed, but in practice
the reconstruction of their Wigner functions  is associated with a measurement
of an infinite number of independent moments of system observables
which is not realistic.
In the experiments by Raymer et al. \cite{Smithey93}
 only a finite number of values
of $\theta$ have been considered, i.e., these types of experiments
are associated with observation level for which the corresponding
generalized canonical density  operator reads  
\begin{equation}
\hat{\sigma} = \frac{1}{Z} \exp\left(\lambda_0 \hat{n} + 
\sum_{l=1}^{N_x}
\sum_{m=1}^{N_{\theta}}
\lambda_{l,m} 
|x_{\theta_m}^{(l)} \rangle \langle x_{\theta_m}^{(l)} | \right).
\label{6.4}
\end{equation}

\subsection{Implementation and numerical examples}
\label{sec6.A}
We want to demonstrate our reconstruction scheme and compare it with
known tomography scheme\footnote{The direct sampling method as described
in Section~\ref{sec3} can be straightfowardly applied also in the
case when the quadrature components $\hat{x}_{\theta}$ are measured
at $N_{\theta}$ discrete phases $\theta_m$. As shown by Leonhardt and Munroe
\cite{LeonhardtMunroe} if it is {\em a priori} known that 
 $\rho_{mn}=0$ for $|m-n|\geq N_{\theta}$
then the density matrix elements $\rho_{mn}$ for $|m-n|<N_{\theta}$
can be precisely reconstructed from the measured distributions
$w(x_{\theta},\theta_m)$ at $N_{\theta}$ phases $\theta_m$. On the
other hand, if the parameter $x$ is discretize (which corresponds
to a measurement of $N_x$ projectors
$|x_{\theta_m}^{(l)}\rangle\langle x_{\theta_m}^{(l)}|$ in the
direction $\theta_m$), then the direct-sampling reconstruction
can be applied as well, but 
may lead to ``pathological'' density operators which are not positively
defined. 
Alternatively, the least-square inversion
method (see for instance \cite{Tan1997}) can be efficiently applied.
The advantage of this method is that it is a linear method which means
that the density matrix can be reconstructed in a real time together
with an estimation of the statistical error. We note that this method
may also lead to non-positive density operators.} 
at four nontrivial examples. One is an incoherent
superposition of two coherent states
\begin{equation} 
\label{6.5}
\hat{\rho}_1 = \frac{1}{2} \left( |\alpha_1 \rangle \langle \alpha_1 | 
+ | \alpha_2 \rangle \langle \alpha_2 | \right),
\end{equation}
the second is a  superposition of two coherent states 
\begin{equation}
\label{6.6}
\hat{\rho}_2 = {\cal N} 
\left( |\alpha_1 \rangle + |\alpha_2 \rangle \right)
\left( \langle \alpha_1 | + \langle \alpha_2 | \right),
\end{equation}
the third is a rectangular state
\begin{equation}
\label{6.7}
\hat{\rho}_3 = |\psi \rangle \langle \psi|
\end{equation}

with
\begin{eqnarray}
\psi(x) = 
\left\{ \matrix{\frac{1}{\sqrt{4 \alpha_1}} \qquad \mbox{for}\, x \in 
[-2 \alpha_1,2 \alpha_1] \cr  0 \qquad \qquad \mbox{elsewhere,}} \right.
\label{6.8}
\end{eqnarray}
and the last one is a Fock state
\begin{equation}
\label{6.9}
\hat{\rho}_4 = |n \rangle \langle n|.
\end{equation}

All calculations were carried out in Fock representation, where  the
projection operators 
\be
\hat{O}_{lm}=|x_{\theta_m}^{(l)}\rangle\langle x_{\theta_m}^{(l)}|
\label{6.10}
\ee
 read
\begin{equation}
\left(\hat{O}_{lm}\right)_{n_1,n_2} = \psi^\star_{n_1}(x_l) \psi_{n_2}(x_l) 
\exp(i \theta_m (n_1-n_2)),
\label{6.11}
\end{equation}
and $\theta_m$ is the quadrature phase and $x_l$ is the eigenvalue of the
operator. In the numerical examples we chose
$\alpha_1 = 1.25$ and $\alpha_2=1.25\,i$ for the first three states and $n=4$
for the Fock state. 

Our numerical approach forces us to truncate the Hilbert space at a finite
value $n_{\rm max}$ and we must insure that an increase of this cut-off does
not change our results significantly. On the other hand the number $N_{\theta}$
of different angles $\theta$ and the number $N_x$ and separation $\Delta x$
of different $x$
is given by the experiment. The error of any reconstruction scheme goes to
zero when all $x$ for all angles $\theta$ are covered, i.e.\ when our
knowledge about the state is complete. On the other hand for incomplete
knowledge the different reconstruction schemes give different results and
in this sense we want to compare the schemes.

As a representation of the state and its reconstruction we show the
Wigner function  
(see Fig.~\ref{fig6}). 
The plots in the upper line
show the Wigner functions as surface plots, whereas the lower line
shows the same functions as grey scale  plots. 
The uniform grey background 
corresponds to the value zero whereas darker areas indicate positive values
of the Wigner function. 
In (a) we show the state $\hat{\rho}_1$ itself as defined in 
(\ref{6.5}) and in (b)
the reconstruction $\hat{\tilde{\rho}}_1$ as obtained via the maximum 
entropy principle. 
For the reconstruction we used
only 4 different angles and 13 points on each axis. Despite this extremely
small number the graphical representation of the state $\hat{\tilde{\rho}}_1$
reveals no difference to the original state $\hat{\rho}_1$. 
For completeness we include (c)
the state as obtained via projection onto pattern functions as described
in \cite{Leonhardt96-oc}. 
Contrary to (a) and (b) we also obtain white areas which
correspond to a negative value of the Wigner function.

Already from
this plot it is obvious that the reconstruction via maximum entropy principle
matches much better the original state. For a quantitative comparison we 
calculated
\begin{equation} 
\label{6.12}
\Delta = \sum_{n_1,n_2} \left[ (\rho_1)_{n_1,n_2} - (\tilde{\rho}_1)_{n_1,n_2} 
\right]^2
\end{equation}
as a measure for the error of the reconstruction. We vary the number 
$N_\theta$ of different
angles and $N_x$ of different values on each angle, their separation is
chosen in such a way that they cover uniformly
the integral $[-2,2]$, where---as can
be seen in Fig.~\ref{fig6}---almost the whole state is located. The
numerical cut-off for the Hilbert space was $n_{\rm max}=30$. 

Whereas the errors $\Delta$ (Table~\ref{table1})
for usual quantum tomography are of the order of one
(thus on average of the order of $10^{-3}$ per matrix element) the
inclusion of the maximum entropy principle reduces the errors by several
orders of magnitudes. 
We want to add that the large errors of usual quantum tomography of course
decrease significantly when increasing the amount of measurement data, i.e.
increasing $N_\theta$ and $N_x$.
On the other hand, the reconstruction
via projection onto pattern function is in general not positive definite 
which reflects some fundamental problems associated with this reconstruction
scheme.

\begin{figure}
\begin{center}
\epsfig{file=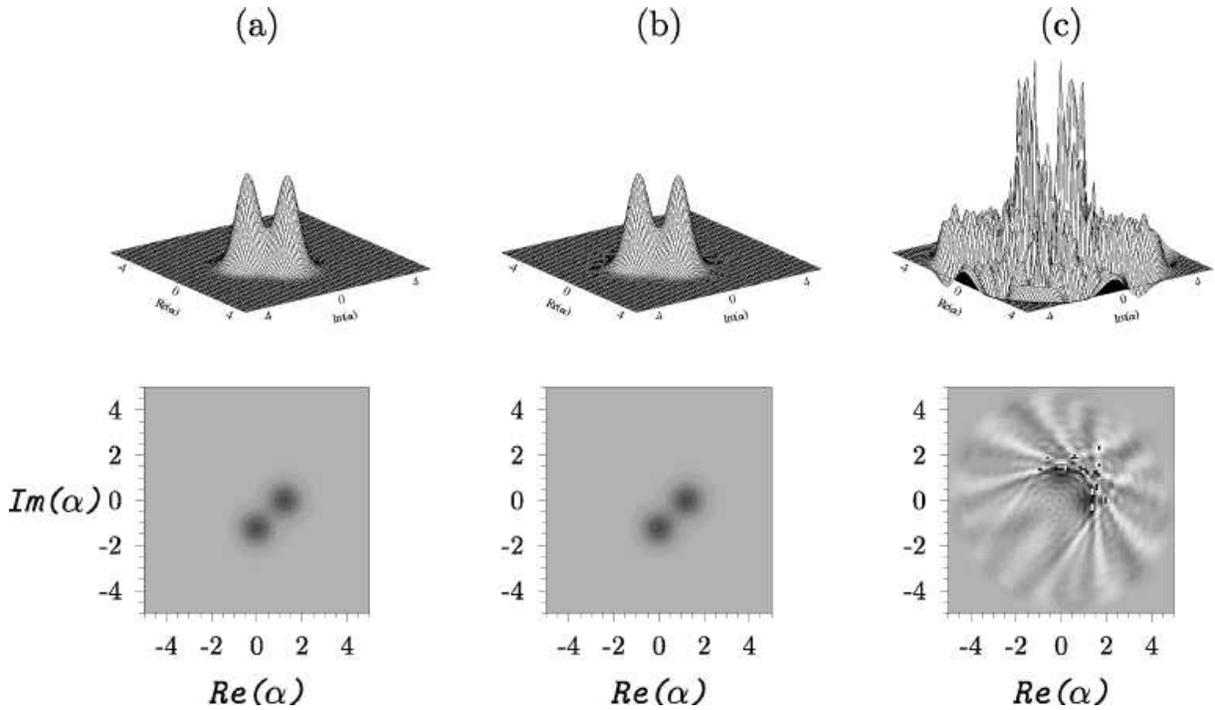, angle=-90, width=16cm}
\end{center}
\caption{
Wigner function of an incoherent superposition of two coherent states
(a), its reconstruction via the maximum entropy principle (b) and its
reconstruction via projection onto pattern functions (c). The upper line
shows the Wigner function as surface plots, whereas the lower line shows
the same functions as a grey scale plot, where dark areas correspond to
higher values and bright areas to lower values. We see that the
reconstruction via the MaxEnt principle is much more reliable then a
straightforward application of direct sampling via pattern functions.
}
\label{fig6}
\end{figure}

\begin{table}
\begin{center}
\epsfig{file=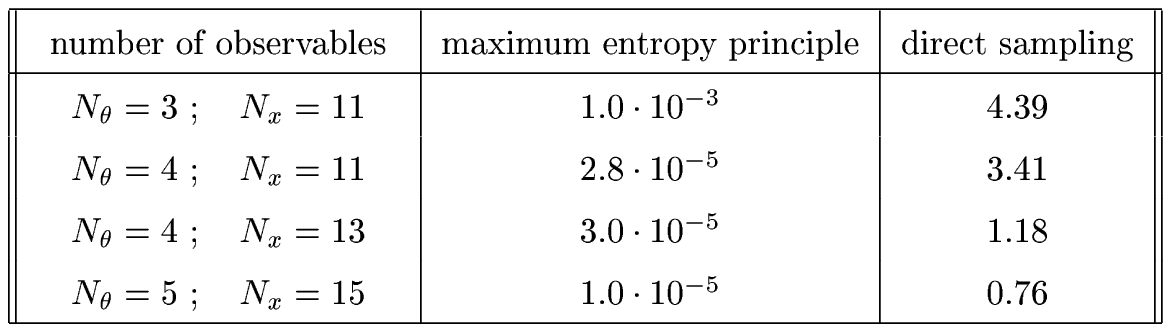, angle=0, width=9cm}
\end{center}
\caption{
Deviation $\Delta$ of the reconstructed state from the incoherent
superposition of two coherent states for reconstruction via the maximum
entropy principle and for reconstruction via projection onto pattern
functions.
}
\label{table1}
\end{table}

One might suspect that the superiority of the reconstruction via the
maximum entropy principle might be a speciality of the selected state,
an incoherent superposition of two coherent states. Therefore we want to
give some more examples, e.g.\ the corresponding coherent superposition, 
as defined in (\ref{6.6}). As in the previous example the values 
of $\alpha_1$ and $\alpha_2$ were chosen to be $1.25$ and $1.25\,i$,
respectively (see Fig.\ref{fig7}).

Surprisingly enough the reconstruction turns out to be simplified by the
quantum interferences apparent in the cat state: the deviations for both
reconstruction schemes are smaller than for the incoherent
superposition (Table~\ref{table2}). Apart from this 
the overall picture remains the same: The reconstruction
with the maximum entropy principle is many orders of magnitudes better than
the reconstruction via pattern functions. Moreover, the reconstruction
via pattern functions may again result in density operators which are not
positive definite.

Next, we discuss the reconstruction of the rectangular state as defined
in (\ref{6.7}). Though this state is of less relevance in quantum optics
it can be easily realized for atomic beams by an aperture. The reason, why
we include this state into our discussion is twofold: on one hand, the
oscillations in its Wigner function (see below) represent a serious difficulty
for quantum tomography, so it is interesting to check, whether other
reconstruction schemes do not have this difficulty. On the other hand
the smoothening character of our reconstruction by selecting the state
with maximum entropy may smooth out just these oscillations and therefore
this state is a critical test of the maximum entropy reconstruction.

\begin{figure}
\begin{center}
\epsfig{file=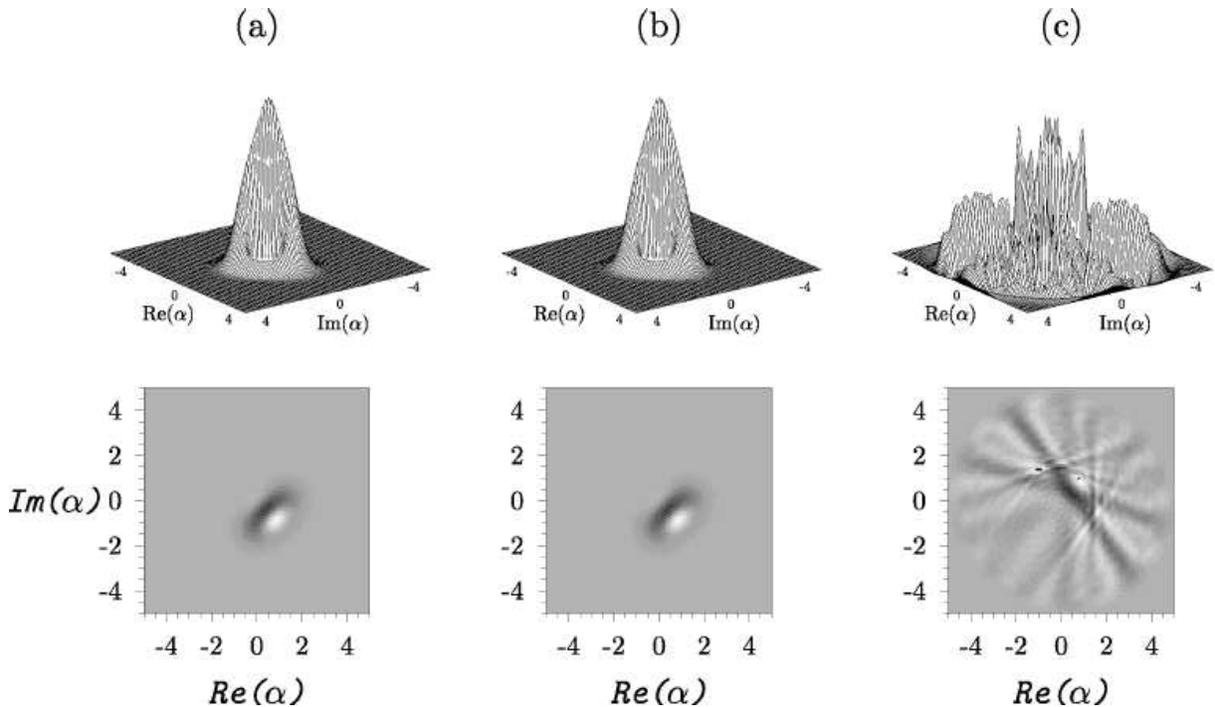, angle=-90, width=16cm}
\end{center}
\caption{
Wigner function of a coherent superposition of two coherent states
given by Eq.(\ref{6.6})
(a), its reconstruction via the maximum entropy principle (b) and its
reconstruction via the projection onto pattern functions (c).
Again the upper line
shows the state as surface plot and the lower line the corresponding grey
scale plots.
}
\label{fig7}
\end{figure}

\begin{table}
\begin{center}
\epsfig{file=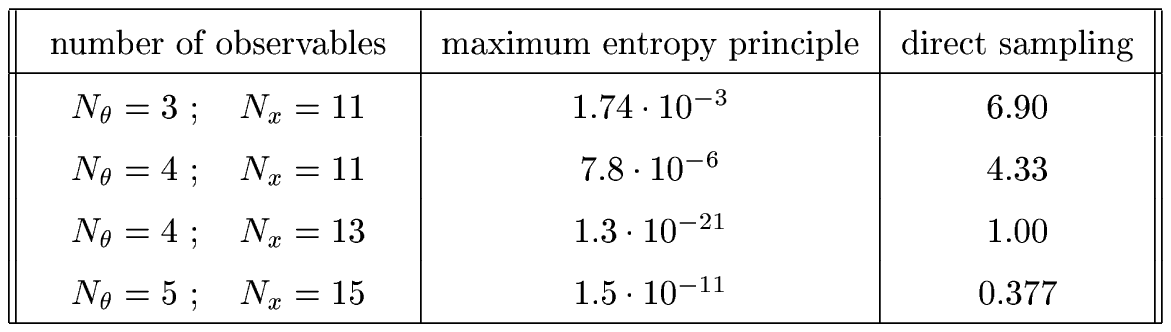, angle=0, width=9cm}
\end{center}
\caption{
Deviation $\Delta$ of the reconstructed state from the coherent
superposition of two coherent states for reconstruction via the maximum
entropy principle and for reconstruction via projection onto pattern
functions.
}
\label{table2}
\end{table}

The overall picture (Fig.~\ref{fig8})
is the same as for the previous two examples: the
reconstruction via the maximum entropy principle gives a deviation (see
also Table~\ref{table3}) from
the original state many orders of magnitude lower than conventional quantum
tomography does. 
Once more we want to stress that the
bad reconstruction by usual quantum tomography is due to the extremely small
number of angles and grid points. Increasing the amount of measurement data
makes this reconstruction scheme working.

Finally, we turn to a state, which is relatively easy to reconstruct via
quantum tomography: a number state (\ref{6.9}) with $n=4$. 
This state can be obtained
even exactly with a finite number of phases $N_\theta$ provided that 
$N_\theta = n+1$ and that each quadrature is measured completely, i.e.\
covering densely the whole axis. 
Since our examples do not and cannot
fulfill the latter condition, we encounter again a situation where
the reconstruction via quantum tomography suffers from too few measurement
data. To allow for a fair comparison we restrict the Hilbert space to
$N=n=4$, otherwise quantum tomography adds additional errors in the higher
density matrix elements.

\begin{figure}
\begin{center}
\epsfig{file=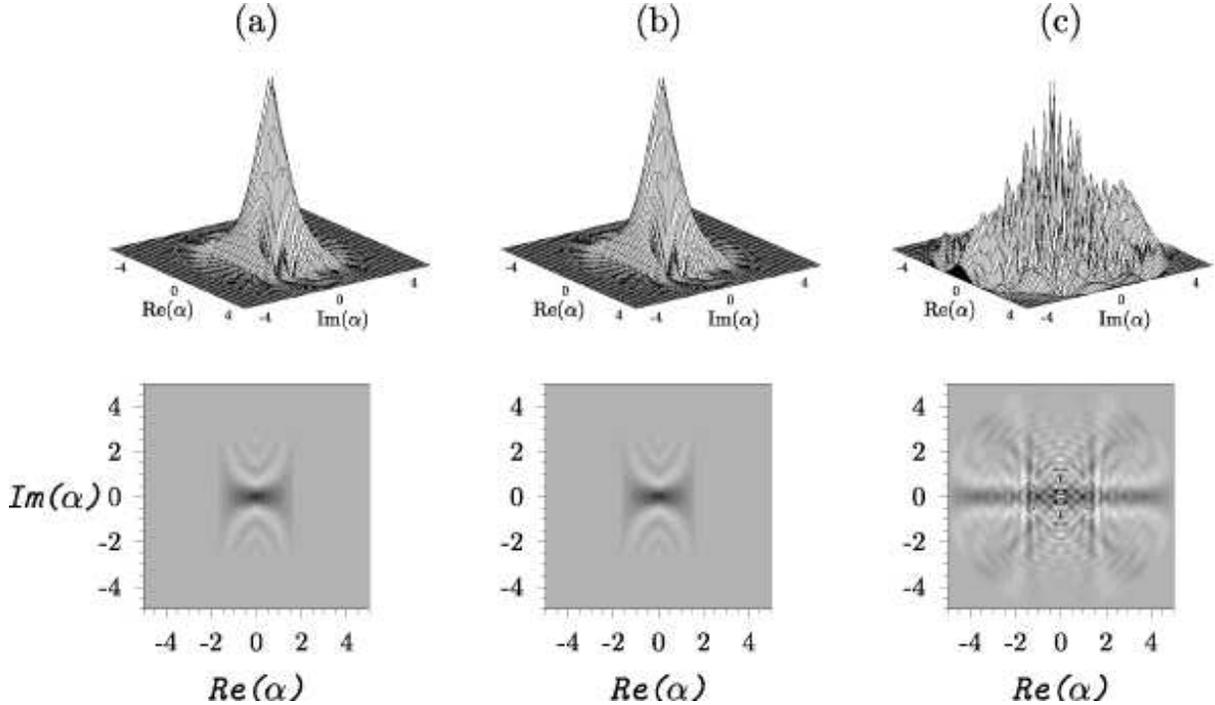, angle=-90, width=16cm}
\end{center}
\caption{
Wigner function of a {\em rectangular} state given by
[see Eq.(\ref{6.7})]
(a), its reconstruction via the maximum entropy principle (b) and its
reconstruction via the projection onto pattern functions (c).
}
\label{fig8}
\end{figure}

\begin{table}
\begin{center}
\epsfig{file=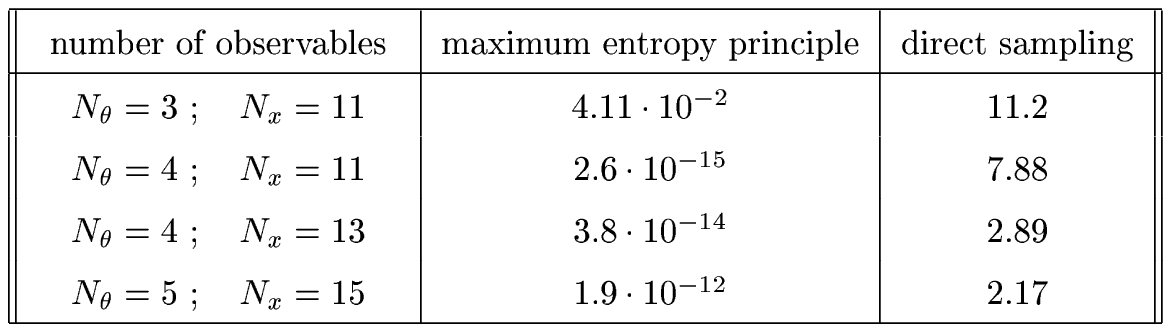, angle=0, width=9cm}
\end{center}
\caption{
Deviation $\Delta$ of the reconstructed state from the rectangular
state for reconstruction via the maximum
entropy principle and for reconstruction via projection onto pattern
functions.
}
\label{table3}
\end{table}

All errors (Table~\ref{table4})
are smaller than their counterparts for the other states
considered so far, which is also due to the smaller Hilbert space under
consideration. For this state the conventional reconstruction also gives
a very good estimate of the state (see Fig.~\ref{fig9}), 
though the absolute values of the
oscillations in the Wigner function (Fig.~\ref{fig9}) are not
completely correct due to the finite number of measurements on each
axis. But even for this state, which is 
advantageous for quantum tomography, the errors of the reconstruction
via the maximum entropy principle are much smaller.

\begin{figure}
\begin{center}
\epsfig{file=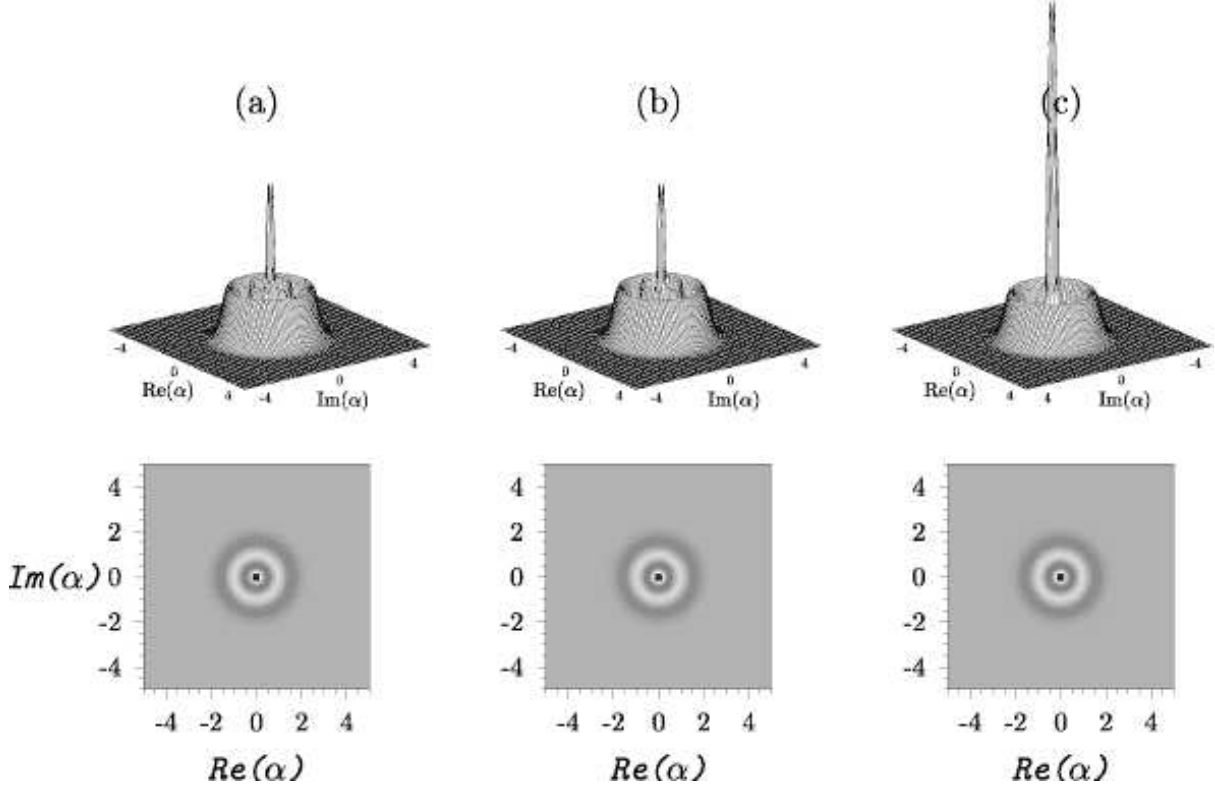, angle=-90, width=16cm}
\end{center}
\caption{
Wigner function of a Fock state
(a), its reconstruction via the maximum entropy principle (b) and its
reconstruction via the projection onto pattern functions (c).
}
\label{fig9}
\end{figure}

\begin{table}
\begin{center}
\epsfig{file=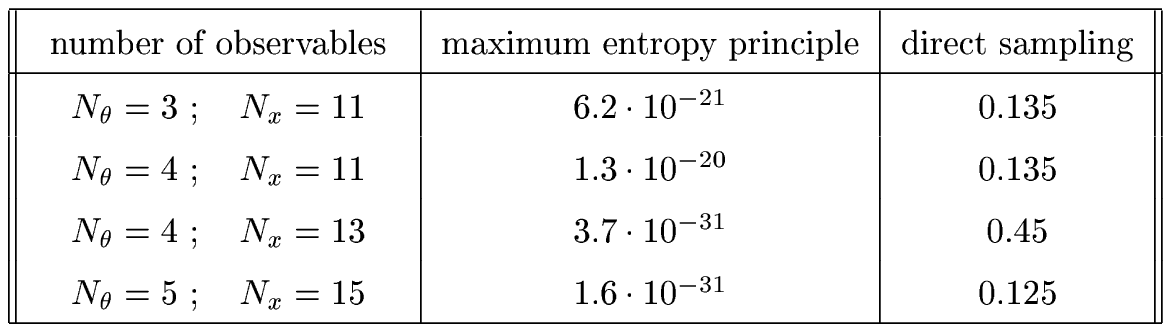, angle=0, width=9cm}
\end{center}
\caption{
Deviation $\Delta$ of the reconstructed state from the Fock
state for reconstruction via the maximum
entropy principle and for reconstruction via projection onto pattern
functions.
}
\label{table4}
\end{table}

\begin{figure}[t]
\begin{center}
\epsfig{file=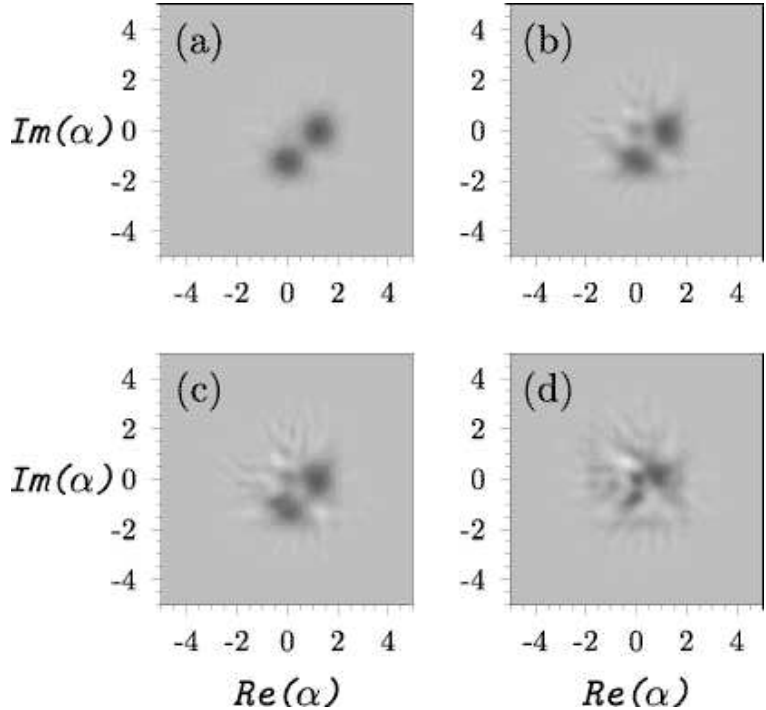, angle=-90, width=10cm}
\end{center}
\caption{
Wigner function as grey-scale plots
of the reconstruction of the incoherent superposition
shown in Fig.~6. Here we assume
 measurement results with errors.
The errors are proportional to a factor $\eta$, which is $10^{-2}$ (a),
$5 \cdot 10^{-2}$ (b), $10^{-1}$ (c) and $5 \cdot 10^{-1}$ (d).
}
\label{fig10}
\end{figure}

\subsection{Measurement errors and incompatible measurement results}
\label{sec6.B}
We want to discuss briefly the influence of measurement errors. Including
measurement errors we encounter a new problem: we cannot guarantee that
there is any state with positive definite density matrix
compatible with all measurement results (\ref{2.8}).
In other words, the set ${\cal C}$ 
as defined by Eq. (\ref{2.8}) is empty. Practically
this means that our numerical procedure to solve numerically the equations
for the Lagrange parameters cannot converge. Fortunately it turns out that
the set of Lagrange parameters minimizing the deviation
\begin{equation}
\left(\bar{n}-{\rm Tr}\left\{\tilde{\rho}\hat{n} \right\} \right)^2 + 
\sum_{lm} \left( o'_{lm} - {\rm Tr}\left\{ \tilde{\rho}\hat{O}_{lm} \right\}
\right)^2
\label{6.13}
\end{equation}
gives generally an excellent estimate for the state to be reconstructed.
To illustrate
 this we take our state and spoil artificially our measurement results by
\begin{equation}
o'_{lm} = o_{lm} + \eta \xi_{lm} \sqrt{o_{lm}}.
\label{6.14}
\end{equation}
$o'_{lm}$ is the result of the measurement with errors, whereas 
$o_{lm}={\rm Tr}(\hat\rho\hat{O}_{lm})$
is the corresponding result one would obtain in an ideal measurement.
The error was chosen to be proportional to the square root of $o_{lm}$
since this quantity is obtained by measuring $x_\theta$ several times
and counting how much results fall into a certain interval. The proportionality
factor $\eta$ characterizes the quality of our measurement and depends on the
number of single measurements made. $\xi_{lm}$ represent independent
Gaussian random numbers with
\begin{eqnarray}
\langle \xi_{lm} \rangle &=& 0 
\nonumber
\\[2pt]
\langle \xi_{lm} \xi_{l'm'} \rangle &=& \delta_{ll'} \delta_{mm'}.
\end{eqnarray}
\begin{figure}[t]
\begin{center}
\epsfig{file=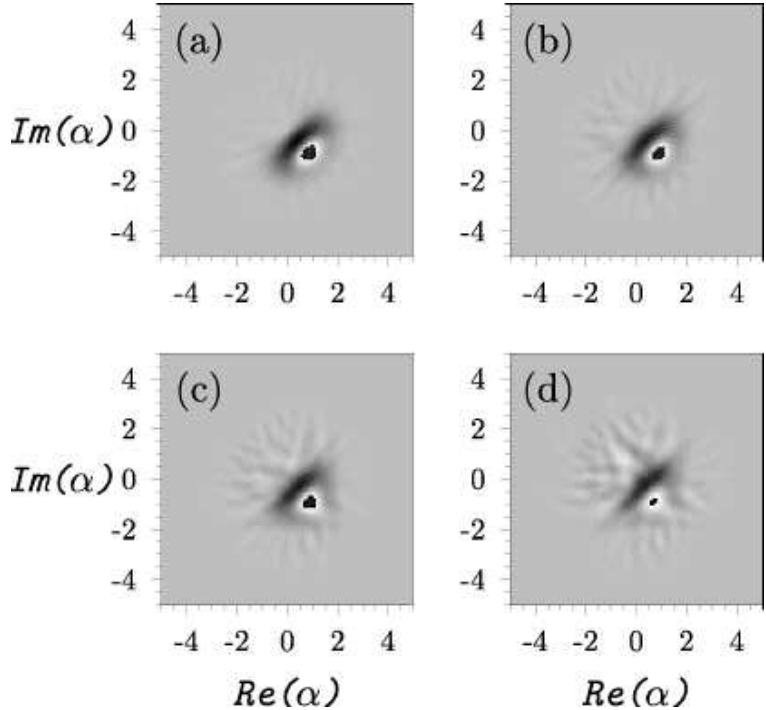, angle=-90, width=10cm}
\end{center}
\caption{
Wigner function as grey-scale plots
of the reconstruction of the cat state
shown in Fig.~7 on the basis of measurement results with errors.
The errors are proportional to a factor $\eta$, which is $10^{-2}$ (a),
$5 \cdot 10^{-2}$ (b), $10^{-1}$ (c) and $5 \cdot 10^{-1}$ (d).
}
\label{fig11}
\end{figure}

With these $o_{lm}$ and various parameters $\eta$ we started again the
reconstruction of the incoherent superposition (\ref{6.5}), the
cat state (\ref{6.6}), the rectangular state and the Fock state
already discussed in the last paragraph. For all reconstructions we used
$N_\theta=4$ different axis' and $N_x=13$ points at each axis, covering the
interval $[-2,2]$ as before.

Fig.~\ref{fig10} shows the Wigner functions of the reconstruction of 
the incoherent superposition for $\eta=10^{-2}$, $5\cdot 10^{-2}$,
$10^{-1}$ and $5\cdot 10^{-1}$. Despite the relative large values of
$\eta$---corresponding to a large error---the reconstruction is very
good. We also calculated the error as defined in (\ref{6.12}) and
obtained $\Delta=4 \cdot 10^{-3}$ for $\eta=10^{-2}$, $\Delta = 4 \cdot
10^{-2}$ for $\eta=5 \cdot 10^{-2}$, $\Delta = 5 \cdot 10^{-2}$ for
$\eta = 0.1$ and $\Delta = 0.3$ for $\eta = 0.5$. For $\eta=0.05$ we 
recognize a slight asymmetry between the two spots, which becomes more
pronounced for the highest value of $\eta$. But still the two dots are
easily distinguishable. Due to the random character of
our calculation these numbers will vary when varying the random numbers---as
the results of a measurements will vary from run to run.

Next, we present the corresponding plots for the cat state 
(Fig.~\ref{fig11}). As in
the previous figure the values for $\eta$ are $10^{-2}$, $5\cdot 10^{-2}$,
$10^{-1}$ and $5\cdot 10^{-1}$ for (a), (b), (c) and (d), respectively.
The obtained errors $\Delta$ are $3 \cdot 10^{-3}$, $3 \cdot 10^{-2}$,
$7 \cdot 10^{-2}$ and $2 \cdot 10^{-1}$. 
We want to emphasize that for both states even 
for the largest value of $\eta$ the reconstruction is better than the
usual quantum tomography without errors.

Now we discuss the influence of measurement errors for the rectangular
state. Fig.~\ref{fig12} shows the Wigner functions of the reconstruction with
an error parameter $\eta=10^{-2}$ (a), $5 \cdot 10^{-2}$ (b), $10^{-1}$
(c) and $5 \cdot 10^{-1}$ (d). As for the previous states only for the
highest error parameter $\eta$ we recognize an asymmetry not present in the
original state (Fig.~\ref{fig8} (a)). 
The resulting errors were $2 \cdot 10^{-3}$,
$2 \cdot 10^{-2}$, $6 \cdot 10^{-2}$ and $3 \cdot 10^{-1}$, respectively.
Despite the relatively large errors the reconstruction is as in the previous 
examples very good. Only for the largest value of $\eta$ we recognize a
qualitative difference to the original state Fig.~\ref{fig8} (a).

\begin{figure}[t]
\begin{center}
\epsfig{file=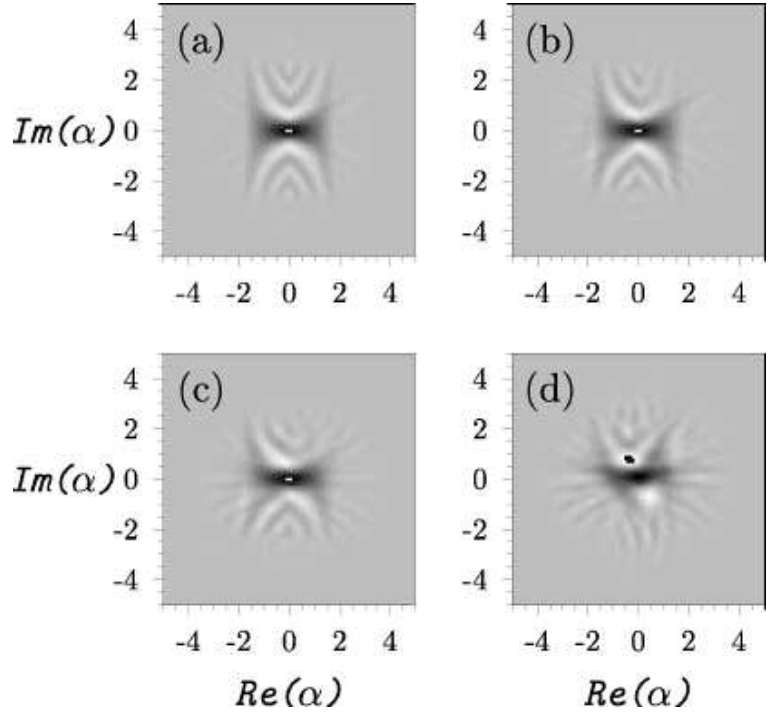, angle=-90, width=10cm}
\end{center}
\caption{
Wigner function as grey-scale plots
of the reconstruction of the rectangular state
shown in Fig.~8 on the basis of measurement results with errors.
The errors are proportional to a factor $\eta$, which is $10^{-2}$ (a),
$5 \cdot 10^{-2}$ (b), $10^{-1}$ (c) and $5 \cdot 10^{-1}$ (d). 
}
\label{fig12}
\end{figure}

\begin{figure}
\begin{center}
\epsfig{file=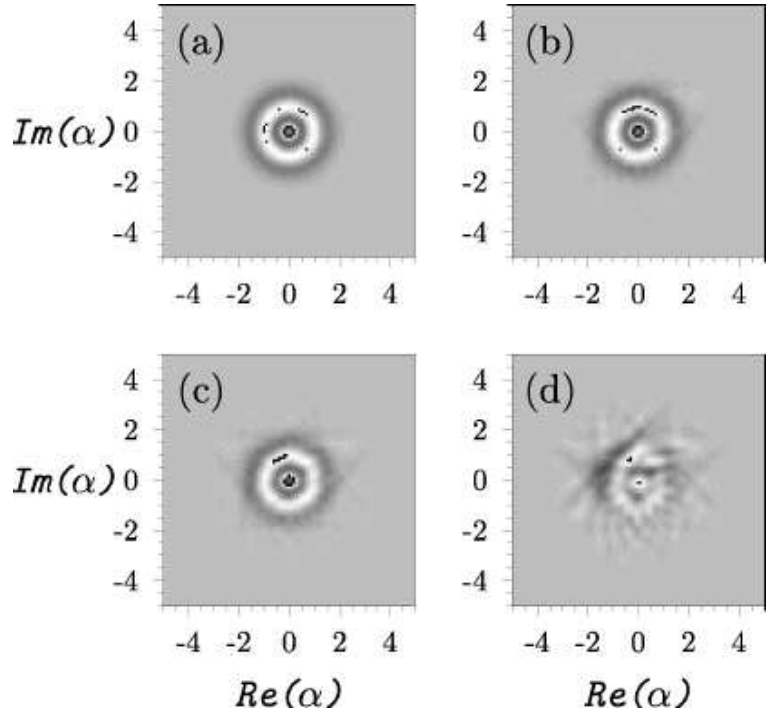, angle=-90, width=10cm}
\end{center}
\caption{
Wigner function as grey-scale plots
of the reconstruction of the Fock state
shown in Fig.~9 on the basis of measurement results with errors.
The errors are proportional to a factor $\eta$, which is $10^{-2}$ (a),
$5 \cdot 10^{-2}$ (b), $10^{-1}$ (c) and $5 \cdot 10^{-1}$ (d). 
}
\label{fig13}
\end{figure}

Finally, we turn to the influence of measurement errors on the reconstruction
of the Fock state. The errors $\Delta$ for the various error parameters
$\eta$ are $\Delta=8 \cdot 10^{-4}$ for $\eta=1\cdot10^{-2}$, $\Delta=1 
\cdot 10^{-2}$ for
$\eta=5\cdot 10^{-2}$, $\Delta=4 \cdot 10^{-2}$ for $\eta=0.1$, and $\Delta=0.7$
for $\eta=0.5$. Also the plots of the corresponding Wigner functions
(Fig.~\ref{fig13})
reveal that the reconstruction is very good and shows the ring-shaped structure
of the original Fig.~\ref{fig9} (a). Only for the $\eta=0.5$ the
reconstruction is not good enough to show clearly this feature.

Usually this kind of measurement results into a very good reconstruction of
Wigner functions (such that the corresponding entropy is close to zero
for pure states). Nevertheless, a certain attention has to be paid for
highly squeezed states, such as the Vogel-Schleich phase states 
 \cite{Vogel1991},
for which the measurement of
distributions $w_{\hat{\rho}}(x_{\theta_j})$ can be problematic.
Namely, $w_{\hat{\rho}}(x_{\theta_j})$ can be very ``wide'', so that
the normalization condition is not fulfilled in a domain of physically
accessible values of $x_{\theta_j}$.

In this Section we have
 presented a numerical application of the reconstruction scheme via
the MaxEnt principle for a reconstruction of Wigner functions
of quantum-mechanical states of light from incomplete tomographic data.
We have shown that when the tomographic data are incomplete, then the
reconstruction via the MaxEnt principle is much more reliable than
the standard inversion 
Radon transformation scheme or the pattern-function scheme.

\section{RECONSTRUCTION OF SPIN STATES VIA MAXENT PRINCIPLE}
\label{sec7}

In the following sections we will apply the Jaynes principle for the
reconstruction of pure spin states (see also \cite{Buzek97}).
 Firstly, for illustrative purposes
we present the simple example of the reconstruction of states of
a single spin-1/2 system with the help of the maximum-entropy principle.
Then we will discuss the partial reconstruction of entangled spin states.
In particular, we will analyze the problem how
to identify incomplete observation levels
on which the complete reconstruction can be performed
for the Bell and the Greenberger--Horne--Zeilinger states (i.e.,
the corresponding entropy is equal to zero
and the generalized canonical density operator is identical to
$\hat\rho_0$).


\subsection{Single spin-1/2}
\label{sec7.A}

Firstly we illustrate the application of the maximum-entropy  principle
for the partial quantum--state reconstruction of single spin--$1/2$ system.
Let us	consider an ensemble of
spins-$1/2$ in an unknown pure state $|\psi_0\rangle$.
In the most general case this unknown state vector $|\psi_0\rangle$
can be parameterized as
\be
|\psi_0\r= \cos\theta/2 |1\r +\mbox{e}^{i\varphi}\sin\theta/2 |0\r ,
\label{7.1}
\ee
where $|0\r$, $|1\r$  are eigenstates of the $z$-component of the
spin operator $\hat{s}_z={1\over 2}\hat{\sigma}_z$
with eigenvalues $-{1\over 2}$, ${1\over 2}$, respectively.
The corresponding density operator $\hat{\rho}_0=|\psi_0\r \l\psi_0|$
can be written in the form
\be
\hat{\rho}_0= {1\over 2}\left( \hat{I} + \vec{n}.\hat{\vec{\sigma}}
\right), 
\label{7.2}
\ee
where $\hat{I}$ is the unity operator,
$\vec{n}=(\sin \theta \cos\varphi,\sin \theta \sin\varphi,\cos \theta)$;
$\hat{\vec{\sigma}}=(\hat{\sigma}_x,\hat{\sigma}_y,\hat{\sigma}_z)$
are the Pauli spin operators which in the matrix representation in the basis
$|0\r$, $|1\r$ read
\be
\hat{\sigma}_x=\left(\begin{array}{cc} 0 & 1 \\ 1 & 0 \end{array}\right),
\quad
\hat{\sigma}_y=\left(\begin{array}{cc} 0 & -i \\ i & 0 \end{array}\right),
\quad
\hat{\sigma}_z=\left(\begin{array}{cc} 1 & 0 \\ 0 & -1 \end{array}\right)
.\label{7.3}
\ee

\begin{table}
\begin{center}
\epsfig{file=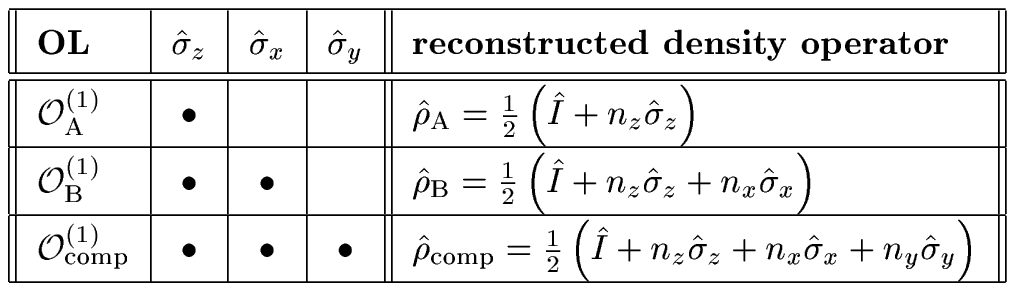, angle=0, width=10cm}
\end{center}
\caption{
In this table we present three observation  levels
${\cal O}_{\rm A}^{(1)}$; ${\cal O}_{\rm B}^{(1)}$,
and ${\cal O}_{\rm comp}^{(1)}$ associated with a measurement of the
particular
spin-1/2  operators.
Bullets ($\bullet$) in the table indicate which observables constitute
a given observation level.
We also present  explicit expressions for the
reconstructed density operators $ \hat{\rho}_{\rm A}$;
$ \hat{\rho}_{\rm B}$ and $ \hat{\rho}_{\rm comp}$.
}
\label{table5}
\end{table}

To determine completely the unknown state one has to measure
three linearly independent (e.g., orthogonal) projections of the spin.
After the measurement of the expectation value of each observable,
a reconstruction
of the generalized canonical density operator (\ref{2.11})
according to the maximum-entropy principle can be performed.
In  Table~\ref{table5}
 we consider three observation levels defined as
${\cal O}_A^{(1)}=\{\hat\sigma_z \}$,
${\cal O}_B^{(1)}=\{ \hat\sigma_z,\hat\sigma_x \}$ and
${\cal O}_C^{(1)}=\{\hat\sigma_z,\hat\sigma_x,\hat\sigma_y \}
\equiv{\cal O}_{comp}$
[the superscript of the observation levels indicates the number of spins--$1/2$
under consideration].

Using algebraic properties of the $\hat\sigma_\nu$-operators, the
generalized canonical density operator (\ref{2.11})
 can be expressed as
\be
\hat\rho_{\cal O}={1\over Z}
\exp (-\vec{\lambda}.\hat{\vec{\sigma}})={1\over Z}
\left[ \cosh|\lambda| \hat{I} - \sinh|\lambda|
{ \vec{\lambda}.\hat{\vec{\sigma}}\over |\lambda| } \right],
\qquad Z=2 \cosh|\lambda| , 
\label{7.4}
\ee
with $\vec{\lambda}=(\lambda_x,\lambda_y,\lambda_x)$ and
$|\lambda|^2=\lambda_x^2+\lambda_y^2+\lambda_z^2$.
The final form of the $\hat\rho_{\cal O}$ on  particular
observation levels is given in Table~\ref{table5}.
The corresponding entropies can be written as
\be
S_{\cal O}=-p_{\cal O} \ln p_{\cal O} -
	   (1-p_{\cal O}) \ln (1-p_{\cal O}), 
\label{7.5}
\ee
where $p_{\cal O}$ is one eigenvalue of $\hat\rho_{\cal O}$
[the other eigenvalue is equal to $(1-p_{\cal O})$] which
reads as
\be
p_A={1+|\l\hat{\sigma}_z\r| \over 2}, \quad
p_B={1+\sqrt{\l\hat{\sigma}_x\r^2+\l\hat{\sigma}_z\r^2} \over 2}, \quad
p_{\rm comp}={1+\sqrt{\l\hat{\sigma}_x\r^2+\l\hat{\sigma}_y\r^2+
\l\hat{\sigma}_z\r^2} \over 2}.
\label{7.6}
\ee
It is seen that the entropy $S_{\cal O}$ is equal to zero if and
only if $p_{\cal O}=1$. From here follows that on ${\cal O}_A^{(1)}$ only
the basis vectors $|0\r$ and $|1\r$ with $|\l\hat{\sigma}_z\r|=1$
can be fully reconstructed. Nontrivial is
${\cal O}_B^{(1)}$, on which a whole set of pure states (\ref{7.1})
with $\l \hat{\sigma}_y\r=0$ (i.e., $\varphi=0$)
can be uniquely determined. For such states $S_B=0$ and further
measurement
of the $\hat\sigma_y$ on ${\cal O}_{comp}$ represents redundant
(useless) information.
\subsection{Two spins--1/2}
\label{sec7.B}

Now we assume a system composed of two {\em distinguishable} spins--$1/2$.
If we are performing only {\em local} measurements
of observables such as $\hat{\sigma}_\mu^{(1)}\otimes\hat{I}^{(2)}$
and $\hat{I}^{(1)}\otimes\hat{\sigma}_\nu^{(2)}$ (here	superscripts label
the particles) which do not reflect correlations between the particles
then the reconstruction of the density operator reduces to an estimation
 of individual (uncorrelated)
spins--1/2, i.e., the reconstruction reduces to the problem discussed
in the previous section.  For each spin--$1/2$
the reconstruction can be performed separately and the resulting
generalized canonical density operator is given as a tensor product of
particular generalized canonical density operators, i.e.,
$\hat\rho=\hat\rho^{(1)}\otimes \hat\rho^{(2)}$.
In this case just the uncorrelated states
$|\psi_0\r=|\psi_0^{(1)}\r \otimes |\psi_0^{(2)}\r$ can be fully
reconstructed.
Nevertheless, the correlated (nonfactorable) states
$|\psi_0\r \neq |\psi_0^{(1)}\r \otimes |\psi_0^{(2)}\r$
are of central interest.
\begin{table}
\begin{center}
\epsfig{file=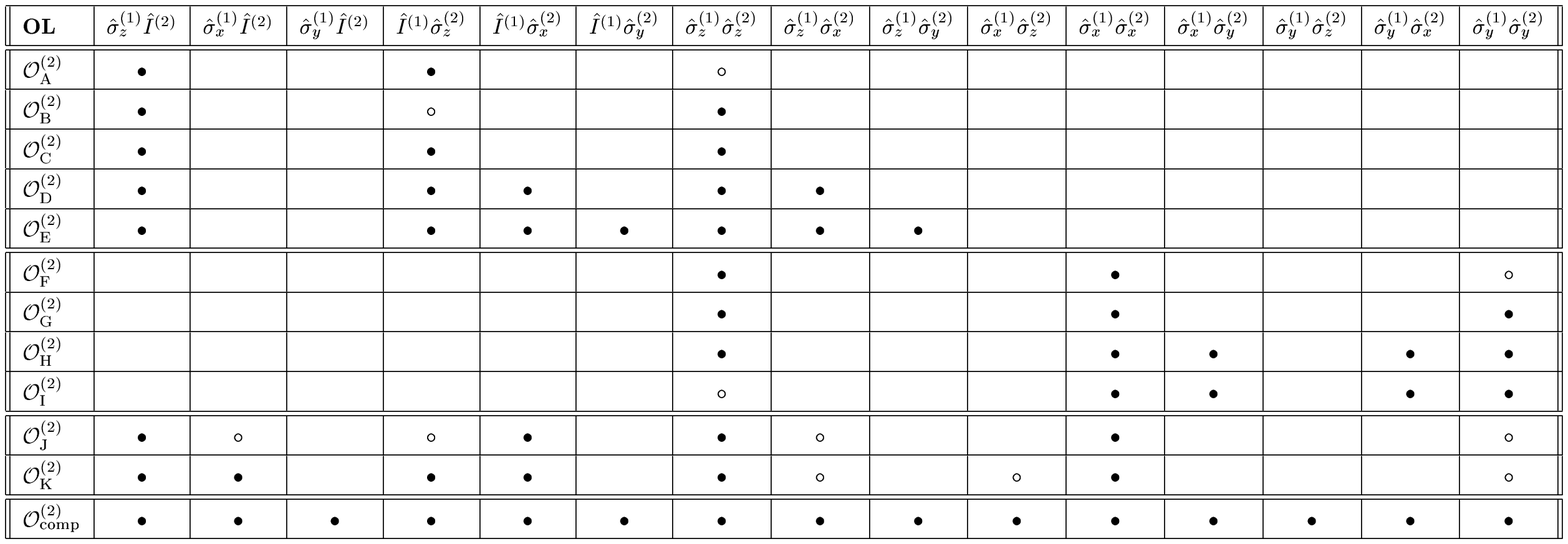, angle=0, width=17.5cm}
\end{center}
\caption{
We present a set of observation levels on which the density operators
of two spins-1/2 can be  partially reconstructed.
Bullets ($\bullet$) in the table indicate which observables constitute
a given observation level while empty circles ($\circ$)
denote
unmeasured observables (i.e., these observables are not included in the
given observation level)
 for which the maximum-entropy principle `predicts''
 nonzero mean values.
}
\label{table6}
\end{table}

In general, any density operator of a system composed of two
distinguishable
spins--$1/2$ can be represented by a $4\times 4$ Hermitian matrix
and $15$ independent numbers are required for its complete
determination. It is worth noticing that $15$ operators (observables)
\be
\{ \hat{\sigma_\mu}^{(1)}\otimes\hat{I}^{(2)},
\hat{I}^{(1)}\otimes\hat{\sigma_\nu}^{(2)},
\hat{\sigma}_\mu^{(1)} \otimes\hat{\sigma}_\nu^{(2)} \}; 
\qquad (\mu,\nu=x,y,z), 
\label{7.7}
\ee
together with the identity operator $\hat{I}^{(1)}\otimes\hat{I}^{(2)}$
form an operator algebra basis in which any operator can be expressed.
In this ``operator'' basis each density operator can be written as
\be
\hat{\rho}= {1\over 4}\left[ \hat{I}^{(1)}\otimes \hat{I}^{(2)} +
\vec{n}^{(1)}. ~\hat{\vec{\sigma}}^{(1)}\otimes\hat{I}^{(2)} +
\vec{n}^{(2)}. ~\hat{I}^{(1)}\otimes\hat{\vec{\sigma}}^{(2)} +
\sum_{\mu,\nu} \xi_{\mu\nu}
\hat{\sigma}_\mu^{(1)} \otimes\hat{\sigma}_\nu^{(2)} \right], 
\label{7.8}
\ee
with $\xi_{\mu\nu}=
\l \hat{\sigma}_\mu^{(1)}\otimes\hat{\sigma}_\nu^{(2)} \r$
($\mu, \nu = x, y, z$).

Using the maximum-entropy principle we can (partially) reconstruct an unknown
density operator $\hat\rho_0$ on various observation levels.
Conceptually the method of maximum entropy is rather straightforward:
one has to express the generalized canonical density operator (\ref{2.11})
for two spins-1/2  in the form (\ref{7.8}) from which a set
of nonlinear equations for Lagrange multipliers $\lambda_\nu$ is obtained.

Due to algebraic properties of the operators under the consideration the
practical realization of this programme can be technically  difficult
(see Appendix A).
In Table~\ref{table6} we define some nontrivial observation
levels.
Measured observables which define a particular observation
level are indicated in Table~\ref{table6} by bullets ($\bullet$)
 while the empty circles ($\circ$) indicate
unmeasured observables (i.e., these observables are not included in the
given observation level)
 for which the maximum-entropy principle ``predicts''
 nonzero mean values.
This means that the maximum-entropy principle
provide us with a nontrivial estimation of mean values of unmeasured
observables. The generalized canonical density operators which correspond
to the observation levels considered in Table~\ref{table6} are presented
in Table~\ref{table7}. The signs ``$\oplus,\ominus$'' are used to indicate
unmeasured observables for which nontrivial information can be obtained
with the help of  the maximum-entropy principle.

\begin{table}
\begin{center}
\epsfig{file=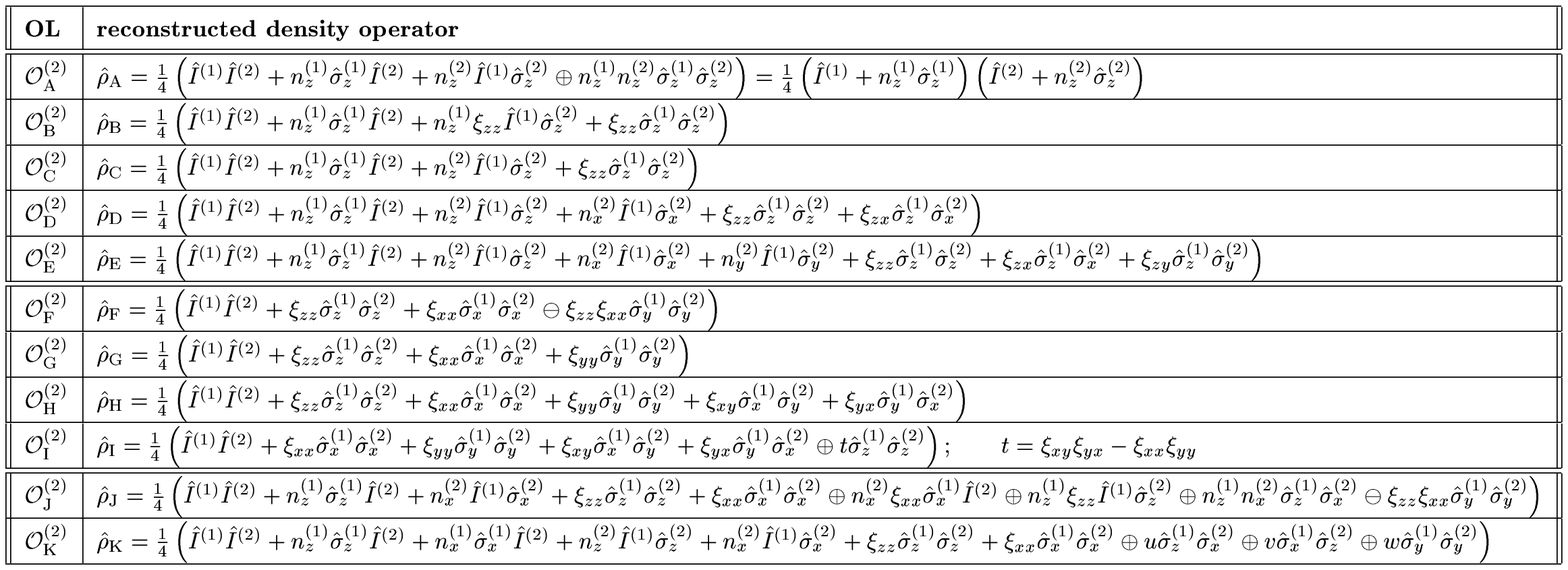, angle=0, width=17.5cm}
\end{center}
\caption{
We present  explicit expressions for the
reconstructed density operators $ \hat{\rho}_{\rm X}$
of two spins-1/2 on the observation levels denoted in Table 2.
We use the notation
$n_\mu^{(p)}=\langle\hat{\sigma}_{\mu}^{(p)}\rangle$
($\mu=z,x,y;\quad p=1,2$) and $\xi_{\mu\nu}=
\langle \hat{\sigma}_{\mu}^{(1)}\otimes \hat{\sigma}_{\nu}^{(2)}\rangle$
with $\mu,\nu=z,x,y$.
 The signs $\oplus$ and $\ominus$ are used to indicate
unmeasured observables for which nontrivial information can be obtained
with the help of  the maximum-entropy principle.
}
\label{table7}
\end{table}

\subsection{Reconstruction of Bell states}
\label{sec7.C}
In what follows we analyze a partial reconstruction of the Bell states
(i.e., the most correlated two particle states)
on observation levels given in Table~\ref{table6}. One  of our
main tasks will be
to find the minimum observation level (i.e., the set of system
observables) on which the complete reconstruction of these states
can be performed. Obviously, if all 15 observables are measured,
then any state of two spins-1/2  can be reconstructed precisely.
Nevertheless, due to the quantum entanglement between the two particles,
measurements of some observables will simply be redundant.
To find the minimal set of observables which  uniquely determine
the Bell state one has to  perform either a sequence of reductions of the
complete observation level, or	a systematic extension
of the most trivial observation level ${\cal O}^{(2)}_{A}$.

Let us consider particular examples of Bell states, of the form
\be
|\Psi_\varphi^{  (Bell)}\rangle={1\over\sqrt{2}}
\left[ |1,1\rangle + \mbox{e}^{i\varphi} |0,0\rangle \right],
\quad \hat\rho_\varphi^{(Bell)}
=|\Psi_\varphi^{ (Bell)}\rangle \langle \Psi_\varphi^{ (Bell)}|,
\label{7.9}
\ee
(other Bell states are discussed later).
These maximally correlated states have the property that the
result of a measurement
performed on one of the two spins--$1/2$ uniquely determines the
state of the
second spin.
Therefore, these states find their applications in quantum communication
systems \cite{Barenco,Ekert91}.
  In addition,
they are suitable for testing fundamental
principles of quantum mechanics \cite{Peres93}
such as the complementarity principle or local hidden--variable theories
\cite{Greenberger}.

Let us analyze now a sequence of successive extensions of the observation
level ${\cal O}_A^{(2)}$
\be
{\cal O}_A^{(2)}\subset{\cal O}_B^{(2)}\subset{\cal O}_C^{(2)}
\subset {\cal O}_D^{(2)}
.\label{7.10}
\ee
The observation level ${\cal O}_A^{(2)}$ (see Table~\ref{table6})
is associated with the measurement
of $\hat{\sigma}_z$ observables of each spin individually, i.e.,
it is insensitive with respect to correlations between the spins.
On ${\cal O}_B^{(2)}$ both $z$-spin components of particular spins
and their correlation have been recorded (simultaneous
measurement of these observables is possible because they commute).
Further extension to the observation level
on ${\cal O}_C^{(2)}$ corresponds to a rotation of
the Stern--Gerlach apparatus such that the $x$-spin component of the
 second spin--$1/2$ is measured.
The observation level ${\cal O}_D^{(2)}$ is associated with
another rotation of the Stern--Gerlach apparatus which would allow us
to measure the $y$-spin
component.
The generalized canonical density operators on the observation levels
${\cal O}_B^{(2)}$, ${\cal O}_C^{(2)}$ and ${\cal O}_D^{(2)}$
predict zero mean values for all the unmeasured observables (\ref{7.7})
(see  Table~\ref{table7}).

In general,    successive extensions (\ref{7.10})
of the observation level
${\cal O}_A^{(2)}$ should be accompanied by a decrease in the
entropy of the reconstructed state which should reflect increase of our
knowledge about the quantum-mechanical system under consideration.
Nevertheless, we note that
 there are states for which the entropy remains
constant when  ${\cal O}_B^{(2)}$ is extended towards ${\cal O}_C^{(2)}$ and
${\cal O}_D^{(2)}$, i.e., the performed measurements are in fact redundant.
For instance, this is the case for the maximally correlated state (\ref{7.9}).
Here entropies associated with given observation levels read
\be
S_A=2\ln 2, \qquad S_B=S_C=S_D=\ln 2
\label{7.11}
,\ee
respectively, which
mean that these observation levels are not suitable for reconstruction
of the Bell states. The reason is that the Bell states
have no ``preferable'' direction for each individual spin,
i.e., $\langle \hat{\sigma}_{\mu}^{(p)} \rangle =0$ for  $\mu=x,y,z$
and $p=1,2$.

From the above it follows that, for a nontrivial  reconstruction
of Bell states, the observables
which reflect correlations between composite spins also have to be included
into the observation level.
Therefore let us now discuss the sequence of observation levels
\be
  {\cal O}_E^{(2)}\subset{\cal O}_F^{(2)}\subset{\cal O}_G^{(2)}
\label{7.12}
\ee
associated with simultaneous measurement of spin components of the two
particles [see Table~\ref{table6}].
 The corresponding generalized canonical
density operators are given  in Table~\ref{table7}.
To answer the question of  which states can be completely reconstructed
on the observation level
${\cal O}_E^{(2)}$ we evaluate
the von Neumann entropy  of the generalized canonical
density operator $\hat{\rho}_E$. For the Bell states we find that
$S_E=-p_E \ln p_E - (1-p_E) \ln (1-p_E)$ where $p_E=(1-\cos\varphi)/2$.
We can also compare directly $\hat{\rho}_\varphi^{(Bell)}$ with
$\hat{\rho}_E$. The density  operator $\hat{\rho}_\varphi^{(Bell)}$
in the matrix form can be written as
\be
\hat{\rho}_\varphi^{(Bell)}={1\over 2}\left(
\begin{array}{cccc}
1 & 0 & 0 & \mbox{e}^{-i\varphi} \\
0 & 0 & 0 & 0 \\
0 & 0 & 0 & 0 \\
\mbox{e}^{i\varphi} & 0 & 0 & 1
\end{array} \right)
,\ee
while the corresponding operator reconstructed on
the observation level ${\cal O}_E^{(2)}$ reads
\be
\hat{\rho}_E={1\over 2}\left(
\begin{array}{cccc}
1 & 0 & 0 & \cos\varphi \\
0 & 0 & 0 & 0 \\
0 & 0 & 0 & 0 \\
\cos\varphi & 0 & 0 & 1
\end{array} \right).
\label{7.13}
\ee
We see that $\hat{\rho}_\varphi^{(Bell)}=\hat{\rho}_E$
and $S[\hat{\rho}_E]=0$ only
if $\varphi=0$ or $\pi$ which means that
the Bell states $|\Psi_{\varphi=0,\pi}\rangle={1\over\sqrt{2}}
\left[ |1,1\rangle \pm |0,0\rangle \right]$ are completely determined
by mean values of two observables
$\hat{\sigma}_z^{(1)}\otimes \hat{\sigma}_z^{(2)}$ and
$\hat{\sigma}_x^{(1)}\otimes \hat{\sigma}_x^{(2)}$
 and that these states can be completely reconstructed on  ${\cal O}_E^{(2)}$.
We note  that two other maximally
correlated states $|\Phi_{\pm}\rangle={1\over\sqrt{2}}
\left[ |0,1\rangle \pm |1,0\rangle \right]$
can also be completely reconstructed on ${\cal O}_E^{(2)}$.

The  extension of ${\cal O}_E^{(2)}$ to ${\cal O}_F^{(2)}$ does not increase
the amount of  information about  the Bell states (\ref{7.9}) with
$\varphi\neq 0,\pi$. For this reason we have to consider
further extension of ${\cal O}_F^{(2)}$
 to the observation level ${\cal O}_G^{(2)}$
(see Table~\ref{table6} and Appendix A). In what follows we will show that
this is an observation level
on which {\em all}  Bell states
(\ref{7.9}) can be completely reconstructed.
To see this one has to realize two facts. Firstly,
the generalized canonical density operator $\hat{\rho}_G$ given by
Eq.(\ref{2.11}) can be expressed as a linear superposition of observables
associated with the given observation level, i.e.:
\be
\hat{\rho}_G=\frac{1}{Z_G}\exp\left(
-\sum_{\mu=x,y,z}\lambda_{\mu\mu}\hat{\sigma}_{\mu}^{(1)}\otimes
\hat{\sigma}_{\mu}^{(2)}
-\lambda_{xy}\hat{\sigma}_x^{(1)}\otimes \hat{\sigma}_y^{(2)}
-\lambda_{yx}\hat{\sigma}_y^{(1)}\otimes \hat{\sigma}_x^{(2)}
\right)
\nonumber
\\
=\frac{1}{4}\left(\hat{1}
-\sum_{\mu=x,y,z}\xi_{\mu\mu}\hat{\sigma}_{\mu}^{(1)}\otimes
\hat{\sigma}_{\mu}^{(2)}
-\xi_{xy}\hat{\sigma}_x^{(1)}\otimes \hat{\sigma}_y^{(2)}
-\xi_{yx}\hat{\sigma}_y^{(1)}\otimes \hat{\sigma}_x^{(2)}
\right), 
\label{7.14}
\ee
where the parameters $\xi_{\mu\nu}$ are functions of
the Lagrange multipliers $\lambda_{\mu\nu}$.
Secondly,
for  Bell states (\ref{7.9}) the
only observables which have nonzero expectation values are those
associated with ${\cal O}_G^{(2)}$.
 Namely,
$\l \hat{\sigma}_z^{(1)} \otimes\hat{\sigma}_z^{(2)} \r=1$,
$\l \hat{\sigma}_x^{(1)} \otimes\hat{\sigma}_x^{(2)} \r=
-\l \hat{\sigma}_y^{(1)} \otimes\hat{\sigma}_y^{(2)} \r=\cos\varphi$
and $\l \hat{\sigma}_x^{(1)} \otimes\hat{\sigma}_y^{(2)} \r=
\l \hat{\sigma}_y^{(1)} \otimes\hat{\sigma}_x^{(2)} \r=\sin\varphi$.
It means that all coefficients in  the generalized
canonical density operator $\hat\rho_G$ given by Eq. (\ref{7.8})
are uniquely determined by the
measurement, i.e., $\hat\rho_G=\hat\rho_\varphi$.

 From the above it follows that	Bell states can be completely
reconstructed on the observation level ${\cal O}_G^{(2)}$.
On the other hand, ${\cal O}_G^{(2)}$ is not the {\em minimum} observation
level on which these states can  be completely reconstructed.
The minimum set of observables which would allow us to
reconstruct  Bell states uniquely can be found
by a {\em reduction} of ${\cal O}_G^{(2)}$.
Direct inspection of a	finite number of possible reductions
reveals that  Bell states can be completely reconstructed
on those observation level which can be obtained from ${\cal O}_G^{(2)}$
when one of the observables
$\hat{\sigma}_\nu^{(1)} \otimes\hat{\sigma}_\nu^{(2)}$ ($\nu=x,y,z$)
is omitted.
As an example, let us consider the observation level
${\cal O}_H^{(2)}$ given in Table~\ref{table6} which
represents a reduction of ${\cal O}_G^{(2)}$ when  the observable
$\hat{\sigma}_z^{(1)} \otimes\hat{\sigma}_z^{(2)}$ is omitted.
Performing the Taylor series expansion of the generalized canonical
density operator $\hat\rho_H$ defined by Eq. (\ref{A.33}) one can find
that the only  new observable
$\hat{\sigma}_z^{(1)} \otimes\hat{\sigma}_z^{(2)}$
 enters the expression for the $\hat\rho_H$
as indicated in Table~\ref{table7}. The coefficient $t$ in front of
$\hat{\sigma}_z^{(1)} \otimes\hat{\sigma}_z^{(2)}$
can either be found explicitly in a closed analytical form
(see Appendix A) or can be obtained from the following variational problem.
Namely, we remind ourselves that
the expression (\ref{A.33}) for $\hat\rho_H$ helps us to identify
those unmeasured observables for which
the Jaynes principle of the maximum entropy ``predicts'' nonzero
mean values. At this stage we
still have  to find the particular value of the parameter $t$
for which the density operator $\hat\rho_H$ in Table~\ref{table7} leads to the
maximum of the von Neumann entropy. To do so
we search through the one-dimensional parametric space which is bounded as
$-1 \le t\le 1$. To be specific,
first of all, for $t\in \l -1,1 \r$  we have to exclude those operators
 which are not true density operators (i.e., any such operators which have
negative eigenvalues). Then we ``pick'' up from a physical parametric
subspace the generalized canonical density operator with the maximum von
Neumann entropy. Direct calculation for  Bell states shows that the
physical parametric subspace is reduced to an isolated ``point''
with $t=\l \hat{\sigma}_z^{(1)} \otimes\hat{\sigma}_z^{(2)} \r=1$.
Therefore we conclude that
  Bell states can completely be reconstructed on ${\cal O}_H$.
Two other minimum observation levels suitable for the complete
reconstruction of  Bell states
can be obtained by a reduction
of ${\cal O}_G^{(2)}$ when either
$\hat{\sigma}_x^{(1)} \otimes\hat{\sigma}_x^{(2)}$
or $\hat{\sigma}_y^{(1)} \otimes\hat{\sigma}_y^{(2)}$ is omitted.
On the other hand, direct inspection shows that
a reduction of ${\cal O}_G^{(2)}$ by exclusion of either
$\hat{\sigma}_x^{(1)} \otimes\hat{\sigma}_y^{(2)}$
or $\hat{\sigma}_y^{(1)} \otimes\hat{\sigma}_z^{(2)}$ leads to
an incomplete observation level with respect to  Bell states.

In what follows we  discuss briefly
two other  observation levels
${\cal O}_I^{(2)}$ and ${\cal O}_J^{(2)}$ which are defined in
Table~\ref{table6}.
The observation level ${\cal O}_I^{(2)}$ serves as an example when one
can find an analytical expression for the Taylor series expansion
of the canonical density operator $\hat\rho_I$ (see Table~\ref{table7}) in
the form (\ref{7.8}). The coefficients (functions of the original Lagrange
multipliers) in front of particular observables in Eq.(\ref{7.8})
can be identified and are given in Table~\ref{table7}. Problems do appear
when  ${\cal O}_I^{(2)}$ is extended towards ${\cal O}_J^{(2)}$.
In this case we cannot simplify the exponential expression 
for $\hat\rho_J$ and rewrite it analytically in the
form (\ref{7.8}) as a linear combination of the observables (\ref{7.7}).
In this situation one should apply the following procedure:
firstly, by performing the Taylor-series expansion of the $\hat\rho_J$
to the lowest orders one can identify
the observables with nonzero coefficients
in the form (\ref{7.8}). Namely, for $\hat\rho_J$
the additional observables
$\hat{\sigma}_z^{(1)}\otimes\hat{\sigma}_x^{(2)}$,
$\hat{\sigma}_x^{(1)}\otimes\hat{\sigma}_z^{(2)}$ and
$\hat{\sigma}_y^{(1)}\otimes\hat{\sigma}_y^{(2)}$
appear in addition to those which form ${\cal O}_H^{(2)}$ 
[see Table~\ref{table7}].
The corresponding coefficients $u,v,w\in \l -1,1 \r$ form a bounded
three--dimensional parametric space $(u,v,w)$.
In the second step one can use constructively the maximum-entropy principle
to choose within this parametric space	the density operator
with the maximum von Neumann entropy. The basic procedure is to scan
the whole three--dimensional parametric space. At the beginning,
one has to select out those density operators (i.e., those
parameters $u,v,w$) which posses negative eigenvalues and do not
represent genuine density operators. Finally,
from a remaining set of  ``physical'' density operators
which are semi--positively defined  the canonical density
operator $\hat\rho_J$ with maximum von Neumann entropy has to be chosen.
For a completeness, let us notice that for  Bell states
the observation levels ${\cal O}_I^{(2)}$ and ${\cal O}_J^{(2)}$
are equivalent to ${\cal O}_E^{(2)}$, i.e.,
$\hat\rho_I=\hat\rho_J=\hat\rho_E$.

In this section we have found  the minimum observation
levels [e.g., ${\cal O}_H^{(2)}$] which are suitable for
the  complete reconstruction  of  Bell states.
These observation levels are associated
with the measurement of two--spin
correlations
$\hat{\sigma}_x^{(1)}\otimes\hat{\sigma}_z^{(2)}$,
$\hat{\sigma}_y^{(1)}\otimes\hat{\sigma}_z^{(2)}$ and
two of the observables
$\hat{\sigma}_\nu^{(1)} \otimes\hat{\sigma}_\nu^{(2)}$
($\nu=x,y,z$).
Once this problem has been solved, it is
interesting then to find a minimum set of observables suitable for
a complete reconstruction of maximally correlated spin states systems
consisting of more than two spins--$1/2$.
In the following section we will investigate the (partial) reconstruction
of Greenberger-Horne-Zeilinger states of three spins--1/2
on various observation levels.


\subsection{Three spins--{1/2}}
\label{sec7.D}

Even though the Jaynes principle of  maximum entropy provides us
with general instructions on how to reconstruct density operators of
quantum-mechanical systems practical applications of this reconstruction
scheme may face serious difficulties. In many cases the reconstruction
scheme fails due to insurmountable technical problems (e.g. the system
of equations for Lagrange multipliers cannot be solved explicitly).
We have  illustrated these problems in the previous section
when we have discussed the reconstruction of a density operator of
two spins--1/2. Obviously, the general problem of reconstruction
of density operators describing a system composed of three spins--1/2
is much more difficult. Nevertheless a (partial) reconstruction of some
states of this system can be performed. In particular, in this section
we will discuss a reconstruction of the maximally correlated three
spin-1/2 states -- the so-called
{\em Greenberger-Horne-Zeilinger} (GHZ) state \cite{Greenberger}:
\be
|\Psi_\varphi^{(GHZ)}\r={1\over\sqrt{2}}
\left[ |1,1,1\r +\mbox{e}^{\mbox{i}\varphi} |0,0,0\r \right], \quad
\hat\rho_\varphi^{(GHZ)}=|\Psi_\varphi \r \l \Psi_\varphi |
.\label{7.15}
\ee
Our main task will be to identify, with the help of the Jaynes principle of
 maximum entropy,  the minimum observation
level on which the GHZ state can be completely reconstructed.

We start with a relatively simple observation level ${\cal O}_B^{(3)}$
such that only
{\em two}-particle correlations of the neighboring spins are measured,
i.e.
\be
{\cal O}_B^{(3)}=\{
\hat{\sigma}_z^{(1)}\otimes \hat{\sigma}_z^{(2)}\otimes \hat{I}^{(3)},
\hat{I}^{(1)}\otimes\hat{\sigma}_z^{(2)}\otimes \hat{\sigma}_z^{(3)}
\}.
\label{7.16}
\ee
The generalized density operator associated with this observation level
reads
\be
\hat{\rho}_B= &{1\over 8}& \left[
\hat{I}^{(1)}\otimes \hat{I}^{(2)}\otimes \hat{I}^{(3)}
+ \l \hat{\sigma}_z^{(1)}\otimes\hat{\sigma}_z^{(2)}\otimes \hat{I}^{(3)} \r
\hat{\sigma}_z^{(1)} \otimes \hat{\sigma}_z^{(2)} \otimes \hat{I}^{(3)}
\right. \nonumber \\
&+&
\l \hat{I}^{(1)}\otimes \hat{\sigma}_z^{(2)}\otimes\hat{\sigma}_z^{(3)}\r
\hat{I}^{(1)} \otimes \hat{\sigma}_z^{(2)}\otimes \hat{\sigma}_z^{(3)}
\nonumber \\
&\oplus& \left.
\l \hat{\sigma}_z^{(1)}\otimes\hat{\sigma}_z^{(2)}\otimes \hat{I}^{(3)} \r
\l \hat{I}^{(1)}\otimes \hat{\sigma}_z^{(2)}\otimes\hat{\sigma}_z^{(3)} \r
\hat{\sigma}_z^{(1)}\otimes \hat{I}_{2} \otimes \hat{\sigma}_z^{(3)}
\right]
.\label{7.17}
\ee
where `$\oplus$'' indicates a prediction for the unmeasured observable.
In particular, for the GHZ states (\ref{7.15}) we obtain the following
generalized canonical density operator
\be
\hat{\rho}_B^{(GHZ)}&=& {1\over 8}\left[
\hat{I}^{(1)}\otimes \hat{I}^{(2)}\otimes \hat{I}^{(3)}
\right. \nonumber \\
&+& \left.
\hat{\sigma}_z^{(1)} \otimes \hat{\sigma}_z^{(2)} \otimes \hat{I}^{(3)}
+ \hat{I}^{(1)}\otimes \hat{\sigma}_z^{(2)} \otimes \hat{\sigma}_z^{(3)}
\oplus \hat{\sigma}_z^{(1)}\otimes \hat{I}_{2} \otimes \hat{\sigma}_z^{(3)}
\right] \nonumber \\
&=& {1\over 2} |1,1,1\r \l 1,1,1| + {1\over 2} |0,0,0\r \l 0,0,0|
.\label{7.18}
\ee
The
reconstructed density operator $\hat{\rho}_B^{(GHZ)}$ describes a mixture
of three-particle states and it
does not
contain any information about the three-particle correlations associated
with the GHZ states. In other words, on ${\cal O}_B^{(3)}$
 the phase information
which plays essential role for a description of quantum entanglement
cannot be reconstructed. This is due to the fact that
 the density
operator $\hat{\rho}_B^{(GHZ)}$  is equal to the phase-averaged
GHZ density operator, i.e.
\be
\hat\rho_B^{(GHZ)}=\frac{1}{2\pi}\int_{-\pi}^{\pi}
\hat\rho_\varphi^{(GHZ)} \, d\varphi.
\label{7.19}
\ee
Because of this loss of information,
the von\,Neumann entropy of  the state $\hat\rho_B^{(GHZ)}$
is equal to $\ln 2$. We note, that when the GHZ states
are reconstructed on the observation levels
$ {\cal O}_{B'}^{(3)}=\{
\hat{\sigma}_\mu^{(1)}\otimes \hat{\sigma}_\mu^{(2)}\otimes \hat{I}^{(3)},
\hat{I}^{(1)}\otimes\hat{\sigma}_\mu^{(2)}\otimes \hat{\sigma}_\mu^{(3)}
\}$ ($\mu=x,y$),
then the corresponding reconstructed operators are again given by
Eq.(\ref{7.18}). These examples illustrate the fact that three-particle
correlation cannot be in general reconstructed via the measurement
of two-particle correlations.

To find the observation level on which the complete reconstruction
of the GHZ states can be performed we recall
the observables which may have nonzero mean values for these states.
Using abbreviations
\be
&\xi_{\mu_1 \nu_2}=\l \hat\sigma_\mu^{(1)} \otimes \hat\sigma_\nu^{(2)}
\otimes \hat{I}^{(3)} \r, \quad
\xi_{\mu_2 \nu_3}=\l \hat{I}^{(1)}\otimes \hat\sigma_\mu^{(2)}
\otimes \hat\sigma_\nu^{(3)} \r, \quad
\xi_{\mu_1 \nu_3}=\l \hat\sigma_\mu^{(1)} \otimes\hat{I}^{(2)}
\otimes \hat\sigma_\nu^{(3)} \r,& \nonumber \\
&\zeta_{\mu_1 \nu_2 \omega_3} = \l  \hat\sigma_\mu^{(1)}
\otimes \hat\sigma_\nu^{(2)} \otimes \hat\sigma_\omega^{(3)} \r,
\qquad (\mu,\nu,\omega=x,y,z), &
\label{7.20}
\ee
we find the nonzero mean values to be
\be
\xi_{z_1 z_2} &=&\xi_{z_2 z_3}=\xi_{z_1 z_3}=1, \nonumber \\
\zeta_{x_1 x_2 y_3} &=& \zeta_{y_1 x_2 x_3}=
\zeta_{x_1 y_2 x_3}= \sin\varphi,   \nonumber \\
\zeta_{y_1 y_2 x_3} &=& \zeta_{x_1 y_2 y_3}=
\zeta_{y_1 x_2 y_3}= -\cos\varphi,  \nonumber \\
\zeta_{x_1 x_2 x_3} &=& \cos\varphi, \nonumber \\
\zeta_{y_1 y_2 y_3} &=&-\sin\varphi.
\label{7.21}\ee
We  see that for arbitrary $\varphi$ there exist non--vanishing
three--particle correlations $\zeta_{\mu_1 \nu_2 \omega_3}$.
The observation level which consists
of all the observables with nonzero mean values is the complete
observation level with respect to the GHZ states.
Our task now is to reduce this set of observables to a minimum
observation level on which the GHZ states can  still be uniquely determined.
In practice it means that each observation level which is suitable
for the detection of  the existing coherence and correlations
should incorporate some of the observables with nonzero mean
values. The other observables of these observation levels
should	result as a consequence of mutual tensor products
which appear in the Taylor series expansion of the generalized canonical
density operator (\ref{2.11}). It can be seen
 by direct inspection of  the finite number of possible reductions
that the minimum set of the observables which matches
these requirements consists of two two--spin observables and
two three--spin observables. For the illustration we consider the
 observation level
\be
{\cal O}_C^{(3)}=\{
\hat{\sigma}_z^{(1)}\otimes\hat{\sigma}_z^{(2)}\otimes\hat{I}^{(3)},
\hat{I}^{(1)}\otimes\hat{\sigma}_z^{(2)}\otimes\hat{\sigma}_z^{(3)},
\hat{\sigma}_x^{(1)}\otimes\hat{\sigma}_x^{(2)}\otimes\hat{\sigma}_x^{(3)},
\hat{\sigma}_y^{(1)}\otimes\hat{\sigma}_y^{(2)}\otimes\hat{\sigma}_y^{(3)}
\}. 
\label{7.22}
\ee
In this case
the exponent $\hat{C}$ of the generalized canonical density operator
$\hat\rho_C=\exp(-\hat{C})/Z_C$  can be rewritten as
$\hat{C}=\hat{C}_1+\hat{C}_2$ with $\hat{C}_1=\gamma_{12}
\hat{\sigma}_z^{(1)}\otimes\hat{\sigma}_z^{(2)}\otimes\hat{I}^{(3)}
+\gamma_{23}
\hat{I}^{(1)}\otimes\hat{\sigma}_z^{(2)}\otimes\hat{\sigma}_z^{(3)}$
and $\hat{C}_2=\alpha
\hat{\sigma}_x^{(1)}\otimes\hat{\sigma}_x^{(2)}\otimes\hat{\sigma}_x^{(3)}
+\beta
\hat{\sigma}_y^{(1)}\otimes\hat{\sigma}_y^{(2)}\otimes\hat{\sigma}_y^{(3)}$.
The operators $\hat{C}_1$, $\hat{C}_2$ commute and further calculations
are straightforward. After some algebra the generalized density operator
$\hat\rho_C$ can be found in the form
\be
&&\hat{\rho}_C=
{1\over 8} \left[
\hat{I}^{(1)}\otimes \hat{I}^{(2)}\otimes \hat{I}^{(3)} +
\xi_{z_1 z_2} \hat{\sigma}_z^{(1)}\otimes\hat{\sigma}_z^{(2)}
\otimes\hat{I}^{(3)} +
\xi_{z_2 z_3} \hat{I}^{(1)}\otimes\hat{\sigma}_z^{(2)}
\otimes\hat{\sigma}_z^{(3)}   \right.
\label{7.23} \\	 
&& + \zeta_{x_1 x_2 x_3} \hat{\sigma}_x^{(1)}\otimes\hat{\sigma}_x^{(2)}
\otimes\hat{\sigma}_x^{(3)} +
\zeta_{y_1 y_2 y_3} \hat{\sigma}_y^{(1)}\otimes\hat{\sigma}_y^{(2)}
\otimes\hat{\sigma}_y^{(3)}
\oplus \xi_{z_1 z_2}  \xi_{z_2 z_3}
\hat{\sigma}_z^{(1)}\otimes\hat{I}^{(2)}\otimes\hat{\sigma}_z^{(3)}
\nonumber \\
&& \ominus
\zeta_{x_1 x_2 x_3} \left(
\xi_{z_1 z_2} \hat{\sigma}_y^{(1)}\otimes\hat{\sigma}_y^{(2)}\otimes
\hat{\sigma}_x^{(3)} +
\xi_{z_2 z_3} \hat{\sigma}_x^{(1)}\otimes\hat{\sigma}_y^{(2)}\otimes
\hat{\sigma}_y^{(3)} +
\xi_{z_1 z_2} \xi_{z_2 z_3}
\hat{\sigma}_y^{(1)}\otimes\hat{\sigma}_x^{(2)}\otimes
\hat{\sigma}_y^{(3)}
\right) \nonumber\\
&& \ominus \left.
\zeta_{y_1 y_2 y_3} \left(
\xi_{z_1 z_2} \hat{\sigma}_x^{(1)}\otimes\hat{\sigma}_x^{(2)}\otimes
\hat{\sigma}_y^{(3)} +
\xi_{z_2 z_3} \hat{\sigma}_y^{(1)}\otimes\hat{\sigma}_x^{(2)}\otimes
\hat{\sigma}_x^{(3)} +
\xi_{z_1 z_2} \xi_{z_2 z_3}
\hat{\sigma}_x^{(1)}\otimes\hat{\sigma}_y^{(2)}\otimes
\hat{\sigma}_x^{(3)}
\right) \right] \nonumber
.\ee
For the GHZ states 
the von\,Neumann entropy of the generalized canonical density operator
$\hat{\rho}_C$ is equal to zero, from which it follows that
$\hat{\rho}_C=\hat{\rho}^{(GHZ)}_\varphi$ [see Eq.(\ref{7.15})], i.e.,
the GHZ states can be completely reconstructed on ${\cal O}_C^{(3)}$.
Moreover, the  observation level ${\cal O}_C$ represents
the {\em minimum} set of observables for complete determination
of the GHZ states.

\section{QUANTUM BAYESIAN INFERENCE}
\label{sec8}

The {\em exact} meanvalue of an arbitrary observable
can only be obtained when a very large (in principle, infinite) number
of measurements on
individual elements of an ensemble is performed. On the other hand,
it is a very legitimate question to ask ``What is the best {\em a posteriori}
estimation of a quantum state when  a measurement is performed on a
{\em finite} (arbitrarily small)  number of elements of the ensemble?''~.
To estimate the state of the system based on an incomplete
set of data, one has to utilize more powerful estimation schemes
such as the quantum Bayesian inference.

The general idea
of the Bayesian reconstruction scheme (see for instance \cite{Derka96})
is based on manipulations with
probability
distributions in parametric state spaces $\Omega$ and $A$
of the measured system and the
measuring apparatus, respectively.
The quantum Bayesian method  as discussed in the literature
\cite{Helstrom76,Holevo82,Jones91}
is based on the assumption
that the reconstructed system is in a pure
state described by a state vector $|\Psi\rangle$,
or equivalently by a pure-state density
operator $\hat{\rho}=|\Psi\rangle\langle\Psi|$.
The manifold of all pure states is a continuum which we denote as
$\Omega$.
The state space $A$ of reading states
of a measuring	apparatus associated with the observable $\hat{O}$
is assumed to be discrete.
These states are intrinsically related to the projectors
$\hat{P}_{{\lambda_i,\hat O}}$, where
$\lambda_i$ are the eigenvalues of the observable $\hat O$.

The Bayesian reconstruction scheme
is formulated as a three-step inversion procedure:\newline
{\bf (1)} As a result of a measurement a conditional probability
\begin{eqnarray}
p(\hat O,\lambda_i \vert \hat{\rho})={\rm Tr}
\left(\hat{P}_{\lambda_i,\hat O}\hat{\rho}\right),
\label{8.1}
\end{eqnarray}
on the discrete space $A$ is obtained.
This conditional probability distribution  specifies the probability of finding
the result $\lambda_i$ if the measured system is in a particular
state $\hat{\rho}$.\newline
{\bf (2)} To perform the second step of the inversion procedure
we have to specify an {\em a priori} distribution
$p_0(\hat{\rho})$ defined on the space	$\Omega$. This distribution
describes our initial knowledge concerning the measured system. Using the
conditional probability distribution
$p(\hat O,\lambda_i \vert \hat{\rho})$ and the {\em a priori} distribution
$p_0(\hat{\rho})$ we can define the {\em joint} probability distribution
$p(\hat O,\lambda_i ; \hat{\rho})$
\begin{eqnarray}
p(\hat O,\lambda_i ; \hat{\rho})= p(\hat O,\lambda_i \vert \hat{\rho})
p_0(\hat{\rho}),
\label{8.2}
\end{eqnarray}
on the space $\Omega\otimes A$. We note that if no initial information
about the measured system is known, then the prior $p_0(\hat{\rho})$
has to be assumed to be constant (this assumption is related to the
Laplace principle of indifference \cite{Jeffreys}).\newline
{\bf (3)} The final step of the Bayesian reconstruction is based on the
well known Bayes rule $p(x|y)p(y)=p(x;y)=p(y|x)p(x)$, with the help
of which we find the conditional probability
$p(\hat{\rho} \vert \hat O,\lambda_i)$ on the state space $\Omega$:
\begin{eqnarray}
p(\hat{\rho} \vert \hat O,\lambda_i)=
{{p(\hat O,\lambda_i ,\hat{\rho})}\over
{\int _\Omega p(\hat O,\lambda_i ,\hat{\rho}) d_\Omega}},
\label{8.3}
\end{eqnarray}
from which the reconstructed density operator can be obtained
[see Eq.(\ref{8.4})].

In the case of the repeated $N$-trial measurement, the reconstruction
scheme consists of an
iterative utilization of the three-step procedure as described above.
 After the $N$-th measurement we
use as an input for the prior distribution the conditional
probability distribution given by the output of the $(N-1)$-st measurement.
However, we can equivalently define the $N$-trial measurement conditional
probability  $p(\{\ \}_{_N}\vert \hat{\rho})=
\prod_{i=1}^N p(\hat O_i,\lambda_j \vert \hat{\rho})$ 
[the so-called likelihood function, which is also denoted as ${\cal
L}(\hat{\rho})$ ]
and
applying the three-step procedure
just once to obtain the
reconstructed density operator
\begin{eqnarray}
\hat{\rho}(\{ \ \}_{N})={\int _\Omega p(\hat{\rho}\vert\{\ \}_{N})
\hat{\rho} d_\Omega \over
 {\int _\Omega p( \hat{\rho}\vert\{\ \}_{N}) d_\Omega}},
\label{8.4}
\end{eqnarray}
where $\hat\rho$ in the r.h.s. of Eq.(\ref{8.4}) is a properly parameterized
density operator in the state space $\Omega$. We note that in general,
the reconstructed 
density operator (\ref{8.4}) corresponds to a mixed state inspite of
the fact that an {\em a priori} assumption is that the system is in a pure
state. This deviation from the purity (let say expressed in terms
of the von Neumann entropy) may serve as a measure of fidelity of the
estimatimation procedure\footnote{We note that Hradil \cite{Hradil}
has recently proposed another statistical quantum-state-reconstruction
method related to the Bayesian scheme considered in this Section.
His method is based on the maximization of the likelihood function
${\cal L} (\hat{\rho})$.}

At this point we should mention one essential problem in the Bayesian
reconstruction scheme, which is the determination of the integration measure
$d_\Omega$.
The integration measure has to be invariant under
unitary transformations in the space $\Omega$. This
requirement uniquely determines the form of the measure.
However, this is no longer valid when
$\Omega $ is considered to be
a  space of mixed states formed by all convex
combinations of elements of the original pure state space
$\Omega$. Although the Bayesian procedure itself does not require any special
conditions imposed on the space $\Omega$, the ambiguity
in determination of the integration measure is
the main obstacle in generalization of the Bayesian inference scheme
for  reconstruction of {\em a priori} impure quantum states.
We will show in later in this Section 
that this problem can be solved with the
help of a purification ansatz. We will also discuss in detail
how to apply the quantum Bayesian inference for a reconstruction
of states of a spin-1/2 when just a finite number of elements of an ensemble
have been measured. Before we do this 
we will analyze the limit of large  number of measurements.

\subsection{Bayesian inference in limit of infinite number of measurements}
\label{sec8.A}

The explicit evaluation of an {\em a posteriori} estimation of the
density operator $\hat{\rho}\{\ \}_{N}$ is significantly limited
by technical difficulties when integration over parametric space
is performed [see Eq.(\ref{8.4})].
Even for the simplest quantum systems  and for	a relatively small number
 of measurements,
the reconstruction procedure can present technically insurmountable
 problems.

On the other hand let us assume that the number of measurements of
observables  $\hat O_i$
approaches  infinity (i.e.  $N\rightarrow \infty$).
It is clear that in this case the
mean values of all projectors $\langle \hat{P}_{\lambda_j,\hat O_i}\rangle$
associated with the observables $\hat O_i$
are {\em precisely} known (measured): i.e.
\begin{eqnarray}
\langle \hat{P}_{\lambda_j,\hat O_i}\rangle=\alpha _j^i,
\label{8.5}
\end{eqnarray}
where $\sum_j \alpha_j^i=1$. In this case the integral in the
right-hand side of Eq.(\ref{8.4}) can be significantly
simplified  with the help of  the following lemma:
\medskip

{\it Lemma:} \newline
Let us define the integral expression
\begin{eqnarray}
I(\alpha_1,\dots,\alpha_{n-1})\equiv
\int_0^1 dx_1 \int_0^{y_2}dx_2\dots
\int_0^{y_{n-1}}
dx_{n-1}\,
F(x_1,\dots,x_{n-1}|\, \alpha_1,\dots,\alpha_{n-1}).
\label{8.6}
\end{eqnarray}
where
\begin{eqnarray}
F(x_1,\dots,x_{n-1}|\, \alpha_1,\dots,\alpha_{n-1})
={1\over B}
x_1^{\alpha_1N}x_2^{\alpha_2N}\dots x_{n-1}^{\alpha_{n-1}N}
(1-x_1\dots -x_{n-1})^{\alpha_{n}N}.
\label{8.7}
\end{eqnarray}
and $\alpha_i$ satisfy condition $\sum_i^n \alpha_i=1$. The integration
boundaries $y_k$ are given by relations:
\begin{eqnarray}
y_k = 1 -\sum_{j=1}^{k-1} x_j;\qquad k=2,\dots, n-1.
\label{8.8}
\end{eqnarray}
and  $B$ equals to
the product of Beta functions $B(x,y)$:
\begin{eqnarray}
B \equiv B (a_{n}+1,a_{n-1}+1)
B (a_{n}+a_{n-1}+1,a_{n-2}+2)\dots
B (a_{n}+a_{n-1}\dots a_2+1,a_1+n-1).
\label{8.9}
\end{eqnarray}

\noindent
{\it i.}
The function $F(x_1,\dots,x_{n-1}|\, \alpha_1,\dots,\alpha_{n-1})$
in the integral (\ref{8.6}) is a normalized probability
distribution in the $(n-1)$-dimensional volume given by integration
boundaries.

\noindent
{\it ii.}
For $N\rightarrow \infty$,
this probability distribution has the
following properties:
\begin{eqnarray}
\langle x_i\rangle \rightarrow \alpha_i\qquad
\langle x_i^2\rangle \rightarrow \alpha_i^2\qquad
i=1,2,3,\dots,n-1,
\label{8.10}
\end{eqnarray}
i.e., this probability density tends to the product of delta
functions:
\begin{eqnarray}
\lim_{N\rightarrow\infty}
F(x_1,\dots,x_{n-1}|\, \alpha_1,\dots,\alpha_{n-1})=
\delta (x_1- \alpha_1)\delta (x_2-\alpha_2)\dots \delta
(x_{n-1}-\alpha_{n-1}).
\label{8.11}
\end{eqnarray}

{\it Proof:}\newline
Statement {\it i.} can be derived by the successive application
of the equation [see for example \cite{Gradstein}, Eqs.(3.191)]
\begin{eqnarray}
\int_0^u x^{\nu-1}(u-x)^{\mu-1}dx=u^{\mu +\nu -1}B(\mu, \nu ).
\label{8.12}
\end{eqnarray}

\noindent
Statement {\it ii.} can be obtained as a result  of  straightforward
calculation  of limits of certain
expressions containing Beta functions with integer-number arguments.
In our calculations we have used
the identity
\begin{eqnarray}
\frac{B(n+1,m)}{B(n,m)}=\frac{n}{n+m},
\label{8.13}
\end{eqnarray}
which is satisfied by Beta functions with integer-number arguments.

\subsection{Conditional density distribution}
\label{sec8.B}
Let us start with the expression for  conditional  probability distribution
$p(\{\ \}_{_{N}}\vert \hat\rho )$
for the $N$-trial measurement of a set of observables $\hat O_i$.
If we assume that the number of measurements of each observable
$\hat O_i$ goes to  infinity then we can write:
\begin{eqnarray}
p(\{\ \}_{_{N\rightarrow\infty}}\vert \hat\rho )=\lim_{N\rightarrow\infty}
\prod_i \Big\lbrack
\prod_{j=1}^{n_i} {\rm Tr}\left(\hat P_{_{\lambda_j,\hat O_i}}
\hat\rho\right)^{\alpha _j^iN}\Big\rbrack.
\label{8.14}
\end{eqnarray}
The first product on the right-hand side (r.h.s.)
 of Eq.(\ref{8.14})
is associated with each measured observable $\hat O_i$ on
a given  observation level.
The second  product
runs over   eigenvalues $n_i$ of each observable $\hat O_i$.

In what follows we formally rewrite the  r.h.s. of Eq.(\ref{8.14}):
we insert in it a set of $\delta$-functions and we perform the following
 integration
\begin{eqnarray}\nonumber
p(\{\ \}_{_{N\rightarrow\infty}}\vert \hat\rho )=\prod_i \left\{
\int_0^1dx^i_1 \int_0^{y_{2}^i} dx^i_2	   \dots
\int_0^{y_{n_i-1}^i}
dx^i_{n_i-1}\right.
 \delta\left[x^i_1-{\rm Tr}\left(\hat P_{_{\lambda_1,\hat O_i}}
\hat \rho\right)
\right]\dots
\end{eqnarray}
\begin{eqnarray}
\times\left.
\delta\left[x^i_{n_i-1}-{\rm Tr}\left(P_{_{\lambda_{n_i-1},\hat O_i}}
\rho\right)
\right]
\prod_{j=1}^{n_i-1}(x^i_j)^{\alpha _j^iN}\ \ \ (1-x_1^i\dots x^i_{n_i-1})^
{\alpha^i_{n_i}N}\right\}.
\label{8.15}
\end{eqnarray}
In Eq.(\ref{8.15}) we perform an integration over a volume
determined by the integration boundaries $y_k^i$ [see
Eq.(\ref{8.8})], i.e., due to the condition
$\sum^{n_i}_{j=1}{\rm Tr}(\hat{P}_{\lambda_j,\hat O_i}\hat{\rho})=1$,
there is no need to perform integration from $-\infty$ to $\infty$.

At this point we utilize our {\em Lemma}. To be specific,  firstly
we separate in	Eq.(\ref{8.15}) the term, which corresponds to the
 function $I$ given by Eq.(\ref{8.6}). Then we replace this term by
its limit expression (\ref{8.11}). After a
straightforward integration over variables $x^i_j$
we finally obtain an explicit expression for the conditional
probability
$p( \hat\rho\vert\{\ \}_{_{N\rightarrow\infty}} )$
which we insert into Eq.(\ref{8.4}), from which we obtain the  expression
for  an {\em a posteriori} estimation of the density operator
$\hat\rho(\{\ \}_{N\rightarrow \infty})$ on the given observation level:
\begin{eqnarray}
\hat{\rho}(\{\ \}_{_{N\rightarrow \infty}})={1 \over {\cal N}}
\int_\Omega \prod_i \left\{
\prod_{j=1}^{n_i-1}\delta \left[{\rm Tr}
\left(\hat{P}_{\lambda_j,\hat O_i}\hat{\rho}\right)-\alpha _j^i\right]
\right\}
\hat{\rho} d_\Omega.
\label{8.16}
\end{eqnarray}
Here ${\cal N}$ is a normalization constant determined
by the condition
${\rm Tr}\left[\hat{\rho}(\{\ \}_{_{N\rightarrow \infty}})\right]=1$.

The interpretation of Eq.(\ref{8.16}) is straightforward.
The reconstructed density operator is equal to the sum of
equally-weighted
pure-state density operators on the manifold $\Omega $, which  satisfy
the conditions given by Eq.(\ref{8.5})
[these conditions
are guaranteed by the presence of the $\delta$-functions in
the r.h.s. of Eq.(\ref{8.16})].
In terms of statistical physics Eq.(\ref{8.16}) can be interpreted as
 an averaging over the generalized microcanonical
ensemble of those {\em pure} states which
satisfy the conditions on the mean values of
the measured observables. Consequently, Eq.(\ref{8.16}) represents the
principle of the ``maximum entropy'' associated with
the generalized microcanonical ensemble which fulfills  the
constraint (\ref{8.5}).

\subsection{Bayesian reconstruction of impure states}
\label{sec8.C}

In classical statistical physics a
mixed state is interpreted as a statistical average over an ensemble in
which any individual realizations is in a pure state. This is also true
in quantum physics, but here
a mixture can also be interpreted as a state of a
quantum system, which  can not be completely
described in terms of its own Hilbert space. That is the system under
consideration	is a nontrivial part
of a larger  quantum system. When we say nontrivial, we mean that
the system under consideration is quantum-mechanically entangled
\cite{Peres93} (see also \cite{Peres96})
with the other parts of the composite system.
Due to the lack of information about
other parts of this complex system, the description of the subsystem
is possible  only in terms of mixtures.

Let assume that the quantum system $P$ is entangled with
another quantum system	$R$ (a reservoir).
Let us assume that the composed system $S$ ($S=P\times R$) itself
is in a pure state $\vert \Psi\rangle$. The density operator $\hat{\rho}_P$ of
the subsystem $P$ is then obtained via tracing over the reservoir degrees of
freedom:
\begin{eqnarray}
\hat{\rho}_P={\rm Tr}_R\,\left[\hat{\rho}_S\right];\qquad
\hat{\rho}_S =|\Psi\rangle\langle \Psi| .
\label{8.17}
\end{eqnarray}
Once the system $S$ is in a pure state, then we can determine
an invariant integration measure on the state space of the composite system
$S$ and then we  can safely apply
the Bayesian reconstruction scheme as described in Section III.
The reconstruction itself is based only on data associated with
measurements performed on the system $P$. When the density operator
$\hat{\rho}_S$ is {\em a posteriori} estimated, then by tracing
over the reservoir degrees of freedom, we obtain the {\em a posteriori}
estimated
density operator $\hat{\rho}_P$ for the system $P$ (with no {\em a priori}
constraint on the purity of the state of the system $P$). These arguments
are intrinsically related
to the ``purification'' ansatz as proposed by Uhlmann \cite{Uhlmann}.

To make our reconstruction scheme for impure states consistent,
we have to chose the reservoir $R$ uniquely. This can be
done with the help of the Schmidt theorem (see Ref.\cite{Peres93,Ekert95})
from which it follows that if the composite system $S$ is in a pure
state $|\Psi\rangle$ then its state vector can be written in the form:
\begin{eqnarray}
|\Psi\rangle=\sum_{i=1}^M c_i \vert\alpha_i\rangle_{_P}\otimes\vert
\beta_i\rangle_{_R},
\label{8.18}
\end{eqnarray}
where $\vert\alpha_i\rangle_{_P}$ and $\vert\beta_i\rangle_{_R}$ are
elements from two specific
orthonormalized bases associated with the subsystems $P$ and $R$,
respectively,
and $c_i$ are appropriate complex  numbers satisfying the
normalization condition  $\sum \vert c_i \vert^2=1$.
The maximal index of summation ($M$) in Eq.(\ref{8.18}) is given by the
dimensionality of the Hilbert space of the system $P$.
In other words, when we apply the Bayesian method  to the case of impure
states
of $M$-level system, it is sufficient to ``couple'' this system  to an
$M$-dimensional ``reservoir''.  In this case the dimensionality of the
Hilbert space of the composite system is $2M$. Using the standard techniques
(see Appendix B) we can then evaluate the invariant integration  measure
on the manifold of pure states and we can apply the quantum Bayesian
inference as discussed above. We stress once again, that using the purification
procedure we have determined the {\em invariant} integration measure on the
space of {\em pure} states of the composite  system.

Concluding this Section  we note that  there also exists
another approach to the problem of the integration measure
on the	space of {\em impure} states. Namely, Braunstein and Caves
\cite{Braunstein} used statistical distinguishability between
neighboring quantum states to define
the Bures metric \cite{Bures} on the space of all (pure and mixed)
states of the original system $S$ (see also recent work
by Slater \cite{Slater}).
The two approaches differ conceptually in understanding what is an
impure quantum-mechanical state. That is, in our approach we assume
that impurity results as a consequence of the fact that the
system under consideration is entangled with some other system.
The other approach  accepts the possibility that an isolated quantum system
can be in a statistically mixed state (we will not discuss consequences
of these two conceptually different approaches here, but this problem
definitely deserves due attention).

\section{RECONSTRUCTION OF SPIN STATES VIA BAYESIAN INFERENCE}
\label{sec9}
We start this
section with the Bayesian reconstruction of spin-1/2
states on various observation levels. That is, we  investigate
how the best {\em a posteriori} estimation of the density operator
of the spin-1/2 system based on an incomplete set of data (in this case
the exact mean values of the spin observables are not available)
can be obtained. We have already stressed the fact that the
Bayesian inference scheme as introduced by Jones \cite{Jones91}
is suitable only for pure states. This means that the completely
reconstructed density operator has to fulfill the purity condition
\begin{eqnarray}
|\langle \hat{\sigma}_x\rangle|^2+
|\langle \hat{\sigma}_y\rangle|^2+
|\langle \hat{\sigma}_z\rangle|^2=1.
\label{8.18a}
\end{eqnarray}

We start our example with a definition of the parametric state space
associated with the spin-1/2.
The rigorous way  to determine	this parametric state space $\Omega $
is   based on  the diffeomorphism between $\Omega $ and the quotient space
$\ ^{SU(n)}|_{U(n-1)}$, where
$n$ is the dimensionality of the  Hilbert space of the
measured quantum system. In a particular case of the spin-1/2 we work with
the commutative group $U(1)$ and the construction of $\Omega$ is very simple.
The space $\Omega$ can be mapped on to the Poincar\'{e} sphere
and the parameterized density operator (i.e. the point on the Poincar\'{e}
sphere) is given by Eq.(\ref{7.2}).
The topology of the Poincar\'{e} sphere determines also the integration measure
for which we have  $d_{_\Omega}=\sin \theta\, d\theta d\phi$
(for more details see Appendix B).

The observables
associated with the spin-1/2 are  spin projections for three
orthogonal directions represented by Hermitian operators 
$\hat{s}_j=\hat{\sigma}_j/2$.
These observables
have spectra equal to  $\pm{1\over2}$. In what follows
we  distinguish between these two possible measurement results
by the sign, i.e.  $s=\pm 1$. The projectors $\hat{P}_{s,\hat{s}_i}$
on to the corresponding eigenvectors are
\begin{eqnarray}
\hat{P}_{s,\hat{s}_i}={{\hat1+s\hat{\sigma}_i}\over2};\qquad i=x,y,z \, ,
\label{9.1}
\end{eqnarray}
and the  conditional probabilities associated with this kind of
measurement  can be written as
\begin{eqnarray}
p(s,\hat{s}_i \vert \hat{\rho}(\theta,\phi))={{1+ s\,r_i}\over 2};
\qquad i=x,y,z,
\label{9.2}
\end{eqnarray}
where we use the parameterization $\hat{\rho}(\theta,\phi)=(\hat{1}+
\vec{r}\hat{\vec{\sigma}})/2$ [see Eq.(\ref{7.2})].
Now using the procedure described in Section~\ref{sec8}, we can construct an
{\em a posteriori} estimation of the  density operator $\hat{\rho}(\{\,\}_N)$
based on a given sequence of measurement outcomes on different observation
levels.

\subsection{Estimation based on results of fictitious measurements}
\label{sec9.A}

In Table~\ref{table8} we present results of an {\em a posteriori} estimation
of density operators based on data obtained from
``experiments'' performed with three Stern-Gerlach devices
oriented along the  axes $x$, $y$, and $z$.  We  first
discuss in detail reconstruction of a single spin-1/2 state
under the {\em a priori} assumption that the system is in a pure
state.
\begin{table}
\begin{center}
\epsfig{file=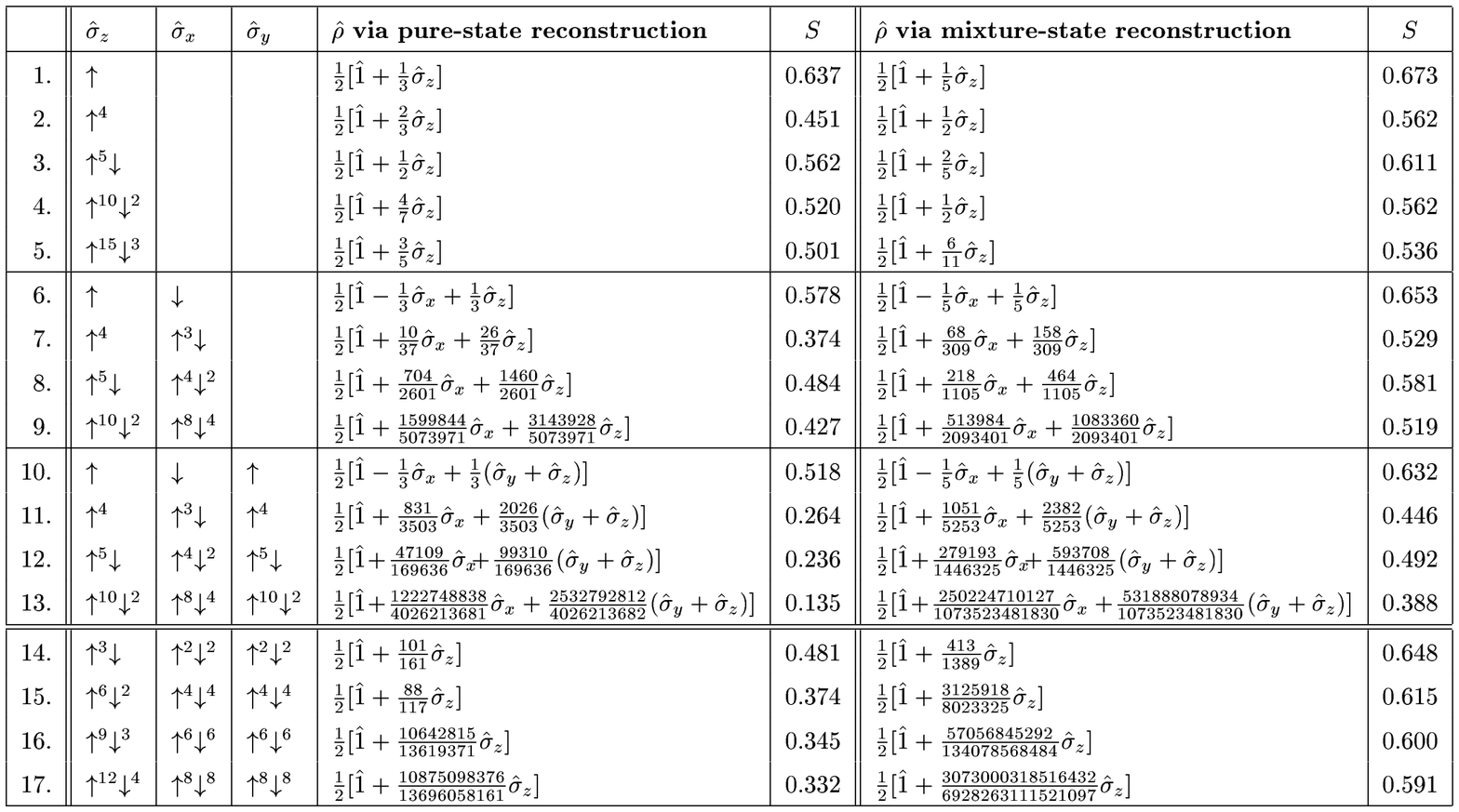, angle=0, width=16cm}
\end{center}
\caption{
Results of {\em a posterior} Bayesian estimation of
density operators of the spin-1/2 are presented for two different cases:
{\bf (1)} when it is {\em a priori} assumed that the spin is in a pure state
and {\bf (2)} when no {\em a priori} constraint on the state is imposed. In
this second case the generalized Bayesian scheme has been applied.
We also present values of von\,Neumann entropy [see Eq.(6.13)]
associated with the
given estimated density operator. In the case of a reconstruction of
pure states, the value of the von\,Neumann entropy reflects the fidelity
of the estimation.
}
\label{table8}
\end{table}

\subsubsection{Observation level  ${\cal O}_A^{(1)}=\{\hat s_z \}$}
\label{sec9.A.1}
The first five lines
in Table~\ref{table8} describe results of a fictitious measurement
of the spin component $\hat s_z$ and the corresponding
estimated density operators. In particular, let us assume that
just one detection event (spin ``up'', i.e. $\uparrow$) is registered
in the given Stern-Gerlach apparatus (associated with the measurement
of $\hat s_z$). Taking into account the parameterization of the single
spin-1/2 density operator expressed by Eq.(\ref{7.2}) we
find for the corresponding conditional probability distribution
$p(s,\hat{s}_i \vert \hat{\rho}(\theta,\phi))$ (\ref{9.2})
the expression
\begin{eqnarray}
p(s,\hat{s}_i \vert \hat{\rho}(\theta,\phi))={{1+ \cos\theta}\over 2}.
\label{9.3}
\end{eqnarray}
Using Eq.(\ref{8.4}) we can express the estimated density operator
based on the registration of just one result (spin ``up'') as
\begin{eqnarray}
\hat\rho=\frac{1}{8\pi}\int_0^\pi \sin\theta \, d\theta \int_0^{2\pi}\,d\phi
(1+\cos\theta)(\hat 1 + \sin\theta \cos\phi \hat\sigma_x
+ \sin\theta \sin\phi \hat\sigma_y + \cos\theta \sigma_z)=
\frac{1}{2}\left(\hat 1 + \frac{1}{3}\hat\sigma_z \right).
\label{9.4}
\end{eqnarray}
We stress that we started our estimation procedure with an a priori
assumption  that the
measured system is in a {\em pure} state, for which the
von\,Neumann entropy $S$ (\ref{7.4}) has to be equal to zero.
But the estimated
density operator (\ref{9.4}) describes a statistical mixture
with the von\,Neumann entropy $S\simeq 0.637$ (see Table~\ref{table8}). 
There is no contradiction here.
In the reconstruction of pure states,
a nonzero value of the von\,Neumann entropy of the estimated density operator
reflects the fidelity with
which the reconstruction is performed. That is, before any measurement
is performed, the ``estimated'' density operator is $\hat\rho=\hat 1/2$,
for which the von\,Neumann entropy takes the maximal value
 $S=\ln 2 \simeq 0.693$.  As soon as the first measurement is performed,
some information about the state of the system is acquired, which
is reflected by the decrease of the entropy and a better estimation
of the density operator. The estimated density operator is
expressed as a statistical mixture because it
is equal to a specifically weighted sum  of a set of {\em pure} states
[see the reconstruction formula (\ref{8.4})] which also
reflects our incomplete knowledge about the state of the measured
system. Obviously, the more measurements we perform, the better the estimation
can be performed (compare lines 2--5 in Table~\ref{table8}). Nevertheless, we
have to stress that the von\,Neumann entropy is not a monotonically
decreasing function of a number of measurements. To be specific, in the
case when just a small number of measurements is performed, the estimation
is very sensitive with respect to the outcome of any additional measurement.
Comparing the lines 2 and 3 in Table~\ref{table8},
we see that the entropy ``locally'' increases in spite of the fact that
more measurements are performed.
Nevertheless,
in the limit of large number of measurements, the entropy approaches
its minimum possible value associated with a given measurement.
Providing the quorum of observables is measured, the entropy tends
to zero and the state is completely reconstructed.

In general,
increasing  the number of measurements	improves
the {\em a posterior} estimation of the density operator on the
given observation level (see lines 2--5 in Table~\ref{table8}).
Using the general results of Section~\ref{sec8} we can evaluate the
{\em a posteriori} estimation of the density operator of
the spin-1/2 system on the observation level ${\cal O}_A^{(1)}$
in the limit of infinite number of measurements of the spin component
$\hat s_z$. We note, that
in this case, when observable has only two eigenvalues,
the information obtained in the spectral distribution (\ref{8.5})
is equivalently given only by  the mean value of this observable.
Once we know the spectral distribution
Eq.(\ref{8.5}) corresponding to the
measurement of the spin projection $\hat s_z $ of single spin-1/2, then
with the help of Eq.(\ref{8.16}) we can express
the reconstructed density operator as
\begin{eqnarray}
\hat\rho=\frac{1}{\cal N}
\int_0^{2\pi}\!\!d\phi\!\!\int_0^\pi\!\!\sin\theta d\theta
\ \ \delta(\langle \hat \sigma_z \rangle -\cos\theta)
(\hat 1 + \sin\theta\cos\phi\,
\hat\sigma_x + \sin\theta\sin\phi\, \hat\sigma_y +
\cos\theta\, \hat\sigma_z),
\label{9.5}
\end{eqnarray}
where ${\cal N}$ is the normalization constant such that ${\rm Tr}\hat\rho =1$.
Integration over the variable $\phi$ in Eq.(\ref{9.5})
cancels all terms in front of the operators
 $\hat\sigma_x$ and $\hat\sigma_y$ and we obtain
\begin{eqnarray}
\hat\rho=\frac{1}{\cal N}
\int_0^\pi\sin\theta d\theta
\ \ \delta(\langle \hat \sigma_z \rangle -\cos\theta)
(\hat 1 +\cos\theta\,  \hat\sigma_z).
\label{9.6}
\end{eqnarray}
The right hand side of this equation suggests a simple geometrical
interpretation of the quantum Bayesian inference in the limit
of infinite number of measurements.
Namely,
the density operator (\ref{9.6}) can be understood as an equally-weighted
average of all {\em pure} states with the same (i.e., measured) mean value of
the operator $\hat s_z$. These states are represented as points on a circle on
the	Poincar\'{e} sphere.
When we perform integration over $\theta$ in  Eq.(\ref{9.6}) we obtain
the final expression
\begin{eqnarray}
\hat\rho=\frac{1}{2}\left(
\hat 1 + \langle \hat \sigma_z \rangle \hat\sigma_z\right).
\label{9.7}
\end{eqnarray}
for the density operator on the given observation level.
Formally this is the same density operator as that reconstructed
with the help of the Jaynes principle [see Tab.~\ref{table5}]. But there is
a difference: the formula (\ref{9.7}) is obtained as a result of
averaging of the generalized {\em microcanonical} ensemble of {\em pure}
states, while the reconstruction via the MaxEnt principle is based
on an averaging over the generalized {\em grand canonical} ensemble
of all states. The two reconstruction schemes
differ by the {\em a priori} assumptions about the possible states
of the measured system. As we will see later,
these different assumptions result in different
estimations (see below).

\subsubsection{Observation level  ${\cal O}_B^{(1)}=\{\hat s_z, \hat s_x \}$}
\label{sec9.A.2}

The results of a numerical reconstruction of the density operator
of the spin-1/2 based on the measurement of two spin components
$\hat s_z$ and $\hat s_x$ are presented in Table~\ref{table8} (lines 6--9).
The lines 1--4 and 6--9 describe estimations
based on the same data for the $\hat s_z$ measurement, but
they differ in the data for the $\hat s_x$ measurement. That is, the lines
1--4 describe the situation for which no results for $\hat s_x$
are available, while lines 6--9 describe the situation with
specific outcomes for the $\hat s_x$ measurements.
Comparing these two cases (i.e., if we compare the values of the von\,Neumann
entropy for pairs of lines $\{x,x+5\};~~x=1,2,3,4$)
we see
that any measurement performed	on the
additional observable ($\hat s_x$) can only improve
our  estimation based on the measurement of the original observable
($\hat s_z$).

In the limit of infinite number of measurements, when
 we have information about the spectral distribution corresponding
to measurement of spin projections $\hat s_x, \hat s_z$ the particular form of
Eq.(\ref{8.16}) reads
\begin{eqnarray}
\hat\rho=\frac{1}{\cal N}
\int_0^{2\pi}\!\!d\phi\!\!\int_0^\pi\!\!\sin\theta d\theta
\ \ \delta(\langle \hat \sigma_z \rangle -\cos\theta)
\delta(\langle \hat \sigma_x \rangle -\sin\theta\cos\phi)
(\hat 1 + \sin\theta\cos\phi
\hat\sigma_x + \sin\theta\sin\phi\, \hat\sigma_y +
\cos\theta\, \hat\sigma_z).
\label{9.8}
\end{eqnarray}
As seen from the right-hand side of Eq.(\ref{9.8})
in this case the reconstructed density operator is represented
by an equally weighted sum of points given by an intersection of
two circles lying on the Poincar\'{e} sphere. These two circles
are specified by the two equations
$\langle \hat \sigma_z \rangle =\cos\theta$ and
$\langle \hat \sigma_x \rangle =\sin\theta\cos\phi$.

With the help of the  identity
\begin{eqnarray}
\delta(f(x))=\sum_{_{x_0,f(x_0)=0}}{\delta(x-x_0) \over |f^{'}(x_0)|},
\label{9.9}
\end{eqnarray}
we can perform the integration over  $\phi$ in Eq.(\ref{9.8}) and
obtain
\begin{eqnarray}
\rho=\frac{1}{\cal N}
\int_{\cal L}
d\theta  \sum_{\phi_0}
{\sin \theta \over \vert \sin \theta \sin \phi_0 \vert}
\delta(\langle \hat \sigma_z \rangle -\cos\theta)
(\hat 1 +  \langle \hat \sigma_x \rangle\, \hat\sigma_x +
\sin\theta\sin\phi_0\, \hat\sigma_y +
\cos\theta\, \hat\sigma_z).
\label{9.10}
\end{eqnarray}
The integration boundaries ${\cal L}$ on the right-hand side of Eq.(\ref{9.10})
are defined as
\begin{eqnarray}
{\cal L} := \, 0\leq \theta \leq \pi ~~~{\mbox and} ~~
\vert \sin\theta \vert \geq \vert \langle \hat
\sigma_x\rangle\vert.
\label{9.11}
\end{eqnarray}
The sum on the right-hand side of Eq.(\ref{9.10}) refers to two values
of the parameter $\phi_0$ which fulfill the condition
$\cos \phi_0=\langle \hat \sigma_x \rangle/\sin \theta$. We note that the
function in front of the operator
 $\hat\sigma_y$ disappears due to the fact that it is proportional to
 $\sin \phi_0 /\vert \sin \phi_0 \vert$, which is an odd function of $\phi_0$.
After we perform the  integration over $\theta$ we obtain
\begin{eqnarray}
\hat\rho=\frac{1}{2}\left(
 \hat 1 + \langle \hat \sigma_x \rangle\hat\sigma_x +
\langle \hat \sigma_z \rangle\hat\sigma_z\right).
\label{9.12}
\end{eqnarray}
What we see again is that in the limit of a large number of measurements
the Bayesian inference formally gives us the same result as the
Jaynes principle of  maximum entropy [see Tab.~\ref{table5}].

\subsubsection{Observation level  ${\cal O}_C^{(1)}=\{\hat s_z, \hat s_x ,
\hat s_y \}$}
\label{sec9.A.3}

Further extension of the observation level ${\cal O}_B^{(1)}$
leads us to the complete observation level, when all three spin
components $\hat s_x, \hat s_y$ and $\hat s_z$ of the spin-1/2
are measured. Results of the numerical reconstruction are presented
in Table~\ref{table8} (lines 10--13). Now we compare the {\em a posteriori}
estimation of density operators based on data presented in lines 6--9.
 The ``experimental data'' in line 10 are equal to those presented
in line 6 except that now some additional knowledge concerning the
spin component $\hat s_y$ is available. We note that this additional
information about $\hat s_y$ improves our estimation of the density
operator  which is clearly seen when we compare
values of the von\,Neumann entropy presented in Table~\ref{table8}.

Providing that we have information concerning the  spectral
distribution associated with  the measurement of a complete
set (i.e. the quorum) of operators $\hat s_x,\hat s_y, \hat s_z$
(i.e., after an infinite number of measurements of the three spin
components have been performed), then  we can express the estimated
density operator as  [see Eq.(\ref{8.16})]
\begin{eqnarray}\nonumber
\hat\rho =\frac{1}{\cal N}
\int_0^{2\pi}\!\!d\phi\!\!\int_0^\pi\!\!\sin\theta d\theta
\ \ \delta(\langle \hat \sigma_z \rangle -\cos\theta)
\delta(\langle \hat \sigma_x \rangle -\sin\theta\cos\phi)
\delta(\langle \hat \sigma_y \rangle -\sin\theta\sin\phi)
\end{eqnarray}
\begin{eqnarray}
\times(\hat 1 + \sin\theta\cos\phi
\hat\sigma_x + \sin\theta\sin\phi\hat\sigma_y +
\cos\theta\hat\sigma_z).
\label{9.13}
\end{eqnarray}
The integral on the right-hand-side of Eq.(\ref{9.13}) can
only be performed if
the purity condition (\ref{8.18a}) is fulfilled, otherwise it simply
does not exist. When the purity condition is fulfilled then
from Eq.(\ref{9.13}) we obtain
\begin{eqnarray}
\hat\rho=\frac{1}{2}\left(
 \hat 1 +  \langle \hat \sigma_x \rangle\hat\sigma_x +
 \langle \hat \sigma_y \rangle\hat\sigma_y +
 \langle \hat \sigma_z \rangle\hat\sigma_z\right).
\label{9.14}
\end{eqnarray}
Here we can again utilize a simple geometrical interpretation of
the limit formula (\ref{9.13})	for the Bayes inference.
The three
$\delta$-functions in Eq.(\ref{9.13}) correspond to three specific orbits
(circles) on the
Poincar\'{e} sphere each of which is associated with a set of pure states
which posses the measured value of a given observable $\hat{s}_i$.
The reconstructed density operator then describes a point on the Poincar\'{e}
sphere which coincides with an intersection of these three orbits.
Consequently, if the three orbits have no intersection the reconstruction
scheme fails, because
there does not exist a {\em pure} state with the given mean values
of the measured observables.

We illustrate this failure of the Bayesian inference scheme in lines 14--17
of Table~\ref{table8}.
Here we present a numerical simulation of the measurement in
which all three observables are measured.
It is assumed that the spin-1/2 is in the
state with $\langle\hat{\sigma}_z\rangle=1/2$ and
$\langle\hat{\sigma}_x\rangle=\langle\hat{\sigma}_y\rangle=0$, which apparently
does not fulfill the purity condition (\ref{8.18a}).
For a given set of measurement outcomes (line 14) the Bayesian inference
scheme provides us with an {\em a posteriori} estimation such that
$\langle\hat{\sigma}_z\rangle=101/161$ which is above the expected mean value
which is equal to 1/2. Moreover if we increase the number of measurements
(lines 15--17) the a posteriori estimation
deviates more and more from what would be a correct  estimation
(i.e., results presented in lines 14--17 correspond to the following
sequence of mean values of $\hat\sigma_z$: $0.481; 0.375; 0.345; 0.332$)
but simultaneously the von\,Neumann entropy $S$ decreases, which should
indicate that our estimation is better and better. This clearly illustrates
the intrinsic conflict in the estimation procedure.

The reason for this contradiction lies in the {\em a priori} assumption about
the purity of the reconstructed  state, i.e. the mean values of the
spin components do not fulfill the condition (\ref{8.18a}) and so the
Bayesian method {\em cannot} be applied safely in the present case.
The larger the number of measurement the more clearly  the inconsistency
is seen and, as follows from Eq.(\ref{9.13}), in the limit of infinite number
of measurements the Bayesian method fails completely.
On the other hand the Jaynes method can be applied safely in this case.
The point is that this method is not based on an {\em a priori}
assumption about the purity of the reconstructed state. The Jaynes
principle is associated with maximization of entropy on the generalized
grand canonical
ensemble, which means that all states (pure and impure) are taken into
account.

In the present example the discrepancy between the {\em a posteriori}
estimations of density operators based on the two different schemes
has appeared only on the complete observation level. For more complex
quantum-mechanical systems the difference between the density operator
reconstructed with the help of the Jaynes principle of maximum entropy
and the density operator obtained via the Bayesian inference scheme
may differ even on incomplete observation levels. To see this
we present in the following sections an example of reconstruction
of density operators describing states of two spins-1/2.

\subsection{Quantum Bayesian inference of states of two spins-1/2}
\label{sec9.B}

In order to apply the general formalism of quantum Bayesian inference
as described in Section~\ref{sec8} we have to properly parameterize the state
space of the quantum system under consideration. Once this is done
we have to find the invariant integration measure $d_{\Omega}$ associated
with the state space and only then  can we effectively use the reconstruction
formula (\ref{8.3}). We start this section with a description of
how the state space of two spins-1/2 has to be parameterized and we show
how the integration measure can be found.

\subsubsection{Parameterization of two-spins-1/2 state space}
\label{sec9.B.1}

One way to determine the state space $\Omega$ of a given
quantum-mechanical system is via a
diffeomorphism $\Omega \equiv \ ^{SU(n)}\vert_{U(n-1)}$. This directly
provides us with  information about the dimensionality of $\Omega$,
which is  $(dim_{SU(n)}-dim_{U(n-1)})=2n-2$. This means that in our
case of two spins-1/2  which are prepared in a {\em pure} state
we need 6 coordinates which parameterize $\Omega$ $(n=4)$.
Unfortunately, it is not very convenient to determine the
state space via  the given diffeomorphism because then we have
to work with noncommutative groups.

It is much simpler to parameterize the state space $\Omega$ utilizing
the idea of the  Schmidt decomposition \cite{Peres93,Ekert95}.
In this case we can represent any pure state
$\vert \Psi\rangle$ describing two spins-1/2 as:
\begin{eqnarray}
\vert\Psi\rangle=A\vert \!\! \uparrow_1\rangle \otimes \vert \!\!
\uparrow_2\rangle
+B\vert \!\! \downarrow_1\rangle \otimes \vert \!\! \downarrow_2\rangle,
\label{9.15}
\end{eqnarray}
where $\vert \!\! \downarrow_j\rangle,\vert \!\! \uparrow_j\rangle,$ are two general orthonormalized
bases in $H^2$ and $A,B$ are two complex numbers satisfying the condition
$\vert A \vert ^2+\vert B \vert ^2=1$.
The corresponding density operator of a pure state  in $\Omega$ then reads
\begin{eqnarray}\nonumber
\hat{\rho} =\vert A \vert^2\vert \!\! \uparrow_1\rangle \langle \uparrow_1\!\!
\vert \otimes
\vert \!\! \uparrow_2\rangle \langle \uparrow_2\!\!\vert +
AB^\ast \vert \!\! \uparrow_1\rangle \langle \downarrow_1\!\!\vert \otimes
\vert \!\! \uparrow_2\rangle \langle \downarrow_2\!\!\vert
\end{eqnarray}
\begin{eqnarray}
\ \ \ +A^\ast B\vert \!\! \downarrow_1\rangle \langle \uparrow_1\!\!\vert
\otimes
\vert \!\! \downarrow_2\rangle \langle \uparrow_2\!\!\vert +
\vert B \vert^2\vert \!\! \downarrow_1\rangle \langle \downarrow_1\!\!\vert
\otimes
\vert \!\! \downarrow_2\rangle \langle \downarrow_2\!\!\vert.
\label{9.16}
\end{eqnarray}
The projectors $\vert\!\!\uparrow_j\rangle \langle \uparrow_j\!\!\vert$ and
$\vert \!\!\downarrow_j\rangle \langle \downarrow_j\!\!\vert$ ($j=1,2$)
are given by
$(\hat 1+\vec r^{_{(j)}}\hat{\vec{\sigma}}^{_{(j)}})$ and
$(\hat 1-\vec r^{_{(j)}}\hat{\vec{\sigma}}^{_{(j)}})$,
respectively [see Eq.(\ref{9.1})],
where $\vec r^{_{(1)}}$ and $\vec r^{_{(2)}}$ are two arbitrary unity vectors.
The operators $\vert\!\!\downarrow_j\rangle\langle \uparrow_j\!\!\vert$ and
their Hermitian conjugates
$\vert\!\!\uparrow_j\rangle\langle \downarrow_j\!\!\vert$
are determined as
\begin{eqnarray}
\vert \!\! \downarrow_j\rangle \langle \uparrow_j\!\!\vert (\hat 1+
\vec r^{_{(j)}}\hat{\vec{\sigma}}^{_{(j)}})
\vert \!\! \uparrow_j\rangle \langle \downarrow_j\!\!\vert=(\hat 1-
\vec r^{_{(j)}}\hat{\vec{\sigma}}^{_{(j)}}),
\label{9.17}
\end{eqnarray}
from which  the relation
\begin{eqnarray}
\vert \!\! \uparrow_j\rangle \langle \downarrow_j\!\!\vert =e^{i\psi _j}
(\vec k^{_{(j)}}\hat{\vec{\sigma}}^{_{(j)}} +i\vec l^{_{(j)}}
\hat{ \vec{\sigma}}^{_{(j)}}), 
\label{9.18}
\end{eqnarray}
follows.
Here the vectors $\vec k^{_{(j)}}$ are two arbitrarily chosen unity vectors
which satisfy the condition $\vec k^{_{(j)}} \perp \vec r^{_{(j)}}$,
and $\vec l^{_{(j)}}$ are
equal to  vector products $\vec l^{_{(j)}}=\vec r^{_{(j)}}\times\vec k^{_{(j)}}$.
A particular choice of vectors $\vec k_j$ is not important because
phase factors $e^{i\psi_j}$ [ $\psi_j \in (0,2\pi)$]
rotate them along all possible directions.
We also note that the phase factors $e^{i\psi_j}$ can be always
incorporated in the phase $\psi $ of a complex number $AB^\ast $.
Using the parameterization $\vert A\vert =\cos(\alpha/ 2)$ and
$\vert B\vert =\sin(\alpha/2) $ we can parameterize
$\hat{\rho}$ as:
\begin{eqnarray}
\hat{\rho}(\alpha,\psi,\phi_1,\theta_1,\phi_2,\theta_2)={\hat 1\otimes\hat 1
\over4}
+{{\vec r^{_{(1)}}\hat{\vec{\sigma}}\otimes\vec r^{_{(2)}}\hat{\vec{\sigma}}}
\over4}
+\cos\alpha\Big\lbrack
{{\vec r^{_{(1)}}\hat{\vec{\sigma}}\otimes\hat 1}\over4}+
{{\hat1\otimes\vec r^{_{(2)}}\hat{\vec{\sigma}}}\over4}\Big\rbrack
\label{9.19}
\end{eqnarray}
\begin{eqnarray}\nonumber
+\sin\alpha\cos\psi\Big\lbrack {{{\vec k^{_{(1)}}\hat{\vec{\sigma}}}\otimes
\vec k^{_{(2)}}\hat{\vec{\sigma}} }\over4}
-{{\vec l^{_{(1)}}\hat{\vec{\sigma}}\otimes\vec l^{_{(2)}}\hat{\vec{\sigma}} }
\over4}\Big\rbrack
-\sin\alpha\sin\psi\Big\lbrack {{\vec k^{_{(1)}}\hat {\vec{\sigma}}\otimes
\vec l^{_{(2)}}\hat{\vec{\sigma}} }\over4}
+{{\vec l^{_{(1)}}\hat{\vec{\sigma}}\otimes\vec k^{_{(2)}}\hat{\vec{\sigma}} }\over4}\Big\rbrack,
\end{eqnarray}
where $\psi,\phi_1,\phi_2 \in (0,2\pi)$;
$\alpha, \theta _1,\theta _2 \in (0,\pi)$ and
\begin{eqnarray}
\vec k^{_{(j)}}&=&(\sin \phi_j,-\cos \phi_j,0);\nonumber \\
\vec l^{_{(j)}}&=&(\cos\theta_j\cos\phi _j,\cos\theta_j
\sin\phi _j,-\sin\theta_j);
\label{9.20}\\
\vec r^{_{(j)}}&=&(\sin\theta_j\cos\phi_j,\sin\theta_j\sin\phi_j, \cos\theta_j).\nonumber
\end{eqnarray}
Once we have parameterized the state space $\Omega$ we can find the
invariant integration measure $d_{\Omega}$ (see Appendix B) which reads
\begin{eqnarray}
d_{\Omega}=\cos^2\alpha \sin\alpha \sin\theta_1 \sin\theta_2 d\alpha d\psi d\phi_1
d\theta_1 d\phi_2 d\theta_2.
\label{9.21}
\end{eqnarray}

\subsection{Quantum Bayesian inference of the state of two-spins-1/2}
\label{sec9.C}

To perform the Bayesian reconstruction of density operators of the
two-spins-1/2 system we introduce a set of projectors associated with
the observables 
\begin{eqnarray}
\hat P_{s,\hat{s}_i^{_{(1)}}}
={({\hat 1+s\hat{\sigma}_i)}\over 2}\otimes \hat 1;\qquad
\hat P_{s,\hat{s}_i^{_{(2)}}}
=\hat 1 \otimes {({\hat 1+s\hat{\sigma}_i)}\over 2};\qquad
\hat P_{s,\hat{s}_i^{_{(1)}}\hat{s}_j^{_{(2)}}}
={\hat 1 \otimes \hat 1 \over 2}+
s{\hat{\sigma}_i\otimes \hat {\sigma}_j\over 2}.
\label{9.22}
\end{eqnarray}
The corresponding conditional probabilities can be expressed as
\begin{eqnarray}
p(s,\hat{s}_i^{_{(1)}} \vert \hat{\rho} (\alpha \dots ))=
{1 \over 2}+s{{\cos(\alpha)}\over 2}r_{i}^{_{(1)}};\qquad
p(s,\hat{s}_i^{_{(2)}} \vert \hat{\rho} (\alpha \dots ))=
{1 \over 2}+s{{\cos(\alpha)}\over 2}r_{i}^{_{(2)}};
\label{9.23}
\end{eqnarray}
\begin{eqnarray}\nonumber
p(s,\hat{s}_i^{_{(1)}}\hat{s}_j^{_{(2)}} \vert \hat{\rho} (\alpha \dots ))=
{1 \over 2}+s{{r_i^{_{(1)}}r_j^{(2}}\over2}+
+s\left[ {{\sin(\alpha)\cos\psi}
\over 2}(k_i^{_{(1)}}k_j^{_{(2)}}-l_i^{_{(1)}}l_j^{_{(2)}})
-{{\sin(\alpha)\sin\psi}
\over 2}(k_i^{_{(1)}}l_j^{_{(2)}}+l_i^{_{(1)}}k_j^{_{(2)}})\right],
\end{eqnarray}
where $s$ is the sign of the measured eigenvalue.
Here we comment briefly on the physical meaning of the projectors defined
by Eq.(\ref{9.22}). Namely, the single-particle projectors
of the form $\hat P_{s,\hat{s}_i^{_{(1)}}}$ are associated  with a measurement
of the spin component of the first particle in the $i$-direction ($i=x,y,z$).
Obviously this spin component can have
only two values, i.e., ``up'' ($s=1$)
and  ``down'' ($s=-1$).

\begin{table}
\begin{center}
\epsfig{file=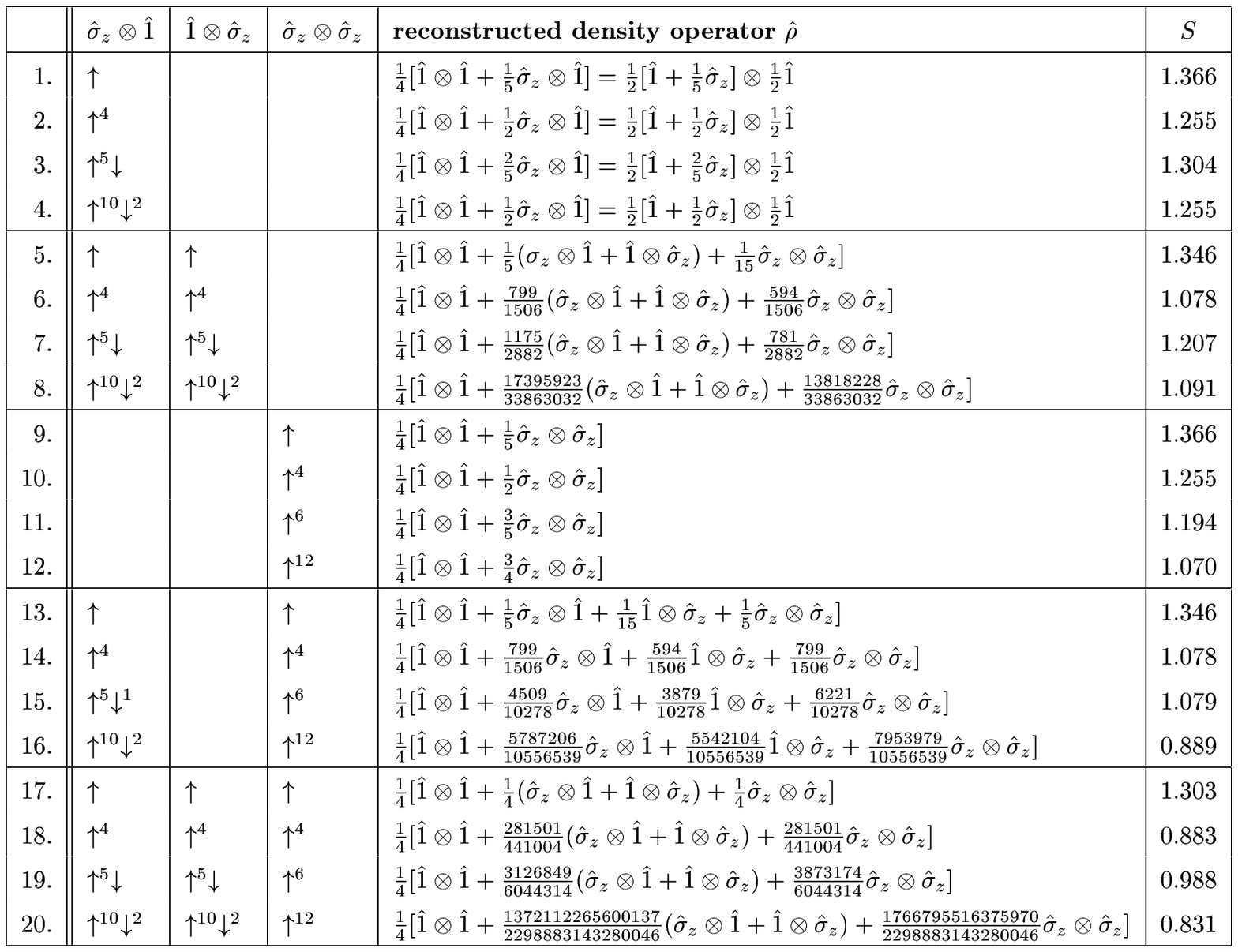, angle=0, width=16cm}
\end{center}
\caption{
Results of {\em a posterior} Bayesian estimation of
density operators of the two-spin-1/2 system. We also present explicit
values of the von\,Neumann entropy associated with given measured data.
}
\label{table9}
\end{table}

In Tables \ref{table8} and \ref{table9} we will denote
outcomes of the measurements ``up'' and ``down'' as $\uparrow$ and
$\downarrow$, respectively. The two-particle projectors
$\hat P_{s,\hat{s}_i^{_{(1)}}\hat{s}_j^{_{(2)}}}$ are associated with
measurements of correlations between the two spin. Namely, if $s=1$,
the two spins are {\em correlated}, which means that
they both are registered in
the same, yet unspecified, state (that is, both spins are registered
either in the state $|\uparrow_1\uparrow_2\rangle$
or $|\downarrow_1\downarrow_2\rangle$).

In Tables \ref{table8} and \ref{table9}
we will denote this outcome of the measurement as $\uparrow$. On the
contrary, if the particles are registered as
{\em anticorrelated}, that is after the measurement they are  in
one of the two states
$|\uparrow_1\downarrow_2\rangle$ or
$|\downarrow_1\uparrow_2\rangle$, then	$s=-1$. In Tables \ref{table8} and 
\ref{table9} we will denote the
outcome of this measurement for $\hat\sigma_i\otimes\hat\sigma_j$
as $\downarrow$.

Now we can apply general rules of Bayesian inference
presented in Section~\ref{sec8}.  for a  two-spins-1/2  system.
We will
consider three specific incomplete observation levels and we will
derive asymptotic expressions for the density operators
in the limit of large number of measurements.
We stress here that we assume the measured system to be prepared in a
pure state. To be specific, let us suppose that the two spins
are prepared in a state  described by the state vector
(obviously, this can be determined only after an infinite number of
measurements on the complete observation level is performed)
\begin{eqnarray}
|\Psi\rangle= A |\!\!\uparrow\rangle\otimes|\!\!\uparrow\rangle+
B|\!\!\downarrow\rangle\otimes|\!\!\downarrow\rangle,
\label{9.24}
\end{eqnarray}
 where
$|\!\!\uparrow\rangle$ and
$|\!\!\downarrow\rangle$ are eigenstates corresponding to the observable of
the spin  projection into the $z$-direction (i.e.,
$\langle \hat\sigma_z\otimes \hat 1\rangle
=\langle\hat 1\otimes \hat\sigma_z\rangle=|A|^2-|B|^2$ and
$\langle \hat{\sigma}_z\otimes \hat{\sigma}_z \rangle=1$).
When we assume the coefficients $|A|$ and $B|$ to be real, then we can rewrite
the density operator associated with the state vector 
(\ref{9.24}) in the form (\ref{9.19}), i.e.
\begin{eqnarray}
\hat{\rho}
=\frac{\hat 1\otimes\hat 1}{4}
+\frac{\hat\sigma_z\otimes\hat\sigma_z}{4}
+\frac{A^2-B^2}{4}\left(\hat\sigma_z\otimes \hat 1 + \hat 1 \otimes\sigma_z
\right)+
\frac{AB}{2}\left(\hat\sigma_x\otimes\hat\sigma_x
-\hat\sigma_y\otimes\hat\sigma_y\right),
\label{9.25}
\end{eqnarray}
with $\psi=0,\phi_1=\pi/2,\theta_1=0,\phi_2=\pi/2,\theta_2=0$ and
$\sin\alpha/2=A$.
In what follows we perform {\em a posteriori} estimation of  the density
operator based on incomplete data obtained from three different fictitious
measurement sequences.

\subsubsection{Observation level ${\cal O}_A^{(2)}=
\{ \hat s_z^{_{(1)}}, \hat s_z^{_{(2)}} \} $}
\label{sec9.C.1}

In the first sequence
of measurements we reconstruct	a density operator  from data which refer
to a measurement of the first spin-1/2 in the direction $z$, i.e.,
only the spin component
$\hat s_z^{_{(1)}}$ is measured (see lines 1--4 in Table \ref{table9}).
We see that if only one
spin is measured, then the reconstructed two-spin density operator can be
factorized, while, as expected, the state of the unmeasured spin is
estimated as $\rho=\hat 1/2$. Obviously, in this kind of measurement,
correlations between the two spins cannot be revealed,
i.e., the estimated value of $\hat\sigma_z\otimes\hat\sigma_z$ is equal to zero.
As in the case of the reconstruction
of a single-spin-1/2 state, the reconstructed density operators
describe statistical mixtures and the corresponding von\,Neumann
entropy is directly related to the fidelity of the reconstruction.
The maximum value of the von\,Neumann entropy is in the case of
two-spins-1/2 equal to $S=\ln 4\simeq 1.386$. This entropy is associated with
the ``total'' mixture of the two-spin-1/2 system and in our case
it reflects
a complete lack of information	about the state of the measured system
(i.e., we have no knowledge about the state before a measurement is
performed). As soon as the first measurement is performed, we gain some
knowledge about the state of the system and the entropy of the
estimated density operator is smaller than $\ln 4$ (see line 1).

Let us assume now that data from the measurement of the spin components
$\hat s_z^{_{(1)}}$ and
$\hat s_z^{_{(2)}}$ of the
first and the second particle (spin-1/2), are available.
In Table~\ref{table9} (lines 5--8)  
we present results of a reconstruction procedure
based on the given ``measured'' data. We see that though correlations
between the two spins have not been measured directly our estimation procedure
provides us with a nontrivial estimation for this observable
(i.e., the density operator cannot be factorized).
Obviously, this estimation is affected by the prior assumption about
the purity of the reconstructed state.	We see that with the increased number
of detected spins the von\,Neumann entropy of the estimated
density operator decreases (we note that it does not decrease
monotonically as a function of the number of measurements).

In the limit of large (infinite) number of measurements
spectral distributions	Eq.(\ref{8.5}) associated with observables
on a given  incomplete observation level are precisely determined
by the measured data.
Using the parameterization introduced earlier in this section
[see Eqs.(\ref{9.19}) and  (\ref{9.21}-\ref{9.23})]
we can write down the expression (\ref{8.16})
for the Bayesian {\em a posteriori} estimation of the
density operator in the limit of large number of measurements.
After we perform some trivial integrations
and when the substitution
$\cos \alpha=x$, $\cos \theta_1=y$, $\cos \theta_2=z$ is performed we
can write the reconstructed density  operator as
\begin{eqnarray}
\hat\rho=\frac{1}{\cal N}
\int_{-1}^{1}x^2dx\int_{-1}^{1}dy\int_{-1}^{1}
dz
 \delta(\langle \hat \sigma_z^{_{(1)}} \rangle -xy)
\delta(\langle \hat \sigma_z^{_{(2)}} \rangle -xz)
(\hat 1\otimes \hat 1+xy\hat\sigma_z \otimes \hat 1
+xz\hat 1\otimes \hat\sigma_z
+yz\hat\sigma_z\otimes\hat\sigma_z),
\label{9.25a}
\end{eqnarray}
The right-hand side of
Eq.(\ref{9.25a}) can easily be integrated over the variables $y$ and $z$
so we can write
\begin{eqnarray}
\hat\rho =\frac{1}{\cal N}
 \int_{\cal L}
 dx \,
\left(\hat 1\otimes \hat 1+  \langle \hat \sigma_z^{_{(1)}} \rangle\,
\hat\sigma_z \otimes \hat 1+
 \langle \hat \sigma_z^{_{(2)}} \rangle\, \hat 1\otimes \hat\sigma_z
+{\langle \hat \sigma_z^{_{(1)}} \rangle  \langle \hat \sigma_z^{_{(2)}}
\rangle \over x^2}\,
\hat\sigma_z\otimes\hat\sigma_z\right),
\label{9.26}
\end{eqnarray}
where the integration is performed over the interval ${\cal L}$
\begin{eqnarray}
{\cal L} := \{ -1, 1\} ~~~{\mbox{\rm  and}~~~}
\vert x \vert \geq s_{max},
\label{9.27}
\end{eqnarray}
with $s_{max}={\rm max}\{\, \vert \langle \hat \sigma_z^{_{(1)}}\rangle \vert,
\vert \langle \hat \sigma_z^{_{(2)}}\rangle \vert\, \}$.
After we perform
the integration over the variable $x$ we find
\begin{eqnarray}
\hat\rho =\frac{1}{4}
\left(\hat 1\otimes \hat 1+
\langle \hat \sigma_z^{_{(1)}}\rangle\, \hat\sigma_z \otimes \hat 1+
\langle \hat \sigma_z^{_{(2)}}\rangle\, \hat 1\otimes \hat\sigma_z
+{ \langle \hat \sigma_z^{_{(1)}}\rangle
\langle \hat \sigma_z^{_{(2)}}\rangle \over s_{max}}\,
\hat\sigma_z\otimes\hat\sigma_z\right).
\label{9.28}
\end{eqnarray}
Comparing results presented in Tab.~\ref{table7} 
 and (\ref{9.28}) we see that on the
observation level  ${\cal O}_A^{(2)}$
the quantum Bayesian inference	and the Jaynes principle of  maximum
entropy provides us with the different {\em a posteriori} estimations
of density operators. To be specific, the density operator 
obtained with the help of the MaxEnt principle can be expressed in a
factorized form while
the density operator $\hat{\rho}_{\rm A}$ cannot be factorized into
a product of two density operators describing each spin separately
[the only exception is when $s_{max}=1$].

\subsubsection{Observation level ${\cal O}_B^{(2)}=
\{ \hat s_z^{_{(1)}}, \hat s_z^{_{(1)}}\hat s_z^{_{(2)}} \} $}
\label{sec9.C.2}

Here we start  our discussion with an assumption that only
correlations between the particles are measured,
while the state of the each individual particle after the measurement is
unknown (see lines 9--12 in Table~\ref{table9}). 
In this case we are not able
to make any nontrivial estimation for the mean values of the
spin components of the individual particles. In order to have a better
estimation we have also to measure  at least one of the spin components
of the first or the second spin.

Let us assume that the $z$-component of the first spin and the
 correlation $\hat s_z^{_{(1)}}\hat s_z^{_{(2)}}$ are measured.
That is the $z$-component of the second spin $\hat s_z^{_{(2)}}$
is not directly observed. The question is what is the estimation
of the density operator on this observation level and in particular,
what is the estimation for the mean value of the observable
$\hat s_z^{_{(2)}}$. In Table~\ref{table9} (lines 13--16)
we present numerical results for
the {\em a posteriori} estimation of the density  based on a finite set
of ``experimental'' data. We see that the Bayesian scheme provides us
with a nontrivial (i.e., nonzero) estimation of the mean value of
$\hat s_z^{_{(2)}}$. But the question is whether in the limit of a large number
of measurements this is equal to the mean value estimated
with the help of the Jaynes principle of  maximum entropy.
The expression for the {\em a posteriori}
Bayes estimation of the density  operator in the limit of infinite
number of measurements on the given observation level [for technicalities
see Appendix C] reads
\begin{eqnarray}
\hat\rho = \frac{1}{4}
\left[\hat 1\otimes \hat 1+
\langle \hat \sigma_z^{_{(1)}}\rangle \hat\sigma_z \otimes \hat 1+
{\langle \hat \sigma_z^{_{(1)}}\rangle \langle \hat \sigma_z^{_{(1)}}
\hat \sigma_z^{_{(2)}}\rangle
\over s_{max}} \hat 1\otimes \hat\sigma_z
+\langle \hat \sigma_z^{_{(1)}}\hat \sigma_z^{_{(2)}}\rangle
\hat\sigma_z\otimes\hat\sigma_z\right],
\label{9.29}
\end{eqnarray}
where $s_{max} ={\rm max} \left\{\, \vert\langle
\hat \sigma_z^{_{(1)}}\rangle\vert,
\vert \langle \hat \sigma_z^{_{(1)}}\hat
\sigma_z^{_{(2)}}\rangle\vert\, \right\}$.
Here again
the Bayesian {\em a posteriori}  estimation  (\ref{9.29})
is in general different from  the estimation  obtained
with the help of the  Jaynes
{\em MaxEnt} principle [see $\hat{\rho}_{\rm B}$ in Tab.~\ref{table7}].
We see that these two results coincide only when
$s_{max}=1$. For instance, if
$\langle {\hat\sigma}_z\otimes {\hat\sigma}_z \rangle=1$, then
$s_{max}$ is equal to unity and the estimated density operators
$\hat{\rho}_{\rm B}$ and $\hat{\rho}$ given by Eq.~(\ref{9.29}) 
are equal and read
\begin{eqnarray}
\hat\rho = \frac{1}{4}
\left[\hat 1\otimes \hat 1+
\langle \hat \sigma_z^{_{(1)}}\rangle \hat\sigma_z \otimes \hat 1+
\langle \hat \sigma_z^{_{(1)}}\rangle
 \hat 1\otimes \hat\sigma_z
+\hat\sigma_z\otimes\hat\sigma_z\right].
\label{9.30}
\end{eqnarray}
In the case when  $\langle \hat \sigma_z^{_{(1)}}\rangle=1$ the von\,Neumann
entropy  is equal to zero,
i.e., the measured state is completely reconstructed, and is described by
the state vector $|\Psi\rangle=|\uparrow_1\uparrow_2\rangle$.

\subsubsection{Observation level ${\cal O}_C^{(2)}=
\{ \hat s_z^{_{(1)}}, \hat s_z^{_{(2)}}, \,
\hat s_z^{_{(1)}}\hat s_z^{_{(2)}} \} $}
\label{sec9.C.3}

Finally, we will consider
a measurement of both the spins projections $\hat s_z^{_{(1)}}$,
$\hat s_z^{_{(2)}}$,
as well as the correlation $\hat s_z^{_{(1)}}\hat s_z^{_{(2)}}$. Results of
an estimation of the density operator based on a sequence of
data associated with this observation level
are given in  Table~\ref{table9} (lines 17--20).
If an infinite number of measurements on the given observation level is
performed  then we can evaluate the {\em a posteriori} density
operator analogously to that of the previous example [see Appendix C]
and after some algebra we find
\begin{eqnarray}
\hat\rho = \frac{1}{\cal N}
\int_{{\cal L}''}\, {x^2\over
\vert x\vert}dx\!\!\int_{z_1}^{z_2}\!\!dz\,
{\delta(\langle \hat \sigma_z^{_{(2)}}\rangle - xz)\over\sqrt{a+bz+cz^2}}
\left[\hat 1\otimes \hat 1+\langle \hat \sigma_z^{_{(1)}}\rangle \hat\sigma_z
\otimes \hat 1+xz\hat 1\otimes \hat\sigma_z+
\langle \hat \sigma_z^{_{(1)}}\hat \sigma_z^{_{(2)}}
\rangle\hat\sigma_z\otimes\hat\sigma_z\right].
\label{9.31}
\end{eqnarray}
Due to	the presence of the  $\delta$-function
the integration over the parameter $z$
on the right-hand side of Eq.(\ref{9.31})  is straightforward and we obtain
\begin{eqnarray}
\hat\rho=\frac{1}{\cal N}
\int_{{\cal L}''}{dx\over\sqrt{a+bz_0+cz_0^2}}
\left[\hat 1\otimes \hat 1+\langle \hat \sigma_z^{_{(1)}}\rangle
\hat\sigma_z \otimes \hat 1+
\langle \hat \sigma_z^{_{(2)}}\rangle \hat 1\otimes \hat\sigma_z+
\langle \hat \sigma_z^{_{(1)}}\hat \sigma_z^{_{(2)}}\rangle
\hat\sigma_z\otimes\hat\sigma_z\right],
\label{9.32}
\end{eqnarray}
where $z_0=\langle \hat \sigma_z^{_{(2)}}\rangle/x$.
From Eq.(\ref{9.32}) we directly obtain the reconstructed density operator
which reads
\begin{eqnarray}
\hat\rho=\frac{1}{4}\left[\hat 1\otimes \hat 1+
\langle \hat \sigma_z^{_{(1)}}\rangle \hat\sigma_z \otimes \hat 1+
\langle \hat \sigma_z^{_{(2)}}\rangle \hat 1\otimes \hat\sigma_z+
\langle \hat \sigma_z^{_{(1)}}\hat \sigma_z^{_{(2)}}\rangle
\hat\sigma_z\otimes\hat\sigma_z\right].
\label{9.33}
\end{eqnarray}
We see that on the present observation level
the density operator (\ref{9.33}) estimated via  Bayesian inference
is {\em equal} to the density operator $\hat{\rho}_{\rm C}$
 estimated
with  Jaynes principle of maximum entropy [see Tab.~\ref{table7}].
 Nevertheless, we have to
remember that the estimation (\ref{9.33}) based on  quantum  Bayesian
inference is
intrinsically related to an averaging over a generalized microcanonical
ensemble of pure states. On the other hand, the MaxEnt-estimation is
associated with  averaging over the grand canonical ensemble.

\subsection{Reconstruction of impure spin states}
\label{sec9.D}

In this section we apply the purification ansatz as shown in
Section~\ref{sec8.C} for a reconstruction (estimation) of an impure state
of a single spin-1/2. To do so, we apply the results of the previous
section where we have discussed the Bayesian estimation of pure
two-spins-1/2 states. In particular, in lines 1--4 of Table~\ref{table9}
we present results of the  estimation of a two-spin density operator
based on  ``results'' of
measurements of  the $\hat\sigma_z$-component of just one spin-1/2.
We see that in this case the two-spin density operator can be written
in a factorized form, $\hat\rho_{ab}=\hat\rho_{a}\otimes\frac{1}{2}\hat 1$.
In this case we can easily trace over the unmeasured spin
and we obtain the estimation for the density operator of the first
spin (compare with lines 1--4 in Table~\ref{table8}).
This estimation is not based on the {\em a priori} purity assumption.

Comparing results of two estimations which differ by the
{\em a priori} assumption about the purity of the reconstructed state
we can conclude the following: \newline
{\bf (1)} In general, under the purity assumption the reconstruction procedure
converges faster (simply compare the two columns in Table~\ref{table8}) to
a particular result. This is easy to understand, because  in the case
when the purity of measured states is {\em a priori} assumed, the state
space of all possible states is much smaller compared to the
state space of all possible (pure and impure) states.\newline
{\bf (2)} When the measured data are inconsistent with an a priori purity
assumption, then estimations based on this assumption  become incorrect.
For instance, for the ``measured'' data presented in
 lines 14--17 of Table~\ref{table8} we find that the estimated mean values of
$\hat\sigma_z$ diverge from the expected mean value  $1/2$ (i.e., this is
the mean value of $\hat\sigma_z$ when we detect in a sequence of
$4N$ measurements  $3N$ spins ``up'' and $N$ spins ``down'').
As we have shown in Section~\ref{sec9.A.3}
 in the limit $N\rightarrow\infty$
the reconstruction can completely fail when the purity condition is
imposed. In the other hand, if it is {\em a priori} assumed
that the measured state can be in a statistical mixture, then
the Bayesian quantum inference provides us with estimations
which in the limit $N\rightarrow\infty$ coincide with
estimations based on the Jaynes principle of maximum entropy.

\subsubsection{Observation level  ${\cal O}_A^{(1)}=\{\hat s_z^{_{(1)}} \}$}
\label{sec9.D.1}

Using the techniques which  have been demonstrated in Section~\ref{sec8}
we can express the estimated density operator on the given observation
level in the limit  $N\rightarrow\infty$ as [see Eq.(\ref{8.16})].
We note
that  on the considered observation level, Eq.(\ref{8.16})  contains
many terms, which are odd functions of the corresponding integration variables.
Therefore
the integration over these parameters ($\theta_2, \phi_2, \psi,\phi_1$) is
straightforward. Moreover, if we perform the trace over
the ``second'' (reservoir) spin we can express
the density operator of the spin-1/2 under consideration as
\begin{eqnarray}
\hat\rho=\frac{1}{\cal N}
\int_{-1}^{1}y^2dy\int_0^\pi\sin\theta_1 d\theta_1
\ \ \delta(\langle \hat \sigma_z^{_{(1)}} \rangle -y\cos\theta_1)
(\hat 1+y\cos\theta_1\hat\sigma_z),
\label{9.33a}
\end{eqnarray}
where the variable $\alpha$ is substituted by $y=\cos\alpha$.
When we perform  integration  over  $y$ we obtain the expression
\begin{eqnarray}
\hat\rho=\frac{2}{\cal N} \int_{{\cal L}}
  d\theta_1
{\sin\theta_1\over \cos^2\theta_1\vert \cos\theta_1\vert}
(\hat 1 +\langle \hat \sigma_z^{_{(1)}} \rangle \hat\sigma_z),
\label{9.34}
\end{eqnarray}
with  ${\cal L}$ defined as
\begin{eqnarray}
{\cal L} := \{0,\pi\}~~~\mbox{ such that }~~
\vert \cos\theta_1 \vert \geq \vert \langle\hat \sigma_z^{_{(1)}}\rangle\vert.
\label{9.35}
\end{eqnarray}
After we perform the integration over $\theta_1$ we obtain the expression
for the density operator identical to that obtained via the
Jaynes principle of  maximum entropy [see Tab.~\ref{table5}].

\subsubsection{Observation level  ${\cal O}_B^{(1)}=\{\hat s_z^{_{(1)}},
\hat s_x^{_{(1)}}   \}$}
\label{sec9.D.2}

In the limit of infinite number of measurements one can express
the Bayesian estimation  of the
density operator of the spin-1/2
on the given observation level as (here the trace over the ``reservoir''
spin has already been performed):
\begin{eqnarray}\nonumber
\hat\rho =\frac{1}{\cal N}
\int_{-1}^{1}\!\!y^2dy\!\!\int_0^\pi\!\!\sin\theta_1 d\theta_1
\!\!\int_0^{2\pi}\!\!d\phi_1
\ \ \delta(\langle \hat \sigma_z^{_{(1)}} \rangle -y\cos\theta_1)
\delta(\langle \hat \sigma_x^{_{(1)}} \rangle -y\sin\theta_1\cos \phi_1)
\end{eqnarray}
\begin{eqnarray}
\times \left(\hat 1+y\sin\theta_1\cos \phi_1\hat\sigma_x+y
\sin\theta_1\sin \phi_1\hat\sigma_y+
y\cos\theta_1\hat\sigma_z\right).
\label{9.36}
\end{eqnarray}
When we perform integration over the variable $y$ we find
\begin{eqnarray}\nonumber
\hat\rho=\frac{1}{\cal N}\int_0^{2\pi}d\phi_1
\int_ {{\cal L}'}
d\theta_1
{\sin\theta_1\over \cos^2\theta_1\vert \cos\theta_1\vert}
 \delta(\langle \hat \sigma_x^{_{(1)}} \rangle -\tan\theta_1\cos \phi_1
\langle \hat \sigma_z^{_{(1)}}\rangle)
\end{eqnarray}
\begin{eqnarray}
\times\left(\hat 1 +\langle \hat \sigma_z^{_{(1)}}\rangle
\tan\theta_1\cos\phi_1\hat\sigma_x+
\langle \hat \sigma_z^{_{(1)}} \rangle \tan\theta_1\sin \phi_1\hat\sigma_y+
\langle \hat \sigma_z^{_{(1)}} \rangle \hat\sigma_z\right).
\label{9.37}
\end{eqnarray}
The integration over the variable $\phi_1$ in the right-hand side of
Eq.(\ref{9.37}) gives us
\begin{eqnarray}
\hat\rho=\frac{1}{\cal N} \int_{{\cal L}''}
\, d\theta_1
\sum_{j=1}^2
{1 \over
\cos^2\theta_1\vert\sin\phi_{1}^{(j)}\vert}
\left(\hat 1 +\langle \hat \sigma_x^{_{(1)}}\rangle \hat\sigma_x+
\langle \hat \sigma_z^{_{(1)}} \rangle \tan\theta_1
\sin \phi_{1}^{(j)}\hat\sigma_y+
\langle \hat \sigma_z^{_{(1)}} \rangle \hat\sigma_z\right),
\label{9.38}
\end{eqnarray}
where the integration is performed over the interval
\begin{eqnarray}
{\cal L}'' := \{0,\pi\}~~~\mbox{ such that }~~
\vert \cos\theta_1 \vert \geq \vert \langle\hat \sigma_z^{_{(1)}}\rangle\vert,
~~~\mbox{and}~~~
\vert \tan \theta_1 \vert \geq \vert \frac{\langle \hat \sigma_x^{_{(1)}} \rangle}{
\langle \hat \sigma_z^{_{(1)}} \rangle \vert}.
\label{9.39}
\end{eqnarray}
The sum in Eq.(\ref{9.38}) is performed over two values $\phi_1^{(j)}$
of the variable $\phi_1$ which are equal to the two solutions
of the equation
\begin{eqnarray}
\cos \phi_{1}=\frac{\langle \hat \sigma_x^{_{(1)}} \rangle}{
\langle \hat \sigma_z^{_{(1)}} \rangle \tan \theta_1}.
\label{9.40}
\end{eqnarray}
Due to the fact that the term in front of the operator
$\hat{\sigma}_y^{_{(1)}}$
is the odd function of $\phi_1^{(j)}$, we can straightforwardly perform
in Eq.(\ref{9.38}) the
integration over $\theta_1$ and we find the expression of the reconstructed
density operator which again is exactly the same as if we perform
the reconstruction with the help of the Jaynes principle
[see Tab.~\ref{table5}].

\subsubsection{Observation level  ${\cal O}_C^{(1)}=\{\hat s_z^{_{(1)}},
\hat s_x^{_{(1)}} ,  \hat s_y^{_{(1)}} \}$}
\label{sec9.D.3}

On the complete observation level, the expression for the Bayesian
estimation of the density operator of the spin-1/2 in the
limit of infinite number of measurements can be expressed as
(here again we have already traced over the ``reservoir'' degrees
of freedoms) [see  Eq.(\ref{9.36})]:
\begin{eqnarray}\nonumber
\hat\rho=\frac{1}{\cal N}
\int_{-1}^{1}\!\!y^2dy\!\!\int_0^\pi\!\!\sin\theta_1 d\theta_1
\!\!\int_0^{2\pi}\!\!d\phi_1
\ \ \delta(\langle \hat \sigma_z^{_{(1)}} \rangle -y\cos\theta_1)
\delta(\langle \hat \sigma_x^{_{(1)}} \rangle -y\sin\theta_1\cos \phi_1)
\delta(\langle \hat \sigma_y^{_{(1)}} \rangle -y\sin\theta_1\sin \phi_1)
\end{eqnarray}
\begin{eqnarray}
\times\left(\hat 1+y\sin\theta_1\cos \phi_1\hat\sigma_x
+y\sin\theta_1\sin \phi_1\hat\sigma_y+
y\cos\theta_1\hat\sigma_z\right).
\label{9.41}
\end{eqnarray}
Performing similar calculations as in the previous subsection we
can rewrite Eq.(\ref{9.41}) as
\begin{eqnarray}\nonumber
\hat\rho\simeq
\int_{{\cal L}''}
\, d\theta_1
\sum_{j=1}^2
{1 \over
\cos^2\theta_1\vert\sin\phi_{1}^{(j)}\vert}
 \delta(\langle \hat \sigma_y^{_{(1)}}\rangle -\tan\theta_1\sin \phi_{1}^{(j)}
\langle \hat \sigma_z^{_{(1)}}\rangle)
\end{eqnarray}
\begin{eqnarray}
\times\left(\hat 1 +\langle \hat \sigma_x^{_{(1)}}\rangle \hat\sigma_x+
\langle \hat \sigma_y^{_{(1)}}
\rangle \tan\theta_1\sin \phi_{1}^{(j)}\hat\sigma_y+
\langle \hat \sigma_z^{_{(1)}}\rangle \hat\sigma_z\right)\otimes \hat 1,
\label{9.42}
\end{eqnarray}
where ${\cal L}''$ and $\phi_1^{(j)}$ are defined by Eqs.(\ref{9.39})
and (\ref{9.40}), respectively. Now the
integration over  $\theta_1$ can be easily performed and for the
density operator of the given spin-1/2 system we find
\begin{eqnarray}
\hat\rho=\frac{1}{2}
\left[\hat 1 + \langle \hat \sigma_x^{_{(1)}}\rangle \hat\sigma_x+
\langle \hat \sigma_y^{_{(1)}}\rangle \hat\sigma_y+
\langle \hat \sigma_z^{_{(1)}}\rangle \hat\sigma_z\right],
\label{9.43}
\end{eqnarray}
where the mean values $\langle \hat \sigma_j^{_{(1)}}\rangle$
do not necessarily satisfy the purity condition (\ref{8.18a}).
In other words the generalized Bayesian scheme provides us
with a possibility of reconstruction of impure quantum-mechanical
states and the results in the limit of infinite number
of measurements are equal to those obtained with the help of
the Jaynes principle of  maximum entropy. Moreover, when the
quorum of observables is measured, a {\em complete} reconstruction
of the measured state is performed.

\section{OPTIMAL ESTIMATION OF QUANTUM STATES FROM FINITE ENSEMBLES}
\label{sec10}

In the previous section we have analyzed how quantum states can be reliably
estimated from the data obtained in a {\it given} measurement performed on a 
finite ensemble of $N$ identically prepared objects in  
 an unknown {\it pure}
quantum state described by a density operator $\hat
\rho=|\psi\rangle \langle \psi|$. In this section we will address the
question  - {\it What kind of
measurement provides the best possible estimation of $\hat \rho$}?\, 
(see also \cite{MasPop,Derka98}).
This leads to
an important problem of the {\it optimal}
 state estimation with limited physical
resources. It is a generic problem, common to many areas of quantum physics
ranging from the ultra-precise quantum metrology to eavesdropping in quantum
cryptography.

Within a framework of an elementary group theory the  problem of the state
estimation can be reformulated as a more general problem of estimating an
unknown unitary operation from a group of transformations acting on 
a given quantum system (i.e., the state estimation follows as a special case).
Holevo \cite{Holevo82} has shown that  this problem can be
solved via the {\it covariant measurement} (CM) approach.
Unfortunately,
the covariant measurement corresponds
to an {\it infinite} (i.e. consisting of infinite continuous set of operators)
and therefore physically non-realizable  positive operator-valued measure
measurement (POVM).
We note that from the logic of the CM it follows
that if any optimal measurement (finite or infinite) does  exist
then using a simple formal construction one can  generate from the
original optimal measurement another measurement
which is covariant and which, in the same time	conserves optimality
of the original solution. 
In the present Section we address the question how to find {\it finite}
optimal generalized measurements if they exist. This is a fundamental
question because only {\it finite} POVM schemes are experimentally 
realizable. 
We propose a universal algorithm how to look for these POVM schemes 
and we apply it explicitly in  two physically
interesting  cases of the state estimation of $N$ 
identically prepared two-level systems (qubits).

In order to set up the scene, let us assume that
state $\hat \rho$ is generated from a reference state
$\hat\rho_0=|\psi_0\rangle\langle \psi_0|$ by a unitary operation
$U({\bf x})$ which is an element of a particular unitary
finite dimensional representation of a compact Lie group $G$.
Different ${\bf x}$ denote different
points of the group (e.g., different angles of rotation in the case of
the $SU(2)$) and we assume that all values of ${\bf x}$ are equally
probable.

Our task is to design the most general POVM, mathematically described
as a set $\{\hat O_r\}_{r=1}^ R$ of positive Hermitian operators such
that $\sum_r \hat O_r = \hat{1}$ \cite{Peres93,Helstrom76}, which when
applied to the {\it combined} system of {\em all} $N$ copies provides
us with the best possible estimation of $\hat \rho$ (and therefore
also of $U({\bf x})$).	We quantify the quality of the state
estimation in terms of the {\it mean} fidelity
\begin{eqnarray}
\bar f  =  \sum_r \int_G \!\! d{\bf x}\ {\bf Tr}[\hat O_r\
\overbrace{U({\bf x})\hat\rho_0 U^\dagger({\bf x})\otimes \dots
\otimes  U({\bf x})\hat\rho_0
U^\dagger({\bf x})}^{\bf N\ times}]  
{\bf Tr}[ U({\bf x}) \hat \rho_0  U^\dagger ({\bf x}) \ U_r
\hat \rho_0 U_r^\dagger],
\label{10.1}
\end{eqnarray}
which corresponds to a particular choice of a cost function
\cite{Helstrom76} used in a context of detection and estimation theory.
The mean fidelity (\ref{10.1}) can be understood as follows: In order to
assess how good a chosen measurement is we apply it many times {\it
simultaneously} on {\it all} $N$ particles each in state $U({\bf x})
\hat \rho_0 U^\dagger ({\bf x})$.  The parameter ${\bf x}$ varies randomly
and isotropically\footnote{
We note that this isotropy condition is equivalent to a ``no a priori
information'' condition and is associated with the specific
integration measure in Eq.(\ref{10.1}). This measure has to be invariant
under the action of all unitary transformations on the state space of
pure states (for details see Appendix B).
}
over all points of the group $G$ during
many runs of the measurement.

For each result $r$ of the measurement, i.e., for each operator $\hat
O_r$, we prescribe the state $|\psi_r\rangle=U_r|\psi_0 \rangle$ 
representing our guess (i.e., estimation) of the original state.	The
probability of the outcome $r$ is equal to $ {\bf Tr}[\hat O_r\ U({\bf
x})\hat\rho_0 U^\dagger({\bf x})\!\otimes\! \dots \! \otimes \! U({\bf
x})\hat\rho_0 U^\dagger({\bf x})]$, while the corresponding fidelity of
our  estimation is ${\bf Tr}[
U\! ({\bf x}) \hat \rho_0 U^\dagger ({\bf x}) \ U_r \hat \rho_0
U_r^\dagger]$.	This fidelity is then averaged over all possible
outcomes and over many independent runs of the measurement with
randomly and isotropically distributed parameters ${\bf x}$. We want
to find the generalized measurement which {\it maximizes} the mean
fidelity $\bar f$ given by Eq.(\ref{10.1}).

The combined system of $N$ identically prepared reference states
always remains within the {\em totally symmetric subspace} of
$H^k\otimes H^k\otimes \dots H^k$, where $H^k$ is $k$-dimensional
Hilbert space of the reference state in which the corresponding
unitary representation $U({\bf x})$ acts. Thus the dimensionality $d$
of the space in which we construct the POVM $\{\hat O_r\}$ is $d=(\
^{N+k-1}_{k-1})$.  In this case
the first trace in Eq.(\ref{10.1}) can be
rewritten as
\begin{eqnarray}
\bar f  = \sum_r \int_G {\bf Tr}[\hat O_r\  U^{N}({\bf x})\hat \Omega_0 U^
{N\dagger}
({\bf x})] 
 {\bf Tr}[ U({\bf  x})\hat \rho_0 U^\dagger({\bf  x})\
U_r\hat \rho_0 U_r^\dagger ] d{\bf x},
\label{10.2}
\end{eqnarray}
where $ U^{N}({\bf x})$ is a new representation of the same group $G$;
it is equivalent to the $N$-fold symmetrized direct product
\cite{BarutRoncka} of the original representation $U({\bf x})$.
Here
$ U^{N}({\bf x})$ transforms the $(\ ^{N+k-1}_{k-1})$-dimensional
reference state denoted as $\hat{\Omega}_0$.

We can insert the identity operator $U_{r}^N U^{N\dagger}_{r}$ into
the first trace in Eq.~(\ref{10.2}) and, taking into account that in
Eq.(\ref{10.2}) we integrate over whole the group $G$ parameterized by
${\bf x}$, we can substitute $U^N({\bf x})U^{N\dagger}_{r}\rightarrow
U^N({\bf x}) $ and $U({\bf x})U^\dagger_{r}\rightarrow U({\bf x})$.
Now, using the linearity of the trace operation as well as the
linearity of the representation of the group $G$ ($U\hat \rho U^\dagger$ is
a linear adjoint representation) we rewrite Eq.(\ref{10.2}) as
\begin{eqnarray}
\bar f=\sum_r\ {\bf Tr}[\hat O_r\  U_{r}^N \hat F U^{N\dagger}_r],
\label{10.3}
\end{eqnarray}
where
\begin{equation}
\hat F= \int_G \
 U^{N}({\bf x})\hat \Omega_0 U^{N\dagger} ({\bf x})\ \ {\bf Tr}[
U({\bf x})\hat \rho_0 U^\dagger ({\bf x})\ \hat \rho_0] d{\bf x},
\label{10.4}
\end{equation}
is a positive Hermitian operator.

Let us now  derive an upper bound on the mean fidelity.  Taking into
account positivity of the operator $\hat F$ (i.e., $\hat
F=\sum_i\lambda_i |\phi_i\rangle\langle \phi_i|;~~\lambda_i\geq 0$)
and the completeness condition for POVM (i.e., $\sum_r \hat O_r=\hat
1$) we obtain
\begin{eqnarray}
\bar f&\!\! = \!\!\! &\sum_r\ \!\!\! {\bf Tr}[\hat O_r U_{r}^N \hat F
U^{N\dagger}_r]
	  \! = \!\! \sum_{i\ r}\!\!\!
	    \ \lambda_i {\bf Tr}[\hat O_r
	    U_{r}^N |\phi_i\rangle\langle \phi_i| U^{N\dagger}_r]
\nonumber \\
     &\leq & \lambda_{max}\sum_{i\ r}
	     \	{\bf Tr}[\hat O_r\
	     U_{r}^N |\phi_i\rangle\langle \phi_i| U^{N\dagger}_r]
\label{10.5} \\
	   &\! = \! & \lambda_{max} \sum_r \!\!\!
	     \	{\bf Tr}[\hat O_r\  U_{r}^N \hat 1 U^{N\dagger}_r]
	      \! = \! \lambda_{max} \ {\bf Tr}[ \hat 1]
	     =\lambda_{max}\ d.\nonumber
\end{eqnarray}
From Eq.(\ref{10.5}) it clearly follows that the upper bound can be
achieved if and only if all operators $\hat O_r$ forming the POVM
satisfy the following conditions:\newline {\it i)} Each $\hat O_r$ is
proportional to a suitably rotated (by some $U_{r}^N$) projector on
the eigenvector of $\hat F$ with the highest eigenvalue, i.e.  for all
$\hat O_r$ there exists $ U_{r}^N$, such that $ \hat O_r=c_r^2\
U_{r}^N|\phi_{max}\rangle\langle \phi_{max}|U^{N\dagger}_r$. This
$U_{r}^N$, or more precisely $U_r |\psi_0\rangle$, is our guess
associated with the result ``$r$''.\newline {\it ii)} All $c_r^2$ are
real and positive, to assure that all $\hat O_r$ are positive
operators.\newline {\it iii)} Finally, the operators $\hat O_r$ have
to satisfy the completeness criterion $\sum_r\ c_r^2\
U_{r}^N|\phi_{max}\rangle\langle \phi_{max}|U^{N\dagger}_r=
\hat 1$.\newline
As shown by Holevo  \cite{Holevo82}
in the case of the {\it infinite} POVM the condition
{\it iii)} is  fulfilled for covariant measurements,   providing
 the representation  $U^N$ of the group $G$ is irreducible (see
Example~A below). This statement  follows from the  Shur lemma.
However, in the general case of reducible representation and
specifically for {\it finite} realizable POVMs this  argument cannot be used
and we have to proceed differently.

To find the solution of the problem
we start with the following observation.  Let us assume that we have
some POVM $\{\hat O_r \}_{r=1}^R$ and the corresponding guesses
$U_{r}^N$ which maximize the mean fidelity $\bar f$. We can always
construct another POVM with more elements which is also optimal. For
example, let us consider a one-parametric subgroup $U(\phi)=\exp
(i\hat X \phi)$ of our original group $G$ and choose a basis $\{
|m\rangle \}_{m=1}^d$ in which the action of this subgroup is
equivalent to multiplication by a factor ${\bf e}^{i \omega_m
\phi}$ (i.e., the operator $U(\phi)$ is diagonal in this basis and
$\omega_m$ are eigenvalues of the generator $\hat X$).	Then we take
$d$ points $\phi_s$ ($s=1,\dots d$) and generate from each
original operator $\hat O_r$ a set of $d$ operators $\hat O_{rs}^,= {1
\over d} U^{N}(\phi_s) \hat O_r U^{N\dagger} (\phi_s)$. In this way we
obtain a new set of $(d\cdot R)$ operators such that the mean fidelity
for this new set of operators, $\bar f=\sum_{r,s}{\bf Tr}[\hat
O_{r,s}^,\ U_{r,s}^N \hat F U_{r,s}^{N\dagger}]$, is {\it equal} to
the mean fidelity of the original POVM $\{\hat O_r\}$ because we
ascribe to each eventual result $[r,s]$ a new guess $U_{r,s}=U(\phi_s)
U_r$.  However, in order to guarantee that the new set of operators
$\hat O_{rs}^,$ is indeed a POVM we have to satisfy the completeness
condition
\begin{eqnarray}
\hat 1  = \sum_{s} \sum_r \hat O^,_{rs}=
\sum_{s} \sum_r  {1 \over d} U^N(\phi_s)\hat O_r
U^{N\dagger} (\phi_s)  
        = \sum_s \sum _{m,n}
{{\bf e}^{i \phi_s(\omega_m-\omega_n)}\over d} \sum_r \left(
\hat O_r \right)_{mn}
|m\rangle \langle n|.
\label{10.8}
\end{eqnarray}
Let us notice that,
by the appropriate choice of $\phi_s$, the sum
$\sum_s {{\bf e}^{i \phi_s(\omega_m-\omega_n)}\over d}$ can {\em
always}
 be made equal to $\delta_{m,n}$ providing all eigenvalues
are non-degenerate\footnote{
In the case when the spectrum of the generator $\hat X$ is degenerate,
i.e., for some $m$ and $n$ we have $\omega_m=\omega_n$, then our algorithm
is still valid, provided we increase a number of Lagrange multipliers
in Eq.(\ref{10.9}) to account for off-diagonal elements $L_{mn}$ and $L_{nm}$
in the definition of the operator $\hat L$ in Eq.(\ref{10.9}).
}
(this is basically a discrete
Fourier transform and we illustrate this point in detail in
Example~A).
In this case, the conditions (\ref{10.8}) for the off-diagonal terms in
the basis $| m\rangle$ are {\em trivially} satisfied whereas the
diagonal terms are equal to unity because the original POVM $\{\hat
O_r\}$ guarantees that $\sum_r (\hat O_r )_{mm}=1$. Moreover, even if
the original set of operators $\{\hat O_r\}$ does not satisfy the full
completeness condition and the conditions for the off-diagonal terms
are {\it not} satisfied (i.e., these operators do not constitute a
POVM) we can, using our extension ansatz, always construct a {\it
proper } POVM $\{\hat O_{r,s}\}$.  This {\it proves} that when we
maximize the mean fidelity (\ref{10.3}) it is enough to assume $d$
diagonal conditions rather than the original complete set of $d^2$
constraints for diagonal {\it and} off-diagonal elements.

Now we turn back to our original problem of how to construct the POVM
which maximizes the mean fidelity.  To do so we first express the
operators $\hat O_r$ in the form $\hat O_r\!= \!c_r^2\ U_{r}^N
|\Psi_r\rangle \langle \Psi_r| U^{N\dagger}_r$, where $|\Psi_r
\rangle$ are general normalized states in the $d$-dimensional space in
which the operators $\hat O_r$ act, and $c_r^2$ are positive
constants.  This substitution is done without any loss of 
generality\footnote{
The most general choice of $\hat O_r$ would be
$\hat O_r=\sum_i c_{r,i}\ |\Psi_{r,i}\rangle \langle
\Psi_{r,i}|$.
 However, from the point of view of optimality
of the POVM these operators
are always less effective than operators
$\hat O_r\!= \!c_r^2\ U_{r}^N |\Psi_r\rangle \langle \Psi_r| U^{N\dagger}_r$
which are proportional to one-dimensional projectors.
}
and it permits us to rewrite Eq.(\ref{10.3}) so that the
mean fidelity $\bar f$ does not explicitly depend on $U_r^N$, i.e.
\begin{equation}
\bar f=\sum_r\ c_r^2\ {\bf Tr}[|\Psi_r\rangle \langle \Psi_r| \hat F ].
\label{10.6}
\end{equation}
Obviously, the completeness condition $\sum_r \hat O_r=\hat 1$ is now
modified and it reads
\begin{equation}
\sum_r	c_r^2\ U_{r}^N |\Psi_r\rangle \langle \Psi_r|
U^{N\dagger}_r\  =\  \hat 1.
\label{10.7}
\end{equation}

From our discussion above it follows that when maximizing the mean
fidelity (\ref{10.6}) it is enough to apply only $d$ constraints $\sum_r
c_r^2\left|\langle m |U_{N,r}|\Psi _r\rangle \right|^2 =1$ (here
$m=1,\dots, d$) out of the $d^2$ constraints (\ref{10.7}). Therefore to
accomplish our task we solve a set Lagrange equations with $d$
Lagrange multipliers $L_m$.  If we express $L_m$ as eigenvalues of the
operator $\hat L =\sum_m L_m\ |m\rangle \langle m|$ then we obtain the
final very compact set of equations determining the optimal POVM
\begin{eqnarray}
\left[ \hat F - U^{N\dagger}_r\! \hat L U_r^N \right]\! |\Psi_r\rangle\!
=\! 0, &\ \ \ \ &
\sum_r \! c_r^2\left|\langle m |U_r^N|\Psi _r\rangle \right|^2\!\!\! =\! 1.
\label{10.9}
\end{eqnarray}
From here it follows that $|\Psi_r\rangle$ are determined as
zero-eigenvalue eigenstates.  More specifically, they are functions of
$d$ Lagrange multipliers $\{L_m\}_{m=1}^d$ and $R$ vectors $\{{\bf
x}_r\}_{r=1}^R$ [where ${\bf x}_r$ determine $U_r$ as $U_r=U({\bf
x}_r)$].  These free parameters are in turn related via $R$ conditions
${\bf Det}[ (\hat  F - U_r^{N\dagger} \hat L U_r^N)]=0 $.  The mean
fidelity now is equal to ${\rm Tr} \hat L$.  At this stage we solve a
system of $d$ linear equations [see the second formula in
Eq.(\ref{10.9})] for $R$ unknown parameters $c_r^2$.  All solutions for
$c_r^2$ parametrically depend on $L_m$ and ${\bf x}_r$ which are
specified above.  We note that the number of free parameters in our
problem depends on $R$ which has not been specified yet. We choose $R$
such that there are enough free parameters so that the mean fidelity
is maximized and simultaneously all $c_r^2$ are positive.  This
freedom in the choice of the value of $R$ also reflects the fact that
there is an infinite number of equivalent (i.e., with the same value
of the mean fidelity) optimal POVMs.  The whole algorithm is completed
by finding $\phi_s$ from Eq.(\ref{10.8}) which explicitly determine the
{\it finite} optimal POVM $\{ \hat O_{rs}^,\}$. 

\subsection{Optimal reconstruction of spin-1/2 states}
\label{sec10.A}

Suppose we have $N$
identical copies of spin
$1/2$ all prepared in the same but unknown pure quantum state. If we
chose the group $G$ to be $U(2)$, i.e.	the complete unitary group
transforming a two-level quantum system, we can straightforwardly
apply the optimal estimation scheme as described above.  To be more
precise, due to the fact that there exist elements of the group $U(2)$
for which the reference state is the fixed point (i.e., it is
insensitive to its action) we have to work only with the coset space
$\ ^{SU(n)}|_{U(n-1)}$~\cite{BarutRoncka}.  In the present case this
is a subset of the $SU(2)$ group parameterized by two Euler angles
$\theta,\psi$ (the third Euler angle $\chi$ is fixed and equal to
zero). This subset  is isomorphic to the Poincare sphere.

The unitary representation $U$ is now the representation $({1\over
2})$ (we use a standard classification of $SU(2)$ representations,
where $(j)$ is the spin number).  Its $N$-fold symmetrized direct
product (we denote this representation as $U^N$) is the representation
classified as $({N \over 2})$ (which transforms a spin-$N/2$
particle).  Choosing the standard basis $|j,m\rangle$ with $m=-j,\dots
j$ in which the coordinate expression for $U(\theta,\psi)$ corresponds
to standard rotation matrices $D^j_{m,n}(\theta,\psi,0)={\rm
e}^{-im\psi}\ d^j_{m,n}(\theta)$~\cite{Zare}, we obtain the matrix
expression for the operator $\hat F$
\begin{eqnarray}
F_{m,n} & = &  \int _0^{2\pi} d\phi \int_0^\pi {\sin (\theta)
d\theta\over 8\pi}  (1 +\cos \theta)
D^{N \over 2}_{m,{N \over 2}}(\theta,\phi)
\ D^{{N \over 2}\ast }_{n,{N \over 2}}(\theta,\phi)
 =  {N/2 + m + 1 \over (N + 2)(N + 1)}\delta_{m,n}.
\label{10.10}  
\end{eqnarray}
When we insert this operator in the Eq.(\ref{10.5}) we immediately find
the upper bound on the mean fidelity to be equal to ${N+1\over N+2}$.

This is the main result of the paper
by Massar and Popescu \cite{MasPop} who
noted  that this upper bound can be attained using the
special POVM which consists of an {\it infinite} continuous set of
operators proportional to isotropically rotated projector
$|{N \over 2},{N \over 2} \rangle\langle {N \over 2},{N \over 2}|$.
This result is closely related to the covariant measurements of Holevo
\cite{Holevo82}.

However, our aim is to construct an  optimal and {\it finite}
POVM. To do so, we have to find a finite set of
pairs of angles $\{ (\theta_r, \psi_r) \}$ such that the completeness
conditions (\ref{10.7}) which now take the form
\begin{equation}
\sum_{r} c_r^2\  {\rm e}^{-i\psi_r(m-n)} d_{m,{N \over 2}}^{N \over 2}
(\theta_r)\
 d_{n,{N \over 2}}^{N \over 2}(\theta_r) = \delta_{m,n},
\label{10.11}
\end{equation}
are fulfilled.	Following our general scheme we first satisfy the
completeness conditions (\ref{10.11}) for diagonal terms [compare with
Eq.(\ref{10.9})]
\begin{eqnarray}
\sum_r	c_r^2\	d_{m,{N \over 2}}^{N \over 2}(\theta_r)^2= 1;\qquad
 m=-N/2,\dots	N/2.
\label{10.12}
\end{eqnarray}
\begin{figure}[t]
\begin{center}
\epsfig{file=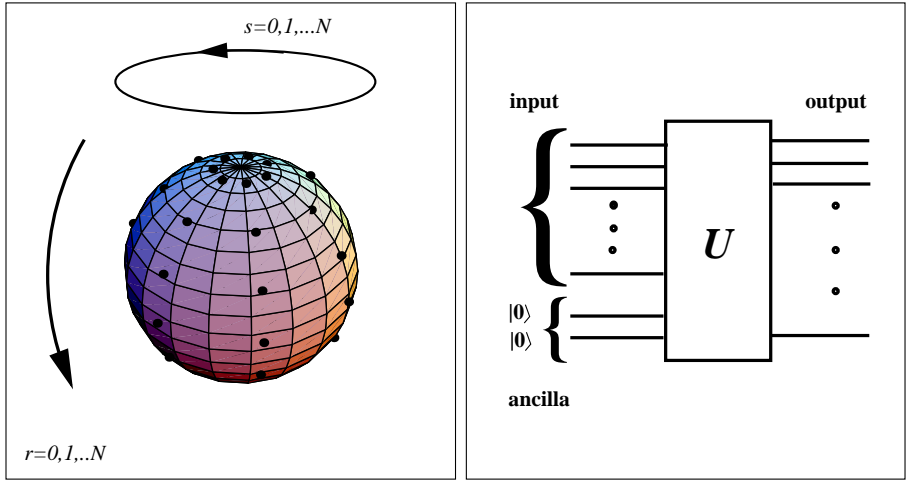, angle=-90, width=12cm}
\end{center}
\caption{
The visualization of the optimal POVM for measurement of a
quantum state of the spin-1/2 from $N$ identical copies as provided by
the solution in Section~\ref{sec10.A}.
On the right  the schematic picture of an
quantum network (for more information see  Section \ref{sec10.C}) as a
physical realization of the  measuring apparatus which embodies
the optimal solution from the left part of the picture. 
}
\label{fig14}
\end{figure}

To satisfy these completeness conditions we choose $N+1$ angles
$\theta_r$ to be equidistantly distributed in the $\langle 0,\pi
\rangle$ (obviously, there are many other choices which may suite the
purpose -- see discussion below Eq.(\ref{10.9})).	Then we solve the
system of linear equations for $N+1$ variables $c_r$.  For this
choice of $\theta_r$ the system~(\ref{10.12}) has non-negative solutions.
Finally we satisfy the off-diagonal conditions by choosing $N+1$
angles $\psi_{s}={2 s\pi \over N+1}$ for each $\theta_r$. In this case
${1 \over N+1} \sum_{s=0}^N {\rm e}^{i \psi_s y}=\delta_{y,0}$ for all
$y=-N/2,\dots N/2$ and the off-diagonal conditions are satisfied
straightforwardly.  This concludes the construction of the {\em
optimal} and {\it finite} POVM for the spin-$1/2$ state estimation.
In Fig.~\ref{fig14} we present a schematic description of the optimal
POVM performed on spin-1/2, while in Fig.~\ref{fig15} the mean fidelity
$\bar{f}$ as a function  of number $N$ of measured spins (initially
prepared in the same state) is presented.
\begin{figure}
\begin{center}
\epsfig{file=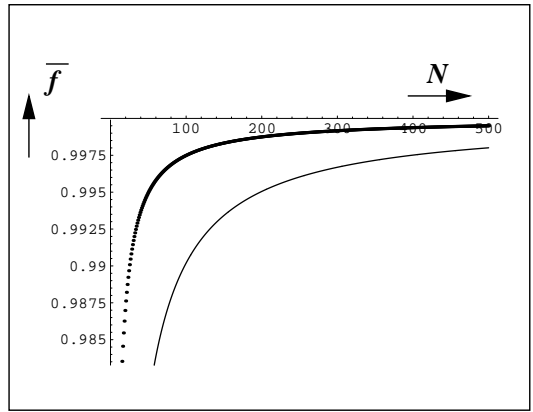, angle=0, width=7cm}
\end{center}
\caption{
The mean fidelity of the estimation of
a quantum state of the spin-1/2 (lower line) and of a phase shift (upper
line) based on the optimal POVM (as discussed in this Section) performed on
$N$ identical copies of quantum objects under consideration. As expected,
for $N$ large enough the fidelity in both cases becomes equal to unity.
}
\label{fig15}
\end{figure}

\subsection{Optimal estimation of phase shifts}
\label{sec10.B}

Consider  a system
of $N$ effectively two level atoms (qubits), all initially prepared in the 
reference state ${1\over \sqrt 2}(|0\rangle + |1\rangle)$  by applying
so called ${\pi \over 2}$ pulse to initially deexcited atoms.
Then the  atoms  undergo the  free evolution effectively
described by the $U(1)$ group, i.e. the  state of the single qubit 
evolves as 
${1\over \sqrt 2}(|0\rangle + \exp \{i\psi(t)\} |1\rangle)$.  
Our task is to find a measurement which provides the optimal estimation 
of the phase  $\psi(t)$ of the $U(1)$ rotation which carries the 
information about the interaction parameters.

In the standard classification 
of representations of the $U(1)$ group the
 single isolated qubit is described by the direct sum of two 
 one-dimensional representations 
$U=(0)\oplus(1)$. The representation $U^N$ transforming
entire system of $N$ qubits is then equal to the
direct sum of representations of the form $(0)\oplus
(1)\oplus \dots (N)$. This  acts in the $N+1$ dimensional space spanned
by basis vectors $|m\rangle$, $m=0,1,\dots N$.	In this basis matrix
elements $\hat F_{m,n}$ of the operator $\hat{F}$ given by Eq.(\ref{10.4})
take the form
\begin{eqnarray}
\hat F_{m,n}  = \int_0^{2\pi}{d\psi \over 2\pi}\ {\sqrt{
(^N_{N-m})
(^N_{N-n})}\over 2^{N+1}}\
{\rm e}^{i\psi (n-m)}\ (1 + \cos\psi) 
 = {\sqrt { (^N_{N-m})
(^N_{N-n})} \over 2^{N+2}} \left( 2 \delta_{m,n} +
 \delta_{m,n+1} +  \delta_{m+1,n} \right).
\label{10.13}
\end{eqnarray}
The upper bound on the fidelity Eq.~(\ref{10.5}) is now too conservative
to be of any use (greater than unity).	We can, however, solve the
system of Eqs.~(\ref{10.9}) which in this particular case of the
commutative group reads
\begin{eqnarray}
\left[ \hat F -\hat L \right ]|\Psi \rangle=0; \qquad
|\langle m|\Psi\rangle|^2=1;
\ \ \ \forall m.
\label{10.14}
\end{eqnarray}
The condition ${\bf Det}(\hat F - \hat L)=0$ now determines the
eigenvector $|\Psi\rangle$ with the zero eigenvalue as a function of
Lagrange multipliers $L_m$.  When we substitute this eigenvector into
the second equation in Eq.(\ref{10.14}) we obtain a set of equations for
$L_m$ from which the  state $|\Psi \rangle$ can be
determined.  The final POVM is then constructed by rotation of
$|\Psi\rangle$ by $N+1$ angles $\phi_s$ in such a way that all
off-diagonal elements of $\sum_s(\hat O_s)_{m,n}$
become equal to zero. This is done
in exactly the same way as in the example presented above.
(see Section~\ref{sec10.A}).
  The resulting POVM
corresponds to the {\it von Neumann measurement} performed on the {\em
composite} system of {\em all} $N$ ions characterized
by the set of orthogonal projectors
\begin{eqnarray}
\hat P_s=|\Psi_s\rangle \langle \Psi_s|; &\ \ \ \ \ &
|\Psi_s\rangle={1 \over \sqrt{N+1}} \sum _{q=0}^N
{\rm e}^{i {2 \pi \over N+1} s q} \ |q\rangle.
\label{10.15}
\end{eqnarray}
and the maximal mean fidelity $\bar f$ is given as the sum:
$\bar f=1/2 + 1/2^{N+1} \sum_{i=0}^{N-1} \sqrt{ (\ ^N_i)(\ ^N_{i+1})}$. 
We plot this fidelity in Fig.~\ref{fig15} (see upper line).

Finally, we note  that
the Hermitian operator $\hat \Phi$ 
constructed from the optimal POVM (\ref{10.15}) 
\begin{eqnarray}
\hat \Phi=\sum_{s-0}^N {2 \pi \over N+1}s \hat P_s,
\label{10.16}
\end{eqnarray}
with the corresponding guesses as eigenvalues is identical to 
  the Pegg-Barnett Hermitian phase operator
\cite{pegg} originally introduced within completely different
context.

In conclusion, we have presented a general algorithm for the optimal state
estimation from finite ensembles. It provides finite POVMs which, following the
Neumark theorem~\cite{Neumark}, can, at least in principle, be implemented as
simple quantum computations. 

\subsection{Neumark theorem and realization of generalized measurements}
\label{sec10.C}
Applicability of many ideas presented  in previous parts of this Section
critically relies on the assumption that the generalized quantum
measurements are in principle realizable. This is not obvious, since
the typical quantum measurements, e.g. measurements which may be performed
with the help of the Stern-Gerlach apparatus,
a photo-detection, or a measurement of an atomic population by means of
photo-ionization, etc.,  are all  orthogonal measurements.\footnote{This is, 
of course,
an  idealization. In practice, we never posses a perfect measuring apparatus. 
In other words, our measurements are always subject of an external and 
uncontrollable
noise. Therefore, in principle, we  always  perform a 
randomized measurements corresponding to  non-orthogonal POVMs.} 
Can we under this conditions hope to perform
a controlled generalized measurement?

To answer this question we start with the following consideration:
Assume a quantum object (${\cal S}$) 
in a state described by a density matrix $\hat \rho$.
Instead of directly  measuring it we subject this object to an interaction  
with another quantum object - the ancilla ${\cal A}$ (see Fig.~\ref{fig14}). 
The ancilla is initially prepared in 
a particular (fixed)  quantum state $|\alpha\rangle$. After some interaction
time  the composite system (${\cal S}+{\cal A}$) evolves into a nontrivial
entangled quantum state   
$U\hat \rho \otimes |\alpha\rangle \langle \alpha| U^\dagger$. The  orthogonal
measurement $\{ \hat P_r \}$
is then performed  on the composite quantum system. In this case
 the conditional probability distribution $p(r|\hat{\rho})$ defined as
\begin{eqnarray}
p(r|\hat{\rho})= 
{\rm Tr_{{\cal S} + {\cal A}}}
\left[\hat P_r\ U\hat\rho \otimes |\alpha\rangle \langle \alpha| U^\dagger\right] =
{\rm Tr_{\cal S}}\left[{\rm Tr_{\cal A}}
\left[U^\dagger \hat P_r U\ \hat 1 \otimes |\alpha\rangle \langle \alpha|
\right]
\ \hat \rho \right]=
{\rm Tr_{\cal S}}\left[\hat O_r \hat \rho\right].
\label{10.17}
\end{eqnarray}
can be specified.
It can be seen, that  each
projector $\hat P_r$ is associated with a new operator 
$\hat O_r={\rm Tr_{\cal A}}
[ \hat P_r \ U\hat 1 \otimes |\alpha\rangle \langle \alpha|U^\dagger]$.
In addition, it is easy to check that the operators $\hat O_r$ 
together compose the nonorthogonal POVM. Therefore, the described procedure 
represents
{\em an  orthogonal realization of a nonorthogonal} POVM.

Relation between the orthogonal and the nonorthogonal POVMs is actually even
more close. As showed by Neumark~\cite{Neumark}, 
not only any particular case of the procedure
we have described realizes an nonorthogonal POVM, 
but also the converse is true:

\vspace*{0.25cm} \noindent
{\bf Theorem  (Neumark)}
{\em Any POVM $\hat O(x)~dx$  defined in the  Hilbert 
space ${\cal H}$ may arise as a
restriction of an orthogonal POVM  $\hat E(x)~dx$ in a larger Hilbert space 
 $\bar{\cal H}$
\begin{eqnarray}
\hat O(x)~dx=\hat P \hat E(x) \hat P~dx,
\label{10.18}
\end{eqnarray}
where $\hat P$ is the projection from  $\bar{\cal H}$ onto  ${\cal H}$.}
\vspace*{0.25cm}

\noindent
In what follows we present  a construction which  proves  
the restricted version of this theorem.
Namely, we will assume only a finite-component POVMs 
$\{ \hat O_r\}_{r=1}^R$, where
each component is of the form $\hat O_r=c_r^2 |\Psi_r\rangle \langle \Psi_r|$.
For this  purpose the following construction is suitable:

\vspace*{0.25cm} \noindent
{\bf Proof} (special case)
We are looking for a unitary
transformation $U$, which satisfies the condition
\begin{eqnarray}
\hat O_r  = {\rm Tr_{\cal A}}\left[ U^\dagger \hat P_r U\ \hat 1 \otimes
|\alpha\rangle \langle \alpha|\right];~~~~~~~~~~~~~~~~~\forall r=1,2,\dots R.
\label{10.19}
\end{eqnarray}
It turns out, that for the  construction only $R$ of the all $d \times a $
($a$ and $d$ is the number of dimensions
of the ancilla and the measured system Hilbert spaces, respectively)
dimensions of the space ${\cal H}_{{\cal S} + {\cal A}}$  are relevant. 
Therefore we will construct
the unitary operation $U$ only on a subspace ${\cal H}^R \subset 
{\cal H}_{{\cal S} + {\cal A}}$.
To have a suitable  notation we
also divide Hilbert space ${\cal H}^R$ into two 
subspaces ${\cal H}^R={\cal H}^d \oplus {\cal H}^{R-d}={\cal H}^d 
\oplus {\cal H}^{k}$, where the first coincides  with  the linear span 
of  the vectors $|\Psi_r\rangle \otimes |\alpha\rangle $.
By inspection, if the unitary operation $U$ is of the form
\begin{eqnarray}
U=\sum_{r=1}^R |p_r\rangle \langle \psi_r | + 
\sum_{r=1}^R |p_r\rangle \langle \chi_r |,
\label{10.20}
\end{eqnarray}
where $|\psi_r\rangle \equiv c_r |\Psi_r\rangle \otimes |\alpha\rangle 
\in {\cal H}^d$  
and  $|\chi_r\rangle$ are  orthogonal to all $|\psi_s\rangle$ (i.e. 
 $|\chi_r\rangle \in H^k$ ), then Eq.~(\ref{10.18}) is 
satisfied. Therefore we need to find a proper set
of vectors $|\chi_r\rangle$, so  that $U$ is indeed a unitary
transformation, i.e.
\begin{eqnarray}
U U^\dagger  =  \sum_{r,s=1}^R \langle \psi_r | \psi_s \rangle
|p_r\rangle \langle p_s | +
\sum_{r,s=1}^R \langle \chi_r | \chi_s \rangle |p_r\rangle \langle p_s|
 = \hat A + \hat B = \hat 1.
\label{10.21}
\end{eqnarray}   
The  operators $\hat A$ and $\hat B$ are not diagonal in the basis 
$|p_r\rangle$. However, a new basis $|\bar p_r \rangle$ can be found in which
both operators $\hat A$ and $\hat B$ are diagonal. This basis  is found by
diagonalization of the matrix $A_{rs}=\langle \psi_r | \psi_s \rangle$ 
using the suitable  unitary matrix $V_{rs}$
[i.e., $|\bar p_r \rangle=\sum_{s=1}^R V_{sr} |p_s \rangle$, $\bar A_{rs}=
\sum_{i,j=1}^R V_{ri}^\dagger A_{ij} V_{js}$]. 
Moreover, since the following relations 
\begin{eqnarray}
(\hat A^2)_{rs}=\sum_{i=1}^R A_{ri} A_{is}=(\hat A)_{rs}, 
&~~~~~~~~~~~~&
{\rm Tr}[\hat A]
=\sum_{s=1}^R A_{ss}=d,
\label{10.22}
\end{eqnarray}
are satisfied (both as the  consequence of the completeness condition of the original POVM) 
we know that the  spectra of the  operators $\hat A$ and $\hat B$ are: 
${\rm Sp} \hat A =\{ \overbrace{1,1,\dots 1}^d,\overbrace{0,0,\dots
0}^{R-d}\}$;
${\rm Sp} \hat B =\{ \overbrace{0,0,\dots 0}^d,\overbrace{1,1,\dots 1}^{R-d}\}$. 
Therefore, if we use the basis $|\bar p_r \rangle$ instead of the basis  
$|p_r \rangle$  we 
can write the  unitary transformation $U$ in the Eq.~(\ref{10.18}) in the form
\begin{equation}
U=\sum_{r=1}^d |\bar p_r\rangle \langle \bar \psi_r | +
\sum_{r=d+1}^R |\bar p_r\rangle \langle \bar \chi_r |.
\label{10.23}
\end{equation}
Because the matrices $\bar A_{rs}=\langle \bar \psi_r |
\bar \psi_s \rangle$ and $\bar B_{rs}=\langle \bar \chi_r |
\bar \chi_s \rangle$ are diagonal 
 only those $|\bar \psi_r \rangle = \sum_{s=1}^R \langle  p_s | 
\bar p_r \rangle
|\psi _s \rangle =\sum_{s=1}^R V_{sr}
|\psi _s \rangle$ and $|\bar \chi_s \rangle = \sum_{r=1}^R V_{rs}|\chi _r 
\rangle$
 are nonzero for  which $r=1,2,\dots d$ and
$s=d+1,\dots R$, respectively. Moreover, this also justifies our 
assumption about the existence of the vectors $|\chi_r\rangle$ orthogonal to
all $|\psi_s\rangle$. 
More exactly, as we know the vectors $ |\bar \psi_r\rangle$, $r=1,2,\dots d$ 
we can
complete them into an orthonormal basis in ${\cal H}^R$  by choosing the set of
vectors $|\bar \chi_s\rangle$,
$s=d+1,\dots R$. Because the vectors with  bar are related to the vectors
without  bar  via  the unitary matrix 
$V_{sr}$, the vectors $|\chi_r\rangle$ 
and  $|\psi_s\rangle$ remain  mutually orthogonal. This concludes our proof.
\vspace*{0.25cm}

\noindent We note, that  there is a  freedom in the 
way how  we choose to complete the set of vectors 
$ |\bar \psi_r\rangle$, $r=1,2,\dots d\ $  
by the vectors  $|\bar \chi_r\rangle$,
$r=d+1,\dots R$, so that they together form an orthonormal basis.
The explicit form of the unitary operation $U$ from  Eq.~(\ref{10.18}) is
\begin{equation}
U=\sum_{r=1}^R |p_r\rangle \langle \psi_r | +
\sum_{r=1}^R |p_r\rangle  \sum_{s=d+1}^R V_{rs} \langle \bar \chi_s |.
\label{10.24}
\end{equation}
Therefore we  conclude: If we can in the system composed  
of the original object
and  the auxiliary system perform  a particular von 
Neumann measurement characterized by a
 set of orthogonal projectors $|p_r\rangle \langle p_r|$ and, in addition,
we can transform this composed system by 
the unitary transformation $U$ [see Eq.~(\ref{10.24})], then we can realize 
a general POVM measurement $\{\hat O_r\}$ (for more details see \cite{MY}).

\section{INSTEAD OF CONCLUSIONS}
\label{sec11}
We conclude this paper by a citation from the  
 Jaynes'  Brandeis lectures 
(see p. 183 of Ref.\cite{Jaynes62}): 
``{\it Conventional quantum theory has provided
an answer to the problem of setting up initial state descriptions only in the
limiting case where measurements of a  ``complete set of commuting
observables'' have been made, the density matrix $\hat{\rho}(0)$ then
reducing to the projection operator onto a pure state $\psi(0)$ which is the
appropriate simultaneous eigenstate of all measured quantities. But there is
almost no experimental situation in which we really have all this information,
and before we have a theory able to treat actual experimental situations,
existing quantum theory must be supplemented with some principle that tells
us how to translate, or encode, the results of measurements into a definite 
state
description $\hat{\rho}(0)$. Note that the problem is not to find
$\hat{\rho}(0)$  which correctly describes ``true physical situation''.
That is unknown, and always remains so, because of incomplete information.
In order to have a usable theory we must ask the much more
modest question: {\em {\bf  What $\hat{\rho}(0)$ best describes our state of
knowledge about the physical situation?}}\,}''.

\acknowledgements
We thank many our friends with whom we had a pleasure to discuss various
aspects of the problem of quantum-state reconstruction. In particular,
we thank Peter Knight, Artur Ekert, Ulf Leonhardt, Zden\v{e}k Hradil,
Tom\'{a}\v{s} Opatrn\'y, and Jason Twamley.
This work was supported in part by the Royal Society and in part by the 
Jubil\"aumsfonds der \"Osterreichischen Nationalbank under Contract No 5968.

\newpage


\section*{Appendix A}
Conceptually the reconstruction scheme based on the Jaynes principle
of the maximum entropy is very simple. On the other hand particular
analytical calculations can be difficult and in many cases cannot
be performed. In this appendix we present explicit calculations
of generalized canonical density operators (GCDO) and corresponding
entropies for two observation levels
${\cal O}_G^{(2)}$ and ${\cal O}_H^{(2)}$ defined in  Table~\ref{table6}.

\subsection*{ A. 1. Observation level ${\cal O}_G^{(2)}$}
Let as assume the observation level  ${\cal O}_G^{(2)}$
given by the set of observables
$\{\hat \sigma _z^{(1)}\otimes\hat\sigma _z^{(2)};
\hat\sigma _x^{(1)}\otimes\hat\sigma _x^{(2)};
\hat \sigma _x^{(1)}\otimes\hat \sigma _y^{(2)};
\hat\sigma _y^{(1)}\otimes\hat\sigma _x^{(2)};
\hat \sigma _y^{(1)}\otimes\hat \sigma _y^{(2)}\}$. In this case
the GCDO reads
$$
\hat\rho_{ G}={1 \over {Z_G}}\exp \left( {-\hat E} \right),
\eqno({\rm A.1})
$$
where
$$
Z_G= \mbox{Tr}\left[ {\exp \left( {-\hat E} \right)} \right],
\eqno({\rm A.2}) 
$$
is the partition function. Here we have used the abbreviation
$$
\hat E=\lambda _{zz}\hat \sigma _z^{(1)}\otimes\hat\sigma _z^{(2)}
+\lambda _{xx}
\hat\sigma _x^{(1)}\otimes\hat\sigma _x^{(2)}+\lambda _{xy}
\hat \sigma _x^{(1)}\otimes\hat \sigma _y^{(2)}
+\lambda _{yx}\hat\sigma _y^{(1)}\otimes\hat\sigma _x^{(2)}
+\lambda _{yy}\hat \sigma _y^{(1)}\otimes\hat \sigma _y^{(2)}
.\eqno({\rm A.3}) $$
The corresponding entropy has the form
$$
S_G=\ln Z_G+\lambda _{zz}\xi _{zz}+\lambda _{xx}\xi _{xx}
+\lambda _{xy}\xi _{xy}
+\lambda _{yx}\xi _{yx}+\lambda _{yy}\xi _{yy}
,\eqno({\rm A.4})
$$
Using the algebraic properties of the operators associated with the given
observation level we find  the GCDO (A.1) to read
$$
\hat\rho_{ G}={1 \over 4}\left[ \hat I^{(1)}\otimes\hat I^{(2)}+
\xi _{zz}\hat\sigma _z^{(1)}\otimes \hat\sigma _z^{(2)}
+\xi _{xx}\hat \sigma _x^{(1)} \otimes\hat \sigma _x^{(2)} \right.
$$ $$
\left. +\xi _{xy}\hat \sigma _x^{(1)} \otimes\hat \sigma _y^{(2)}
+\; \xi _{yx}\hat \sigma _y^{(1)}\otimes \hat\sigma _x^{(2)}
+\xi _{yy} \hat\sigma _y^{(1)}\otimes \hat\sigma _y^{(2)}\right],
\eqno({\rm A.5})
$$
where we use the notation
$$
\xi _{\mu\nu}\equiv \left\langle {\hat \sigma _\mu^{(1)}
\otimes\hat \sigma _\nu^{(2)}} \right\rangle, \qquad (\mu,\nu=x,y,z).
\eqno({\rm A.6}) $$

Now we	express the entropy as a function of expectation
values of operators associated with the observation level ${\cal O}_G^{(2)}$.
With the help of this entropy function we can perform reductions of
${\cal O}_G^{(2)}$ to the observation levels
${\cal O}_H^{(2)}$, ${\cal O}_F^{(2)}$ and ${\cal O}_E^{(2)}$.
In order to perform this reduction we
express $\lambda _{\mu\nu}$ in Eq. ({\rm A.4}) as functions of
the expectation values $\xi _{\mu\nu}$. To do so we utilize the relation
$$
\xi _{\mu\nu}=-{{\partial \ln Z_G} \over {\partial \lambda _{\mu\nu}}}
.\eqno({\rm A.7})
$$
The partition function $Z_G$ can be found when we rewrite the operator
$\hat{E}$ in Eq.(A.4) as a 4$\times$4 matrix:
$$
\hat E=\left( {\matrix{a&0&0&{d^*}\cr
0&{-a}&{b^*}&0\cr
0&b&{-a}&0\cr
d&0&0&a\cr
}} \right)
,\eqno({\rm A.8})
$$
where we used the abbreviations
$$
  a=\lambda _{zz} , \quad
  b=\lambda _{xx}+\lambda _{yy}-i(\lambda _{xy}-\lambda _{yx}), \quad
  d=\lambda _{xx}-\lambda _{yy}+i(\lambda _{xy}+\lambda _{yx})
.\eqno({\rm A.9})
$$
The powers of the operator $\hat{E}$ can be written as
$$
\hat E^n=\left( {\matrix{{E_{11}^{(n)}}&0&0&{E_{14}^{(n)}}\cr
0&{E_{22}^{(n)}}&{E_{23}^{(n)}}&0\cr
0&{E_{32}^{(n)}}&{E_{33}^{(n)}}&0\cr
{E_{41}^{(n)}}&0&0&{E_{44}^{(n)}}\cr
}} \right)
,\eqno({\rm A.10})
$$
with the matrix elements given by the relations
$$
\begin{array}{rcl}
E_{11}^{(n)} &= & E_{44}^{(n)}=
{1 \over 2}\left[ {\left( {a+\left| d \right|} \right)^n
+\left( {a-\left| d \right|} \right)^n} \right] \, ,\\
E_{14}^{(n)} & = & {1 \over 2}\left[ {\left( {a+\left| d \right|} \right)^n
-\left( {a-\left| d \right|} \right)^n} \right]{{d^*}
\over {\left| d \right|}} \, , \\
E_{22}^{(n)} & = & E_{33}^{(n)}=
{1 \over 2}\left[ {\left( {-a+\left| b \right|} \right)^n
+\left( {-a-\left| b \right|} \right)^n} \right] \, ,\\
E_{23}^{(n)} & = & {1 \over 2}\left[ {\left( {-a+\left| b \right|} \right)^n
-\left( {-a-\left| b \right|} \right)^n} \right]{{b^*}
\over {\left| b \right|}} \, ,\\
   E_{32}^{(n)} & = & E_{23}^{(n)*} \, ,\\
   E_{41}^{(n)}& = & E_{14}^{(n)*} \, .
\end{array}
\eqno({\rm A.11})
$$
Now we find
$$
\exp \left( {-\hat E} \right)=\left( {\matrix{{e^{-a}\cosh\left| d
\right|}&0&0&
{-e^{-a}\sinh(\left| d \right|){{d^*} \over {\left| d \right|}}}\cr
0&{e^a\cosh\left| b \right|}&{-e^a\sinh(\left| b \right|){{b^*}
\over {\left| b \right|}}}&0\cr
0&{-e^a\sinh(\left| b \right|){{b^{}} \over {\left| b \right|}}}&
{e^a\cosh\left| b \right|}&0\cr
{-e^{-a}\sinh(\left| d \right|){{d^{}} \over {\left| d \right|}}}&
0&0&{e^{-a}\cosh\left| d \right|}\cr
}} \right), 
\eqno({\rm A.12})
$$
from which we obtain the expression for  the partition function $Z_G$
$$
Z_G=2e^{-a}\cosh\left| d \right|+2e^a\cosh\left| b \right| .
\eqno({\rm A.13})
$$
For the  expectation values given by Eq.(A.7) we obtain
$$
\begin{array}{rcl}
\xi _{zz} & = & {1 \over {Z_G}}\left[ {2e^{-a}\cosh\left| d \right|
-2e^a\cosh\left| b \right|} \right] \, ;\\
\xi _{xx}& =& -{1 \over {Z_G}}\left[ {2e^{-a}\sinh(\left| d \right|){{1^{}}
\over
{\left| d \right|}}\left( {\lambda _{xx}-\lambda _{yy}}
\right)+2e^a\sinh(\left| b \right|){{1^{}} \over {\left| b \right|}}
\left(
{\lambda _{xx}+\lambda _{yy}} \right)} \right] \, ; \\
\xi _{xy} & = &-{1 \over {Z_G}}\left[ {2e^{-a}\sinh(\left| d \right|)
{{1^{}}
  \over {\left| d \right|}}\left( {\lambda _{xy}+\lambda _{yx}}
\right)
  +2e^a\sinh(\left| b \right|){{1^{}} \over {\left| b \right|}}\left(
  {\lambda _{xy}-\lambda _{yx}} \right)} \right] \, ; \\
\xi _{yx} & = &-{1 \over {Z_G}}\left[ {2e^{-a}\sinh(\left| d \right|)
{{1^{}}
  \over {\left| d \right|}}\left( {\lambda _{xy}+\lambda _{yx}} \right)
  -2e^a\sinh(\left| b \right|){{1^{}} \over {\left| b \right|}}\left(
{\lambda _{xy}
  -\lambda _{yx}} \right)} \right] \, ;\\
\xi _{yy} & = & -{1 \over {Z_G}}\left[ {-2e^{-a}\sinh(\left| d \right|)
{{1^{}}
  \over {\left| d \right|}}\left( {\lambda _{xx}-\lambda _{yy}} \right)
  +2e^a\sinh(\left| b \right|){{1^{}} \over {\left| b \right|}}\left(
{\lambda _{xx}
  +\lambda _{yy}} \right)} \right] \,.
\end{array}
\eqno({\rm A.14})
$$
If we introduce the abbreviations
$$
B=\xi _{xx}+\xi _{yy}-i\left( {\xi _{xy}-\xi _{yx}} \right) \, , \quad
D=\xi _{xx}-\xi _{yy}+i\left( {\xi _{xy}+\xi _{yx}} \right), 
\eqno({\rm A.15})
$$
then with the help of Eq. (A.14) we obtain
$$
B=-{4 \over {Z_G}}e^a\sinh(\left| b \right|){{b^{}} \over {\left| b
\right|}}
\, , \quad
D=-{4 \over {Z_G}}e^{-a}\sinh(\left| d \right|){{d^{}} \over
{\left| d \right|}}
.\eqno({\rm A.16})
$$
Taking into account that
$$
\left| B \right|={4 \over {Z_G}}e^a\sinh(\left| b \right|) \, , \quad
\left| D \right|={4 \over {Z_G}}e^{-a}\sinh(\left| d \right|), 
\eqno({\rm A.17})
$$
we find
$$
{B \over {\left| B \right|}}=-{b \over {\left| b \right|}}
 \, , \quad
{D \over {\left| D \right|}}=-{d \over {\left| d \right|}}
.\eqno({\rm A.18})
$$
Now we introduce four new parameters $M_i$
$$
M_1=1+\xi _{zz}+\left| D \right|  \, , \quad
  M_2=1+\xi _{zz}-\left| D \right|  \, , \\
$$
$$
  M_3=1-\xi _{zz}+\left| B \right|  \, , \quad
  M_4=1-\xi _{zz}-\left| B \right|, 
\eqno({\rm A.19})
$$
in terms of which we can express the von\,Neumann entropy on the
given observation level.
Using Eqs. (A.13), (A.14) and (A.17) we obtain
$$
\begin{array}{rl}
M_1={4 \over {Z_G}}\exp \left( {-a+\left| d \right|} \right)  \, , &
\quad
    M_2={4 \over {Z_G}}\exp \left( {-a-\left| d \right|} \right)
\, ,  \\
  M_3={4 \over {Z_G}}\exp \left( {a+\left| b \right|} \right)	\,
, &
\quad
  M_4={4 \over {Z_G}}\exp \left( {a-\left| b \right|} \right).
\end{array}
\eqno({\rm A.20})
$$
The Lagrange multipliers
$\lambda _{kl}$ can be expressed
as functions of the expectation values $\xi _{kl}$:
$$
  \exp \left( a \right)=\left( {{{M_3M_4}
  \over {M_1M_2}}} \right)^{{1 \over 4} }  \, , \quad
\exp \left( {\left| b \right|} \right)=
\left( {{{M_3} \over {M_4}}} \right)^{{1 \over 2}}    \, , \quad
  \exp \left( {\left| d \right|} \right)=
  \left( {{{M_1} \over {M_2}}} \right)^{{1 \over 2}} \, .
\eqno({\rm A.21})
$$
After inserting these expressions into Eq. (A.13) we obtain
for the  partition  function
$$
Z_G={4 \over {\left( {M_1M_2M_3M_4} \right)^{{1 \over 4}}}}
.\eqno({\rm A.22})
$$
When we insert Eqs. (A.18), (A.21) and (A.22) into Eqs.
(A.1), (A.4) and (A.12) then we find  both the entropy
$$
S_G=-\sum\limits_{i=1}^4 {{{M_i} \over 4}}\ln \left( {{{M_i} \over 4}}
\right), 
\eqno({\rm A.23})
$$
and the GCDO
$$
\hat{\rho}_G ={1 \over 4}\left( {\matrix{{1+\xi _{zz}}&0&0&{D^*}\cr
0&{1-\xi _{zz}}&{B^*}&0\cr
0&B&{1-\xi _{zz}}&0\cr
D&{}&{}&{1+\xi _{zz}}\cr
}} \right), 
\eqno({\rm A.24})
$$
as functions of the expectation values $\xi _{kl}$. Finally,
we can rewrite the reconstructed density operator (A.24)
in terms of the spin operators (see Table~\ref{table7}).

\subsection*{ A. 2.  Observation level ${\cal O}_H^{(2)}$}

The GCDO on the  ${\cal O}_H^{(2)}$ can be obtained as a result of a reduction
of the observation level ${\cal O}_G^{(2)}$. The difference between these
two observation levels is  that the ${\cal O}_H^{(2)}$ does not contain the operator
$\hat \sigma _z^{(1)}\otimes \hat\sigma _z^{(2)}$, i.e.,
the corresponding mean value is unknown from the measurement.

According to the maximum--entropy principle,
the observation level ${\cal O}_H^{(2)}$ can be obtained from ${\cal O}_G^{(2)}$
by setting the Lagrange multiplier
$\lambda_{zz}$ equal to zero.
With the help of the relation [see Eq.(A.7)]
$$
\lambda_{zz}={{\partial S_G} \over {\partial \xi_{zz}}}
=-{1 \over 4}\ln \left( {{{M_1 M_2} \over {M_3 M_4}}} \right) = 0, 
\eqno({\rm A.25})
$$
we obtain
$$
M_1M_2=M_3M_4
.\eqno({\rm A.26})
$$
From this equation we find the ``predicted'' mean value of the
operator $\hat \sigma _z^{(1)}\otimes \hat\sigma _z^{(2)}$ (i.e.,
the parameter $t$ in Table~\ref{table3})
$$
\xi _{zz}={1 \over 4}\left( {\left| D \right|^2
-\left| B \right|^2} \right) \equiv t .
\eqno({\rm A.27})
$$
Taking into account that the parameters $| B|$ and $|D|$ read
$$
\left| B \right|^2=\left( {\xi _{xx}+\xi _{yy}} \right)^2
+\left( {\xi _{xy}-\xi _{yx}} \right)^2  \, , \quad
  \left| D \right|^2=\left( {\xi _{xx}-\xi _{yy}} \right)^2
+\left( {\xi _{xy}+\xi _{yx}} \right)^2 \, ,
\eqno({\rm A.28})
$$
we can express the predicted mean value
$\xi _{zz}$ as a function of the
measured mean values $\xi _{xx}, \,
\xi _{xy},  \, \xi _{yx} \, $ and $ \, \xi _{yy}$:
$$
\xi _{zz}=\left( {\xi _{xy}\xi _{yx}-\xi _{xx}\xi _{yy}} \right) \, .
\eqno({\rm A.29})
$$

When we insert
Eq. (A.27) into Eq. (A.19) we obtain:
$$
  M_1=N_1 N_2  \, , \quad
  M_2=N_3 N_4  \, , \quad
  M_3=N_1 N_3  \, , \quad
  M_4=N_2 N_4\, ,
  \eqno({\rm A.30})
$$
where the parameters $N_i$ are defined as
$$
N_1=1+{1 \over 2}\left( {\left| D \right|+\left| B \right|} \right)
\, , \quad
  N_2=1+{1 \over 2}\left( {\left| D \right|-\left| B \right|} \right)
\, , \\
$$
$$
  N_3=1-{1 \over 2}\left( {\left| D \right|-\left| B \right|} \right)
\, , \quad
  N_4=1-{1 \over 2}\left( {\left| D \right|+\left| B \right|} \right)
\, .
\eqno({\rm A.31})
$$
In addition, from
Eqs. (A.30) and (A.23) we obtain the expression for the
von\,Neumann entropy of the density operator reconstructed on
the observation level ${\cal O}_H^{(2)}$:
$$
S_H=-\sum\limits_{i=1}^4 {{{N_i} \over 2}}
\ln \left( {{{N_i} \over 2}} \right) \, .
\eqno({\rm A.32})
$$
Finally, from Eqs. (A.28) and (A.24) we find the expression
for the GCDO on the observation level  ${\cal O}_H^{(2)}$
(see Table~\ref{table7}):
\begin{eqnarray}
\begin{array}{rcl}
\hat\rho_{ H}&=&{1 \over 4}{[ \hat I^{(1)}\otimes \hat I^{(2)}+
\left( {\xi _{xy}\xi _{yx}-\xi _{xx}\xi _{yy}} \right)
\hat \sigma _z^{(1)}\otimes \hat \sigma _z^{(2)}} \\
&+& \; {\xi _{xx}\hat \sigma _x^{(1)}\otimes \hat \sigma _x^{(2)}
+\xi _{xy} \hat \sigma _x^{(1)}\otimes \hat \sigma _y^{(2)}}
+\; \xi _{yx}\hat \sigma _y^{(1)}\otimes \hat\sigma _x^{(2)}
+\xi _{yy} \hat\sigma _y^{(1)}\otimes \hat\sigma _y^{(2)}] \, .
\end{array}
\eqnum{A.33}
\label{A.33}
\end{eqnarray}

\section*{APPENDIX B: INVARIANT INTEGRATION MEASURE}
In differential geometry the
integration measure is a global object - the so called invariant volume
form $\omega$.
The condition that $d_{\Omega}$ is invariant under the action of each group
element $U\in SU(n)$ is equivalent to the requirement
\begin{eqnarray}
d_{_\Omega}=d_{_{U\Omega U^{-1}}}\ \ \Longleftrightarrow \ \
L_{V_i}\omega=0\qquad
i=1,\dots,n^2-1,
\eqnum{B.1}
\label{B.1}
\end{eqnarray}
that the Lie derivative of
$\omega$ with respect to the fundamental field $V_i$ of action of the group
$SU(n)$ in the space $\Omega $ is zero.
The vector fields
\begin{eqnarray}
V_i=V^b_i(x_1,\dots ,x_{(2n-2)})
{\partial\over \partial x_b};~~~ b=1,2\dots (2n-2), 
\eqnum{B.2}
\label{B.2}
\end{eqnarray}
are defined
via the actions of one-parametric subgroups
$\exp (it\hat S_i)\subset SU(n),\ t\in R$ (one
action for each generator $\hat S_i$).
On the other hand the elements of the space $\Omega$  [see Eq.(\ref{9.19})]
have a structure
\begin{eqnarray}
\hat\rho(x_1,\dots ,x_{(2n-2)})={\hat 1\over n}+f^i(x_1,\dots ,x_{(2n-2)})\hat S_i,
\eqnum{B.3}
\label{B.3}
\end{eqnarray}
where $\hat S_i$ are $n^2-1$ linearly independent, zero-trace, Hermitian,
$n\times n$ matrixes, i.e. they  are  generators of the $SU(n)$ group.
Due to this  we can  express the vector fields	$V_i$
\begin{eqnarray}
V^b_i{\partial \over \partial x_b}\hat \rho=
{\partial \over \partial t} \left.\left[\exp (it\hat S_i)\hat \rho
\exp (-it\hat S_i)\right]\right|_{t=0}, 
\eqnum{B.4}
\label{B.4}
\end{eqnarray}
as the solutions of the  following equation:
\begin{eqnarray}
V^b_i{\partial \over \partial x_b}f^k=ic^k_{ij}f^j.
\eqnum{B.5}
\label{B.5}
\end{eqnarray}
The complex numbers $c^k_{ij}$ are the	coefficients in  commutation relations
$[\hat S_i,\hat S_j]=c^k_{ij}\hat S_k$. We note, that Eq.~(\ref{B.5})
represents  for each fixed index
$i$ an overdetermined system of $n^2-1$ linear equations for
$2n-2$ unknown functions $V^b_i$ (the fact that this system is
consistent confirms the  correctness of our  parameterization of the state
space  $\Omega $).
Finally, we present an explicit coordinate form of Eq.(\ref{B.1}), which
determines the invariant volume form
$\omega=m(x_1,\dots,x_{(2n-2)})\land dx_1 \land \dots \land dx_{(2n-2)}$
as the solution of a system of partial differential equations:
\begin{eqnarray}
{\partial \over \partial x_b}(m V_i^b)=0. 
\eqnum{B.6}
\label{B.6}
\end{eqnarray}
Here we note, that  $m \vec V_i$ in Eq.(\ref{B.6}) has the
meaning of a ``flow'' of the density of states generated by
unitary transformations  associated  with the  $i$-th generator.
From  the physical point of view Eq.(\ref{B.6}) means
that the divergence of
this flow is zero, i.e. the number of states in each (confined) volume element
is constant.

As an illustration of the above discussion
 we firstly evaluate the invariant measure for the state space
of a single spin-1/2.
Using   the
definition (\ref{B.5}) we find the fundamental field of action $V_i$
($i=1,2,3$) for the three generators  of  the $SU(2)$ group:
\begin{eqnarray}
V_1=\cos (\phi)\cot (\theta) \partial_\phi+\sin (\phi)\partial_\theta;\ \ \ \
V_2=\sin (\phi)\cot (\theta) \partial_\phi-\cos (\phi)\partial_\theta; \ \ \ \
V_3= -\partial_\phi.
\eqnum{B.7}
\label{B.7}
\end{eqnarray}
We substitute these generators into Eq.(\ref{B.6}) and after some
algebra we obtain the  system of differential equations:
\begin{eqnarray}
{\partial \over \partial \phi}m=0\qquad
{\partial \over \partial \theta}m=m\cot(\theta ),
\eqnum{B.8}
\label{B.8}
\end{eqnarray}
which can be easily solved,
\begin{eqnarray}
m(\theta,\phi )=const\ \sin(\theta ). 
\eqnum{B.9}
\label{B.9}
\end{eqnarray}
The multiplicative factor is given by the normalization condition.
This is the route to
derive the integration measure of the Poincar\'{e} sphere.
Analogously we evaluate the invariant integration measure for a state space
of two spins-1/2.
The calculations are technically more involved, but the
result is simple see Eq.~(\ref{9.21}).

\section*{Appendix C: Bayesian inference on ${\cal O}_B^{(2)}$
in the limit of infinite number of measurements}
On the given observation level we can express the estimated
density operator in the limit of infinite number of measurements
as 
\begin{eqnarray}\nonumber
\hat\rho= \frac{1}{\cal N}
\int_0^{2 \pi}\!\!d\psi\!\!\int_{-1}^{1}\!\!x^2dx\!\!
\int_{-1}^{1}\!\!dy\!\!\int_{-1}^{1}\!\!dz
\delta\left(\langle \hat \sigma_z^{_{(1)}}\rangle -xy)
\delta(\langle \hat \sigma_z^{_{(1)}}\hat \sigma_z^{_{(2)}}\rangle -yz+
\left[
(1-x^2)(1-y^2)(1-z^2)\right]^{1/2}\cos\psi\right)
\end{eqnarray}
\begin{eqnarray}
\times
\left\{ \hat 1\otimes \hat 1+xy\hat\sigma_z \otimes \hat 1+xz\hat 1\otimes
\hat\sigma_z
+\left[ yz-
\left((1-x^2)(1-y^2)(1-z^2)\right)^{1/2}\cos\psi\right]
\hat\sigma_z\otimes\hat\sigma_z\right\}.
\eqnum{C.1}
\label{C.1}
\end{eqnarray}
When  we integrate Eq.(\ref{C.1}) over the variable $y$ and we obtain
\begin{eqnarray}\nonumber
\hat\rho = \frac{1}{\cal N}
\int_0^{2 \pi} \, d\psi \int_{{\cal L}'}
{x^2\over
\vert x\vert}dx\!\!\int_{-1}^{1}\, dz
 \delta\left(\langle \hat \sigma_z^{_{(1)}}\hat \sigma_z^{_{(2)}} \rangle -
\langle \hat \sigma_z^{_{(1)}}\rangle
{z\over x}+\left[
(1-x^2)(1-z^2)\left(1-{\langle \hat \sigma_z^{_{(1)}}
\rangle^2\over x^2}\right)\right]^{1/2}
\cos\psi\right)
\end{eqnarray}
\begin{eqnarray}
\times\left\{
\hat 1\otimes \hat 1+\langle \hat \sigma_z^{_{(1)}}\rangle \hat\sigma_z
\otimes \hat 1+xz\hat 1\otimes \hat\sigma_z+
\left[\langle \hat \sigma_z^{_{(1)}}\rangle
{z\over x}\right.\right.
\nonumber
\end{eqnarray}
\begin{eqnarray}\left.\left.
-\left( (1-x^2)(1-z^2)\left(1-{\langle \hat \sigma_z^{_{(1)}}\rangle^2\over x^2}\right)\right)^{1/2}\cos\psi)\right]
\hat\sigma_z\otimes\hat\sigma_z\right\},
\eqnum{C.2}
\label{C.2}
\end{eqnarray}
where the integration boundaries are defined as
\begin{eqnarray}
{\cal L}'' := \{ -1, 1\} ~~~{\mbox {\rm and}}~~~
\vert x \vert \geq
\vert \langle \hat \sigma_z^{_{(1)}} \rangle \vert.
\eqnum{C.3}
\label{C.3}
\end{eqnarray}
Now we will integrate Eq.(\ref{C.2}) over the variable $\psi$. There are two
values $\psi_0^{(j)}$ ($j=1,2$) of  $\psi$, such that
\begin{eqnarray}
\cos\psi_0=
{\langle \hat \sigma_z^{_{(1)}}\hat \sigma_z^{_{(2)}}\rangle -
\langle \hat \sigma_z^{_{(1)}}\rangle
{z/x}\over\left[
(1-x^2)(1-z^2)(1-(\langle \hat \sigma_z^{_{(1)}}\rangle/x)^2)\right]^{1/2}},
\eqnum{C.4}
\label{C.4}
\end{eqnarray}
providing that inequality
\begin{eqnarray}
1 \geq \left| {
\langle \hat \sigma_z^{_{(1)}}\hat \sigma_z^{_{(2)}}\rangle -
\langle \hat \sigma_z^{_{(1)}}\rangle
z/x}\over{\left[
(1-x^2)(1-z^2)(1-(\langle \hat \sigma_z^{_{(1)}}\rangle/x)^2)\right]^{1/2}}
\right|.
\eqnum{C.5}
\label{C.5}
\end{eqnarray}
holds.
The last
relation
can be rewritten as the condition $a+bz+cz^2\geq0$, where  the explicit
forms of  the coefficients $a,b$, and $c$ are:
\begin{eqnarray}
a=1-\langle \hat \sigma_z^{_{(1)}}\rangle^2/x^2
+\langle \hat \sigma_z^{_{(1)}}\rangle^2-
\langle \hat \sigma_z^{_{(1)}}\hat \sigma_z^{_{(2)}}\rangle^2-x^2;
\nonumber
\end{eqnarray}
\begin{eqnarray}
b=2\langle \hat \sigma_z^{_{(1)}}\rangle
\langle \hat \sigma_z^{_{(1)}} \hat \sigma_z^{_{(2)}}\rangle/x;\qquad
c=x^2-\langle \hat \sigma_z^{_{(1)}}\rangle^2-1.
\eqnum{C.6}
\label{C.6}
\end{eqnarray}
The coefficient $c$ is
always negative, which means that we have a new condition for the parameter
$z$,  that is  $z\in \langle z_1, z_2 \rangle$, where $z_1$ and $z_2$ are
two roots of
the quadratic equation $a+bz+cz^2=0$. However, these roots exist only providing
the discriminant $b^2-4 a c\geq 0$ is nonnegative. Taking into account
Eq.(\ref{C.6})	we see that the last relation  is a cubic
equation with respect to the variable $x^2$, which imposes a
new condition on the integration parameter $x$.
That is, the interval ${\cal L}''$
through which  the integration over $x$ in
Eq.(\ref{C.2}) is performed is defined as
\begin{eqnarray}
{\cal L}'' :=\left\{
\begin{tabular}{ll}
$\left\{ \vert \langle \hat \sigma_z^{_{(1)}}\rangle\vert,1 \right\}$ &
$\mbox{\rm ~~for~~}  \vert \langle \hat \sigma_z^{_{(1)}}\hat
\sigma_z^{_{(2)}}\rangle \vert \leq
\vert \langle \hat \sigma_z^{_{(1)}}\rangle \vert;$
\\
$\left\{ \vert \langle \hat \sigma_z^{_{(1)}}\rangle\vert,
\sqrt{1+\langle \hat \sigma_z^{_{(1)}}\rangle^2-
\langle \hat \sigma_z^{_{(1)}}\hat \sigma_z^{_{(2)}}\rangle^2}\right\}$ &
$\mbox{\rm ~~~for~~~} \vert \langle \hat \sigma_z^{_{(1)}}\hat
\sigma_z^{_{(2)}}\rangle
\vert \geq \vert \langle \hat \sigma_z^{_{(1)}}\rangle \vert.$
\end{tabular}
\right.
\eqnum{C.7}
\label{C.7}
\end{eqnarray}
Taking into account all conditions imposed on parameters of integration
we can	rewrite Eq.(\ref{C.2}) as
\begin{eqnarray}
\hat\rho=\frac{1}{\cal N}
\int_{{\cal L}''}\, {x^2\over
\vert x\vert}dx\!\!\int_{z_1}^{z_2}{dz\over\sqrt{a+bz+cz^2}}
(\hat 1\otimes \hat 1+\langle \hat \sigma_z^{_{(1)}}\rangle \hat\sigma_z
\otimes \hat 1+xz\hat 1\otimes \hat\sigma_z+
\langle \hat \sigma_z^{_{(1)}}\hat \sigma_z^{_{(2)}}\rangle\hat\sigma_z
\otimes\hat\sigma_z).
\eqnum{C.8}
\label{C.8}
\end{eqnarray}
Using standard formulas [see, for example, \cite{Gradstein},
 Eq.(2.261) and Eq.(2.264)] the integration over parameter $z$
in Eq.(\ref{C.8})
can now be performed and we obtain
\begin{eqnarray}\nonumber
\hat\rho = \frac{1}{\cal N}
\int_{{\cal L}''}\, dx{x^2\over
\vert x\vert}{1\over(1+\langle \hat \sigma_z^{_{(1)}}\rangle^2-x^2)^{1\over2}}
(\hat 1\otimes \hat 1+\langle \hat \sigma_z^{_{(1)}}\rangle \hat\sigma_z
\otimes \hat 1+
\langle \hat \sigma_z^{_{(1)}}\hat \sigma_z^{_{(2)}}\rangle\hat\sigma_z
\otimes\hat\sigma_z)
\end{eqnarray}
\begin{eqnarray}
+\int_{{\cal L}''}\, dx{x^2\over
\vert x\vert}{\langle \hat \sigma_z^{_{(1)}}\rangle
\langle \hat \sigma_z^{_{(1)}}\hat \sigma_z^{_{(2)}}
\rangle\over(1+\langle \hat \sigma_z^{_{(1)}}\rangle^2-x^2)^{3\over2}}
(\hat 1\otimes\hat\sigma_z).
\eqnum{C.9}
\label{C.9}
\end{eqnarray}
After performing integration over $x$ in Eq.(\ref{C.9})
we obtain final the expression (\ref{9.29}) for the {\em a  posteriori}
estimation of the density operator on the given observation level.


\end{document}